\documentclass[11pt]{article}
\usepackage{amsmath}
\usepackage{amssymb}
\usepackage{amssymb}
\usepackage{enumerate}
\usepackage{textcomp}
\usepackage{graphicx}
\usepackage{tabu}
\usepackage{tikz}

\usepackage[T1]{fontenc}

\usepackage{cite}
\usetikzlibrary{decorations.markings}
\usepackage[a4paper,left=2.5cm,right=2.5cm,top=3cm,bottom=3cm]{geometry}
\usepackage[onehalfspacing]{setspace}
\usepackage{fancyhdr}
\usepackage[colorlinks=true,allcolors=blue]{hyperref}
\counterwithin{equation}{section}
\counterwithin{figure}{section}

\usepackage{url}

\usepackage[a-1a]{pdfx}

\newcommand{\eq}[1]{(\ref{eq:#1})}
\newcommand{\fig}[1]{\ref{fig:#1}}
\newcommand{\sect}[1]{\ref{sec:#1}}

\begin{document}
	
\thispagestyle{empty}
\begin{center}
	\Large
	Dissertation\\
	submitted to the\\
	Combined Faculty of Mathematics, Engineering and Natural Sciences\\
	of Heidelberg University, Germany\\
	for the degree of\\
	Doctor of Natural Sciences\\
	
	\vspace{14cm}
	
	Put forward by\\
	Philipp Heinen\\
	born in: Bonn\\
	Oral examination: 8 May 2024
\end{center}	
	
\clearpage

\thispagestyle{empty}
\begin{center}
	\vspace{2cm}
	\Huge Simulation of ultracold Bose gases\\ with the complex Langevin method\\
\end{center}

\vspace{18cm}

{\Large \noindent Referees:
	
\hspace*{0.5cm} Prof. Dr. Thomas Gasenzer

\vspace{0.3cm}

\hspace*{0.5cm} Prof. Dr. Tilman Enss}

\clearpage
\thispagestyle{empty}

\section*{Abstract} 
Evaluating the field-theoretic path integral of interacting non-relativistic bosons is not feasible with standard Monte Carlo techniques because the kinetic term in the action is purely imaginary, rendering the path integral weight a complex quantity. The complex Langevin (CL) method attempts to circumvent this problem by recasting the path integral into a stochastic differential equation and by complexifying originally real degrees of freedom. We explore the applicability of the algorithm in numerous scenarios and demonstrate that it is a viable tool for the numerically exact simulation of weakly interacting Bose gases. We first review the construction of the path integral of non-relativistic bosons and discuss in depth its discretization on a computational lattice as well as the extraction of observables; furthermore, the CL method is reviewed. We then perform benchmark studies of CL simulations against approximate analytical descriptions of the three-dimensional Bose gas in the condensed and thermal phase and against literature predictions for the critical temperature. In a two-dimensional gas, we study the Berezinskii-Kosterlitz-Thouless transition, recovering its known hallmarks and extracting the critical temperature. We also compute density profiles in a two-dimensional harmonic trap and compare them to experiment. Finally, we employ the CL method to simulate dipolar Bose gases. We consider both the stable phase above the roton instability, where we compare excitation energies to experiment, as well as the unstable one, where we demonstrate the stabilizing effect of quantum fluctuations.

\clearpage
\thispagestyle{empty}

\section*{Zusammenfassung}
Die Auswertung des feldtheoretischen Pfadintegrals wechselwirkender, nicht-relativistischer Bosonen ist mit gewöhnlichen Monte-Carlo-Methoden nicht möglich, da der kinetische Term der Wirkung rein imaginär ist und dadurch der Gewichtungsfaktor im Pfadintegral zu einer komplexen Größe wird. Die Complex-Langevin-Methode (CL) versucht dieses Problem zu umgehen, indem das Pfadintegral zu einer stochastischen Differentialgleichung umgeschrieben wird und ursprünglich reelle Freiheitsgrade komplexifiziert werden. Wir untersuchen die Anwendbarkeit des Algorithmus in zahlreichen Szenarien und zeigen, dass dieser ein nützliches Werkzeug für die numerisch exakte Simulation schwach wechselwirkender Bosegase darstellt. Wir behandeln zunächst die Konstruktion des Pfadintegrals für nicht-relativistische Bosonen und diskutieren ausführlich dessen Diskretisierung auf dem Gitter und die Extraktion von Observablen; außerdem wird die CL-Methode besprochen. Anschließend unterziehen wir CL-Simulationen einem Vergleich mit approximativen analytischen Beschreibungen des dreidimensionalen Bosegases in der kondensierten und thermischen Phase sowie mit Literaturwerten für die kritische Temperatur. In einem zweidimensionalen Gas untersuchen wird den Beresinski-Kosterlitz-Thouless-Übergang, wobei wir dessen bekannte Charakteristika reproduzieren und die kritische Temperatur extrahieren. Außerdem berechnen wir Dichteprofile in einer zweidimensionalen harmonischen Falle und vergleichen mit dem Experiment. Schließlich verwenden wir die CL-Methode zur Simulation dipolarer Bosegase. Wir betrachten sowohl die stabile Phase oberhalb der Roton-Instabilität, wo wir Anregungsenergien mit dem Experiment vergleichen, als auch die instabile Phase, wo wir den stabilisierenden Effekt der Quantenfluktuationen nachweisen.
	
\clearpage
\thispagestyle{empty}

\section*{Publications}
Parts of this thesis are based on the following publications:
\begin{itemize}
	\item Philipp Heinen and Thomas Gasenzer, Complex Langevin approach to interacting Bose
	gases, Phys. Rev. A, 106, 063308, 2022 \cite{heinen2022complex}.
	\item Philipp Heinen and Thomas Gasenzer, Simulating the Berezinskii-Kosterlitz-Thouless
	transition with the complex Langevin algorithm, Phys. Rev. A, 108, 053311, 2023 \cite{heinen2023simulating}.
\end{itemize}
In particular, chapter \sect{3D} is based on \cite{heinen2022complex} and chapter \sect{2D_gas} is based on \cite{heinen2023simulating} (as well as the respective appendices \sect{3D_supp} and \sect{2D_supp}). Also section \sect{observ} on the extraction of observables on the lattice, section \sect{disc_eff} on imaginary-time dicretization errors as well as the discussion of multi-component Bogoliubov theory in \sect{Bog_theo} are in part based on \cite{heinen2022complex}. Chapter \sect{dipolars} is partly based on so far unpublished manuscripts that are coauthored by Wyatt Kirkby, Lauriane Chomaz and Thomas Gasenzer. 

Furthermore, the work presented in the following publications was in large part done during my doctoral studies but the material therein contained is not included in this thesis:  
\begin{itemize}
	\item Philipp Heinen, Aleksandr N. Mikheev, and Thomas Gasenzer, Anomalous scaling at
	nonthermal fixed points of the sine-Gordon model, Phys. Rev. A, 107, 043303, 2023 \cite{heinen2023anomalous}.
	\item Philipp Heinen, Aleksandr N. Mikheev, Christian-Marcel Schmied, and Thomas Gasenzer,
	Non-thermal fixed points of universal sine-Gordon coarsening dynamics, ArXiv preprint, 2212.01162, 2022\cite{heinen2022nonthermal}.
\end{itemize}

\clearpage

\thispagestyle{empty}

\vspace*{2cm}

\begin{minipage}{0.6\textwidth}
	
	It is through wonder that humans now begin and at first began to philosophize; initially wondering at the oddities in their everyday life, and then, proceeding little by little, raising questions about the greater matters such as the moon phases, the phenomena of the sun and the stars and the genesis of the universe. [...] Hence, since they philosophized in order to escape from ignorance, it is evident that they were pursuing science for the sake of knowledge alone and not for any practical purpose. \\\vspace{-0.5cm}
	\begin{flushright}Aristotle, Metaphysics\end{flushright}
\end{minipage}

\clearpage	
\thispagestyle{plain}
{\hypersetup{linkcolor=black}\tableofcontents}
\clearpage

\thispagestyle{plain}
\section{Introduction\label{sec:intro}}
The simulation of quantum systems on classical computers is of high theoretical and practical importance and at the same time highly challenging. They are of great relevance to numerous fields of pure and applied natural sciences, such as atomic and solid state physics, chemistry, material sciences and quantum technology. Even with the advent of quantum computers, which by the time of this writing have not reached full technical maturity yet, there will always be the need to benchmark and validate their outcomes on a classical computational device, which, due to its classicality, is more accessible to human comprehension than a quantum computer.  The enormous challenge of such simulations, however, lies in the basic quantum mechanical principle of superposition, which sets it fundamentally apart from classical physics. This can already be seen just from the amount of information required to store the state of a system. In classical mechanics, this amount grows linearly with the number of its constituent particles; in classical field theory, it also grows linearly with the extent of the simulated space. In contrast, the amount of information to store a quantum state grows exponentially with the number of constituent particles in quantum mechanics and with the number of lattice points in (lattice) quantum field theory, as the size of the underlying Hilbert space increases exponentially. While there are also classical physical systems that are notoriously difficult to simulate (non-linear and chaotic systems, complex systems, turbulence), quantum systems are thus unique in that their complexity is built-in and is present even in the most simple models. 

Within the most fundamental formulation of quantum theory in terms of Hilbert spaces and operators, ``solving'' a model amounts to diagonalizing a matrix of the size of its Hilbert space or, if only specific observables are required, to evaluating matrix exponentials of this matrix. This remains feasible numerically for matrices up to sizes of $\mathcal{O}(10^6)$ to $\mathcal{O}(10^7)$ such that e.g. spin or Hubbard models with a few sites may be studied by exact diagonalization \cite{lin1988pairing,lin1990exact,weisse2008exact}. However, exact diagonalization becomes unpractical for realistic continuum bosonic or fermionic field theories, the simulation of which requires thousands or millions of lattice sites.

The consideration of the Hilbert space teaches one important lesson: There is no chance of realistically simulating a quantum system that is able to explore the entire complexity of its theory space, unless it is sufficiently tiny. However, in a broad range of scenarios, this whole complexity is never fully explored in practice. This may enable either approximate simulation methods that permit calculations of the relevant quantities to a sufficient degree of precision or even numerically exact approaches that run in polynomial time. Several scenarios where the full complexity of the Hilbert space is not explored are rather obvious. These include, inter alia, the classical limit of quantum theory; cases where different sections of a Hilbert space decouple such that one may study these smaller Hilbert spaces separately; systems of interacting particles that can be well described as an ensemble of non-interacting quasi-particles; and weakly interacting systems that can be treated perturbatively. In many other cases of interest, it requires substantial ingenuity to detect such a reduction of complexity by devising a suitable algorithm that exploits the reduced complexity.

As reference \cite{berger2021complex} puts it, one may divide these numerical algorithms into two classes, memory and statistics intensive algorithms. Into the former class fall such diverse algorithms as the Hartree-Fock method \cite{slater1951a}, density functional theory \cite{hohenberg1964inhomogeneous} or dynamical mean-field theory \cite{georges1996dynamical}. The latter class consists of algorithms that are commonly known as quantum Monte Carlo (QMC) algorithms. While finding an algorithm of the first class demonstrates that the problem of simulating the given quantum system falls in the complexity class P of problems that are solvable deterministically in polynomial time, the existence of a QMC algorithm demonstrates it to be in BPP, the class of problems that can be solved in polynomial time probabilistically (i.e. the algorithm gives the right answer with a certain probability that can be made arbitrarily small). In practice, QMC algorithms are as powerful as deterministic algorithms, as the statistical error inherent to any quantity that they can calculate can be made arbitrarily small in polynomial time. Apart from few exceptions (e.g. variational QMC), the framework of quantum Monte Carlo methods is the path integral (PI) approach to quantum theory.

The path-integral formalism provides a tremendously useful and powerful reformulation of the original problem, which avoids the Hilbert space and operators but requires the evaluation of very high-dimensional integrals of the type $\int d^nx \exp[-S(\mathbf{x})] O(\mathbf{x})$. It is unfeasible to compute such integrals with exact and deterministic integration schemes but they appear as tailor-made for powerful Monte Carlo integration methods such as the Metropolis-Hastings algorithm \cite{metropolis1953equation}. These are based on interpreting the path integral weighting factor $\exp[-S(\mathbf{x})]$ as a probability density and construct a Markov chain of configurations $\mathbf{x}$ that sample this weighting factor. Thereby the (statistically) exact simulation  of quantum systems can be achieved in polynomial instead of exponential run time as long as $\exp[-S(\mathbf{x})]$ can be interpreted as a probability density.  

Within the path integral formalism, the exponential complexity of quantum physics is thus blurred to a certain extent and not so apparent as in the operator formalism. However, it is not removed. Namely, the weight $\exp[-S(\mathbf{x})]$ is not necessarily real and positive-definite as a probability density in actual physical problems but in the most general case complex and thus not suitable for the application of standard Monte Carlo algorithms. Scenarios in which this is the case include: any quantum system in non-equilibrium; relativistic bosonic and fermionic theories at nonzero chemical potential; spin-imbalanced non-relativistic fermions; and non-relativistic bosons in the field-theoretic formulation. 

This obstacle to the application of Monte Carlo methods is known in the literature as \textit{sign problem}. The naive way of avoiding it is known as \textit{reweighting}: One splits off the imaginary part of $S$ and pulls $\exp(-i \text{Im} S)$ into the observable $O$. In the most extreme case of a vanishing real part of $S$ this amounts to randomly drawing configurations $\mathbf{x}$ with equal probability. While reweighting can be successful in some cases of a mild sign problem \cite{nakamura1992reweighting,fodor2002a}, it will in general have the problem that the computational cost grows exponentially with the system size, such that the advantage over exact diagonalization is lost. Namely, for the evaluation of observables, one needs to normalize $\int d^nx \exp[-S(\mathbf{x})] O(\mathbf{x})$ by the partition function $\int d^nx \exp[-S(\mathbf{x})]$. If the oscillatory factor $\exp(-i \text{Im} S)$ has been pulled into the observable, this in general leads to a division of two very tiny numbers, as the oscillatory behavior causes huge cancellations between positive and negative domains. Thus, an exponentially growing number of samples is required for reaching a fixed accuracy.  

It can be shown that the sign problem in its most general form is NP hard \cite{troyer2005computational} such that a generic solution can be considered unlikely. Nonetheless, there may still be solutions for all cases of physical interest. Numerous approaches have thus been proposed to overcome or ameliorate the sign problem for simulations of physical systems \cite{berger2021complex}. Some of these are model specific to a certain extent, such as the dual variables \cite{endres2007method,gattringer2016approaches} or the density of states method \cite{fodor2007the,gattringer2016approaches}.

A method that is completely model-independent instead is the complex Langevin (CL) method \cite{ parisi1983complex,klauder1985spectrum,berger2021complex}, which relies on the well-known relation between path integrals and stochastic differential equations \cite{wiener1921average,parisi1981perturbation} as well as a complexification of the original degrees of freedom. It has already proven successful in a wide range of physical scenarios \cite{ganesan2001field,berges2007lattice,aarts2008stochastic,aarts2009can,aarts2010on,sexty2014simulating,hayata2015complex,loheac2017third,rammelmuller2018finite,nishimura2019complex,kogut2019applying,ito2020complex,attanasio2020complex,attanasio2020thermodynamics,delaney2020numerical,berger2021complex,attanasio2022density,boguslavski2023stabilizing,mcgarrigle2023emergence} and has the advantage of being easy to implement. Nonetheless, by the time of this writing, the method is still not very widespread and its use is still to a large extent confined to the lattice gauge theory community, with ultracold atoms applications slowly emerging \cite{loheac2017third,attanasio2020thermodynamics,mcgarrigle2023emergence,attanasio2023harmonically}. However, its popularity has been steadily rising in recent years, and with this work we aim at contributing to this progress by further widening the range of physical scenarios that have been successfully simulated with the complex Langevin method. 

The physical system studied throughout this thesis are condensates of ultracold bosonic atoms, which nowadays can be routinely created by numerous experimental groups worldwide. The reason for the great and ongoing interest that these systems have attracted since the first experimental demonstration of Bose-Einstein condensation by Cornell, Wieman and Ketterle in 1995 \cite{anderson1995observation,davis1995bose} lies in the high degree of experimental control that these systems offer, as well as the plethora of intriguing physical phenomena they exhibit, including superfluidity \cite{onofrio2000observation} and supersolidity \cite{li2017stripe,tanzi2019observation,bottcher2019transitient,chomaz2019long}, thermal and quantum phase transitions \cite{hadzibabic2006berezinskii,clade2009observation,bookjans2011quantum,zhang2013tunable}, topological objects such as solitons and vortices \cite{burger1999dark,matthews1999vortices,madison2000vortex}, pattern formation \cite{engels2007observation,pollack2010collective,zhang2020pattern} and scaling behavior \cite{prufer2018observation,erne2018universal,glidden2021bidirectional}. This renders them an ideal testbed for concepts, theories and numerical methods from quantum many-body theory. Furthermore, they have found numerous applications in the analog simulation of models that describe more intricate or less controllable quantum systems, ranging from black hole physics and cosmology \cite{lahav2010realization,viermann2022quantum} to Hubbard models \cite{greiner2002quantum}.

Bose-Einstein condensates (BECs) of ultracold atoms are also a prime example of the aforementioned reduction of Hilbert space complexity. While the bosonic Hilbert space grows exponentially with the number of particles, it is possible to simulate interacting BECs to good accuracy with algorithms that do not only run in polynomial time but are also comparatively cheap. It is clear that for high temperature, far above the Bose-Einstein transition, a gas of bosonic atoms loses its quantum characteristics and transitions into a gas of classical particles. Nonetheless, a description within a classical formalism is also possible far below the transition, namely within classical \textit{field theory}, in contrast to the high-temperature limit where the gas can be described by classical \textit{mechanics}. Ironically, it is precisely the genuinely quantum mechanical effect that bosons tend to batch together in the same state that enables a description in terms of classical field theory, similarly to photons that can be very well described by classical electrodynamics once they occupy the same macroscopic state in large numbers (however, in contrast to bosonic atoms, they do not feature a classical particle limit due to their vanishing mass). 

Approximate numerical methods for describing BECs that are based on their field-theoretic  classicality in the condensed phase are numerous. For non-equilibrium scenarios, a highly successful approach is to simulate the classical equation of motion of the underlying field theory, the Gross-Pitaevskii equation (GPE), a non-linear partial differential equation \cite{gross1961structure,pitaevskii1961vortex}. Several semi-classical approaches have been developed that go beyond the purely classical GPE: The truncated Wigner approximation (TWA) adds noise to the initial condition in order to capture quantum effects in the initial state (but neglects them during the subsequent evolution) \cite{polkovnikov2010phase}; the stochastic Gross-Pitaevskii (SGPE) equation promotes the ordinary GPE to a stochastic differential equation in order to capture the coupling of the  condensate mode to the bath of thermally excited particles \cite{stoof2001dynamics,gardiner2002the,gardiner2003the}; the extended Gross-Pitaevskii equation (EGPE) attempts to include the effect of quantum fluctuations by adding an additional term to the original GPE \cite{wachtler2016quantum}. Most of these approaches possess some sort of thermal equilibrium counterpart, typically involving a projection or cutoff procedure that removes the high-momentum modes \cite{davis2001simulations,davis2002simulations}. Furthermore, it is possible to employ Monte Carlo simulations of classical thermal field theory \cite{arnold2001bec}. Apart from such approximate numerical methods, there exist also several analytical approximate descriptions, including Bogoliubov theory, the Popov and Hartree-Fock approximation \cite{andersen2004theory,pitaevskii2016bose} as well as several renormalization group schemes \cite{bijlsma1996renormalization,floerchinger2009superfluid,rancon2012universal}.

It is important to note, however, that none of these approaches provides a fully exact, ab initio simulation in the sense that the numerical errors can be made arbitrarily small by increasing the resolution of the numerical grid or decreasing the numerical time step, because the correct description of interacting bosons is quantum field theory and not classical field theory. However, the path integral representation of non-relativistic bosonic quantum field theory features a sign problem and is thus inaccessible to standard Monte Carlo algorithms. In fact, the only major well-established numerical method for simulating a gas of interacting bosons in a fully exact manner is the path integral Monte Carlo method (PIMC) \cite{pollock1984simulation,ceperley1986path,ceperley1995path}, which precisely avoids the quantum field theoretic description of interacting bosons. Instead, it recasts the problem into its original, quantum mechanical formulation of single atoms. In the quantum mechanical path integral one must then not only sample the positions of the single atoms but also take care of their bosonic nature, i.e. sample symmetric permutations thereof. This approach has enabled a plethora of high-precision computations of various properties of liquid helium, in extraordinarily good agreement with experiment \cite{ceperley1995path}, and has also variously been employed for the simulation of ultracold atomic systems \cite{nho2004bose,nho2005bose,pilati2006equation,pilati2008critical,bombin2019berezinskii,spada2021thermodynamics}. Nonetheless, it has the disadvantage that the computational cost scales (polynomiallly) with the number of particles that are simulated, with most publications at the time of this writing reaching a few thousand atoms, i.e. one to two orders of magnitude smaller than typical experimental values. 

It is thus an attractive possibility to employ the complex Langevin algorithm designed to tackle the sign problem to simulate the field-theoretic path integral of interacting bosons, as quantum field theory (second quantization) is the most suitable framework for condensed bosons, which occupy the same modes in high numbers. This has not only the advantage that the computational cost becomes independent of the particle number, but it is also of importance for systems genuinely requiring a second-quantization formulation, most notably spinor condensates where spin-changing collisions can change the species of an atom \cite{ stamper-kurn2013spinor}. Furthermore, it is convenient to dispose of an alternative, equally exact method apart from path-integral Monte Carlo that is formulated in a completely different framework, because it enables benchmark studies between the two methods.

The approach to make use of the complex Langevin algorithm for simulating ultracold bosonic atoms has been pioneered by references \cite{hayata2015complex,attanasio2020thermodynamics,delaney2020numerical} but is still in its infancy at the time of this writing. We hope that the present work will help to better establish this very promising approach in the ultracold atoms community. In comparison to the mentioned seminal publications, we extract numerous new observables with the method, including momentum spectra, dispersions, structure factors, superfluid densities and vortex numbers; we perform systematic benchmarks to approximate descriptions and known results; and we extend the application to trapped, low-dimensional and dipolar bosons. 

Still, the application of the CL method requires some justification in view of the tremendous success of semi-classical, GPE-based algorithms and other approximate descriptions as well as the fact that it demands a computational cost that is one to two orders of magnitude higher than that of GPE simulations (as we need to evolve a four-dimensional instead of three-dimensional lattice). In the course of this thesis, we will discuss some scenarios where the difference between a full quantum simulation and approximate descriptions becomes relevant, mostly close to phase transitions. Nonetheless, deviations are often only in the few percent range. In our opinion, the main prospect of the approach is thus that future experimental improvements will enable measurements in ultracold atomic systems with far greater precision than nowadays, such that CL can be employed for high-precsion comparisons between theory and experiment, as they have for long been known from particle physics.

This thesis is organized as follows. Chapter \sect{bose_gas_and_pi} reviews the physics of ultracold bosonic atoms. Further, it discusses in detail the path integral representation of Bose gases as well as approximate approaches such as Bogoliubov and Hartree-Fock theory. Chapter \sect{sign_prob_and_cl} reviews the sign problem and the complex Langevin approach to tackling it. Additionally, we provide expressions for the Langevin equations of the non-relativistic Bose gas. The subject of chapter \sect{3D} is the homogeneous, three-dimensional Bose gas. As this is a comparatively simple setting, it serves mainly as a testbed for the application of the complex Langevin method to ultracold bosonic atoms and we show that well-known theoretical descriptions are recovered by the method. Chapter \sect{2D_gas} is dedicated to the two-dimensional Bose gas, which is subject to much stronger fluctuation effects than the three-dimensional Bose gas. We employ the CL method to study in detail the Berezinskii-Kosterlitz-Thouless (BKT) transition in a two-dimensional Bose gas. We show that well-known hallmarks of this transition can be correctly reproduced by the method. Furthermore, we use it to compute the critical density of the gas and demonstrate that it deviates from results obtained by purely classical methods. In chapter \sect{trap} we show how CL can be applied to computing density profiles of bosons in a harmonic trapping potential. We compare our simulations to experimental results of the Heidelberg BECK experiment. While chapters \sect{3D} to \sect{trap} study bosons that interact only locally, chapter \sect{dipolars} explores the physics of dipolar atoms, which feature long-range, anisotropic interactions. It is precisely this peculiar type of interaction that renders dipolar atoms the most susceptible to quantum fluctuations, which play a crucial role in the formation of exotic types of matter in dipolar gases such as supersolids. Apart from benchmark studies, we compute excitation energies close to the point of the mean-field instability, which are compared to experimental results from the Innsbruck Erbium experiment, and demonstrate the stabilizing effect of quantum fluctuations beyond this point.  We close with a summary and outlook (chapter \sect{conclusion}). Details on our unit conventions can be found in appendix \sect{units}. 
\clearpage

\thispagestyle{plain}
\section{Ultracold Bose gases and the coherent state path integral \label{sec:bose_gas_and_pi}}	

\subsection{Ultracold Bose gases and their description by quantum field theory\label{sec:physics_bose}}

As famously predicted by Einstein, based on work by Bose, in 1925 \cite{einstein2005quantentheorie}, a gas of bosonic particles possesses a critical temperature below which the state of lowest energy is occupied by a macroscopic fraction of particles. This temperature is given for an ideal gas by 
\begin{align}
\label{eq:T_BEC}
T_\mathrm{c}=\frac{2\pi\hbar^2\rho^{2/3}}{mk_B\zeta(3/2)^{2/3}}\,,
\end{align}
where $m$ is the mass of the atoms and $\rho$ their density.
It is not until this temperature is approached that a gas of atoms begins to exhibit quantum-mechanical effects. For typical densities of atomic gases, this formula predicts critical temperatures of the order of $10^{-7} \mathrm{K}$, which are very challenging to reach experimentally.

While superfluid helium had long been considered to be a kind of Bose-Einstein condensate (for which, however, the Bose-Einstein theory is applicable only qualitatively due to its strongly interacting nature), the first experimental demonstration of Bose-Einstein condensation in an ultracold atomic gas was achieved by Cornell and Wieman as well as independently by Ketterle \cite{anderson1995observation,davis1995bose} in 1995, in a gas of rubidium and sodium atoms, respectively. The necessary extremely low temperatures were obtained by a combination of laser cooling and evaporative cooling. Since then, ultracold bosonic atoms have become the subject of intensive experimental and theoretical investigations and can nowadays be prepared in the laboratories of numerous research groups worldwide. One of the main reasons for this interest is the high degree of experimental control that these systems provide, enabling a very direct study of concepts from quantum many-body theory. 

There are two in principle equivalent theoretical descriptions of quantum Bose gases, known as first and second quantization. In first quantization, one describes the quantum state of the atomic ensemble by a multi-particle wave function that must be totally symmetrized according to the bosonic statistics. For a large number of particles, this approach is not longer a very suitable description~\footnote{It is the framework in which some Monte Carlo algorithms such as path integral Monte Carlo (PIMC) are formulated, but apart from these first quantization is rarely employed for practical computations.}. Instead, practical computations are typically performed in second quantization. 

Consider an arbitrary complete and orthonormal basis of the space of one-particle wave functions $\{\phi_l\}$. For a system in a periodic box with volume $\mathcal{V}$, these can e.g. be the plane waves
\begin{align}
\phi_\mathbf{k}(\mathbf{x})=\frac{1}{\sqrt{\mathcal{V}}}e^{i\mathbf{k}\cdot\mathbf{x}}
\end{align}
with $\{\mathbf{k}\}$ being the set of discrete momenta that fulfill the periodic boundary conditions in the box. Another possible choice are e.g. the eigenfunctions of the harmonic oscillator. Note, however, that the $\phi_l$ do not need to be eigenfunctions of some Hamiltonian.

One now introduces a new type of physical state $|n_l\rangle_l$, characterized by $n_l$ particles occupying mode $l$. For fermions, $n_l$ can only be either $0$ or $1$ due to the Pauli exclusion principle while it can take any non-negative integer value for bosons. These states are known as \textit{Fock states}. We furthermore introduce \textit{creation and annihilation operators} $a_l^\dagger$ and $a_l$, respectively, which are characterized by their action onto the Fock states:
\begin{align}
a_l^\dagger|n_l\rangle_l&=\sqrt{n_l+1}\,|n_l+1\rangle_l\\
a_l|n_l\rangle_l&=\sqrt{n_l}\,|n_l-1\rangle_l\,,
\end{align}
i.e. they create and annihilate a particle in state $\phi_l$, respectively. The operator $a_l^\dagger a_l$ is known as \textit{number operator}, as the Fock states are its eigenstates with their respective number of particles as eigenvalue, $a^\dagger_l a_l |n_l\rangle_l=n_l|n_l\rangle_l$. Furthermore, the creation and annihilation operators fulfill the commutation relation
\begin{align}
[a_l,a_{l'}^\dagger]=\delta_{ll'}\,.
\end{align}
The basis states of the full multi-particle Hilbert space are now formed by the Cartesian product of the single-mode Fock states, i.e.
\begin{align}
|n_1, n_2,\dots\rangle=|n_1\rangle_1 |n_2\rangle_2 \dots 
\end{align} 
such that an arbitrary physical state $|\Psi\rangle$ can be represented as 
\begin{align}
|\Psi\rangle=\sum_{\{n_l\}} c(n_1,n_2,\dots) |n_1, n_2,\dots\rangle\,.
\end{align}
The way that usual first-quantization operators translate into second quantization is the following: Suppose we have a first quantization operator $\Omega^{(1)}$ that acts on single particles. Then it is represented in second quantization as 
\begin{align}
\Omega^{(1)}=\sum_{ij} t_{ij}\, a_i^\dagger a_j
\end{align} 
where
\begin{align}
t_{ij}=\int d^3 x \, \phi_i(\mathbf{x})^*\,\Omega^{(1)}(\mathbf{x})\phi_j(\mathbf{x})\,.
\end{align}
Similarly, for an operator that acts on two particles $\Omega^{(2)}$ we have
\begin{align}
\Omega^{(2)}=\frac{1}{2}\sum_{ijkl} V_{ijkl}\, a_i^\dagger a_j^\dagger a_k a_l
\end{align}
where
\begin{align}
V_{ijkl}=\int d^3 x\int d^3y \, \phi_i(\mathbf{x})^*\phi_j(\mathbf{y})^*\,\Omega^{(2)}(\mathbf{x},\mathbf{y})\phi_k(\mathbf{x})\phi_l(\mathbf{y})\,.
\end{align}
The one-particle-operator contribution to the Hamiltonian is typically the kinetic energy of a particle and its energy in an external potential. The two-particle contribution stems from the interaction between the particles. Without the latter, the problem of describing a quantum many-body system reduces to that of describing a single particle.

Let us now consider bosonic particles of mass $m$ that are enclosed in a periodic box of volume $\mathcal{V}$. They may be subject to an external potential $U(\mathbf{x})$ and interact with a potential $V(\mathbf{x},\mathbf{y})$ that we assume to be dependent only on the relative distance, $V(\mathbf{x},\mathbf{y})=V(\mathbf{x}-\mathbf{y})$. Let us take the single-particle basis functions to be plane waves numbered by their momentum $\mathbf{k}$. Then the Hamiltonian of the system reads 
\begin{align}
H=
\sum_\mathbf{k}\frac{\mathbf{k}^2}{2m}a^\dagger_{\mathbf{k}}a_{\mathbf{k}}
+\sum_{\mathbf{k},\mathbf{q}}U_{\mathbf{q}}\,a^\dagger_{\mathbf{k}}a_{\mathbf{k}+\mathbf{q}}+\frac{1}{2\mathcal{V}}\sum_{\mathbf{k}\mathbf{k}'\mathbf{q}}
V_\mathbf{q}\,a^\dagger_{\mathbf{k}+\mathbf{q}}a^\dagger_{\mathbf{k}'-\mathbf{q}}
a_{\mathbf{k}}a_{\mathbf{k}'}
\end{align}
Here $U_{\mathbf{q}}$ is the Fourier transform of the external potential $U(\mathbf{x})$ and $V_{\mathbf{q}}$ is the Fourier transform of $V(\mathbf{x})$. It is also a common experimental setting to have a system composed of several components, e.g. atoms in different hyperfine-levels \cite{myatt1997production} or of different species \cite{modugno2002two}. Say the different components are numbered by $\alpha$ and $a^\dagger_{\mathbf{k},\alpha}$ and $a_{\mathbf{k},\alpha}$ create and annihilate a particle of component $\alpha$ with momentum $\mathbf{k}$. Then the multi-component Hamiltonian reads
\begin{align}
\nonumber H=
&\sum_\mathbf{k}\sum_\alpha\frac{\mathbf{k}^2}{2m}a^\dagger_{\mathbf{k},\alpha}a_{\mathbf{k},\alpha}
+\sum_{\mathbf{k},\mathbf{q}}\sum_\alpha U_{\mathbf{q},\alpha}\,a^\dagger_{\mathbf{k},\alpha}a_{\mathbf{k}+\mathbf{q},\alpha}\\&+\frac{1}{2\mathcal{V}}\sum_{\mathbf{k}\mathbf{k}'\mathbf{q}}\sum_{\alpha\alpha'\beta\beta'}
V_{\mathbf{q},\alpha\alpha'\beta\beta'}\,a^\dagger_{\mathbf{k}+\mathbf{q},\alpha'}a^\dagger_{\mathbf{k}'-\mathbf{q},\beta'}
a_{\mathbf{k},\alpha}a_{\mathbf{k}',\beta}\,.
\end{align}
In this most general form, the non-linear term can also make particles switch between components upon interaction. While there exist experimental realizations of such a setting \cite{kawaguchi2012spinor}, in this thesis we will restrict ourselves to the case where the interaction is of the form $V_{\mathbf{q},\alpha\alpha'\beta\beta'}=V_{\mathbf{q},\alpha\beta}\delta_{\alpha\alpha'}\delta_{\beta\beta'}$, i.e. particles maintain their component upon interaction. In the even more special case that the interaction between all components is equal, $V_{\mathbf{q},\alpha\alpha'\beta\beta'}=V_{\mathbf{q}}\delta_{\alpha\alpha'}\delta_{\beta\beta'}$, and the external potential is independent of the component, $U_{\mathbf{q},\alpha}=U_{\mathbf{q}}$, the Hamiltonian acquires
a $U(\mathcal{N})$ symmetry with $\mathcal{N}$ the number of components. In this case, it simplifies to
\begin{align}
\label{eq:H_UN}
H^{U(\mathcal{N})}=
\sum_\mathbf{k}\sum_\alpha\frac{\mathbf{k}^2}{2m}a^\dagger_{\mathbf{k},\alpha}a_{\mathbf{k},\alpha}
+\sum_{\mathbf{k},\mathbf{q}}\sum_\alpha U_{\mathbf{q}}\,a^\dagger_{\mathbf{k},\alpha}a_{\mathbf{k}+\mathbf{q},\alpha}+\frac{1}{2\mathcal{V}}\sum_{\mathbf{k}\mathbf{k}'\mathbf{q}}\sum_{\alpha\beta}
V_{\mathbf{q}}\,a^\dagger_{\mathbf{k}+\mathbf{q},\alpha}a^\dagger_{\mathbf{k}'-\mathbf{q},\beta}
a_{\mathbf{k},\alpha}a_{\mathbf{k}',\beta}\,.
\end{align}
This will be the only type of multi-component system that we will consider in this thesis, see chapter \sect{3D}. 

Let us also introduce the field operator $\psi(\mathbf{x})$ 
as
\begin{align}
\psi(\mathbf{x})=\frac{1}{\sqrt{\mathcal{V}}}\sum_\mathbf{k} e^{i\mathbf{k}\cdot\mathbf{x}} a_\mathbf{k}\,.
\end{align}
Then it is possible to write the Hamiltonian also in the following form
\begin{align}
H=\int d^3x\, \psi^\dagger(\mathbf{x})\left(-\frac{1}{2m}\Delta+U(\mathbf{x})\right)\psi(\mathbf{x})+\frac{1}{2}\int d^3x\int d^3y\,V(\mathbf{x}-\mathbf{y})\psi^\dagger(\mathbf{x})\psi^\dagger(\mathbf{y})\psi(\mathbf{x})\psi(\mathbf{y})\,.
\end{align}
The dynamics of the system is described by the usual Schrödinger equation $i\partial_t|\Psi\rangle=H|\Psi\rangle$ also in second quantization. In this work, however, we will not be concerned with dynamics but rather with thermal equilibrium. Thermal equilibrium is commonly described either in the canonical or grand-canonical formalism, which both allow the total energy of the system $E_\text{tot}$ to fluctuate but impose a fixed temperature $T$ that determines the probability distribution of $E_\text{tot}$ according to the Boltzmann factor $\exp(-E_\text{tot}/T)$. However, while in the canonical formulation the number of particles is fixed to $N_\text{tot}$, the grand-canonical formalism allows the total particle number to fluctuate too and imposes only a chemical potential $\mu$ as the energy cost of adding one particle to the system. A central object of study in thermodynamics is the partition function $Z_\mathrm{c}$ (canonical formalism) or $Z_\mathrm{gc}$ (grand-canonical formalism), defined as 
\begin{align}
Z_\mathrm{c}&=\text{Tr}|_{\sum_l n_l=N_\text{tot}}\left\{\exp\left(-\beta H\right)\right\}\\
Z_\mathrm{gc}&=\text{Tr}\left\{\exp\left(-\beta[ H-\mu N]\right)\right\}\,,
\end{align}  
where we have introduced the inverse temperature $\beta=1/T$ and the total number operator $N$, the latter being defined as
\begin{align}
N\equiv \sum_l a^\dagger_l a_l\,.
\end{align}
While the trace runs over all Fock states in the grand-canonical case, it is restricted to those fulfilling $\sum_l n_l=N_\text{tot}$ in the canonical case. From the canonical and grand-canonical partition function, one can read off the free energy $F$ and grand-canonical potential $\Omega$, respectively:
\begin{align}
F&=-T\ln Z_\mathrm{c}\\
\Omega&=-T\ln Z_\mathrm{gc}\,,
\end{align} 
from which in turn all thermodynamic quantities can be obtained. More generally, it is possible to obtain the expectation value of an arbitrary operator $\mathcal{O}$ in thermal equilibrium as 
\begin{align}
\langle\mathcal{O}\rangle&=Z_\mathrm{c}^{-1}\text{Tr}|_{\sum_l n_l=N_\text{tot}}\left\{\exp\left(-\beta H\right)\mathcal{O}\right\}\qquad\text{(canonical)}\\
\langle\mathcal{O}\rangle&=Z_\mathrm{gc}^{-1}\text{Tr}\left\{\exp\left(-\beta [ H-\mu N]\right)\mathcal{O}\right\}\qquad\text{(grand-canonical)}\,.
\end{align} 
For large enough systems, the actual fluctuation of the total particle number in the grand-canonical formalism becomes negligible. Then it becomes a mere question of convenience whether one rather wants to impose a total particle number or a chemical potential, as for every total particle number $N_\text{tot}$ there is a chemical potential that leads to the expectation value $\langle N\rangle=N_\text{tot}$. Throughout this thesis, we will exclusively work in the grand-canonical formulation, as this formulation turns out to be much more suitable for the construction of a path integral, cf. section \sect{pi_constr}. For later convenience, let us introduce the generalized (grand-canonical) Hamiltonian $K$ as
\begin{align}
K\equiv H-\mu N
\end{align}
such that the grand-canonical Boltzmann operator simply becomes $\exp(-\beta K)$.

Let us now have a closer look at the interaction potential $V(\mathbf{x}-\mathbf{y})$ between the atoms. Neutral, non-magnetic atoms typically feature only short-range interactions~\footnote{\label{shortrange}``Short-range'' here means decaying faster than $1/r^3$. For short-range interactions according to this definition, the potential energy felt by a particle in a homogeneous gas that is caused by particles closer than $R$ converges for $R\to\infty$.} among each other. At short distances, they repel each other due to the repulsion of the electrons in the outer shells, at larger distances they feel a weak attractive van-der-Waals potential that decays as $\sim1/r^6$. Let us consider some numbers. Typical temperatures in ultracold atoms experiments are of the order of $\sim 10^{-7}\,\mathrm{K}$. Taking as an example sodium ($m=23 \,\mathrm{u}$), we obtain a typical momentum $p\sim\sqrt{2mk_BT}\approx 3.2\cdot 10^{-28}\,\mathrm{kg}\,\mathrm{m}/\mathrm{s}$, corresponding to a wavelength of $\lambda=2\pi\hbar/p\approx 3.9\cdot 10^{4}\,a_0$ with the Bohr radius $a_0$. A characteristic length scale for the extension of the van-der-Waals potential is $R_{\text{vdW}}\equiv 1/2\left(mC_6/\hbar^2\right)^{1/4}$ with $C_6$ the prefactor of the van-der-Waals potential (i.e. $V_\text{vdW}(r)=C_6/r^6$), which for sodium amounts to $R_{\text{vdW}}=44.9\,a_0$ \cite{chin2010feshbach}. I.e. the characteristic momentum of the atoms in a Bose-Einstein condensate is typically by around three orders of magnitude too small to resolve the characteristic form of the interaction potential. 

In such a scenario, scattering between two atoms can be described in excellent approximation as pure $\mathrm{s}$-wave scattering, i.e. the scattering amplitude $f(\mathbf{k},\mathbf{k}')$ can be approximated as
\begin{align}
f(\mathbf{k},\mathbf{k}')\approx -a
\end{align}
with $a$ the $\mathrm{s}$-wave scattering length, such that the scattering cross section reads $\sigma=4\pi a^2$. This means that for a theoretical description we do not need to bother about the precise form of $V(\mathbf{r})$. Instead, we may take the value of the scattering length $a$ as the only input parameter from experiment and then employ for $V(\mathbf{r})$ an arbitrary short-range pseudo-potential that reproduces the correct value of $a$. The simplest and and most widely employed choice is a Dirac delta function, 
\begin{align}
\label{eq:contact_pot}
V(\mathbf{r})=g\,\delta(\mathbf{r})\,.
\end{align}
In first-order Born approximation, the scattering amplitude is given by
\begin{align}
f(\mathbf{k},\mathbf{k}')\approx -\frac{m}{4\pi} \,V_{\mathbf{k}-\mathbf{k}'}
\end{align}
with $V_\mathbf{k}$ the Fourier transform of $V(\mathbf{r})$, such that $g$ must be chosen as 
\begin{align}
g=\frac{4\pi a}{m}\,.
\end{align}
In this contact interaction approximation, the Hamiltonian becomes
\begin{align}
\label{eq:H}
H=\int d^3x\,\left\{ \psi^\dagger(\mathbf{x})\left(-\frac{1}{2m}\Delta+U(\mathbf{x})\right)\psi(\mathbf{x})+\frac{g}{2}\psi^\dagger(\mathbf{x})\psi^\dagger(\mathbf{x})\psi(\mathbf{x})\psi(\mathbf{x})\right\}\,.
\end{align}
Experimentally it is possible to vary the value of the scattering length $a$ and thus $g$ in a highly controlled manner by employing atomic Feshbach resonances \cite{chin2010feshbach}. A measure for how strongly a gas of bosonic atoms is coupled is the \textit{diluteness} parameter $\eta\equiv\sqrt{\rho a^3}$ with $\rho$ the density of atoms. Typical experimental values are of the order of $\eta\sim 10^{-3}$.

While the contact interaction approximation has the advantage of considerably simplifying the Hamiltonian, the unphysical nature of the Dirac function causes also some mathematical difficulties. Namely, performing practical computations with the Hamiltonian \eq{H}, one finds numerous quantities, e.g. the ground state energy, to be ultraviolet divergent, i.e. dependent on the imposed momentum cutoff. These divergences can be removed by going to higher orders in the Born approximation. Including also the next-to-leading and next-to-next-to-leading order, the Born series for the scattering length reads
\begin{align}
\label{eq:Born1}
\frac{4\pi a}{m}=V_0-m \int \frac{d^3p}{(2\pi)^3}\frac{V_{\mathbf{p}}V_{-\mathbf{p}}}{\mathbf{p}^2}+m^2\int \frac{d^3p}{(2\pi)^3}\int \frac{d^3q}{(2\pi)^3}\frac{V_{-\mathbf{p}}V_{\mathbf{q}}V_{\mathbf{p}-\mathbf{q}}}{\mathbf{p}^2\mathbf{q}^2}+\mathcal{O}\left(V^4\right)\,,
\end{align} 
i.e. specifying to the contact potential \eq{contact_pot}
\begin{align}
\label{eq:Born4}
\nonumber\frac{4\pi a}{m}&=g-m g^2\int \frac{d^3p}{(2\pi)^3}\frac{1}{\mathbf{p}^2}+m^2g^3\left(\int \frac{d^3p}{(2\pi)^3}\frac{1}{\mathbf{p}^2}\right)^2+\mathcal{O}\left(g^4\right)\\
&=g-m g^2\frac{1}{2\pi^2}\Lambda +m^2g^3\left(\frac{1}{2\pi^2}\Lambda\right)^2+\mathcal{O}\left(g^4\right)\,,
\end{align} 
where we have introduced a cutoff $\Lambda$ to the momentum integrals. For this simple case of a contact interaction, we can also formally resum the entire Born series as it becomes a geometrical series,
\begin{align}
\label{eq:Born_resummed}
\frac{4\pi a}{m}=\frac{g}{1+\frac{mg}{2\pi^2}\Lambda}\,.
\end{align} 
It turns then out that the UV divergent contributions to $a$ exactly cancel divergences that appear in computations with the Hamiltonian \eq{H}. Thus if we express all physical quantities that we compute as functions of the scattering length $a$ or the \textit{renormalized coupling} $g_R\equiv 4\pi a/m$, which we can obtain via \eq{Born4} or \eq{Born_resummed} from the bare coupling $g$, the result will be independent of $\Lambda$. 

So far we have considered only Bose gases in three spatial dimensions. By tightly confining a gas of ultracold atoms in one direction by an external harmonic potential that is sufficiently strong, it is possible to experimentally create also effectively two-dimensional Bose gases. In order to theoretically describe such a two-dimensional gas one assumes that the quantum field $\psi$ factorizes into an in-plane part and a Gaussian (i.e. the harmonic oscillator ground state) in the tight direction. By integrating out the latter one finds that the two-dimensional gas may be described theoretically by the same Hamiltonian \eq{H}, specified to two spatial dimensions and with a new 2D coupling constant $g_\text{2D}$.

However, if we try to compute a two-dimensional scattering length $a_\text{2D}$ as in three dimensions via the Born series from the bare coupling $g_\text{2D}$, additionally to the ultraviolet divergence we encounter also an infrared divergence, which must be regularized by an IR cutoff $\Lambda_0$. $\Lambda_0$ can be seen as a momentum scale at which the two-dimensional renormalized coupling $g_\text{2D,R}(\Lambda_0)$ is defined. In contrast to the three-dimensional case where we can simply define the renormalized coupling at $\Lambda_0=0$, we need to specify a convention at which (nonzero) scale we want to define $g_\text{2D,R}$ in 2D. In the literature, different conventions can be found \cite{rancon2012universal,franca2017two}. One possibility is e.g. to set $\Lambda_0=\sqrt{\rho}$ with $\rho$ the two-dimensional particle density.
Once we have chosen a $\Lambda_0$, the relation between $g_\text{2D,R}$ and $g_\text{2D}$ can be written as 
\begin{align}
\label{eq:renormalization_general}
g_\text{2D,R}=\frac{g_\text{2D}}{1+\frac{mg_\text{2D}}{2\pi}\,\log\left(\Lambda/\Lambda_0\right)}
\,.
\end{align}
One can show that $g_\text{2D,R}$ can be computed from the experimental parameters as follows \cite{petrov2000bose}:
\begin{align}
\label{eq:full_g2D}
g_\text{2D,R}=\frac{\sqrt{8\pi}}{m}\frac{1}{l_z/a+\frac{1}{\sqrt{2\pi}}\ln\left(\frac{1}{\pi l_z^2\Lambda_0^2 }\right)}
\end{align}
with $a$ the three-dimensional scattering length and $l_z$ the harmonic oscillator length of the strong harmonic confinement, i.e. $l_z=1/\sqrt{m\omega_z}$ with $\omega_z$ the frequency of the  harmonic potential. For large enough $l_z/a$ one may neglect the logarithmic term, leading to the simplified and widely employed expression
\begin{align}
g_\text{2D,R}=\frac{\sqrt{8\pi}}{m}\frac{a}{l_z}\,.
\end{align}

Finally, let us also remark that in two dimensions, $mg_\text{2D,R}$ is a purely dimensionless quantity (at least in natural units). Thus, while in three dimensions we also need the particle density $\rho$ in order to define a dimensionless measure for how strongly the system is coupled, in two dimensions the coupling constant and the mass suffice for this purpose. Typical experimental values are of the order of $mg_\text{2D,R}\sim 0.1$. In the following, we will drop again the subscript ``2D'' since it will be clear from the context whether the three-dimensional or two-dimensional coupling is meant. 

To conclude this section, let us have a brief look at characteristic length scales in an interacting Bose gas. For $U=0$, the Hamiltonian \eq{H} can be brought into dimensionless form by rescaling $\psi\to \sqrt{\rho} \tilde{\psi}$ with $\rho$ the particle density and $\mathbf{x}\to\xi_\mathrm{h}\tilde{\mathbf{x}}$ with the \textit{healing length}~\footnote{The term derives from the fact that the healing length is a measure for the size of topological defects such as vortices that are characterized by a dip in the density, i.e. it sets the length scale on which density defects are ``healed''.} $\xi_\mathrm{h}\equiv1/\sqrt{2mg\rho}$, which thus constitutes a fundamental, temperature-independent characteristic length scale that is of particular relevance in the condensed phase where temperature effects play a minor role. In the non-condensed phase or in the vicinity of the transition, however, the latter are of crucial importance such that here the most relevant length scale is set by the thermal de Broglie wavelength $\lambda_T\equiv\sqrt{2\pi/mT}$ of the atoms. As soon as  $\lambda_T$ becomes of the order of the interparticle distance, Bose-Einstein condensation sets in. Finally, the scattering length $a$ sets the length scale at which the effective description \eq{H} with a contact interaction breaks down. In numerical computations, one must make sure that $\xi_\mathrm{h}$ and $\lambda_T$ are properly resolved while $a$ must \textit{not} be resolved. 
\subsection{The coherent state path integral\label{sec:pi_constr}}
While the canonical (operator) formulation of quantum field theory is both the historically earliest and most fundamental one (the standard axioms of quantum theory being formulated in the language of Hilbert spaces and operators), the alternative path integral approach is often much more suitable for practical calculations. This applies to both analytical tools, the path integral being the suitable framework for techniques such as effective action methods, the functional renormalization group or variational perturbation theory, as well as numerical computations, since the formulation of quantum theory in terms of the path integral allows the application of powerful Monte Carlo methods.

In this section, we will review the construction of the path integral of bosons in thermal equilibrium and discuss some of its general properties, as well as the extraction of observables. Furthermore, we will examine the effect and systematic bias introduced by a finite discretization of the path integral on a lattice, which is necessary for understanding the systematic errors of numerical path integral computations. Finally, we will discuss the renormalization of the coupling constant on a discrete computational lattice.
\subsubsection{Construction\label{sec:constr_pi}}
In order to transform the thermal partition function
\begin{align}
\label{eq:partfun} Z=\text{Tr}\left\{\exp\left[-\beta H\left(\{a^\dagger_i\},\{a_i\}\right)+\beta\mu\sum_i a_i^\dagger a_i\right]\right\}
\end{align}
of a Bose gas with Hamiltonian $H\left(\{a^\dagger_i\},\{a_i\}\right)$ into a path integral expression, one follows the standard route of dividing the imaginary time interval $\beta$ into $N_\tau$ slices of size $a_\tau=\beta/N_\tau$ and inserting resolutions of the identity in terms of eigenstates of the operators $\{a_i\}$ in between the slices. 

The eigenstates of the annihilation operator $a$ are the coherent states~\footnote{In this thesis, we use a convention where the coherent states are normalized, $\langle\alpha|\alpha\rangle=1$, while in some textbooks a convention is followed where they are not normalized, i.e. $\langle\alpha|\alpha\rangle=e^{|\alpha|^2}$.} $|\alpha\rangle$, $\alpha\in \mathbb{C}$, which are defined as
\begin{align}
|\alpha\rangle\equiv e^{-|\alpha|^2/2}\sum_{n=0}^\infty\frac{\alpha^n}{\sqrt{n!}}|n\rangle\,,
\end{align}
where the Fock states $|n\rangle$ are the eigenstates of the number operator $a^\dagger a$. As one easily checks, these are indeed eigenstates of $a$,
\begin{align}
a|\alpha\rangle=\alpha|\alpha\rangle\,.
\end{align}
However, they do not form an orthonormal set, their overlap reading
\begin{align}
\langle \alpha|\beta\rangle=\exp\left(\alpha^*\beta-\frac{1}{2}[|\alpha|^2+|\beta|^2]\right)\,.
\end{align}
The identity may be written in terms of the coherent states as 
\begin{align}
\label{eq:resid}
\mathbb{I}=\int d\alpha d\alpha^* \,|\alpha\rangle\langle\alpha|\,,
\end{align}
where the integral measure $d\alpha d\alpha^*$ has to be read as $d\alpha d\alpha^*=d\text{Re}(\alpha) d\text{Im}(\alpha)/\pi$. \eq{resid} can be proven as follows: 
\begin{align}
\label{eq:tr_cs}
\nonumber\int d\alpha d\alpha^* \,|\alpha\rangle\langle\alpha|&=\sum_{n=0}^\infty\sum_{m=0}^\infty\int \limits_0^\infty \frac{dr\,r}{\pi}\int\limits _0^{2\pi}d\theta \, \frac{e^{-r^2} r^{n+m} e^{i(n-m)\theta}}{\sqrt{n!m!}}|n\rangle\langle m|\\\nonumber
&=\sum_{n=0}^\infty\int \limits_0^\infty dr\,2r \frac{e^{-r^2} r^{2n}}{n!}|n\rangle\langle n|\\
&=\sum_{n=0}^\infty|n\rangle\langle n|=\mathbb{I}\,.
\end{align}
From this it follows immediately that 
\begin{align}
\label{eq:reptr}
\nonumber\text{tr}\{\Omega\}&=\sum_{n=0}^\infty\langle n|\Omega|n\rangle=\int d\alpha d\alpha^* \,\sum_{n=0}^\infty\langle n|\Omega|\alpha\rangle\langle\alpha|n\rangle\\&=\int d\alpha d\alpha^* \,\sum_{n=0}^\infty\langle\alpha|n\rangle\langle n|\Omega|\alpha\rangle=\int d\alpha d\alpha^* \,\langle\alpha|\Omega|\alpha\rangle\,.
\end{align}
We are now in the position to construct the path integral representation of the partition function for a bosonic Hamiltonian $H$. Let the generalized Hamiltonian $K\equiv H-\mu N$ with $N$ the number operator be of the generic form
\begin{align}
\label{Hamiltonian}K=H-\mu N=\sum_{kl}\left(t_{kl}-\mu\delta_{kl}\right)a^\dagger_k a_l+\frac{1}{2}\sum_{klmn}V_{klmn}a^\dagger_k a^\dagger_l a_m a_n\,,
\end{align} 
where the indices  $k,l,m,n$ number arbitrary modes of the system, $t_{kl}$ is the one-particle and $V_{klmn}$ the two-body interaction. Dividing $\beta$ into $N_\tau$ intervals of length $a_\tau=\beta/N_\tau$, we can write $\exp(-\beta K)=[\exp(-a_\tau K)]^{N_\tau}$. We thus write the partition function as
\begin{align}
\nonumber Z=\text{tr}\{\exp(-\beta K)\}=\text{tr}\left\{ [\exp(-a_\tau K)]^{N_\tau} \right\}\,.
\end{align}
By inserting resolutions of the identity, equation \eq{resid}, and the representation of the trace \eq{reptr}, in terms of the coherent states, this can be brought to the form
\begin{align}
Z=\int \prod \limits_{i=0}^{N_\tau-1} d\boldsymbol{\psi}_i^* d\boldsymbol{\psi}_i\,\prod\limits_{i=0}^{N_\tau-1}\langle\boldsymbol{\psi}_{i+1}|\exp(-a_\tau K)|\boldsymbol{\psi}_{i}\rangle
\end{align}
where we define a vector coherent state $|\boldsymbol{\psi}_i\rangle$ as the product over coherent states for the single modes, $|\boldsymbol{\psi}_i\rangle\equiv \prod_j|\psi_{i,j}\rangle_j$, and $i=N_\tau$ is to be identified as $i=0$. 
Now we use that 
\begin{align}
\langle\psi|\exp(\epsilon\Omega)|\chi\rangle=\exp\left(\epsilon\frac{\langle\psi|\Omega|\chi\rangle}{\langle\psi|\chi\rangle}\right)\langle\psi|\chi\rangle+\mathcal{O}(\epsilon^2)
\end{align}
for arbitrary states $|\phi\rangle$, $|\chi\rangle$ and operator $\Omega$. This enables us to derive the following approximate expression for the partition function, which becomes exact in the limit $a_\tau\to0$:
\begin{align}
\label{eq:disc_pi_repr_Z}
Z\approx \int \prod \limits_{i=0}^{N_\tau-1} d\boldsymbol{\psi}_i^* d\boldsymbol{\psi}_i\,\exp(-S_{N_\tau}[\boldsymbol{\psi},\boldsymbol{\psi}^*])\,,
\end{align} 
with action 
\begin{align}
\nonumber S_{N_\tau}[\boldsymbol{\psi},\boldsymbol{\psi}^*]=\sum_{i=0}^{N_\tau-1}\Bigg\{&\sum_k\psi^*_{i+1,k}(\psi_{i+1,k}-\psi_{i,k})\\&+a_\tau\left[\sum_{kl}\left(t_{kl}-\mu\delta_{kl}\right)\psi^*_{i+1,k} \psi_{i,l}+\frac{1}{2}\sum_{klmn}V_{klmn}\psi^*_{i+1,k} \psi^*_{i+1,l} \psi_{i,m} \psi_{i,n}\right]\Bigg\}\,,
\end{align}
where $i=N_\tau$ has to be interpreted as $i=0$.
In this form, the action is ready to be used in numerical path integral computations. One may, however, suggestively write down a continuum form of the above discretized action:
\begin{align}
S[\psi,\psi^*]=\int\limits_0^\beta d\tau\left\{\psi^*\partial_\tau\psi+\sum_{kl}\left(t_{kl}-\mu\delta_{kl}\right)\psi^*_{k} \psi_{l}+\frac{1}{2}\sum_{klmn}V_{klmn}\psi^*_{k} \psi^*_{l} \psi_{m} \psi_{n}\right\}
\end{align}
where $\psi_k(\tau)$ is now a continuous function of the ``imaginary time'' $\tau$ and fulfills periodic boundary conditions, $\psi_k(\tau=0)=\psi_k(\tau=\beta)$. Since in the discretized version of the action the conjugated fields appear evaluated one lattice point later in imaginary time than the non-conjugated ones, $\psi^*$ has to be read as $\psi^*(\tau+0^+)$ and $\psi$ as $\psi^*(\tau)$ in the continuum. While this may seem trifling at first sight, it is important for correctly recovering the results of the operator formalism from the path integral, cf. section \sect{disc_eff}. The high-dimensional but ordinary integral measure in \eq{disc_pi_repr_Z} may be promoted to a path integral measure, such that \eq{disc_pi_repr_Z} becomes
\begin{align}
\label{eq:cont_pi_repr_Z}
Z=\int \mathcal{D}\boldsymbol{\psi}^* \mathcal{D}\boldsymbol{\psi}\,\exp(-S[\boldsymbol{\psi},\boldsymbol{\psi}^*])\,.
\end{align} 
Crucially, the ``Berry phase'' term in the action, $\psi^*\partial_\tau\psi$, yields a purely imaginary contribution, as can be seen from a partial integration~\footnote{This argument does not hold for the discretized path integral, as it requires $\psi$ to be a function of a continuous variable and to be differentiable. Hence $\sum_i \psi^*_{i+1}(\psi_{i+1}-\psi_{i})$ can also have a non-vanishing real part, but it it is clear that it is in general a complex quantity and not purely real.}:
\begin{align}
\left(\int\limits_0^\beta d\tau\,\psi^*\partial_\tau\psi\right)^*=\int\limits_0^\beta d\tau\,\psi\partial_\tau\psi^*=-\int\limits_0^\beta d\tau\,\psi^*\partial_\tau\psi\,.
\end{align}
As a consequence, the entire action $S$ will in general be a complex quantity, $S\in\mathbb{C}$. This makes the numerical evaluation of the path integral unfeasible with standard Monte Carlo methods, as they rely on interpreting the path integral weight $e^{-S}$ as a (real and positive-definite) probability distribution. This problem is known as \textit{sign problem} in the literature and will be discussed in detail in section \ref{sec:sign_prob}. 

\begin{figure}	
	\centering\includegraphics[width=0.6\textwidth]{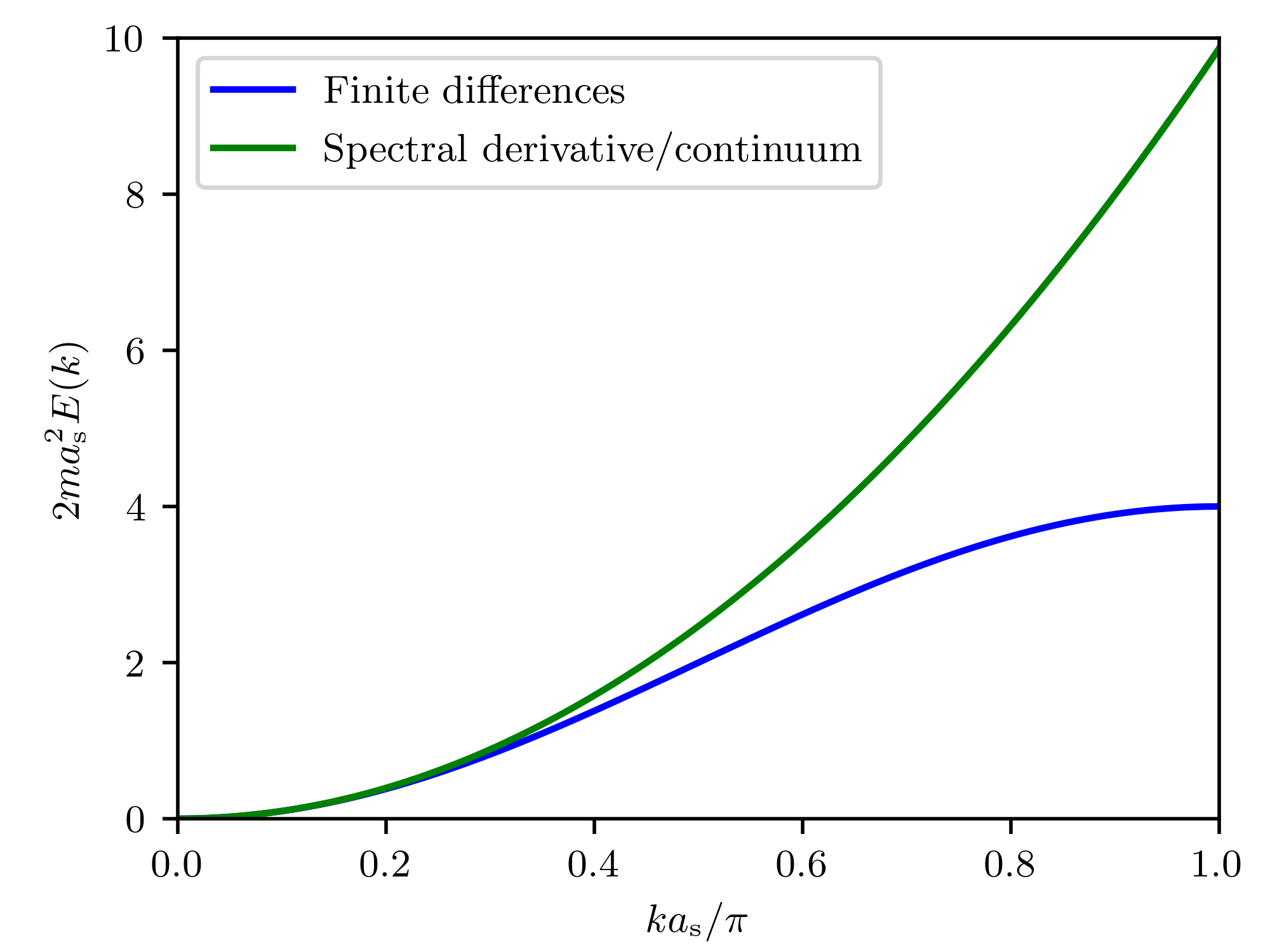}
	\caption{Kinetic energy $E(k)$ as a function of momentum $k$ in one spatial dimension for either the finite-difference discretization of the Laplacian (blue) or the spectral derivative, which corresponds to the correct quadratic continuum dispersion (green). While both agree for small momenta as they have to, the deviation at the edges of the Brillouin zone is substantial, with a maximum relative deviation of $\pi^2/4-1\approx 147\%$.}
	\label{fig:FD_vs_spect}
\end{figure}

For lattice computations of interacting bosons, we do not only have to discretize in imaginary time but also need to employ a lattice version of the continuum Hamiltonian. For a homogeneous one-component gas with contact interaction the latter reads, cf. the previous section \sect{physics_bose}:
\begin{align}
H-\mu N=\int d^3x\left\{-\frac{1}{2m}\psi^\dagger(\mathbf{x})\Delta\psi(\mathbf{x})-\mu\psi^\dagger(\mathbf{x})\psi(\mathbf{x})+\frac{g}{2}\psi^\dagger(\mathbf{x})\psi^\dagger(\mathbf{x})\psi(\mathbf{x})\psi(\mathbf{x})\right\}\,.
\end{align} 
The continuous operator $\psi(\mathbf{x})$ is replaced by operators $\psi_\mathbf{j}$ on lattice sites numbered by $\mathbf{j}$, such that $\mathbf{x}=a_\mathrm{s}\mathbf{j}$ with spatial lattice spacing $a_\mathrm{s}$. For our numerical computations, we will set $2ma_\mathrm{s}=1$, cf. appendix \sect{units} about numerical units and their conversion to experimental ones. The representation of the Laplacian operator $\Delta$ on the lattice is not unique. The simplest choice are first-order finite differences, i.e. to approximate
\begin{align}
\label{eq:FD}
\Delta\psi(\mathbf{x})\approx a_\mathrm{s}^{-2}\left(\psi_{\mathbf{j}+\mathbf{e}_x}+\psi_{\mathbf{j}-\mathbf{e}_x}+\psi_{\mathbf{j}+\mathbf{e}_y}+\psi_{\mathbf{j}-\mathbf{e}_y}+\psi_{\mathbf{j}+\mathbf{e}_z}+\psi_{\mathbf{j}-\mathbf{e}_z}-6\psi_{\mathbf{j}}\right)\,.
\end{align}
This definition has several advantages: It is easy to implement, fast to compute and, most importantly, it maintains the locality of the continuum action. It has, however, also a serious disadvantage: The kinetic energy is no longer given by its correct continuum version, $E_\text{kin}(\mathbf{k})=\mathbf{k}^2/2m$ but takes the typical sine-like form of lattice models:
\begin{align}
E_\text{kin}^\text{FD}(\mathbf{k})=\frac{2}{ma_\mathrm{s}^2}\left[\sin^2\left(\frac{k_xa_\mathrm{s}}{2}\right)+\sin^2\left(\frac{k_ya_\mathrm{s}}{2}\right)+\sin^2\left(\frac{k_za_\mathrm{s}}{2}\right)\right]\,.
\end{align} 
As a consequence, numerical computations are substantially flawed by the ``wrong'' dispersion unless the lattice spacing is taken to be very small, cf. figure \fig{FD_vs_spect}. For example, the occupation number in a non-interacting Bose gas will be $1/[\exp(E_\text{kin}^\text{FD}(\mathbf{k}))-1]$ instead of $1/[\exp(\mathbf{k}^2/2m)-1]$. In part, this flaw can be corrected for within in the evaluation of observables, by \textit{defining} the physical momentum to be the square root of the kinetic energy, cf. the subsequent section \sect{observ}. This becomes, however, unfeasible for more complicated observables than momentum spectra, dispersions and particle numbers in homogeneous systems. 

An alternative choice for the evaluation of the Laplacian that avoids the wrong dispersion is to employ a spectral derivative. That is, one Fourier transforms the operators (fields), multiplies with $\mathbf{k}^2$ and Fourier transforms back. While this is numerically more expensive than just calculating finite differences, it is not \textit{much} more expensive because Fourier transforms can be computed very efficiently by means of fast Fourier transform (FFT) algorithms. The Laplacian is then approximated by the following expression:  
\begin{align}
\label{eq:SD}
\Delta\psi(\mathbf{x})\approx  -\frac{1}{N_xN_yN_z}\sum_{\mathbf{k}\mathbf{j}'}
\exp\left(i\mathbf{k}\cdot[a_\mathrm{s}\mathbf{j}-a_\mathrm{s}\mathbf{j}']\right)
\left|\mathbf{k}\right|^2\psi_{\mathbf{j}'}\,,
\end{align}
where $N_{x,y,z}$ denotes the extension of the lattice in the respective direction, the sum over $\mathbf{j}'$ runs over all lattice points and $k_{x,y,z}=2\pi N_{x,y,z}^{-1}a_\mathrm{s}^{-1}(-N_{x,y,z}/2+1\dots N_{x,y,z}/2)$.
The advantage of the spectral derivative is that the only spatial discretization error stems from the fact that there is a maximum momentum $\Lambda=\pi/a_\mathrm{s}$ where the lattice theory is cut off. Up to this cutoff, the kinetic energy takes its correct continuum form, in contrast to the finite differences, which lead to a bias already far below the momentum cutoff of the theory. It must be noted that the spectral derivative leads to a highly non-local lattice representation of the local continuum theory: In fact, every lattice point now interacts with every other lattice point, although this interaction decreases rapidly with distance. However, in practice we never found any unphysical behavior or artifacts as a consequence of the non-local lattice action. Therefore and because of its only slightly increased numerical cost, we consider the spectral derivative the method of choice for path integral computations of Bose-Einstein condensates. Within this thesis, we employ the finite-difference scheme only in the comparatively simple setting of chapter \sect{3D}, while the simulations in all subsequent chapters were performed with the spectral derivative representation of the Laplacian. 

In conclusion, the discretized lattice action for a three-dimensional, one-component Bose gas with contact interaction reads:
\begin{align}
\label{eq:S_discr}
\nonumber S^\text{lat}=a_\mathrm{s}^3\sum_{i=0}^{N_\tau-1}\sum_{\mathbf{j}}\Bigg\{&\psi^*_{i+1,\mathbf{j}}(\psi_{i+1,\mathbf{j}}-\psi_{i,\mathbf{j}})\\&+a_\tau\left[-\frac{1}{2m}\psi^*_{i+1,\mathbf{j}}\Delta^\text{lat}\psi_{i,\mathbf{j}}-\mu\psi^*_{i+1,\mathbf{j}} \psi_{i,\mathbf{j}}+\frac{g}{2}\psi^*_{i+1,\mathbf{j}} \psi^*_{i+1,\mathbf{j}} \psi_{i,\mathbf{j}} \psi_{i,\mathbf{j}}\right]\Bigg\}\,,
\end{align}
with $\Delta^\text{lat}$ given by \eq{FD}, \eq{SD} or some other approximation for the Laplacian (e.g. higher order finite differences), while the multi-component generalization with $U(\mathcal{N})$-symmetric interaction reads 
\begin{align}
\label{eq:S_discr_UN}
\nonumber S^\text{lat}=a_\mathrm{s}^3\sum_{i=0}^{N_\tau-1}\sum_{\mathbf{j}}\Bigg\{&\psi^*_{a,i+1,\mathbf{j}}(\psi_{a,i+1,\mathbf{j}}-\psi_{a,i,\mathbf{j}})\\&+a_\tau\left[-\frac{1}{2m}\psi^*_{a,i+1,\mathbf{j}}\Delta^\text{lat}\psi_{a,i,\mathbf{j}}-\mu\psi^*_{a,i+1,\mathbf{j}} \psi_{a,i,\mathbf{j}}+\frac{g}{2}\left(\psi^*_{a,i+1,\mathbf{j}}\psi_{a,i,\mathbf{j}}\right)^2\right]\Bigg\}\,,
\end{align}
where summation over the component index $a$ is implied.
\subsubsection{Observables\label{sec:observ}}
There are numerous observables that can be straightforwardly extracted in ultracold atom experiments and even more that are accessible within numerical path integral computations. They can be broadly divided into two classes, \textit{equal-time} and \textit{unequal-time} observables. Equal-time observables (correlators) are composed of operators defined (in the Heisenberg picture) at the same time, i.e. are of the form $\langle\mathcal{O}\rangle (t)=\langle\Omega_1(t)\dots\Omega_n(t)\rangle$, while unequal-time observables are unrestricted with this regard, i.e. can be of the more general form $\langle\mathcal{O}\rangle (t_1,\dots,t_n)=\langle\Omega_1(t_1)\dots\Omega_n(t_n)\rangle$. In thermal equilibrium, equal-time correlators become completely independent of $t$. In this thesis, we will almost exclusively consider such ``static'' observables in thermal equilibrium, i.e. observables of the type
\begin{align}
\langle\mathcal{O}\rangle=\frac{\text{Tr}\left\{e^{-\beta (H-\mu N)}\mathcal{O}\right\}}{\text{Tr}\left\{e^{-\beta (H-\mu N)}\right\}}\,.
\end{align} 
However, it is also possible to compute \textit{unequal-imaginary-time} correlators~\footnote{As the complex Langevin method can in principle deal with any complex action, it is \textit{in principle} possible to also compute actual unequal-real-time correlators. However, in practice the method breaks down for any case beyond short-time dynamics \cite{berges2007lattice}.} of the type   
\begin{align}
\langle\mathcal{O}\rangle(\tau_1,\dots,\tau_n)=\frac{\text{Tr}\left\{e^{-\beta (H-\mu N)}\Omega_1(\tau_1)\dots\Omega_n(\tau_n)\right\}}{\text{Tr}\left\{e^{-\beta (H-\mu N)}\right\}}\,,
\end{align} 
where the imaginary-time Heisenberg operator $\Omega(\tau)$ is defined as $\Omega(\tau)=e^{\tau (H-\mu N)}\Omega e^{-\tau (H-\mu N)}$, as long as $\mathcal{O}$ is \textit{time-ordered}, i.e. $\tau_1\ge \dots\ge\tau_n$, and $\tau_1-\tau_n<\beta$. 

The above construction of the path integral representation of the partition function can be straightforwardly translated to a path integral representation of this type of observables. Namely, one finds
\begin{align}
\langle\mathcal{O}\rangle(\tau_1,\dots,\tau_n)=Z^{-1}\int \prod \limits_{i=0}^{N_\tau-1} d\boldsymbol{\psi}_i^* d\boldsymbol{\psi}_i\,\exp(-S_{N_\tau}[\boldsymbol{\psi},\boldsymbol{\psi}^*])\,\Omega_1\left(\boldsymbol{\psi}_{i_1+1}^*,\boldsymbol{\psi}_{i_1}\right)\dots\Omega_n\left(\boldsymbol{\psi}_{i_n+1}^*,\boldsymbol{\psi}_{i_n}\right)\,,
\end{align}
where $i_m=\tau_m/a_\tau$, $Z$ is defined as in \eq{disc_pi_repr_Z} and the recipe for obtaining from an operator $\Omega(\{a^\dagger_j\},\{a_j\})$ its path integral representation $\Omega(\boldsymbol{\psi}^*,\boldsymbol{\psi})$ is to bring $\Omega$ into normal-ordered form and then to replace $\{a^\dagger_j\}$ by $\boldsymbol{\psi}^*$ and $\{a_j\}$ by $\boldsymbol{\psi}$. Here we assumed that the time-ordering is strict, i.e. $\tau_1> \dots>\tau_n$. If this is not the case, i.e. $\tau_{i}=\tau_{i+1}$ for some $i$, one must normal-order the entire operator $\Omega_{i}\Omega_{i+1}$. 

As equal-imaginary-time correlators are time-independent in thermal equilibrium, one may evaluate a static observable $\mathcal{O}$ at any point in the interval $[0,\beta]$. In a numerical simulation, it is most convenient to exploit this time-translational symmetry and to average over the interval $[0,\beta]$ in order to gain statistics. Hence, we write the expectation value of a static observable $\langle\mathcal{O}\rangle$ as
\begin{align}
\langle\mathcal{O}\rangle=Z^{-1}\frac{1}{N_\tau}\sum_{i'=0}^{N_\tau-1}\int \prod \limits_{i=0}^{N_\tau-1} d\boldsymbol{\psi}_i^* d\boldsymbol{\psi}_i\,\exp(-S_{N_\tau}[\boldsymbol{\psi},\boldsymbol{\psi}^*])\,\mathcal{O}\left(\boldsymbol{\psi}_{i'+1}^*,\boldsymbol{\psi}_{i'}\right)\,.
\end{align}
Several static observables will be the subject of interest in this thesis. Due to its simplicity and the high degree of information that can be read off from it, we will most frequently consider the single-particle momentum spectrum $f(\mathbf{k})$, which is defined as 
\begin{align}
f(\mathbf{k})=\langle a^\dagger_\mathbf{k} a_\mathbf{k}\rangle\,,
\end{align}
translating to  
\begin{align}
a^\dagger_\mathbf{k} a_\mathbf{k} \to \psi_{i+1,\mathbf{k}}^*\psi_{i,\mathbf{k}}
\end{align}
in the path integral. $f(\mathbf{k})$ thus measures the number of particles that occupy one particular momentum mode $\mathbf{k}$. The lattice Fourier transform $\psi_{i,\mathbf{k}}$ is defined in the usual way as 
\begin{align}
\psi_{i,\mathbf{k}}=\sqrt{\frac{a_\mathrm{s}^3}{N_xN_yN_z}}\sum_\mathbf{j} e^{i\mathbf{k}\cdot a_\mathrm{s}\mathbf{j}} \psi_{i,\mathbf{j}}
\end{align}
with $k_{x,y,z}=2\pi N_{x,y,z}^{-1}a_\mathrm{s}^{-1}(-N_{x,y,z}/2+1\dots N_{x,y,z}/2)$, and can be computed efficiently by employing standard FFT algorithms. If one chooses for the discretization of the Laplacian in the action a spectral derivative (cf. section \sect{constr_pi}), $\mathbf{k}$ corresponds to the actual physical momentum. If the simpler finite-difference scheme is employed, it must be taken into account that the kinetic energy assumes a sine-like dependence on the index of the momentum mode. As the most reasonable definition of a momentum is via the square root of the kinetic energy, the definition of the physical momentum assumes a sine-like functional dependence as well in this case. Thus, for the momentum mode with index vector $\mathbf{n}$ (i.e. $k_{x,y,z}=2\pi N_{x,y,z}^{-1}a_\mathrm{s}^{-1} n_{x,y,z}$), we have for the physical momentum $\mathbf{k}_\text{phys}$:
\begin{align}
\label{eq:momentum_sd}
\mathbf{k}_\text{phys}&=\mathbf{k}=2\pi a_\mathrm{s}^{-1}\begin{pmatrix}n_x/N_x\\n_y/N_y\\n_z/N_z\end{pmatrix}\qquad\text{(spectral derivative)}\\
\label{eq:momentum_fd}
\mathbf{k}_\text{phys}&=2a_\mathrm{s}^{-1}\begin{pmatrix}\sin\left(\pi n_x/N_x\right)\\\sin\left(\pi n_y/N_y\right)\\\sin\left(\pi n_z/N_z\right)\end{pmatrix}\qquad\text{(finite differences)}\,.
\end{align}  
For isotropic systems, $f$ depends only on $|\mathbf{k}_\text{phys}|$ and it is thus customary to perform angular averages over momentum shells $k_\text{phys}+\Delta k$, with a suitably chosen binning width $\Delta k$. We will in the following drop the distinction between $\mathbf{k}_\text{phys}$ and $\mathbf{k}$ again, as they are equal for spectral derivatives and for finite differences (which we employ only in chapter \sect{3D}) it is usually clear from the context what is meant. 

Similarly to the occupation number in momentum space, one may define the density in real space $\rho(\mathbf{r})\equiv \rho(a_\mathrm{s}\mathbf{j})\equiv\rho_\mathbf{j}$ as $\rho_\mathbf{j}=\langle\psi^\dagger_\mathbf{j}\psi_\mathbf{j}\rangle$, translating to $\psi^*_{i+1,\mathbf{j}}\psi_{i,\mathbf{j}}$ in the path integral.

On several occasions, it is also necessary to extract the total particle number $N_\text{tot}$, which in the grand-canonical formalism is an observable and not a parameter. As the Fourier transform conserves particle number, it can be equivalently extracted in real and momentum space:
\begin{align}
\label{eq:Ntot}
N_\text{tot}=a_\mathrm{s}^3\sum_\mathbf{j} \rho_\mathbf{j}=\sum_\mathbf{k}f(\mathbf{k})\,.
\end{align} 
If one employs the spectral derivative representation of the Laplacian in the action, this expression typically yields only very tiny deviations from the correct continuum value of $N_\text{tot}$ even if the thermal wave length and healing length are taken to be only a few lattice points, because the only error stems from the cutoff of the momentum sum and $f(\mathbf{k})$ decays exponentially. However, for the finite difference discretization, one finds \eq{Ntot} rather inaccurate because already well below the maximum momentum, the dispersion deviates from the true continuum dispersion and becomes sine-like. The corresponding physical momenta are much more densely spaced at the edges of the Brillouin zone than in the middle, while in the continuum they must be equidistant. One may correct for this effect by introducing a Jacobi determinant into \eq{Ntot} that takes into account the unequal spacing of the momenta at the Brillouin zone edges:
\begin{align}
\label{eq:Ntot_corr}
N_\text{tot,corr}=\sum_\mathbf{n}\cos\left(\frac{\pi n_x}{N_x}\right)\cos\left(\frac{\pi n_y}{N_y}\right)\cos\left(\frac{\pi n_z}{N_z}\right)f_\mathbf{n}\,.
\end{align} 

More specific observables such as superfluid densities or structure factors will be introduced in subsequent chapters when they will be needed. The only further observable that we want to discuss in this introductory chapter is the dispersion relation $\omega_\mathbf{k}$ that measures the energy of a stable (quasi-)particle with momentum $\mathbf{k}$. For unstable quasi-particles, $\omega_\mathbf{k}$ turns complex and it is the real part that measures the energy, while the imaginary part is the decay rate. Thus, $\omega(\mathbf{k})$ makes only sense as an observable as long as there are well-defined quasi-particles in the system. The more generic quantity measuring the excitations in a system, which can be defined in all circumstances, is the spectral function $\rho(\omega,\mathbf{k})$, defined as
\begin{align}
\label{eq:spectr_func}
\rho(\omega,\mathbf{k})=\frac{1}{2\pi}\int \limits_{-\infty}^\infty dt\,e^{i\omega t} \, \left\langle\left[a_\mathbf{k}(t),a_\mathbf{k}^\dagger(0)\right]\right\rangle\,.
\end{align}   
For example, for a non-interacting Bose gas this definition amounts to 
\begin{align}
\rho(\omega,\mathbf{k})=\delta\left(\omega-\frac{\mathbf{k}^2}{2m}\right)\,.
\end{align}
As long as the system can be described by well-defined, stable quasi-particles, the spectral function is of the generic form
\begin{align}
\rho(\omega,\mathbf{k})=A_\mathbf{k}\delta\left(\omega-\omega_\mathbf{k}\right)+B_\mathbf{k} \delta\left(\omega+\omega_\mathbf{k}\right)\,,
\end{align}
with dispersion $\omega_\mathbf{k}$ and amplitudes $A_\mathbf{k}$ and $B_\mathbf{k}$. For unstable quasi-particles, this generalizes to 
\begin{align}
\label{eq:ansatz_rho}
\rho(\omega,\mathbf{k})=\frac{A_\mathbf{k}}{\pi}\frac{\text{Im}(\omega_\mathbf{k})}{\left(\omega-\text{Re}(\omega_\mathbf{k})\right)^2+\text{Im}(\omega_\mathbf{k})^2}+ \frac{B_\mathbf{k}}{\pi}\frac{\text{Im}(\omega_\mathbf{k})}{\left(\omega+\text{Re}(\omega_\mathbf{k})\right)^2+\text{Im}(\omega_\mathbf{k})^2}\,,
\end{align}
with the delta functions replaced by finite-width Lorentz curves. We want to show that as long as the spectral function is of this form and hence the notion of a dispersion $\omega_\mathbf{k}$ makes sense, and assuming $\text{Im}(\omega_\mathbf{k})=0$, it can be extracted from the following, numerically easily accessible formula:
\begin{align}
\label{eq:disp}
\omega_\mathbf{k}^2=-\frac{\partial_\tau\partial_{\tau'}\langle a^\dagger_{\mathbf{k}}(\tau)a_{\mathbf{k}}(\tau')\rangle|_{\tau=\tau'}}
{\langle a^\dagger_{\mathbf{k}}(\tau)a_{\mathbf{k}}(\tau')\rangle|_{\tau=\tau'}}\,.
\end{align}  
The first step in proving \eq{disp} is to relate the spectral function as defined in \eq{spectr_func} to a correlation function that does not involve a commutator:
\begin{align}
\nonumber\rho(\omega,\mathbf{k})&=\frac{1}{2\pi}\int \limits_{-\infty}^\infty dt\,e^{i\omega t} \, \left\langle\left[a_\mathbf{k}(t),a_\mathbf{k}^\dagger(0)\right]\right\rangle\\\nonumber
&=\frac{1}{2\pi}\int \limits_{-\infty}^\infty dt\,e^{i\omega t} \, \text{Tr}\left\{e^{-\beta H}\left(e^{it H}a_\mathbf{k}e^{-it H}a_\mathbf{k}^\dagger-a_\mathbf{k}^\dagger e^{it H}a_\mathbf{k}e^{-it H}\right)\right\}
\\\nonumber&=\frac{1}{2\pi}\int \limits_{-\infty}^\infty dt\,e^{i\omega t} \, \left[\text{Tr}\left\{a_\mathbf{k}^\dagger e^{(it-\beta) H}a_\mathbf{k}e^{-it H}\right\}-\text{Tr}\left\{a_\mathbf{k}^\dagger e^{it H}a_\mathbf{k}e^{(-it-\beta) H}\right\}\right]
\\\nonumber&=\frac{1}{2\pi}\int \limits_{-\infty}^\infty dt\,e^{i\omega t} \, \left[e^{\beta\omega}\text{Tr}\left\{a_\mathbf{k}^\dagger e^{it H}a_\mathbf{k}e^{(-it-\beta) H}\right\}-\text{Tr}\left\{a_\mathbf{k}^\dagger e^{it H}a_\mathbf{k}e^{(-it-\beta) H}\right\}\right]
\\&=\left(e^{\beta\omega}-1\right)\frac{1}{2\pi}\int \limits_{-\infty}^\infty dt\,e^{i\omega t} \, \left\langle a_\mathbf{k}^\dagger (0)a_\mathbf{k}(t)\right\rangle\,.
\end{align}
Here, we have exploited that the trace is invariant under cyclic permutations and performed a shift of the integration variable in the first term, $t\to t-i\beta$. This relationship is known as \textit{fluctuation-dissipation theorem}. Now we insert \eq{ansatz_rho} for $\rho(\omega,\mathbf{k})$, perform a Fourier transform and exploit that due to time-translational invariance we can go over from times $0$ and $t$ to $t$ and $t'$:
\begin{align}
\nonumber\left(e^{\beta\omega}-1\right) \left\langle a_\mathbf{k}^\dagger (t)a_\mathbf{k}(t')\right\rangle&=\int \limits_{-\infty}^\infty d\omega\,e^{-i\omega(t'-t)}\rho(\omega,\mathbf{k})\\&=e^{-\text{Im}(\omega_\mathbf{k})|t'-t|}\left[A_\mathbf{k}\,e^{-i\text{Re}(\omega_\mathbf{k})(t'-t)}+B_\mathbf{k}\,e^{i\text{Re}(\omega_\mathbf{k})(t'-t)}\right]\,,
\end{align}
where the Fourier transform of the Lorentz functions is straightforwardly computed with the residue theorem. For the case $\text{Im}(\omega_\mathbf{k})=0$ it is straightforward to verify then that 
\begin{align}
\label{eq:disp_re}
\omega_\mathbf{k}^2=\frac{\partial_t\partial_{t'}\langle a^\dagger_{\mathbf{k}}(t)a_{\mathbf{k}}(t')\rangle|_{t=t'}}
{\langle a^\dagger_{\mathbf{k}}(t)a_{\mathbf{k}}(t')\rangle|_{t=t'}}\,,
\end{align}  
from which \eq{disp} follows by analytical continuation.   

Evaluating the numerator of \eq{disp} on a lattice is a bit tricky due to the property of the path integral to put every observable into time-ordered form. 
Since $\langle a^\dagger_{\mathbf{k}}(\tau)a_{\mathbf{k}}(\tau')\rangle|_{\tau=\tau'}$ translates to $\langle \psi_{i+1,\mathbf{k}}^*\psi_{i,\mathbf{k}}\rangle$ in the path integral, it is tempting to write $\langle (\psi_{i+2,\mathbf{k}}^*-\psi_{i+1,\mathbf{k}}^*)(\psi_{i+1\mathbf{k}}-\psi_{i,\mathbf{k}})\rangle/a_\tau^2$ for the path integral representation of $\partial_\tau\partial_{\tau'}\langle a^\dagger_{\mathbf{k}}(\tau)a_{\mathbf{k}}(\tau')\rangle|_{\tau=\tau'}$. 
This, however, does \textit{not} correspond to the operator finite-time differences
\begin{align}
a_\tau^{-2}\langle [a^\dagger_{\mathbf{k}}(\tau+a_\tau)-a^\dagger_{\mathbf{k}}(\tau)]
[a_{\mathbf{k}}(\tau+a_\tau)-a_{\mathbf{k}}(\tau)]\rangle
\,,
\end{align}
which contain an anti-time-ordered term, and thus do not translate to $\langle (\psi_{\mathbf{k},i+2}^*-\psi_{\mathbf{k},i+1}^*)(\psi_{\mathbf{k},i+1}-\psi_{\mathbf{k},i})\rangle/a_\tau^2$.
Hence, a better choice is to discretize the product of derivatives as
\begin{align}
\partial_\tau\partial_{\tau'}\langle a^\dagger_{\mathbf{k}}(\tau)a_{\mathbf{k}}(\tau')\rangle|_{\tau=\tau'}
\approx a_\tau^{-2}\langle [a^\dagger_{\mathbf{k}}(\tau+a_\tau)-a^\dagger_{\mathbf{k}}(\tau)]
[a_{\mathbf{k}}(\tau)-a_{\mathbf{k}}(\tau-a_\tau)]\rangle
\,,
\end{align}
which contains time-ordered terms only and implies the discretization 
\begin{align}
\label{eq:omegataudiscretized1}
\omega(\mathbf{k})
=\sqrt{-\frac{\langle [\psi_{\mathbf{k},i+2}^*-\psi_{\mathbf{k},i+1}^*][\psi_{\mathbf{k},i}-\psi_{\mathbf{k},i-1}]\rangle}
	{a_\tau^2\,\langle\psi_{\mathbf{k},i+1}^*\psi_{\mathbf{k},i}\rangle}}\,.
\end{align}
\subsubsection{Discretization effects\label{sec:disc_eff}}
For numerical Monte Carlo simulations, the action must be discretized on a lattice, which inevitably introduces systematic biases into any such computation. Since one of the main goals of path integral simulations of ultracold bosons is to enable precision determinations of physical quantities, it is essential to examine the effect of these biases.

The action being discretized on a $d+1$-dimensional spacetime lattice, there are two types of discretization errors: spatial and temporal ones. The former are discretization effects in the Hamiltonian itself, as there is necessarily a maximum momentum that can be represented in a simulation. If the Laplacian is represented by a spectral derivative, the error stems only from this momentum cutoff; if it is represented by finite differences, additionally the lattice dispersion relation deviates from the correct continuum one already below the momentum cutoff. This type of discretization errors has in part been discussed in the previous section and will be further considered in subsequent chapters such that we will refrain from a discussion in this section. Instead, we will consider the temporal discretization effects, which stem from the trotterization procedure in the construction of the path integral. 

We restrict ourselves to the case of a non-interacting Hamiltonian, in which case it suffices to treat a single mode, i.e. we consider $H=\omega a^\dagger a$. The results will partly apply also to the (weakly) interacting Bose gas, since the latter can be described to a certain extent as a system of non-interacting quasi-particles, cf. the discussion in section \ref{sec:approx_desc}. Let us consider the expectation value of the number operator, given by
\begin{align}
\label{eq:fsinglemode}
f\equiv\langle a^\dagger a\rangle
=\frac{\text{Tr}\left\{e^{-\beta\omega a^\dagger a}a^\dagger a\right\}}{\text{Tr}\left\{e^{-\beta\omega a^\dagger a}\right\}}
\,.
\end{align}
Evaluating this expression in the Fock basis $|n\rangle$ gives the Bose-Einstein distribution,
\begin{align}
\langle a^\dagger a\rangle=\frac{1}{e^{\beta\omega}-1}
\,.
\end{align}
Within the path integral formulation, this expectation value is represented by
\begin{align}
\label{eq:pi-number}
\langle a^\dagger a\rangle
=\lim\limits_{{N_\tau}\to\infty}\frac{1}{{N_\tau}}\sum_j
\frac{\int \prod_i d\psi_i^* d\psi_i\,e^{-S_{N_\tau}[\boldsymbol{\psi}]}\,\psi_{j+1}^*\psi_j}
{\int \prod_i d\psi_i^* d\psi_i\,e^{-S_{N_\tau}[\boldsymbol{\psi}]}}
\,,
\end{align}
with the time-discretized action 
\begin{align}
S_{N_\tau}[\boldsymbol{\psi}]
=\sum_{i=0}^{N_\tau-1}\left(\psi_i^*\psi_i-\psi_{i+1}^*\psi_i+a_{\tau}\omega\, \psi_{i+1}^*\psi_i\right)
\,,
\end{align}
where the index $N_\tau$ is identified with the index $0$ and $a_{\tau}=\beta/{N_\tau}$.

\begin{figure}	
	\includegraphics[width=0.495\textwidth]{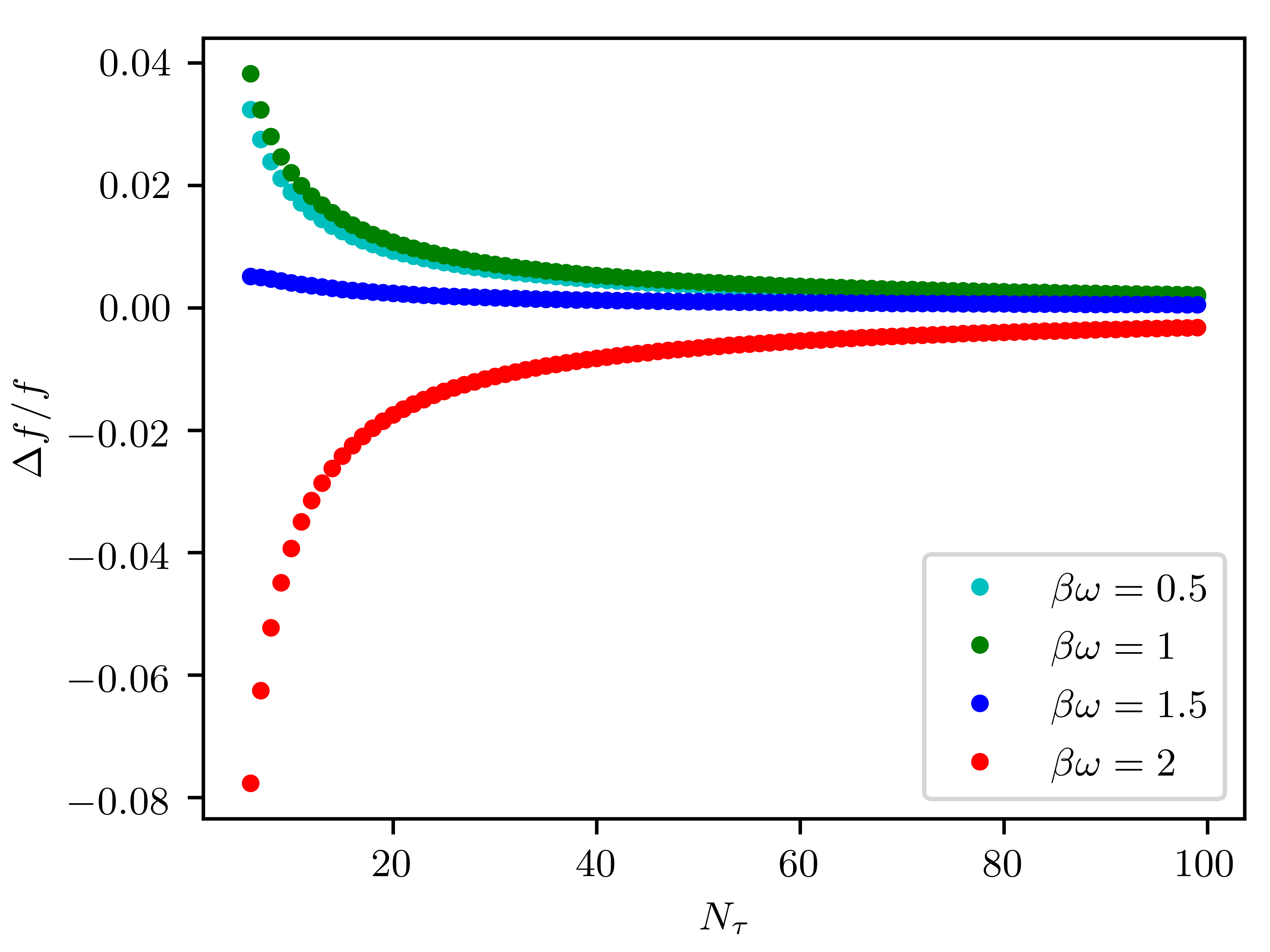}
	\includegraphics[width=0.495\textwidth]{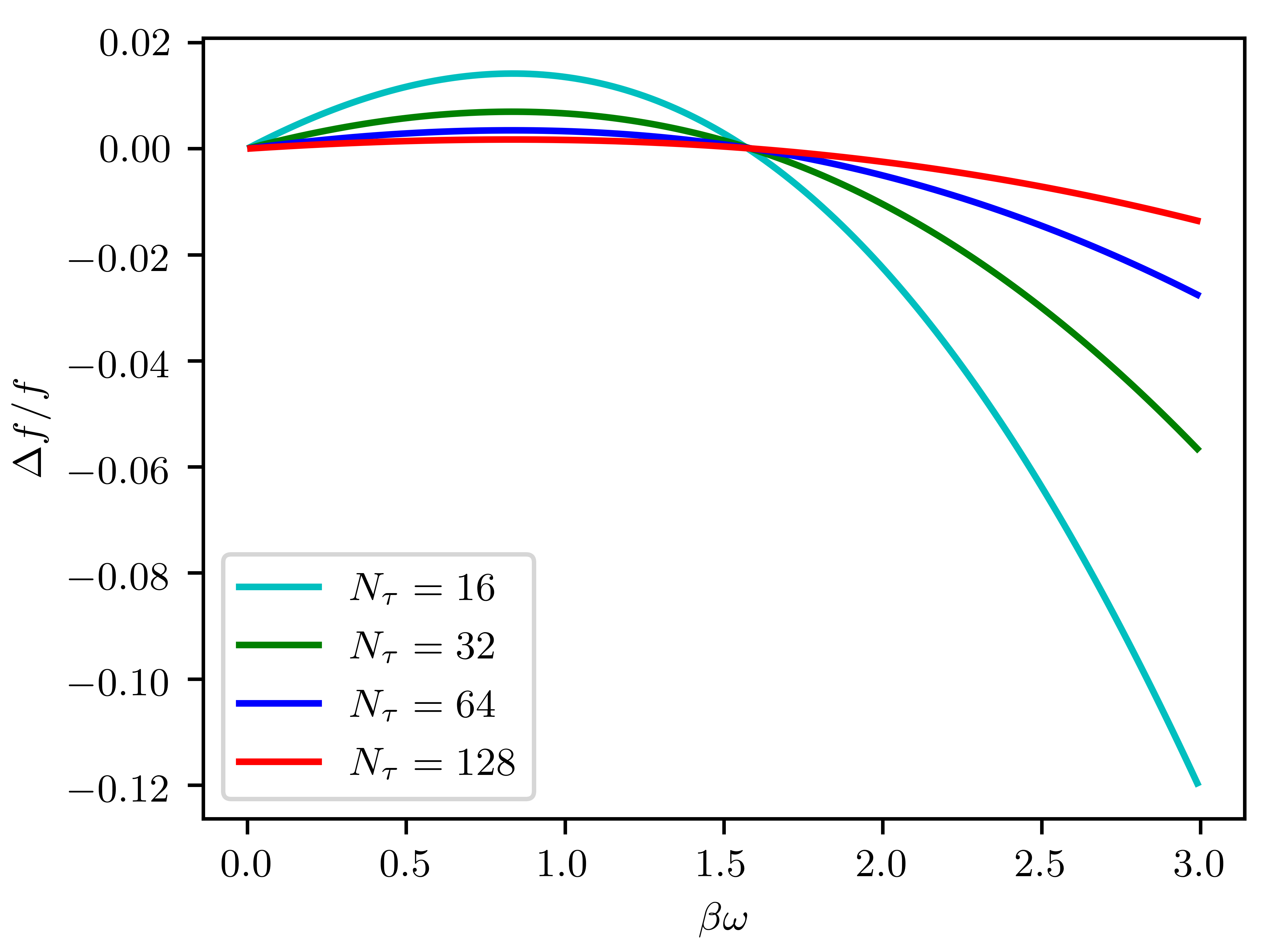}
	\caption{Relative error $\Delta f/f=[f(N_{\tau})-f]/f$ of the particle number $f(N_{\tau})$, \eq{pi3}, with respect to the continuum value $f$, \eq{fsinglemode}, (left panel) for different values of $\beta\omega$ in dependence of the number of temporal lattice points (Matsubara modes) ${N_\tau}$ and (right panel) for different  ${N_\tau}$ as a function of $\beta\omega$.}
	\label{fig:Matsubara}
\end{figure}

Expanding the fields on the finite time interval $[0,\beta]$ in terms of Matsubara modes,
\begin{align}
\psi_j=\frac{1}{\sqrt{\beta}}\sum_{n=0}^{N_{\tau}-1} \tilde{\psi}_n\,e^{i\omega_n j a_{\tau}}
\,,
\end{align}
with the Matsubara frequencies
\begin{align}
\omega_n=\frac{2\pi n}{\beta}, \qquad n=0,\dots, {N_\tau}-1,
\end{align}
the path integral \eq{pi-number} becomes
\begin{align}
\label{eq:pi2}
\langle a^\dagger a\rangle
=\lim\limits_{{N_\tau}\to\infty}\frac{1}{\beta}\sum_{m=0}^{N_{\tau}-1}
\frac{\int \prod_n d\tilde{\psi}_n^* d\tilde{\psi}_n\,e^{-S_{N_\tau}[\tilde{\boldsymbol{\psi}}]}\,
	\tilde{\psi}_m^*\tilde{\psi}_m\,e^{-i\omega_ma_{\tau}}}
{\int \prod_n d\tilde{\psi}_n^* d\tilde{\psi}_n\,e^{-S_{N_\tau}[\tilde{\boldsymbol{\psi}}]}}
\end{align}
with 
\begin{align}
S_{N_\tau}[\tilde{\boldsymbol{\psi}}]
=\sum_{n=0}^{N_{\tau}-1}\left(\frac{1-e^{-i\omega_na_{\tau}}}{a_{\tau}}
+\omega e^{-i\omega_na_{\tau}}\right)\tilde{\psi}_n^*\tilde{\psi}_n
\,.
\end{align}
Performing the Gaussian integrals yields
\begin{align}
\label{eq:pi3}
\langle a^\dagger a\rangle
&=\lim\limits_{{N_\tau}\to\infty}\frac{1}{\beta}\sum_n
\frac{e^{-i\omega_na_{\tau}}}
{a_{\tau}^{-1}({1-e^{-i\omega_na_{\tau}}})+\omega\, e^{-i\omega_na_{\tau}}}
\nonumber\\
&=\lim\limits_{{N_\tau}\to\infty}\sum_n\frac{1}{{N_\tau}(e^{2\pi i n/{N_\tau}}-1)+\beta\omega }
\nonumber\\
&\equiv\lim\limits_{{N_\tau}\to\infty}f(N_\tau)\,.
\end{align}
In the continuum limit, ${N_\tau}\to\infty$, the series \eq{pi3}  converges to the Bose-Einstein distribution $1/(e^{\beta\omega}-1)$. This can be proven with the help of complex analysis, cf. e.g. \cite{altland2010condensed}.  
On the lattice, however, ${N_\tau}$ obviously cannot be taken to infinity but has to be set to a finite value. This causes a numerical error that depends on ${N_\tau}$ but also on the value of $\beta\omega$. 
Figure \ref{fig:Matsubara} shows the numerically computed truncation error for several combinations of ${N_\tau}$ and $\beta\omega$. 
In the right panel, one sees that the relative error for small $\beta\omega$ is positive but turns negative at some point and quickly becomes rather large (while the absolute error still decreases). This demonstrates two things: On the one hand, it is mainly the high-energy modes $\beta\omega(\mathbf{k})\gg1$ in the UV that are affected by temporal discretization effects. On the other hand, the analysis demonstrates that it can be computationally very demanding to resolve all these modes properly. For equidistantly spaced momenta and in three spatial dimensions, one finds for the maximum lattice energy $\beta\omega_\text{max}$:
\begin{align}
\beta\omega_\text{max}=\frac{3}{4}\pi\left(\frac{\lambda_T}{a_\mathrm{s}}\right)^2\,,
\end{align}
with $\lambda_T=\sqrt{2\pi/mT}$ the thermal de Broglie wave length. For typical $\lambda_T$ of a few $a_\mathrm{s}$ this yields $\beta\omega_\text{max}\sim 30$. In this case, one would need $N_\tau\sim 4000$ lattice points in temporal direction if one wanted to resolve all lattice modes with an accuracy of less than $10\%$, which would greatly limit the number of lattice points available for the spatial grid. However, this is not necessary in practice. Since the occupation number of the UV modes decreases exponentially with their energy, it is typically safe to refrain from resolving them beyond a certain momentum. With the theory outlined here, it is possible to estimate a suitable $N_\tau$ by computing the error on the total particle number. Under the assumption that only infrared modes are subject to interaction effects and ultraviolet modes are effectively free in a weakly interacting Bose gas, it is even possible to compute finite-$N_\tau$ corrections to a simulation performed at a certain $N_\tau$ and thereby to extrapolate to $N_\tau\to\infty$. Nonetheless, as the latter assumption is not fully applicable to the condensed phase and even in the uncondensed phase is only an approximation, it is advisable to systematically vary $N_\tau$  additionally and to check whether the relevant observables display convergence. 

One may also study the effect of the temporal discretization on the dispersion that has been defined in section \ref{sec:observ}. For a single mode system we have 
\begin{align}
\label{eq:omegataudiscretized}
\omega(N_\tau)
=\sqrt{-\frac{1}{a_\tau^2\,f(N_\tau)} \frac{1}{{N_\tau}}\sum_i\langle [\psi_{i+2}^*-\psi_{i+1}^*][\psi_{i}-\psi_{i-1}]\rangle}\,,
\end{align}
where $\omega(N_\tau)$ is the energy of the single mode as extracted from the discretized path integral, with $\omega=\lim_{N_\tau\to\infty}\omega(N_\tau)$.
In a similar manner as for he occupation number one shows:
\begin{align}
\omega(N_\tau)
=\frac{N_\tau}{\beta}\sqrt{-\frac{1}{f(N_\tau)}\sum_n
	\frac{2e^{-2\pi i n/{N_\tau}}-e^{-4\pi i n/{N_\tau}}-1}
	{{N_\tau}(e^{2\pi i n/{N_\tau}}-1)+\beta\omega }}
\,.
\end{align}
\subsubsection{Coupling renormalization\label{sec:coupling_renormalization}}
As outlined in section \sect{physics_bose}, the modeling of the inter-atomic interaction in Bose-Einstein condensates by a Dirac delta function causes ultra-violet divergences in the quantum field theoretic computations. These require a renormalization of the coupling constant, which under certain circumstances becomes relevant also for numerical path integral computations on the lattice. What is plugged into these numerical path integral simulations is always the ``bare'' coupling constant $g$. The actual physical quantity that determines the macroscopic behavior of the system~\footnote{E.g. the ground state energy density $\mathcal{E}_0$ as a function of particle density $\rho$ in the limit of vanishing diluteness is given by $\mathcal{E}_0=\frac{1}{2}g_R \rho^2$ and \textit{not} by $\frac{1}{2}g \rho^2$.} is instead the s-wave scattering length $a$, which is defined as the negative of the scattering amplitude between two atoms in vacuum in the limit of zero momentum. Conveniently one introduces also the renormalized coupling constant $g_R\equiv4\pi a/m$. Recall from \sect{physics_bose} that the scattering length is related to the bare coupling as 
\begin{align}
\label{eq:Born2}
\frac{4\pi a}{m}\equiv g_R=g-mg^2\int \frac{d^3p}{(2\pi)^3}\frac{1}{p^2}+\mathcal{O}(g^3)\,.
\end{align}
On a lattice, one introduces a momentum cutoff $\Lambda=\pi a_\mathrm{s}^{-1}$ (or $\Lambda=2 a_\mathrm{s}^{-1}$ if a finite-difference discretization is chosen), such that the second term in \eq{Born2} takes a finite value:
 \begin{align}
 \label{eq:Born3}
 g_R=g-mg^2\frac{1}{2\pi^2}\int \limits_0^\Lambda dp\,\frac{p^2}{p^2}+\mathcal{O}(g^3)=g-\frac{m}{2\pi^2}g^2\Lambda+\mathcal{O}(g^3)\,.
 \end{align}
Note that we have here neglected that a cubic lattice does not introduce a cutoff on the total momentum but rather on the single components thereof (cf. the discussion below). Inserting $\Lambda=\pi a_\mathrm{s}^{-1}$ we obtain for the relative deviation between renormalized and bare coupling:
\begin{align}
\label{eq:diff_renorm_bare_coupling}
\frac{g-g_R}{g}=2\frac{a}{a_\mathrm{s}}+\mathcal{O}(g^2)\,.
\end{align}
We thus see that it depends on the ratio between the scattering length $a$ and the lattice spacing $a_\mathrm{s}$ (as well as the desired accuracy) whether the renormalization of the coupling must be taken into account when comparing numerical path integral computations to analytical or experimental results (or when comparing two simulations with different cutoff among each other). As long as the ratio $a/a_\mathrm{s}$ is sufficiently small, one need not bother about coupling renormalization and can take the bare coupling to be equal to the renormalized one. However, in practice it is not always possible to make this ratio arbitrarily small because the computational cutoff must also be chosen such that relevant physical momentum scales, e.g. the healing momentum $p_\mathrm{h}=\sqrt{2mg\rho}$, are properly resolved. The ratio between inverse scattering length and healing momentum $a^{-1}/p_\mathrm{h}$ is set by the diluteness $\eta=\sqrt{\rho a^3}$, $a^{-1}/p_\mathrm{h}=\eta^{-1}/\sqrt{8\pi}$. For small diluteness of the order of $\sim10^{-3}$, which is both necessary in experimental settings and for the contact interaction approximation to be valid, there is enough room for placing the cutoff $\Lambda$ both substantially apart from $p_\mathrm{h}$ and from $a^{-1}$, but still the effect of the coupling renormalization is not always negligible.

In \eq{Born3} we have made several approximations: On a computational lattice, momenta are not continuous but discrete; it is not the total momentum but rather its single components that are restricted by a cutoff (i.e. we have discarded the ``corners'' of the momentum lattice); for the case of a finite-difference discretization of the Laplacian, the sine-spacing of the momenta was neglected. If even higher precision is required, it is straightforward to evaluate \eq{Born2} in an exact manner by computing the following sum numerically:
\begin{align}
g_R=g-mg^2\sum_\mathbf{n}\frac{1}{\mathbf{p}_\mathbf{n}^2}+\mathcal{O}(g^3)
\end{align}
with $\mathbf{n}$ numbering the discrete momentum modes and $\mathbf{p}_\mathbf{n}$ defined as in \eq{momentum_sd} (for the spectral derivative discretization of the Laplacian) or \eq{momentum_fd} (for the finite difference discretization).

As we recall from \sect{physics_bose}, in two-spatial dimensions, it is impossible to define a renormalized coupling in the macroscopic limit as in the three-dimensional case. Instead, we have to specify a momentum scale $\Lambda_0$ at which we define the renormalized coupling. Once we have chosen such a scale, it is straightforward to compute the renormalized coupling $g_R$ from the bare one $g$ that has been plugged into the simulation by inserting $\Lambda=\pi a_\mathrm{s}^{-1}$ in \eq{renormalization_general}. 

The considerations so far applied to a simulation setup on a discrete spatial lattice, but with the imaginary-time discretization taken to zero, $a_\tau\to 0$. In practice, we have to retain a finite $a_\tau$ of course. As it turns out, a finite $a_\tau$ yields corrections to the continuous-imaginary-time formula \eq{Born2}. While one could make $a_\tau$ sufficiently small to reach convergence to the $a_\tau\to 0$ limit, this can often be quite expensive numerically and it is helpful to dispose of a formula for the scattering length for a discretized imaginary time. In order to derive such a formula, we have to go through the quantum field theoretic derivation for the Born series of the scattering length. The following presentation relies on concepts and techniques from quantum field theory such as propagators, perturbation theory, scattering theory and the LSZ theorem that have not been introduced here but can be found in standard text books such as \cite{altland2010condensed} and \cite{peskin1995an}.

In field theory, the scattering length $a$ is given by the negative of the vacuum (i.e. at $T=\mu=0$) scattering amplitude $f$ of two particles in the limit of vanishing momentum, i.e. $a=-\lim_{\mathbf{k},\mathbf{k}'\to 0}f(\mathbf{k},\mathbf{k}')$. Via the LSZ reduction theorem, $f$ can be related to correlation functions of the fields with truncated external propagators. The scattering length $a$ can thus be represented by Feynmann diagrams as:
\begin{align}
a=\begin{tikzpicture}[baseline={([yshift=-0.75ex]current bounding box.center)},vertex/.style={anchor=base,
	circle,fill=black!25,minimum size=18pt,inner sep=2pt}]
\draw[black,fill=black] (0,0) circle (0.5ex);
\draw[black] (0,0) -- (0.2,0.2);
\draw[black] (0,0) -- (-0.2,0.2);
\draw[black] (0,0) -- (0.2,-0.2);
\draw[black] (0,0) -- (-0.2,-0.2);
\end{tikzpicture}
\quad+\quad
\begin{tikzpicture}[baseline={([yshift=-0.75ex]current bounding box.center)},vertex/.style={anchor=base,
	circle,fill=black!25,minimum size=18pt,inner sep=2pt}]
\draw[black,fill=black] (0,0) circle (0.5ex);
\draw[black,fill=black] (1,0) circle (0.5ex);
\draw[black] (0,0) -- (-0.2,-0.2);
\draw[black] (0,0) -- (-0.2,0.2);
\draw[black] (1,0) -- (1.2,-0.2);
\draw[black] (1,0) -- (1.2,0.2);
\draw[black](0,0) .. controls (0.3,0.3) and (0.7,0.3) .. (1,0);
\draw[black](0,0) .. controls (0.3,-0.3) and (0.7,-0.3) .. (1,0);
\end{tikzpicture}
\quad+ \dots
\end{align}
The leading-order expression for the scattering length $a^\text{(LO)}$ is simply given by 
\begin{align}
\frac{4\pi a^\text{(LO)}}{m}=V_0\,.
\end{align}
For the next-to-leading order $a^\text{(NLO)}$ we have
\begin{align}
\label{eq:NLO}
\frac{4\pi a^\text{(NLO)}}{m}=-\int \frac{d\omega}{2\pi}\int \frac{d^3p}{(2\pi)^3}\,V_{\mathbf{p}}V_{-\mathbf{p}}\,G(\omega,\mathbf{p})G(-\omega,-\mathbf{p})\,.
\end{align}
For a continuum imaginary time, $a_\tau\to 0$, the propagator $G(\omega,\mathbf{p})$ reads
\begin{align}
G(\omega,\mathbf{p})=\frac{1}{i\omega+\frac{\mathbf{p}^2}{2m}} 
\end{align}
and the integral over $\omega$ runs from $-\infty$ to $+\infty$. For a finite $a_\tau$, we obtain a discretized action and thus the propagator reads
\begin{align}
G^\text{disc}(\omega,\mathbf{p})=\frac{1}{a_\tau^{-1}(e^{ia_\tau \omega}-1)+\frac{\mathbf{p}^2}{2m}}\,,
\end{align}
and the integral over $\omega$ runs from $-\pi/a_\tau$ to $+\pi/a_\tau$. 

Let us evaluate the frequency integral in \eq{NLO} for the discretized case:
\begin{align}
\nonumber&\int \limits_{-\pi/a_\tau}^{+\pi/a_\tau}\frac{d\omega}{2\pi}\,G(\omega,\mathbf{p})^\text{disc}G(-\omega,-\mathbf{p})^\text{disc}\\\nonumber
=&\int \limits_{-\pi/a_\tau}^{+\pi/a_\tau}\frac{d\omega}{2\pi}\,\frac{1}{a_\tau^{-1}(e^{ia_\tau \omega}-1)+\frac{\mathbf{p}^2}{2m}}
\frac{1}{a_\tau^{-1}(e^{-ia_\tau \omega}-1)+\frac{\mathbf{p}^2}{2m}}\\\nonumber
=&\frac{2m}{\mathbf{p}^2}\int \limits_{-\pi}^{+\pi}\frac{dx}{2\pi}\frac{1}{2\left[1-\left(a_\tau\frac{\mathbf{p}^2}{2m}\right)^{-1}\right]\cos(x)+2\left(a_\tau\frac{\mathbf{p}^2}{2m}\right)^{-1}-2+a_\tau\frac{\mathbf{p}^2}{2m}}\\
=&\frac{m}{\mathbf{p}^2}\frac{1}{1-a_\tau\frac{\mathbf{p}^2}{4m}}\,.
\end{align}
Here, we have implicitly assumed that $a_\tau\frac{\mathbf{p}^2}{4m}<1$. Inserting this result into \eq{NLO}, we obtain for the next-to-leading contribution to the scattering length for discretized imaginary time 
\begin{align}
\label{eq:NLO_disc}
\frac{4\pi a^\text{(NLO),disc}}{m}=-m\int \frac{d^3p}{(2\pi)^3}\,\frac{V_{\mathbf{p}}V_{-\mathbf{p}}}{\mathbf{p}^2}\frac{1}{1-a_\tau\frac{\mathbf{p}^2}{4m}}\,.
\end{align}
Specifying to $V_{\mathbf{p}}=g$, we thus find as modification of \eq{Born2} for finite $a_\tau$:
\begin{align}
\label{eq:Born_mod}
\frac{4\pi a}{m}\equiv g_R=g-mg^2\int \frac{d^3p}{(2\pi)^3}\frac{1}{p^2}\frac{1}{1-a_\tau\frac{p^2}{4m}}+\mathcal{O}(g^3)\,.
\end{align}
For fixed momentum cutoff and $a_\tau\to 0$, this expression goes over to \eq{Born2}, as it has to. Whether this finite-$a_\tau$ correction should be taken into account in the computation of the renormalized coupling depends on the ratio $a_\tau/a_\mathrm{s}$, on the ratio $a/a_\mathrm{s}$ and, again, on the desired accuracy. 

\subsection{Approximate descriptions\label{sec:approx_desc}}
This section reviews some of the most important approximate descriptions for interacting Bose gases in equilibrium that will later serve as benchmarks for the fully exact numerical path integral simulations. Extensive reviews of these approximate descriptions can also be found e.g. in \cite{dalfovo1999theory,pitaevskii2016bose}.
\subsubsection{Mean field (saddle point of classical action)}
The simplest approximate description of the interacting Bose gas is the mean-field approach~\footnote{The term ``mean field'' is one of the least consistently used expressions in theoretical physics. In this thesis we will follow the convention to restrict it to the saddle point approximation outlined here.}, which consists in computing observables from the saddle point of the classical action $S[\psi^*,\psi]$, i.e. we determine $\psi$ such that
\begin{align}
\frac{\delta S}{\delta \psi}=0\,.
\end{align}
This requires $\psi$ to fulfill the Gross-Pitaevskii equation as the classical equation of motion for the field $\psi$, either in imaginary time in thermal equilbrium or in real time in non-equilibrium. However, there is an important difference between thermal equilibrium and non-equilibrium. In non-equilbrium, the initial condition for the field $\psi$ at $t=0$ is specified, such that there can only be one single saddle point fulfilling this boundary condition. In contrast, thermal quantum field theory does not impose a Dirichlet boundary condition on the field $\psi$ but only requires it to be periodic in imaginary time. As a consequence, there will typically be multiple saddle points of the classical action. We will here understand by ``mean field'' in thermal equilibrium the saddle point with the smallest real part, which contributes the most to the partition function. For simple scenarios, it is typically straightforward to find this global minimum by a steepest-descent method, i.e. by evolving
\begin{align}
\frac{\partial\psi}{\partial\vartheta}=-\frac{\delta S}{\delta\psi^*}\,
\end{align}
until convergence is reached. However, there are also more complicated scenarios, such as dipolar atoms in a trap, for which such a steepest-descent algorithm can get trapped in local minima \cite{roccuzzo2019supersolid}. For a homogeneous gas, the prediction of the mean-field approximation in thermal equilibrium is almost trivial, as it predicts
\begin{align} 
\rho=\frac{\mu}{g}=\text{const}\,.
\end{align}
In the presence of an external trapping potential, the mean-field approximation leads to a non-trivial density distribution $\rho(\mathbf{r})$, which for sufficiently low temperatures is in approximate agreement with the real one, cf. chapter \sect{trap}. It neglects, however, all effects of finite temperature and quantum fluctuations, nor is it able to capture the elementary excitations of a Bose gas.  

There is a further crucial difference between the stationary action approximation in thermal equilibrium and non-equilibrium. In non-equilibrium, demanding the classical action to be stationary and imposing the initial condition as boundary condition at $t=0$ leads directly to the equation of motion of the corresponding classical field theory, which in the case of non-relativistic bosons is the Gross-Pitaevskii equation. In thermal equilibrium instead, the saddle point with the smallest real part is the \textit{minimum} of the classical energy functional. This is not equivalent to classical thermal field theory at finite temperature: Also the classical thermal partition function is defined as a sum over \textit{all} configurations of the classical fields, not just the local minima of the energy. The stationary action approximation corresponds, however, to classical field theory at zero temperature. This is summarized in the following table: 

\vspace{0.5cm}

\noindent
\begin{tabu}{|c|[1.5pt]c|c|}
	\hline
	&Minimal action approximation&Classical approximation\\\tabucline[1.5pt]{-}
	Thermal equilibrium& Thermal classical field theory at $T=0$&Thermal classical field theory\\\hline
	Real-time&\multicolumn{2}{c|}{Gross-Pitaevskii equation}\\\hline
\end{tabu}

\vspace{0.5cm}

\noindent Thermal classical field theory at non-zero temperature will be the subject of section \sect{class_field_theory}.

\subsubsection{Hartree-Fock approximation\label{sec:HF}}
For the Bose gas in the thermal, non-condensed phase, a simple but accurate description of the interacting Bose gas is provided by the Hartree-Fock (HF) approximation. The idea is to replace the interaction term in the Hamiltonian by a quadratic approximation
\begin{align}
\label{eq:HF}
\left(\psi^{\dagger}\psi\right)^2
\to 4\left\langle \psi^{\dagger}\psi\right\rangle \psi^{\dagger}\psi+\left(\left\langle \psi^{\dagger}\psi^{\dagger}\right\rangle \psi\psi+\mathrm{h.c.}\right)
\,.
\end{align}
Above the transition, the anomalous averages $\left\langle \psi^{\dagger}\psi^{\dagger}\right\rangle$ and $\left\langle \psi\psi\right\rangle$ can be assumed to vanish. Then the Hamiltonian becomes of the form of a free Hamiltonian, albeit with a shifted effective chemical potential
\begin{align}
\label{eq:mueffHF}
\mu'=\mu-2g\left\langle \psi^{\dagger}\psi\right\rangle=\mu-2g\rho\,.
\end{align}
In this approximation, the density $\rho$ must be determined self-consistently from the equation
\begin{align}
\rho=\int\frac{\mathrm{d}^{d}k}{(2\pi)^{d}}\,\frac{1}{\exp\left(\beta\left[\mathbf{k}^2/2m-\mu+2g\rho\right]\right)-1}\,.
\end{align}
If a small system is considered where finite-size effects play a role, the momentum integral may be replaced by a momentum sum that then has to be evaluated numerically. The integral version can be computed analytically in all three dimensions and yields the self-consistency equation
\begin{align}
\label{eq:HF_self_cons}
\rho=\lambda_{T}^{-d}\,g_{d/2}\left(e^{\,\beta(\mu-2g\rho)}\right)
\end{align}
with $d$ the dimension, $\lambda_T=\sqrt{2\pi/mT}$ the thermal de Broglie wave length and $g_s(z)$ the polylogarithm. For the special case of $d=2$, the polylogarithm reduces to an ordinary logarithm, $g_1(z)=-\ln(1-z)$, and the equation can be expressed as:
\begin{align}
\rho=-\lambda_{T}^{-2}\,\ln\left(1-e^{\,\beta(\mu-2g\rho)}\right)\qquad\text{(2D)}\,.
\end{align}
The numerical solution of \eq{HF_self_cons} is straightforward. Once we have solved for $\rho$, we can determine the effective chemical potential $\mu'$, from which the momentum spectrum follows, which is just a free Bose-Einstein distribution with $\mu'$ replacing $\mu$, i.e. $f(\mathbf{k})=f_\mathrm{BE}(\mathbf{k},\mu')$. 

We will show in chapter \sect{3D} that for the three-dimensional, non-condensed Bose gas the Hartree-Fock approximation provides an excellent description of the momentum spectrum. It must be noted, however, that the HF approximation as a perturbative approximation cannot capture the non-perturbative interaction effects that lead to a shift of the critical temperature: In fact, the critical temperature $T_c$ is predicted to have the same value $T_c=2\pi \rho^{2/3}/(m\zeta(3/2)^{2/3})$ as in the non-interacting case, while it is known from more sophisticated theoretical studies that they actually differ \cite{andersen2004theory}, cf. also chapter \sect{3D}.

In principle, one can also extend the Hartree-Fock approximation into the condensed phase. The left-hand side of \eq{HF_self_cons} must then be replaced by the thermal part of the density, $\rho_\mathrm{th}$. The full density $\rho$ instead is the sum of the thermal density and the condensate density $\rho_0$, $\rho=\rho_0+\rho_\mathrm{th}$. By demanding the derivative of the free energy by the condensate density to vanish one can show that $\mu=g(\rho_0+2\rho_\mathrm{th})$, cf. section \sect{popov} about the Popov approximation. Hence we obtain for the condensed phase the self-consistency equation
\begin{align}
\label{eq:HF_self_cons2}
\frac{\mu}{g}-\rho=\lambda_{T}^{-d}\,g_{d/2}\left(e^{\,\beta(\mu-2g\rho)}\right)\,.
\end{align}
It must be noted, however, that the Hartree-Fock approximation is not very useful in the condensed phase, because the very assumption that the anomalous averages vanish is not satisfied any more below the transition point. In the condensed phase, they become of crucial importance and lead to the modification of the free dispersion $\varepsilon(\mathbf{k})=\mathbf{k}^2/2m$ to the Bogoliubov dispersion $\omega(\mathbf{k})=\sqrt{\varepsilon(\mathbf{k})(\varepsilon(\mathbf{k})+2g\rho_0)}$, cf. the next section \sect{Bog_theo}. Hence, the Hartree-Fock approximation for the condensate phase can at most provide a useful description not too far below the transition point.

\subsubsection{Bogoliubov theory\label{sec:Bog_theo}}
Arguably the most important approximate description of the weakly interacting Bose gas in the condensed phase, which allows to compute accurately the elementary excitations and momentum spectra, is Bogoliubov theory. The idea is to assume that the vast majority of particles occupies the zero-momentum mode, such that it seems reasonable to replace the operator $a_0$ by the number of particles in this mode, $a_0\to N_0\equiv \rho_0\mathcal{V}$ with $\rho_0$ the condensate density and $\mathcal{V}$ the volume, and expand in the operators of the other modes, $a_{\mathbf{k}\neq 0}$, up to second order. The generalized Hamiltonian, $K\equiv H-\mu N$, of a one-component Bose gas with contact interaction reads in momentum space
\begin{align}
K\equiv H-\mu N=
\sum_\mathbf{k}\left(\frac{\mathbf{k}^2}{2m}-\mu\right)a^\dagger_{\mathbf{k}}a_{\mathbf{k}}
+\frac{g}{2\mathcal{V}}\sum_{\mathbf{k}\mathbf{k}'\mathbf{q}}
a^\dagger_{\mathbf{k}+\mathbf{q}}a^\dagger_{\mathbf{k}'-\mathbf{q}}
a_{\mathbf{k}}a_{\mathbf{k}'}
\,.
\end{align}
If we make the approximation described above, we obtain
\begin{align}
\label{eq:Bog_expanded}
K=
\left(\frac{g}{2}\rho_{0}-\mu\right) \mathcal{V}\rho_0
+\sum_{\mathbf{k}\neq 0}\left(\frac{\mathbf{k}^2}{2m}+ 2g\rho_0-\mu\right)
a^\dagger_{\mathbf{k}}a_{\mathbf{k}}
+\frac{g\rho_0}{2}\sum_{\mathbf{k}\neq 0}
\left(a_{\mathbf{k}}a_{-\mathbf{k}}
+a^\dagger_{\mathbf{k}}a^\dagger_{-\mathbf{k}}\right)
+\mathcal{O}(a_{\mathbf{k}}^3)
\,.
\end{align}
If we discard even the terms containing two operators, $K$ becomes just a number, $K=\mathcal{V}(g\rho_0^2/2-\mu\rho_0)$. From demanding that $\partial \langle K\rangle/\partial\rho_0=0$, we obtain in lowest order that
\begin{align}
\label{eq:mu}
\mu=g\rho_0\,.
\end{align}
The Hamiltonian expanded up to second order \eq{Bog_expanded} is still not diagonal. It can be diagonalized by means of the Bogoliubov transform
\begin{align} 
\label{eq:Bog_traf}
a_{\mathbf{k}} = u_{\mathbf{k}}b_{\mathbf{k}}-v_{\mathbf{k}}b^{\dagger}_{-\mathbf{k}}
\end{align}
where
\begin{align}
\label{eq:bogol_v1}
u_{\mathbf{k}}
&=\sqrt{\frac{1}{2}\left[\frac{\varepsilon(\mathbf{k})+2g\rho_0-\mu}{\omega(\mathbf{k})}+1\right]}\\
\label{eq:bogol_u1}
v_{\mathbf{k}}
&=\sqrt{\frac{1}{2}\left[\frac{\varepsilon(\mathbf{k})+2g\rho_0-\mu}{\omega(\mathbf{k})}-1\right]}
\,,
\end{align}
and 
\begin{align}
\label{eq:omegaBog1}
\omega(\mathbf{k})
&=\sqrt{\left[\varepsilon(\mathbf{k})+g\rho_0-\mu\right]\left[\varepsilon(\mathbf{k})+3g\rho_0-\mu\right]}
\\
\varepsilon(\mathbf{k})
&=\frac{\mathbf{k}^2}{2m}
\,.
\end{align}
Upon inserting \eq{mu} in \eq{bogol_v1}, \eq{bogol_u1} and \eq{omegaBog1}, we obtain
\begin{align}
\label{eq:bogol_v2}
u_{\mathbf{k}}
&=\sqrt{\frac{1}{2}\left[\frac{\varepsilon(\mathbf{k})+g\rho_0}{\omega(\mathbf{k})}+1\right]}\\
\label{eq:bogol_u2}
v_{\mathbf{k}}
&=\sqrt{\frac{1}{2}\left[\frac{\varepsilon(\mathbf{k})+g\rho_0}{\omega(\mathbf{k})}-1\right]}
\\
\label{eq:omegaBog2}
\omega(\mathbf{k})
&=\sqrt{\varepsilon(\mathbf{k})\left[\varepsilon(\mathbf{k})+2g\rho_0\right]}\,.
\end{align}
The generalized Hamiltonian becomes
\begin{align}
\label{eq:K_Bog}
K=K_0 + \sum_{\mathbf{k}\neq 0}\omega({\mathbf{k}}) b^{\dagger}_{\mathbf{k}} b_{\mathbf{k}}\,,
\end{align} 
where the generalized ground state energy reads
\begin{align}
\label{eq:K0}
K_0=-\mu\mathcal{V}\rho_0+\frac{\mathcal{V}}{2}g\rho_0^2+\frac{1}{2}\sum_{\mathbf{k}\neq 0}\left[\omega(\mathbf{k})-\varepsilon(\mathbf{k})-g\rho_0\right]\,.
\end{align}
Note that in the $\mathbf{k}=0$ contribution to \eq{K0} we have retained $\mu$ and not inserted \eq{mu}. Namely, the higher-order corrections to 
\eq{mu} give contributions to the $\mathbf{k}=0$ contribution in \eq{K0} that are of the same order of magnitude as the momentum sum and therefore cannot be neglected.

The Bogoliubov dispersion relation \eq{omegaBog2} takes at small $\mathbf{k}$ the form of a sound wave dispersion, $\omega(k)\sim ck$ with the speed of sound $c=\sqrt{g\rho_0/m}$. Indeed it is possible to show that Bogoliubov quasi-particles created by $b^{\dagger}_{\mathbf{k}}$ correspond to waves in the density of particles. At larger momenta, the dispersion goes over into that of a non-interacting Bose gas instead, albeit shifted by $g\rho_0$. The momentum scale at which the behavior of the dispersion changes from linear to quadratic is the \textit{healing momentum} $k_\mathrm{h}=\xi_\mathrm{h}^{-1}=\sqrt{2mg\rho_0}$, which sets a characteristic momentum scale in an interacting Bose gas.  

At finite temperature $T\equiv1/\beta>0$, the occupation numbers of Bogoliubov quasi-particles follow a Bose-Einstein distribution. From this one can then derive, via the transformation \eq{Bog_traf}, the particle occupation numbers $f(\mathbf{k})=\langle a^\dagger_\mathbf{k} a_\mathbf{k}\rangle$ for $\mathbf{k}\neq 0$:
\begin{align}
\label{eq:bogol_spectrum1}
f(\mathbf{k})
&=\frac{1+2v_{\mathbf{k}}^{2}}{e^{\,\beta\omega(\mathbf{k})}-1}
+v_{\mathbf{k}}^{2}
\,.
\end{align}
The term $v_{\mathbf{k}}^{2}$ arises from commuting a particle creation operator past an annihilation operator and can thus be considered a genuine quantum contribution to the particle occupation number. This leads to a depletion of the condensate mode even at $T=0$, an effect that is absent from a non-interacting Bose gas where all particles condense at zero temperature. One can calculate the condensate depletion by integrating over $v_{\mathbf{k}}^{2}$:
\begin{align}
\delta\rho(T=0)\equiv \rho(T=0)-\rho_0(T=0)=\int\frac{d^3k}{(2\pi)^3} v_{\mathbf{k}}^{2}=\frac{1}{3\pi^2}(mg \rho_0)^{3/2}\,.
\end{align}
The relative depletion $\delta\rho/\rho_0$ can be expressed in terms of the dimensionless diluteness $\eta=\sqrt{\rho a^3}$ with $a=mg/4\pi$ as $\delta\rho/\rho_0=8/(3\sqrt{\pi})\eta+\mathcal{O}(\eta^2)$. Let us now have a closer look at the ground state energy $E_0= K_0+\mu\rho\mathcal{V}$. First one notes that it is possible to replace $\rho_0$ by $\rho$ in \eq{K0} because the terms of order $\delta \rho$ that arise from this replacement cancel each other in the $\mathbf{k}=0$ contribution and the $\mathbf{k}\neq0$ contribution is already of order $\mathcal{O}(\delta\rho)$ such that the replacement $\rho_0\to\rho$ in the momentum sum generates only terms of order $\mathcal{O}(\delta\rho^2)$. I.e. we have 
\begin{align}
\label{eq:E0}
E_0=\frac{\mathcal{V}}{2}g\rho^2+\frac{1}{2}\sum_{\mathbf{k}\neq 0}\left[\omega(\mathbf{k})-\varepsilon(\mathbf{k})-g\rho\right]\,.
\end{align}
This expression turns out to be UV-divergent, which is an artifact of the contact potential. The divergence can be removed by expressing $g$ in terms of the renormalized coupling $g_R=4\pi a/m$, cf. \eq{Born4}. This yields
\begin{align}
\label{eq:E02}
\nonumber E_0&=\frac{\mathcal{V}}{2}g_R\rho^2+\frac{1}{2}\sum_{\mathbf{k}\neq 0}\left[\omega(\mathbf{k})-\varepsilon(\mathbf{k})-g_R\rho+\frac{m g_R^2\rho^2}{\mathbf{k}^2}\right]\\
&\simeq \frac{\mathcal{V}}{2}g_R\rho^2+\frac{\mathcal{V}}{2}\int\frac{d^3k}{(2\pi)^3}\left[\omega(\mathbf{k})-\varepsilon(\mathbf{k})-g_R\rho+\frac{m g_R^2\rho^2}{\mathbf{k}^2}\right]\\
&=\frac{\mathcal{V}}{2}g_R\rho^2\left[1+\frac{128}{15\sqrt{\pi}}\eta\right]\,.
\end{align}
The second term in the brackets is a correction to the mean-field ground state energy $\mathcal{V}g_R\rho^2/2$, which is known as Lee-Huang-Yang (LHY) correction. It plays a crucial role in the treatment of dipolar Bose gases beyond the point of classical instability, cf. chapter \sect{dipolars}.

Let us now discuss the generalization of Bogoliubov theory to the $\mathcal{N}$-component Bose gas with $U(\mathcal{N})$-symmetric interactions. Here, the generalized Hamiltonian reads:
\begin{align}
\label{eq:K_UN}
K\equiv H-\mu N=
\sum_\mathbf{k}\sum_\alpha\left(\frac{\mathbf{k}^2}{2m}-\mu\right)a^\dagger_{\mathbf{k},\alpha}a_{\mathbf{k},\alpha}
+\frac{g}{2\mathcal{V}}\sum_{\mathbf{k}\mathbf{k}'\mathbf{q}}\sum_{\alpha\alpha'}
a^\dagger_{\mathbf{k}+\mathbf{q},\alpha}a^\dagger_{\mathbf{k}'-\mathbf{q},\alpha'}
a_{\mathbf{k},\alpha}a_{\mathbf{k}',\alpha'}
\,,
\end{align}
where $\alpha,\alpha'=1,\dots,\mathcal{N}$ enumerate the different field components. 
Expanding in complete analogy to the one-component case to quadratic order in the non-zero mode operators, we obtain
\begin{align}
\label{eq:coupled}
K=
&\left(\frac{\mathcal{N}g}{2}\rho_{0}-\mu\right)\mathcal{N} \mathcal{V}\rho_0
+\sum_{\mathbf{k}\neq 0}\sum_\alpha\left(\frac{\mathbf{k}^2}{2m}+\mathcal{N} g\rho_0-\mu\right)
a^\dagger_{\mathbf{k},\alpha}a_{\mathbf{k},\alpha}
\nonumber\\
&+g\rho_0\sum_{\mathbf{k}\neq 0}\sum_{\alpha\alpha'}
\left[a^\dagger_{\mathbf{k},\alpha}a_{\mathbf{k},\alpha'}
+\frac12\left(a_{\mathbf{k},\alpha}a_{-\mathbf{k},\alpha'}
+a^\dagger_{\mathbf{k},\alpha}a^\dagger_{-\mathbf{k},\alpha'}\right)
\right]+\mathcal{O}(a_{\mathbf{k},\alpha}^3)
\,,
\end{align}
where $\rho_0$ is the condensate density in a single component, such that the total condensate density is $\mathcal{N}\rho_0$. 

In order to diagonalize in the component degree of freedom, we introduce
\begin{align}
\label{eq:trafo1}
B_{\mathbf{k}}
&=\frac{1}{\sqrt{\mathcal{N}}}\sum_\alpha a_{\mathbf{k},\alpha}
\\
\label{eq:trafo2}
b_{\mathbf{k},\beta}
&=\sum_\alpha w_\beta^\alpha a_{\mathbf{k},\alpha}
\qquad \beta=1,\dots,\mathcal{N}-1
\,,
\end{align}
where the vectors $\mathbf{w}_\beta$ form an arbitrary orthonormal basis of the $\mathcal{N}-1$ dimensional orthogonal complement of the $\mathcal{N}$-component vector $(1,1,\dots,1)^T/\sqrt{\mathcal{N}}$. 
Thereby we obtain
\begin{align}
\label{eq:decoupled}
K=
&\left(\frac{\mathcal{N}g}{2}\rho_{0}-\mu\right)\mathcal{N} \mathcal{V}\rho_0
+\sum_{\mathbf{k}\neq 0}\left(\frac{\mathbf{k}^2}{2m}+2\mathcal{N} g\rho_0-\mu\right)
B^\dagger_{\mathbf{k}}B_{\mathbf{k}}
\nonumber\\
&+\frac{\mathcal{N}g\rho_0}{2}\sum_{\mathbf{k}\neq 0}\left(
B_{\mathbf{k}}B_{-\mathbf{k}}+B^\dagger_{\mathbf{k}}B^\dagger_{-\mathbf{k}}
\right)
+\sum_{\mathbf{k}\neq 0}\sum_{\beta}\left(\frac{\mathbf{k}^2}{2m}+\mathcal{N} g\rho_0-\mu\right)
b^\dagger_{\mathbf{k},\beta}b_{\mathbf{k},\beta}
+\mathcal{O}(a_{\mathbf{k},\alpha}^3)
\,.
\end{align}
The equivalence of \eq{coupled} and \eq{decoupled} can be easily seen by plugging the transformation \eq{trafo1} and \eq{trafo2} into \eq{decoupled} and exploiting the fact that $(1,1,\dots,1)^T/\sqrt{\mathcal{N}}$ and the $\mathbf{w}_{\beta}$ form an orthonormal set.
In the leading-order mean-field approximation the energy is minimized at the chemical potential $\mu=\mathcal{N}g\rho_0$. 
Upon inserting this, \eq{decoupled} becomes
\begin{align}
K=&
-\frac{g}{2}\mathcal{N}^2\rho_0^2\mathcal{V}
+\sum_{\mathbf{k}\neq 0}\left(\frac{\mathbf{k}^2}{2m}+\mathcal{N} g\rho_0\right)
B^\dagger_{\mathbf{k}}B_{\mathbf{k}}\\\nonumber&
+\frac{\mathcal{N}g\rho_0}{2}\sum_{\mathbf{k}\neq 0}\left(
B_{\mathbf{k}}B_{-\mathbf{k}}+B^\dagger_{\mathbf{k}}B^\dagger_{-\mathbf{k}}
\right)+\sum_{\mathbf{k}\neq 0}\sum_{\beta}\frac{\mathbf{k}^2}{2m}
b^\dagger_{\mathbf{k},\beta}b_{\mathbf{k},\beta}
+\mathcal{O}(a_{\mathbf{k},\alpha}^3)
\,,
\end{align}
which describes Bogoliubov modes $B_\mathbf{k}$ and $\mathcal{N}-1$ free Goldstone excitations $b_{\mathbf{k},\beta}$. 
The total occupation number $f(\mathbf{k})=\sum_a\langle a^\dagger_{\mathbf{k},\alpha}a_{\mathbf{k},\alpha}\rangle$ for $\mathbf{k}\neq 0$ can then be written as 
\begin{align}
f(\mathbf{k})
=&\sum_\alpha \langle a^\dagger_{\mathbf{k},\alpha}a_{\mathbf{k},\alpha}\rangle
=\langle B^\dagger_\mathbf{k} B_\mathbf{k}\rangle +\sum_\beta \langle b^\dagger_{\mathbf{k},\beta}b_{\mathbf{k},\beta}\rangle
\nonumber\\
=&\ \frac{1}{2}\left[\frac{\varepsilon(\mathbf{k})+\mathcal{N}g\rho_0}{\omega(\mathbf{k})}-1\right]
+\frac{\varepsilon(\mathbf{k})+\mathcal{N}g\rho_0}{\omega(\mathbf{k})(e^{\beta\omega(\mathbf{k})}-1)}
+\frac{\mathcal{N}-1}{e^{\beta\varepsilon(\mathbf{k})}-1}
\,,
\end{align}
with the Bogoliubov dispersion
\begin{align}
\omega(\mathbf{k})
=\sqrt{\varepsilon(\mathbf{k})\left(\varepsilon(\mathbf{k})+2\mathcal{N}g\rho_0\right)}
\,.
\end{align}

\subsubsection{Popov and modified Popov approximation\label{sec:popov}}
In pure Bogoliubov theory, one includes the effect of temperature in a non-self-consistent way, assuming that the approximation $\mu=g\rho_0$ remains valid at $T>0$ and temperature leads only to a thermal population of the Bogoliubov modes, neglecting the back-reaction onto the condensate mode. Such approximation is only valid at very low temperatures when the thermal depletion remains negligible. The Popov approximation aims at including the effect of temperature in a self-consistent manner. Thereby it extends the range of validity of pure Bogoliubov theory to higher temperatures, though it will equally break down once the thermal phase transition (Bose-Einstein or Berezinskii-Kosterlitz-Thouless transition) is approached. 

Consider the free energy $F$ of the system, given by
\begin{align}
F=-T\ln Z_\text{gc}+\mu\mathcal{V}\rho\,,
\end{align}
which for the Bogoliubov Hamiltonian \eq{K_Bog} yields:
\begin{align}
F=-\mu\mathcal{V}\rho_0+\frac{\mathcal{V}}{2}g\rho_0^2+\frac{1}{2}\sum_{\mathbf{k}\neq 0}\left[\omega(\mathbf{k})-\varepsilon(\mathbf{k})-2g\rho_0+\mu\right]+T\sum_{\mathbf{k}\neq 0}\ln\left(1-e^{-\beta\omega(\mathbf{k})}\right)+\mu\mathcal{V}\rho\,.
\end{align}
From demanding that the derivative of $F$ by $\mu$ vanish, $\partial F/\partial \mu=0$, we obtain
\begin{align}
\rho=\rho_0+\frac{1}{\mathcal{V}}\sum_{\mathbf{k}\neq 0}\left[\frac{\varepsilon(\mathbf{k})+2g\rho_0-\mu}{\omega(\mathbf{k})}\left(\frac{1}{e^{\beta\omega(\mathbf{k})}-1}+\frac{1}{2}\right)-\frac{1}{2}\right]
\end{align}
and from demanding the derivative of of $F$ with respect to $\rho_0$ to vanish, $\partial F/\partial \rho_0=0$:
\begin{align}
\mu=g\rho_0+\frac{g}{\mathcal{V}}\sum_{\mathbf{k}\neq 0}\left[\frac{2\varepsilon(\mathbf{k})+3g\rho_0-2\mu}{\omega(\mathbf{k})}\left(\frac{1}{e^{\beta\omega(\mathbf{k})}-1}+\frac{1}{2}\right)-1\right]\,.
\end{align}
For given chemical potential and coupling strength, these equations must then be solved self-consistently for $\rho$ and $\rho_0$.  

For one-dimensional systems and two-dimensional systems at finite temperature, the Popov equations are plagued by infrared divergences. It has therefore been suggested to modify them in such a way as to include phase fluctuations to all orders \cite{andersen2002phase}, thereby making them applicable also to lower dimensions. These modified Popov equations read:
\begin{align}
\label{eq:mod_Popov1}
\rho&=\rho_0+\frac{1}{\mathcal{V}}\sum_{\mathbf{k}\neq 0}\left[\frac{\varepsilon(\mathbf{k})}{\omega(\mathbf{k})}\left(\frac{1}{e^{\beta\omega(\mathbf{k})}-1}+\frac{1}{2}\right)-\frac{1}{2}+\frac{g\rho_0}{2\varepsilon(\mathbf{k})+2\mu}\right]\\
\label{eq:mod_Popov2}
\mu&=g\left(2\rho-\rho_0\right)\,.
\end{align}
While the modified Popov theory is tailored for the low-dimensional case, it can be equally well applied to the three-dimensional Bose gas, in which case it gives only small deviations from the original Popov theory \cite{andersen2002phase}. 

Both the Popov and modified Popov approximation do not make sense any more once $\rho_0$ becomes zero, i.e. in the normal phase. For this regime one has to use the Hartree-Fock approximation instead, cf. section \sect{HF}. However, the latter does not smoothly transition into Popov theory, as both break down close to the thermal phase transition.
\subsubsection{Classical field theory\label{sec:class_field_theory}}
As mentioned in section \ref{sec:constr_pi}, the Berry phase term $\psi^*\partial_\tau\psi$ renders the action of interacting bosons in the field-theoretic representation inaccessible to straightforward Monte Carlo simulations. Furthermore, even if one employs more sophisticated approaches as the complex Langevin method, the computational cost is rather high because a four-dimensional lattice must be simulated. It is thus a very convenient approximation to drop the $\tau$ dependence of the field $\psi$, setting $N_\tau=1$. The thermal action then becomes
\begin{align}
\int\limits_0^\beta d\tau \int d^dx\left\{\psi^*(\tau)\partial_\tau\psi(\tau)+\mathcal{H}\left[\psi^*(\tau),\psi(\tau)\right]\right\}\to \beta\,\int d^dx\, \mathcal{H}\left[\psi^*,\psi\right]
\end{align}
with $\mathcal{H}$ the Hamiltonian density such that the path integral measure of classical field theory $\exp\left(-\beta \int d^dx\,\mathcal{H}\right)$ is purely real and positive. It can then be subjected to ordinary Monte Carlo approaches, such as the worm or the heat bath algorithm \cite{arnold2001bec,kashurnikov2001critical}. This approximation becomes good if the extension of the interval $[0,\beta]$ is in some sense sufficiently small, but in practice it is difficult to establish criteria in which scenario and for which purpose this may be assumed. For a gas of non-interacting (quasi-)particles with dispersion $\omega(\mathbf{k})$, the approximation amounts to replacing the Bose-Einstein distribution by the classical Rayleigh-Jeans distribution
\begin{align}
\frac{1}{\exp\left[\beta(\omega(\mathbf{k})-\mu)\right]-1}\to \frac{1}{\beta(\omega(\mathbf{k})-\mu)}\,,
\end{align}
which is valid for $\beta(\omega(\mathbf{k})-\mu)\ll 1$. In a regime far above the Bose-Einstein condensation, $-\beta\mu\gg 1$, this condition cannot be satisfied for any $\mathbf{k}$, such that the approximation is not applicable here. In fact, the limit of $-\beta\mu\gg 1$ is the limit of classical particles, not of classical fields. But even close to the transition point or in the condensed phase, $\beta(\omega(\mathbf{k})-\mu)$ will become on the order of $1$ at the thermal momentum $k\gtrsim \lambda_T^{-1}=\sqrt{mT/2\pi}$. The reasoning why the classical field approximation remains still useful is that the interesting physics typically happens at small momenta, in particular that interaction effects take place only within the long-wavelength modes below the healing momentum $k_\text{h}=\sqrt{2mg\rho}$. As long as healing momentum and thermal momentum are sufficiently apart from each other, $k_\text{h}\ll\lambda_T^{-1}$, the classical approximation is justified.

However, even in this case one has to deal in some way with the ultraviolet catastrophe of classical field theory. If one assumes $\omega(\mathbf{k})$ to be quadratic at large momenta, the product of integral measure and occupation number behaves as
\begin{align}
\frac{k^{d-1}}{\beta(\omega(\mathbf{k})-\mu)}\sim k^{d-3}\qquad k\to\infty\,,
\end{align}
which in three dimensions yields a linear and in two dimensions a logarithmic divergence of the particle density. Other observables involving higher powers of the momentum display even worse divergence. In order to cure this, one must assume that at large $k$, the gas becomes effectively non-interacting and the occupation number follows a pure Bose-Einstein distribution. Then, one may correct the density produced by Monte Carlo simulations by subtracting the Rayleigh-Jeans distribution and adding the Bose-Einstein distribution:
\begin{align}
\rho_\text{true}=\rho_\text{class}+\frac{1}{\mathcal{V}}\sum_\mathbf{k} \left\{\frac{1}{\exp\left[\beta\omega(\mathbf{k})\right]-1}-\frac{1}{\beta \omega(\mathbf{k})}\right\}\,.
\end{align}
Employing the Hartree-Fock approximation, one can also correct the chemical potential that has to be plugged into the classical simulations as
\begin{align}
\mu_\text{true}=\mu_\text{class}-\frac{2g}{\mathcal{V}}\sum_\mathbf{k} \left\{\frac{1}{\exp\left[\beta\omega(\mathbf{k})\right]-1}-\frac{1}{\beta \omega(\mathbf{k})}\right\}\,.
\end{align}
It should also be noted that genuine quantum effects, such as the quantum depletion of the condensate (cf. section \ref{sec:Bog_theo}), which are typically tiny but not necessarily neglectable in all circumstances, cannot be captured by a purely classical simulation and require the evaluation of the full quantum path integral.

Due to the numerous assumptions that have to be made and the difficulty to estimate the systematic bias introduced by the classical approximation, it is desirable to compare simulations of classical field theory to simulations of the full quantum theory in order to establish the range of validity of the former. This will be one of the subjects of chapter \ref{sec:2D_gas}.  
\clearpage

\thispagestyle{plain}
\section{Sign problem and the complex Langevin method\label{sec:sign_prob_and_cl}}
\subsection{Sign problem\label{sec:sign_prob}}
The way that the exponential complexity of quantum physics appears within its path integral formulation is known as \textit{sign problem}. That is, that it is in general very hard to compute integrals of the type
\begin{align}
\label{eq:pi}
\int d^n\phi \,\exp\left(-S[\boldsymbol{\phi}]\right)\mathcal{O}[\boldsymbol{\phi}]
\end{align}
if the dimension $n$ is sufficiently large and $S$ is a complex quantity. Physical scenarios in which the latter is the case include:
\begin{itemize}
	\item Any non-equilibrium quantum system.
	\item Relativistic theories with non-vanishing chemical potential.
	\item Non-relativistic fermions with spin imbalance or repulsive interactions.
	\item Non-relativistic bosons in second quantization.
\end{itemize}

To see why the evaluation of \eq{pi} is extremely challenging for complex $S$, let us first have a look at how one can compute such integrals efficiently with Monte Carlo methods if only $n$ is large but $S$ is real. The first naive approach would be to randomly draw configurations $\boldsymbol{\phi}$, take the average of the integrand on this sample and hope that it converges to the right value in a reasonable amount of time. However, for actions $S$ that describe physical systems, the vast majority of configurations is highly suppressed by the weight $\exp\left(-S[\boldsymbol{\phi}]\right)$ such that this approach is doomed to fail. Instead, we must find a way to draw the configurations according to the probability distribution $\exp\left(-S[\boldsymbol{\phi}]\right)$ and evaluate the observable $\mathcal{O}[\boldsymbol{\phi}]$ on this sample. To this purpose, there exist indeed numerous \textit{Markov chain} algorithms that construct a sequence of configurations that in the long run are distributed according to the desired probability distribution. The historically earliest and simplest is the \textit{Metropolis algorithm} \cite{metropolis1953equation}. One starts from some arbitrary configuration $\boldsymbol{\phi}_0$. In every step, one proposes a new configuration $\boldsymbol{\phi}'$ close to the previous one $\boldsymbol{\phi}$, e.g. by adding a Gaussian random number on the old configuration; this proposal is always accepted if $S[\boldsymbol{\phi}']>S[\boldsymbol{\phi}]$, otherwise only with a probability $\exp(S[\boldsymbol{\phi}']-S[\boldsymbol{\phi}])$; if the proposal is not accepted, a new proposal is drawn until acceptance. One can show that this algorithm will indeed produce samples distributed according to $\exp\left(-S[\boldsymbol{\phi}]\right)$.

One immediately sees that this approach has a problem once $S$ is not a real quantity any more, as for a complex $S$ the 
weight $\exp\left(-S[\boldsymbol{\phi}]\right)$ becomes complex as well, and both its real and its imaginary part will have both positive and negative domains. Thus, the weight cannot be considered a probability density any more, which is the crucial ingredient of every Markov chain Monte Carlo algorithm. A very straightforward approach to this problem would be to absorb $\exp\left(-\text{Im}S[\boldsymbol{\phi}]\right)$ into the observable and then to sample only with $\exp\left(-\text{Re}S[\boldsymbol{\phi}]\right)$, a strategy that is known as \textit{reweighting}. Namely, one can rewrite the expectation value of $\mathcal{O}$ as 
\begin{align}
\nonumber\langle\mathcal{O}\rangle_S&=\frac{\int d^n\phi \,\exp\left(-S[\boldsymbol{\phi}]\right)\mathcal{O}[\boldsymbol{\phi}]}{\int d^n\phi \,\exp\left(-S[\boldsymbol{\phi}]\right)}\\
\nonumber&=\frac{\int d^n\phi \,\exp\left(-\text{Re}S[\boldsymbol{\phi}]\right)\exp\left(-\text{Im}S[\boldsymbol{\phi}]\right)\mathcal{O}[\boldsymbol{\phi}]}{\int d^n\phi \,\exp\left(-\text{Re}S[\boldsymbol{\phi}]\right)\exp\left(-\text{Im}S[\boldsymbol{\phi}]\right)}
\\
\nonumber&=\frac{\int d^n\phi \,\exp\left(-\text{Re}S[\boldsymbol{\phi}]\right)\exp\left(-\text{Im}S[\boldsymbol{\phi}]\right)\mathcal{O}[\boldsymbol{\phi}]/\int d^n\phi \,\exp\left(-\text{Re}S[\boldsymbol{\phi}]\right)}{\int d^n\phi \,\exp\left(-\text{Re}S[\boldsymbol{\phi}]\right)\exp\left(-\text{Im}S[\boldsymbol{\phi}]\right)/\int d^n\phi \,\exp\left(-\text{Re}S[\boldsymbol{\phi}]\right)}
\\\label{eq:reweighting}
&=\frac{\langle \exp\left(-\text{Im}S[\boldsymbol{\phi}]\right)\mathcal{O}[\boldsymbol{\phi}]\rangle_{\text{Re}S}}{\langle \exp\left(-\text{Im}S[\boldsymbol{\phi}]\right)\rangle_{\text{Re}S}}\,.
\end{align}
Then, one only has to sample $\exp\left(-\text{Im}S[\boldsymbol{\phi}]\right)\mathcal{O}[\boldsymbol{\phi}]$ and $\exp\left(-\text{Im}S[\boldsymbol{\phi}]\right)$ with the probability density $\exp\left(-\text{Re}S[\boldsymbol{\phi}]\right)$ and divide the results. It is immediately clear that this cannot work in general: If $\text{Re} S$ vanishes, this strategy amounts to the naive approach of completely randomly drawing configurations that, as mentioned above, is doomed to fail. However, even if this is not the case, the reweighting approach performs rather poorly. To see why this is the case, consider the denominator of \eq{reweighting}:
\begin{align}
\langle \exp\left(-\text{Im}S[\boldsymbol{\phi}]\right)\rangle_{\text{Re}S}=\frac{\int d^n\phi \,\exp\left(-S[\boldsymbol{\phi}]\right)}{\int d^n\phi \,\exp\left(-\text{Re}S[\boldsymbol{\phi}]\right)}=\frac{Z}{Z_R}=e^{-\beta (\Omega-\Omega_R)}\,,
\end{align}
where we have introduced the partition functions $Z$ and $Z_R$ corresponding to the actions $S$ and $\text{Re}S$, respectively, as well as the corresponding grand-canonical potentials $\Omega$ and $\Omega_R$, defined as $\Omega\equiv -T\ln Z$ and $\Omega_R\equiv -T\ln Z_R$, with $T$ the temperature and $\beta=1/T$ the inverse temperature. The grand-canonical potential is an extensive thermodynamic quantity, i.e. in the thermodynamic limit it scales with the volume of the system, $\Omega,\Omega_R\propto \mathcal{V}$.  Furthermore, $Z_R>Z$ for non-vanishing $\text{Im} S$. Therefore, there exists some constant $c > 0$ (which depends on intensive parameters but not on the volume) such that 
\begin{align}
\langle \exp\left(-\text{Im}S[\boldsymbol{\phi}]\right)\rangle_{\text{Re}S}\sim e^{-c\mathcal{V}}\,.
\end{align}
The \textit{absolute} uncertainty in a Monte Carlo simulation decreases as $1/\sqrt{n}$ with the number $n$ of drawn configurations. That is, in order to maintain a constant \textit{relative} accuracy in evaluating $\langle \exp\left(-\text{Im}S[\boldsymbol{\phi}]\right)\rangle_{\text{Re}S}$, we must draw a number of configurations that increases exponentially with system size. The stronger the fluctuations in the phase of the weight, the more cancellations of positive and negative domains will occur, i.e. the smaller is the resulting value of the integral and hence the more configurations must be drawn. While there are some cases of a mild sign problems and small system sizes where reweighting is actually applicable, it will fail in the majority of cases of interest.

We thus have to come up with some better approach to evaluating high-dimensional integrals with complex weight. The sign problem has been proven to be NP-hard \cite{troyer2005computational}. That means that finding a generic algorithm that could solve the sign problem in every instance would amount to being able to solve in polynomial time every problem in the complexity class NP of problems whose solution can be verified (but not necessarily found) in polynomial time. In general, it is suspected that $\mathrm{P}\neq \mathrm{NP}$, i.e. that NP contains problems whose solution can be verified but not found in polynomial time. Thus, the existence of such an algorithm is  unlikely. This does not exclude that an algorithm can be conceived that solves the sign problem in all or most cases of physical interest, although even this is unlikely, given the exponential complexity inherent to quantum physics. However, as mentioned in the introduction, chapter \sect{intro}, this exponential complexity in many cases is not fully explored by a system, such that the complexity reduction can be exploited by devising an algorithm that solves the sign problem in the particular case. 

Therefore, several methods have been proposed to tackle the sign problem, which can be divided into model-specific and generic methods. Approaches that are, at least to a certain extent, model-specific, include:
\begin{itemize}
	\item Dual variables \cite{endres2007method,gattringer2016approaches}.
	\item Fixed node Monte Carlo \cite{vanbemmel1994fixed,foulkes2001quantum}.
	\item Diagrammatic Monte Carlo \cite{boninsegni2006worm,vanhoucke2010diagrammatic}.
	\item Density of states algorithm \cite{fodor2007the,gattringer2016approaches}.
\end{itemize}
Algorithms that are completely generic and can \textit{in principle} be used to sample any integral are
\begin{itemize}
	\item Lefschetz thimbles \cite{cristoforetti2012new,pawlowski2021simulating}.
	\item Complex Langevin.
\end{itemize}
Both involve some sort of analytic continuation of the action and observables, i.e. a complexification of the originally real fields $\boldsymbol{\phi}$. While the Lefschetz thimble method attempts to deform the integration contour of $\boldsymbol{\phi}$ away from the real axis in such a way as to reduce the phase oscillations, the complex Langevin approach first rewrites the path integral as a stochastic differential equation and then complexifies $\boldsymbol{\phi}$. This will be the subject of the following section \sect{cl}.

\subsection{Complex Langevin method\label{sec:cl}}
The complex Langevin approach to tackling the sign problem consists of two ingredients: The reformulation of path integrals in terms of stochastic differential equations (stochastic quantization) as well as an artificial doubling of the number of degrees of freedom that transforms the hard problem of sampling from a complex probability distribution to the more accessible problem of sampling from a real and positive-definite distribution with twice as many degrees of freedom. 
The close relationship between functional integrals and stochastic motion stood at the very cradle of the path integral: In fact, the introduction of the path integral in the context of Brownian motion by Wiener in 1921 \cite{wiener1921average} preceded its later application to quantum theory by Wentzel, Dirac and Feynman. However, it was not until the seminal work of Parisi and Wu in 1981 \cite{parisi1981perturbation} that this relationship was exploited to conceive a practical algorithm to simulate a quantum field theory on the lattice and thus the method of stochastic quantization was born. Shortly thereafter, it was recognized that, at least formally, stochastic quantization is not restricted by the crucial requirement that the action be real, in contrast to standard Monte Carlo algorithms known at that time \cite{ parisi1983complex,klauder1985spectrum}. Namely, nothing prevents us from inserting a complex action $S$ into a Langevin equation. It has, however, as its consequence, that also the originally fields $\phi$ must become complex quantities. This is the second ingredient to the complex Langevin algorithm: Since we artificially complexify real fields, we double the number of degrees of freedom but sample from a positive-definite probability distribution in this extended space.

This section is organized as follows: We will first introduce and derive stochastic quantization for the case of a real action. Subsequently, we discuss its generalization to the complex case and the intricate question of the validity of this generalization. Finally, we briefly illustrate the algorithm on a simple toy model. For a comprehensive and pedagogical review the reader is referred to reference \cite{berger2021complex}.

\subsubsection{Real Langevin}
Consider a very high-dimensional integral of the type
\begin{align}
\label{eq:def_pi}
\langle \mathcal{O}\rangle\equiv Z^{-1}\int d^n\phi \,\exp\left(-S[\boldsymbol{\phi}]\right)\mathcal{O}[\boldsymbol{\phi}]
\end{align}
with 
\begin{align}
Z=\int d^n\phi \,\exp\left(-S[\boldsymbol{\phi}]\right)
\end{align}
and $S$ assumed to be real for now. We want to show that the integral \eq{def_pi} can be evaluated as 
\begin{align}
\label{eq:stochastic_obs}
\langle \mathcal{O}\rangle=\lim_{\vartheta\to\infty}\left\langle \mathcal{O}[\boldsymbol{\phi}(\vartheta)]\right\rangle_\eta\,,
\end{align}
where $\boldsymbol{\phi}(\vartheta)$ obeys the stochastic differential equation
\begin{align}
\label{eq:Langevin_eq}
\frac{\partial\boldsymbol{\phi}}{\partial \vartheta}=-\nabla S[\boldsymbol{\phi}]+\boldsymbol{\eta}(\vartheta)\,,
\end{align}
the white noise $\boldsymbol{\eta}$ fulfills 
\begin{align}
\langle\boldsymbol{\eta}(\vartheta)\rangle=0\qquad \langle \eta_i(\vartheta)\eta_j(\vartheta')\rangle=2\delta_{ij}\delta(\vartheta-\vartheta')
\end{align}	
and $\langle\cdot\rangle_\eta$ denotes an average with respect to the noise. The first step in deriving the equivalence of \eq{def_pi} and the stochastic formulation is to derive the Fokker-Planck equation that corresponds to the Langevin equation \eq{Langevin_eq}. 

Consider the probability distribution $P(\boldsymbol{\phi},\vartheta)$ that a field $\boldsymbol{\phi}'(\vartheta)$ whose dynamics is governed by \eq{Langevin_eq} takes the value $\boldsymbol{\phi}$ at time $\vartheta$:
\begin{align}
P(\boldsymbol{\phi},\vartheta)=\left\langle\delta\left(\boldsymbol{\phi}-\boldsymbol{\phi}'(\vartheta)\right)\right\rangle_\eta\,.
\end{align}
Let us derive an equation of motion for $P$ by computing its time derivative:
\begin{align}
\nonumber\frac{\partial P}{\partial\vartheta}&=\left\langle\frac{\partial}{\partial\vartheta}\delta\left(\boldsymbol{\phi}-\boldsymbol{\phi}'(\vartheta)\right)\right\rangle_\eta\\\nonumber
&=-\nabla_{\boldsymbol{\phi}}\cdot\left\langle\delta\left(\boldsymbol{\phi}-\boldsymbol{\phi}'(\vartheta)\right)\partial_\vartheta\boldsymbol{\phi}'\right\rangle_\eta
\\\nonumber&=-\nabla_{\boldsymbol{\phi}}\cdot\left\langle\delta\left(\boldsymbol{\phi}-\boldsymbol{\phi}'(\vartheta)\right)\left[-\nabla S[\boldsymbol{\phi}']+\boldsymbol{\eta}(\vartheta)\right]\right\rangle_\eta
\\\nonumber&=-\nabla_{\boldsymbol{\phi}}\cdot\left\langle\delta\left(\boldsymbol{\phi}-\boldsymbol{\phi}'(\vartheta)\right)\left[-\nabla S[\boldsymbol{\phi}]+\boldsymbol{\eta}(\vartheta)\right]\right\rangle_\eta
\\&=\nabla\cdot\left[P\nabla S\right]-\nabla_{\boldsymbol{\phi}}\cdot\left\langle\delta\left(\boldsymbol{\phi}-\boldsymbol{\phi}'(\vartheta)\right)\boldsymbol{\eta}(\vartheta)\right\rangle_\eta\,.
\end{align}
The evaluation of $\left\langle\delta\left(\boldsymbol{\phi}-\boldsymbol{\phi}'(\vartheta)\right)\boldsymbol{\eta}(\vartheta)\right\rangle_\eta$ requires a closer inspection. The averaging over the noise $\langle\cdot\rangle_\eta$ may be replaced by an averaging over the distribution function $P(\boldsymbol{\phi}',\vartheta)$. However, the problem is that the noise at time $\vartheta$ and $P(\boldsymbol{\phi}',\vartheta)$ are not independent. A solution is to replace the averaging over the noise at all times by an averaging over the noise in the time interval $[\vartheta-\Delta\vartheta,\vartheta]$ and over $P(\boldsymbol{\phi}',\vartheta-\Delta\vartheta)$, which by causality is independent of the noise between $\vartheta-\Delta\vartheta$ and $\vartheta$. $\Delta \vartheta$ will ultimately chosen to be infinitesimally small. We have
\begin{align}
\boldsymbol{\phi}'(\vartheta)=\boldsymbol{\phi}'(\vartheta-\Delta\vartheta)+\int\limits_{\vartheta-\Delta\vartheta}^\vartheta d\vartheta'\,\left[-\nabla S[\boldsymbol{\phi}'(\vartheta')]+\boldsymbol{\eta}(\vartheta')\right]
\end{align}
and thus 
\begin{align}
\nonumber\left\langle\delta\left(\boldsymbol{\phi}-\boldsymbol{\phi}'(\vartheta)\right)\boldsymbol{\eta}(\vartheta)\right\rangle_\eta=&\int d^n\phi''\,P(\boldsymbol{\phi}'',\vartheta-\Delta\vartheta)\\\nonumber
&\times\left\langle\delta\left(\boldsymbol{\phi}-\boldsymbol{\phi}''-\int_{\vartheta-\Delta\vartheta}^\vartheta d\vartheta'\,\left[-\nabla S[\boldsymbol{\phi}'(\vartheta')]+\boldsymbol{\eta}(\vartheta')\right]\right)\boldsymbol{\eta}(\vartheta)\right\rangle_{\eta[\vartheta-\Delta\vartheta,\vartheta]}\\
=&\left\langle P\left(\boldsymbol{\phi}-\int_{\vartheta-\Delta\vartheta}^\vartheta d\vartheta'\,\left[-\nabla S[\boldsymbol{\phi}'(\vartheta')]+\boldsymbol{\eta}(\vartheta')\right],\vartheta-\Delta\vartheta\right)\boldsymbol{\eta}(\vartheta)\right\rangle_{\eta[\vartheta-\Delta\vartheta,\vartheta]}\,.
\end{align}
If we take $\Delta\vartheta$ to be infinitesimal, we may expand $P$ to lowest order:
\begin{align}
\nonumber&\left\langle P\left(\boldsymbol{\phi}-\int_{\vartheta-\Delta\vartheta}^\vartheta d\vartheta'\,\left[-\nabla S[\boldsymbol{\phi}'(\vartheta')]+\boldsymbol{\eta}(\vartheta')\right],\vartheta-\Delta\vartheta\right)\boldsymbol{\eta}(\vartheta)\right\rangle_{\eta[\vartheta-\Delta\vartheta,\vartheta]}\\
=&\left\langle\left[ P\left(\boldsymbol{\phi},\vartheta-\Delta\vartheta\right)-\nabla P\left(\boldsymbol{\phi},\vartheta-\Delta\vartheta\right)\cdot\int_{\vartheta-\Delta\vartheta}^\vartheta d\vartheta'\,\left[-\nabla S[\boldsymbol{\phi}'(\vartheta-\Delta\vartheta)]+\boldsymbol{\eta}(\vartheta')\right]\right]\boldsymbol{\eta}(\vartheta)\right\rangle_{\eta[\vartheta-\Delta\vartheta,\vartheta]}\,.
\end{align}
Here, we have replaced $\vartheta'$ by $\vartheta-\Delta\vartheta$ in the argument of $\boldsymbol{\phi}'$ as the difference amounts to terms of higher order in $\Delta\vartheta$. Now we only need to perform the averages over the noise. Using $\langle\boldsymbol{\eta}\rangle=0$ as well as
\begin{align}
\int\limits_{\vartheta-\Delta\vartheta}^\vartheta d\vartheta'\, \eta_{i}(\vartheta')\eta_{j}(\vartheta)=\delta_{ij}\int\limits_{\vartheta-\Delta\vartheta}^\vartheta d\vartheta'\,2\delta(\vartheta-\vartheta')=\delta_{ij}\,,
\end{align}
and assuming that $P$ is continuous in $\vartheta$, we finally arrive at
\begin{align}
\left\langle\delta\left(\boldsymbol{\phi}-\boldsymbol{\phi}'(\vartheta)\right)\boldsymbol{\eta}(\vartheta)\right\rangle_\eta=-\nabla P(\boldsymbol{\phi},\vartheta)\,.
\end{align}
Putting all pieces together, we obtain the \textit{Fokker-Planck equation}
\begin{align}
\frac{\partial P}{\partial\vartheta}=\nabla\cdot\left[P\nabla S\right]+\nabla^2P\,,
\end{align}
which, by introducing the Fokker-Planck operator
\begin{align}
\mathcal{F}\equiv (\nabla S)\cdot\nabla+(\nabla^2S)+\nabla^2
\end{align}
can be written as
\begin{align}
\frac{\partial P}{\partial\vartheta}=\mathcal{F}P\,.
\end{align}
Let us introduce the modified probability density $\tilde{P}$ and Fokker-Planck operator $\tilde{\mathcal{F}}$ as 
\begin{align}
\tilde{P}&\equiv e^{S/2}P\\
\tilde{\mathcal{F}}&\equiv   e^{S/2}\mathcal{F}e^{-S/2}=\frac{1}{2}(\nabla^2S)-\frac{1}{4}(\nabla S)^2+\nabla^2
\end{align} 
fulfilling
\begin{align}
\frac{\partial \tilde{P}}{\partial\vartheta}=\tilde{\mathcal{F}}\tilde{P}\,.
\end{align}
As one easily checks, $\tilde{\mathcal{F}}$ is a self-adjoint and negative semi-definite operator. Furthermore
\begin{align}
\tilde{\mathcal{F}}e^{-S/2}=0
\end{align}
such that $e^{-S/2}$ is an eigenvector with eigenvalue $0$. Since $\tilde{\mathcal{F}}$ is negative semi-definite, this is also the eigenvector with largest eigenvalue and hence
\begin{align}
\tilde{P}(\vartheta\to\infty)\sim e^{-S/2}\,.
\end{align}
Therefore the original probability distribution behaves as:
\begin{align}
P(\vartheta\to\infty)\sim e^{-S}\,,
\end{align}
which is exactly the desired distribution from which we want to sample. 

For the case of a functional $S[\phi(\mathbf{x})]$, the above considerations can be straightforwardly generalized by replacing the nabla operators by functional derivatives, i.e. the Langevin equation becomes
\begin{align}
\frac{\partial \phi(\mathbf{x})}{\partial\vartheta}=-\frac{\delta S}{\delta \phi(\mathbf{x})}+\eta(\vartheta,\mathbf{x})\,,
\end{align} 
with $\langle\eta(\vartheta,\mathbf{x})\eta(\vartheta',\mathbf{x}')\rangle=2\delta(\vartheta'-\vartheta)\delta(\mathbf{x}-\mathbf{x}')$. However, since we need to discretize the action on a lattice for numerical computations, we will not deal with actual functionals in practice. 

For practical numerical computations, the Langevin equation \eq{Langevin_eq} can be solved with several integration schemes. The simplest and most widely employed one is the (explicit) Euler-Maruyama method, which amounts to
\begin{align}
\boldsymbol{\phi}_{i+1}=\boldsymbol{\phi}_{i}-\Delta\vartheta\, \nabla S[\boldsymbol{\phi}_i]+\sqrt{2\Delta\vartheta}\,\boldsymbol{\eta}_i
\end{align}
with step size $\Delta \vartheta$ and $\boldsymbol{\eta}_i$ Gaussian random vectors with zero mean and standard deviation one. Higher order as well as implicit schemes have also been described and employed in some works \cite{aarts2012complex,alvestad2021stable}. 

Also, for practical computations one typically assumes ergodicity, i.e. that the noise average in \eq{stochastic_obs} is equal to a long-time average, i.e. one computes observables as 
\begin{align}
\langle\mathcal{O}\rangle=\frac{1}{\Theta-\Theta_0}\int\limits_{\Theta_0}^\Theta d\vartheta\, \mathcal{O}[\boldsymbol{\phi}(\vartheta)]
\end{align}
with $\Theta_0$ an equilibration time and $\Theta$ large enough for the average to converge. Statistical uncertainties can be estimated from the variance of several statistically independent runs.

\subsubsection{Complex Langevin}

So far, we have only considered the case of a real action $S\in\mathbb{R}$. However, as Parisi put it ``Nothing forbids to write a Langevin equation also for complex $H$ [i.e. $S$]'' \cite{parisi1983complex}. One sees that if we make $S$ complex, evolving a field $\boldsymbol{\phi}$ with \eq{Langevin_eq} immediately renders it complex also if it is originally real. Thus, we must promote the real field $\boldsymbol{\phi}$ to a complex one~\footnote{This procedure makes of course only sense if the action $S[\boldsymbol{\phi}]$ is holomorphic (which is typically the case for bosonic actions) or at least meromorphic (which is the case for fermionic actions after integrating out the Grassmann fields) in the field $\boldsymbol{\phi}$.}, 
\begin{align}
\boldsymbol{\phi}\to\boldsymbol{\phi}_R+i\boldsymbol{\phi}_I\,.
\end{align}
Upon this complexification, the number of real Langevin equations doubles
\begin{align}
\label{eq:Langevin_eq_compl}
\frac{\partial\boldsymbol{\phi}_R}{\partial \vartheta}&=-\text{Re}\left\{\nabla S[\boldsymbol{\phi}_R+i\boldsymbol{\phi}_I]\right\}+\boldsymbol{\eta}_R(\vartheta)\\
\frac{\partial\boldsymbol{\phi}_I}{\partial \vartheta}&=-\text{Im}\left\{\nabla S[\boldsymbol{\phi}_R+i\boldsymbol{\phi}_I]\right\}+\boldsymbol{\eta}_I(\vartheta)\,,
\end{align}
where the real and imaginary noise fulfill $\langle\eta_{R,i}(\vartheta)\eta_{R,j}(\vartheta')\rangle=2N_R\delta_{ij}\delta(\vartheta-\vartheta')$ and $\langle\eta_{I,i}(\vartheta)\eta_{I,j}(\vartheta')\rangle=2N_I\delta_{ij}\delta(\vartheta-\vartheta')$, respectively. It will soon turn out that $N_R$ and $N_I$ must fulfill $N_R-N_I=1$ but can be chosen arbitrarily apart from this constraint. The expectation value of a (holomorphic) observable  can be computed in complete analogy to the case of a real action
\begin{align}
\label{eq:stochastic_obs_compl}
\langle \mathcal{O}\rangle=\lim_{\vartheta\to\infty}\left\langle \mathcal{O}[\boldsymbol{\phi}_R(\vartheta)+i\boldsymbol{\phi}_I(\vartheta)]\right\rangle_\eta\,.
\end{align}
If $\mathcal{O}$ is hermitian, its expectation value must be real and hence the imaginary part of $\mathcal{O}[\boldsymbol{\phi}_R(\vartheta)+i\boldsymbol{\phi}_I(\vartheta)]$ must average out to zero. 

It now arises the obvious question whether this purely formal generalization of the well-defined real Langevin algorithm will indeed reproduce observables correctly. The probability distribution $P(\boldsymbol{\phi}_R,\boldsymbol{\phi}_I,\vartheta)$ that the complexified Langevin equation produces obeys a Fokker-Planck with twice as many degrees of freedom as before,
\begin{align}
\frac{\partial P}{\partial\vartheta}=\mathcal{F}_2P\,,
\end{align}
with 
\begin{align}
\nonumber\mathcal{F}_2=&\text{Re}\left\{\nabla S[\boldsymbol{\phi}_R+i\boldsymbol{\phi}_I]\right\}\cdot\nabla_R+\left(\nabla_R\cdot\text{Re}\left\{\nabla S[\boldsymbol{\phi}_R+i\boldsymbol{\phi}_I]\right\}\right)+N_R\nabla^2_R\\
&+\text{Im}\left\{\nabla S[\boldsymbol{\phi}_R+i\boldsymbol{\phi}_I]\right\}\cdot\nabla_I+\left(\nabla_I\cdot\text{Im}\left\{\nabla S[\boldsymbol{\phi}_R+i\boldsymbol{\phi}_I]\right\}\right)+N_I\nabla^2_I\,,
\end{align}
where $\nabla_{R/I}$ denotes a derivative by $\boldsymbol{\phi}_{R/I}$ and $\nabla$ a complex derivative by $\boldsymbol{\phi}_R+i\boldsymbol{\phi}_I$. The question of the correctness of the complex Langevin algorithm amounts to whether
\begin{align}
\label{eq:cl_correctness}
\lim_{\vartheta\to\infty}\int d^n\phi_R d^n\phi_I\,P(\boldsymbol{\phi}_R,\boldsymbol{\phi}_I,\vartheta)\mathcal{O}[\boldsymbol{\phi}_R+i\boldsymbol{\phi}_I]=\frac{\int d^n\phi_R\,e^{-S[\boldsymbol{\phi}_R]}\mathcal{O}\left[\boldsymbol{\phi}_R\right]}{\int d^n\phi_R\,e^{-S[\boldsymbol{\phi}_R]}}\,.
\end{align}
An approach for tackling this question was first proposed in \cite{aarts2010complex}. Let us first introduce the operator 
\begin{align}
\mathcal{F}_1\equiv\nabla S[\boldsymbol{\phi}_R+i\boldsymbol{\phi}_I]\cdot\nabla_R+\left(\nabla_R\cdot\nabla S[\boldsymbol{\phi}_R+i\boldsymbol{\phi}_I]\right)+\nabla^2_R\,.
\end{align}
If the constraint $N_R-N_I=1$ is fulfilled, one easily shows that the action of $\mathcal{F}_2$ and $\mathcal{F}_1$ onto a holomorphic function $f(\boldsymbol{\phi}_R,\boldsymbol{\phi}_I)=f(\boldsymbol{\phi}_R+i\boldsymbol{\phi}_I)$ is the same,
\begin{align}
\mathcal{F}_1 f=\mathcal{F}_2 f,
\end{align}
because for a holomorphic function we have $\nabla_I=i\nabla_R$, and equivalently for the transposed operators, $\mathcal{F}_1^T f=\mathcal{F}_2^T f$ . Note that we may \textit{not} assume $P$ to be holomorphic~\footnote{Due to Liouville's theorem, holomorphicity is not even reconcilable with $P$ being a probaility distribution.}. The observable $\mathcal{O}(\boldsymbol{\phi}_R,\boldsymbol{\phi}_I)\equiv \mathcal{O}[\boldsymbol{\phi}_R+i\boldsymbol{\phi}_I]$ instead was assumed to be holomorphic.
$\mathcal{F}_1$ and $\mathcal{F}_2$ both act on the space of functions defined on $\mathbb{R}^{2n}$. Let us also introduce
\begin{align}
\mathcal{F}_1'\equiv\nabla S[\boldsymbol{\phi}_R]\cdot\nabla_R+\left(\nabla_R\cdot\nabla S[\boldsymbol{\phi}_R]\right)+\nabla^2_R\,,
\end{align}
which acts on functions defined on $\mathbb{R}^{n}$. 

Say the initial condition of the complex Langevin evolution is chosen as  $\boldsymbol{\phi}_R(\vartheta=0)=\boldsymbol{\phi}_{R,0}$ and $\boldsymbol{\phi}_I(\vartheta=0)=0$~\footnote{$\boldsymbol{\phi}_I$ does not necessarily have to be zero initially but this assumption simplifies the argument.} such that $P(\boldsymbol{\phi}_R,\boldsymbol{\phi}_I,\vartheta=0)=\delta(\boldsymbol{\phi}_R-\boldsymbol{\phi}_{R,0})\delta(\boldsymbol{\phi}_I)$. Then we define a function $\rho(\boldsymbol{\phi}_R,\vartheta)$ as the solution of the equation
\begin{align}
\label{eq:ev_eq_rho}
\frac{\partial \rho}{\partial\vartheta}=\mathcal{F}_1'\rho\,
\end{align}
with initial condition $\rho(\boldsymbol{\phi}_R,\vartheta=0)=\delta(\boldsymbol{\phi}_R-\boldsymbol{\phi}_{R,0})$. By means of partial integration, it is straightforward to show that \eq{ev_eq_rho} is norm-conserving, i.e. $\partial_\vartheta \int d^n\phi\,\rho=0$. Furthermore, one easily demonstrates that $e^{-S[\boldsymbol{\phi}_R]}$ is a fixed point of \eq{ev_eq_rho}, $\mathcal{F}_1'e^{-S[\boldsymbol{\phi}_R]}=0$. If we assume that $\rho$ approaches this fixed point in the long-time limit (we will comment on this assumption below), i.e. 
\begin{align}
\rho(\boldsymbol{\phi}_R,\vartheta\to\infty)=\frac{e^{-S[\boldsymbol{\phi}_R]}}{\int d^n\phi_R\,e^{-S[\boldsymbol{\phi}_R]}}\,,
\end{align}
we have proven \eq{cl_correctness} and thus the correctness of complex Langevin if we can show that
\begin{align}
\label{eq:P_vs_rho}
\int d^n\phi_R d^n\phi_I\,P(\boldsymbol{\phi}_R,\boldsymbol{\phi}_I,\vartheta)\mathcal{O}[\boldsymbol{\phi}_R+i\boldsymbol{\phi}_I]=\int d^n\phi_R \,\rho(\boldsymbol{\phi}_R,\vartheta)\mathcal{O}[\boldsymbol{\phi}_R]
\end{align} 
for all $\vartheta$. Formally, the evolution equations for $P$ and $\rho$ are solved by
\begin{align}
P(\vartheta)&=\exp\left(\vartheta \mathcal{F}_2\right)P(0)\\
\rho(\vartheta)&=\exp\left(\vartheta \mathcal{F}_1'\right)\rho(0)\,.
\end{align}
While $P$ and $\rho$ have the same initial condition, the problem is that they evolve with different operators. The trick in  proving \eq{P_vs_rho} is to employ a partial integration to move the time evolution from the distribution function to the observable, on which the two different operators agree due to its holomorphicity, and then to move it back to the distribution function, i.e.
\begin{align}
\nonumber\int d^n\phi_R d^n\phi_I\,P(\boldsymbol{\phi}_R,\boldsymbol{\phi}_I,\vartheta)\mathcal{O}[\boldsymbol{\phi}_R+i\boldsymbol{\phi}_I]&=\int d^n\phi_R d^n\phi_I\,\exp(\vartheta \mathcal{F}_2)P(\boldsymbol{\phi}_R,\boldsymbol{\phi}_I,0)\mathcal{O}[\boldsymbol{\phi}_R+i\boldsymbol{\phi}_I]\\\nonumber
&=\int d^n\phi_R d^n\phi_I\,P(\boldsymbol{\phi}_R,\boldsymbol{\phi}_I,0)\exp(\vartheta \mathcal{F}_2^T)\mathcal{O}[\boldsymbol{\phi}_R+i\boldsymbol{\phi}_I]
\\\nonumber
&=\int d^n\phi_R d^n\phi_I\,\rho(\boldsymbol{\phi}_R,0)\delta(\boldsymbol{\phi}_I)\exp(\vartheta \mathcal{F}_1^T)\mathcal{O}[\boldsymbol{\phi}_R+i\boldsymbol{\phi}_I]
\\\nonumber
&=\int d^n\phi_R d^n\phi_I\,\exp(\vartheta \mathcal{F}_1)\rho(\boldsymbol{\phi}_R,0)\delta(\boldsymbol{\phi}_I)\mathcal{O}[\boldsymbol{\phi}_R+i\boldsymbol{\phi}_I]
\\\nonumber
&=\int d^n\phi_R \,\exp(\vartheta \mathcal{F}_1')\rho(\boldsymbol{\phi}_R,0)\mathcal{O}[\boldsymbol{\phi}_R]
\\
&=\int d^n\phi_R \,\rho(\boldsymbol{\phi}_R,\vartheta)\mathcal{O}[\boldsymbol{\phi}_R]\,.
\end{align}
In this proof, we have swept under the rug three potential mathematical difficulties that in practice can spoil the correctness of the method:
\begin{enumerate}[(i)]
	\item The question whether the fixed point $e^{-S}$ of \eq{ev_eq_rho} is unique and whether it is reached in the long-time limit.
	\item The convergence of the matrix exponentials $\exp(\vartheta\mathcal{F}_{2/1})$.
	\item The validity of the performed partial integrations.
\end{enumerate} 
As it turns out, it is very hard to check for the fulfillment of these conditions \textit{a priori}, i.e. it is difficult to establish general and precise criteria for deciding whether complex Langevin will work for a certain model and scenario. However, it is easy to check them \textit{a posteriori}, i.e. after running a simulation of the model in question. 

With regard to (i), it will be immediately clear from the simulation if $\rho$ does not converge to a fixed point as it has as its consequence that $P$ will not converge either, as follows from \eq{P_vs_rho}. Thus if our simulation does not even yield temporal convergence of the observables, it is definitely spoilt. There is still the possibility that the fixed point is not unique, i.e. $\mathcal{F}_1'\rho=0$ for some $\rho$ apart from $\rho(\boldsymbol{\phi}_R)\propto e^{-S[\boldsymbol{\phi}_R]}$. Say there are two distinct (normalized) fixed points $\rho_1$ and $\rho_2$. Then every linear combination thereof, $\lambda\rho_1+(1-\lambda)\rho_2$ with $\lambda\in\mathbb{C}$, will also be a fixed point, i.e. we have an infinitely extended line of continuously connected fixed points. Thus it will continuously depend on the initial condition which fixed point is ultimately reached and hence to which values the observables converge, i.e. ergodicity would be broken. This would also be apparent from a simulation. In practice, it is likely that in such a case the observables would not even display convergence at all, because arbitrarily small perturbations (which are inevitably present in a numerical simulation) could make $\rho$ flow freely along the infinitely extended line of fixed points, each yielding different values for the observables.

Concerning (ii) and (iii), it has been demonstrated in \cite{nagata2016argument} that these conditions are fulfilled if the distribution function of the drift magnitude $p(u)$, defined as 
\begin{align}
\label{eq:pu}
p(u)=\int d^n\phi_R d^n\phi_I\,P(\boldsymbol{\phi}_R,\boldsymbol{\phi}_I)\,\delta\left(u-\max_i \left|\partial_i S[\boldsymbol{\phi}_R+i\boldsymbol{\phi}_I]\right|\right)
\end{align} 
decays at infinity faster than every power law. This condition is also easy to check from numerical simulations.

In conclusion, we can decide rather straightforwardly if a complex Langevin simulation for a given model is correct \textit{after} running the simulation. It is highly likely that the more generic question for which models and under which circumstances this is the case can ultimately be answered only from experience and empirical evidence.  Within this work, we found that complex Langevin works excellently for weakly coupled ultracold Bose gases in thermal equilibrium. Both strong coupling in equilibrium and real-time scenarios in general lead to a breakdown instead. 

The numerical implementation of the complexified Langevin equation works exactly as in the real case, as we can simply consider the $n$-dimensional complex Langevin equation as a $2n$-dimensional real Langevin equation. The only difference is that in the complexified case we have the freedom to choose $N_R$ and $N_I$ arbitrarily under the constraint $N_R-N_I=1$. It is common practice to choose $N_R=1$, $N_I=0$, which has been demonstrated to be numerically most convenient \cite{aarts2010complex}. We will adhere to this choice throughout this thesis. 

\begin{figure}	
	\centering\includegraphics[width=0.47\textwidth]{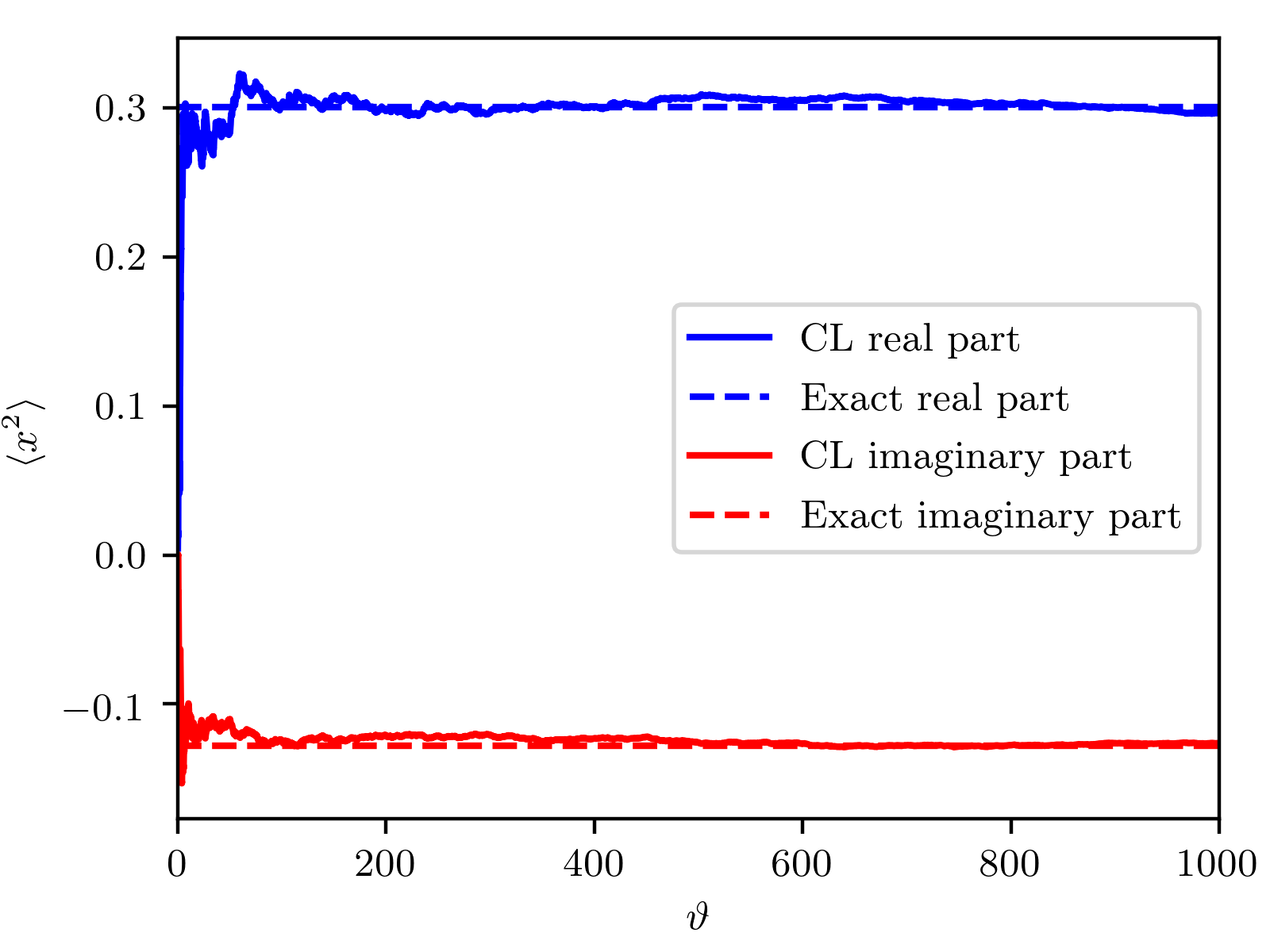}
	\includegraphics[width=0.47\textwidth]{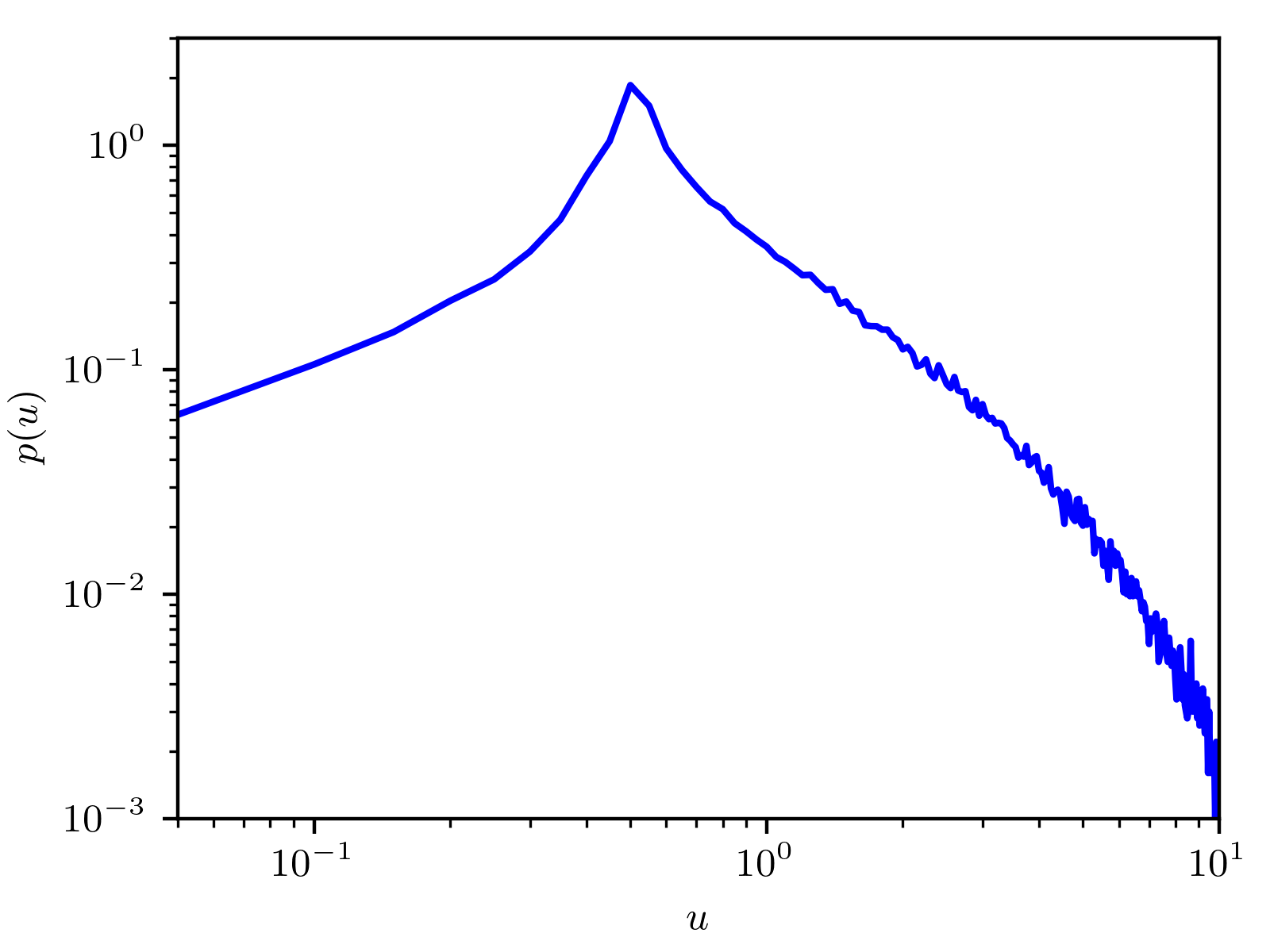}
	\caption{Left panel: Running averages of $x^2$ for $x$ evolving with the complexified Langevin equations, \eq{toymodel_cl_re} and \eq{toymodel_cl_im}. The exact real and imaginary part (dashed lines) are approached in the long-time limit. Right panel: Histogram of the drift magnitude $p(u)$, as defined in \eq{pu}. In the double logarithmic plot, $p(u)$ bends downwards at large $u$, indicating that it falls off faster than every power law.}
	\label{fig:Langevin_toymodel}
\end{figure}

To conclude this section, let us demonstrate the power of the complex Langevin algorithm on a simple toy model. Consider the ``action''
\begin{align}
S(x)=ix^2+x^4\,,
\end{align}
which could mimic a $\phi^4$ type theory with complex mass. We want to compute the expectation value
\begin{align}
\left\langle x^2\right\rangle=\frac{\int dx\, e^{-S(x)} x^2}{\int dx\, e^{-S(x)}}\,.
\end{align}
These integrals can be evaluated with a direct integration scheme (or by expressing them in terms of Bessel functions), which yields $\left\langle x^2\right\rangle=0.3-0.13 i$. The complex Langevin equations with $N_R=1$ read 
\begin{align}
\label{eq:toymodel_cl_re}
\frac{\partial x_R}{\partial\vartheta}&=-\text{Re}\left\{2i(x_R+ix_I)+4(x_R+ix_I)^3\right\}+\eta(\vartheta)\\
\label{eq:toymodel_cl_im}
\frac{\partial x_I}{\partial\vartheta}&=-\text{Im}\left\{2i(x_R+ix_I)+4(x_R+ix_I)^3\right\}\,.
\end{align}
We discretize these equations with a Euler-Maruyama scheme with $\Delta\vartheta=0.01$ and evolve them up to $\vartheta=1000$. The running average of $(x_R+ix_I)^2$ is shown in the left panel of figure \fig{Langevin_toymodel}. In the long-time limit it approaches the correct value. Indeed, the histogram of the drift magnitude $p(u)$, which is shown in the right panel of figure \fig{Langevin_toymodel}, falls off faster than any power law, as it bends downwards in a double logarithmic plot, such that the correctness criterion is fulfilled.

\subsection{Langevin equations for the non-relativistic Bose gas\label{sec:Langevin_eq}}
There are two equivalent formulations of the complex Langevin equations of the bosonic action, which are related to each other by a simple change of variables. As the theory outlined in the previous section applies to an action $S[\phi]$ of a real field $\phi\in \mathbb{R}$, the most straightforward way of deriving Langevin equations for the bosonic action is to consider the complex Bose field $\psi\in\mathbb{C}$ as composed of two real fields, $\psi\equiv \varphi+i\chi$, with $\varphi,\chi\in\mathbb{R}$, and then to derive Langevin equations in the standard way,
\begin{flalign}
\label{eq:cleq_phi} (\text{I}) && \qquad\frac{\partial \varphi}{\partial\vartheta}&=-\frac{\delta S}{\delta \varphi}+\eta_\varphi(\vartheta) &\\
\label{eq:cleq_chi} (\text{II}) && \frac{\partial \chi}{\partial\vartheta}&=-\frac{\delta S}{\delta \chi}+\eta_\chi(\vartheta)\,. &
\end{flalign}
As $S[\varphi,\chi]$ is complex due to the Berry phase term, \eq{cleq_phi} and \eq{cleq_chi} become complex equations and both $\varphi$ and $\chi$ must be complexified, $\varphi\to\varphi_R+i\varphi_I$ and $\chi\to\chi_R+i\chi_I$, leaving us with four equations for four real fields, $\varphi_R,\varphi_I,\chi_R,\chi_I$.

This formalism, however, can sometimes be a bit cumbersome when deriving Langevin equations and actually implementing them in programming code, as one has to keep track of lots of imaginary units. A more convenient choice is to take the complex Bose field and its conjugate field as independent variables. We will denote these two independent complex variables by $\psi\equiv\varphi+i\chi$ and $\bar{\psi}\equiv\varphi-i\chi$ here while reserving the symbol $^*$ to the actual mathematical operation of complex conjugating a variable. While this distinction is superfluous in the original coherent state path integral as $\bar{\psi}=\psi^*$, it will be of importance in the complexified Langevin equations. Taking $(\text{I})\pm i (\text{II})$, using $\psi\equiv\varphi+i\chi$ and $\bar{\psi}\equiv\varphi-i\chi$ and the definition of the complex derivative (Wirtinger calculus)
\begin{align}
\frac{\partial}{\partial \psi}\equiv \frac{1}{2}\left(\frac{\partial}{\partial\varphi}-i\frac{\partial}{\partial\chi}\right)\qquad\qquad \frac{\partial}{\partial \bar{\psi}}\equiv \frac{1}{2}\left(\frac{\partial}{\partial\varphi}+i\frac{\partial}{\partial\chi}\right)\,,
\end{align}
we obtain the two new equations
\begin{align}
\frac{\partial\psi}{\partial\vartheta}&=-2\frac{\delta S}{\delta \bar{\psi}}+\eta(\vartheta)\\
\frac{\partial\bar{\psi}}{\partial\vartheta}&=-2\frac{\delta S}{\delta \psi}+\eta(\vartheta)^*\,,
\end{align}
where the new noise $\eta(\vartheta)\equiv\eta_\varphi(\vartheta)+i\eta_\chi(\vartheta)$ fulfills $\langle\eta\rangle=\langle\eta^*\rangle=\langle\eta^2\rangle=\langle{\eta^*}^2\rangle=0$ and $\langle\eta(\vartheta)\eta(\vartheta')^*\rangle=4\delta(\vartheta-\vartheta')$. A simple rescaling of the Langevin time, $\vartheta\to\vartheta/2$, brings the equations to the even simpler form 
\begin{align}
\frac{\partial\psi}{\partial\vartheta}&=-\frac{\delta S}{\delta \bar{\psi}}+\eta(\vartheta)\\
\frac{\partial\bar{\psi}}{\partial\vartheta}&=-\frac{\delta S}{\delta \psi}+\eta(\vartheta)^*\,,
\end{align}
where now the noise must fulfill $\langle\eta(\vartheta)\eta(\vartheta')^*\rangle=2\delta(\vartheta-\vartheta')$. As one sees, if $S\in\mathbb{R}$ the two equations would just be the complex conjugate of each other and thus $\bar{\psi}=\psi^*$. However, for a complex action $S\in\mathbb{C}$, this is no longer true and $\bar{\psi}\neq\psi^*$, such that $\psi$ and $\bar{\psi}$ are two independent complex variables.

We will here provide, for conciseness, the explicit form of the Langevin equations for a three-dimensional, $\mathcal{N}$-component gas with $U(\mathcal{N})$-symmetric interactions within both formalisms outlined above. For the case of additional dipolar interactions, the reader is referred to chapter \ref{sec:dipolars}. From deriving of \eq{S_discr_UN} we obtain the Langevin equations for independent variables $\varphi$ and $\chi$:
\begin{align}
\label{eq:deriv1}
\frac{\partial\varphi_{a,i,\mathbf{j}}}{\partial \vartheta}
=&\ a_\mathrm{s}^3\,\left(
\varphi_{a,i+1,\mathbf{j}}
+\varphi_{a,i-1,\mathbf{j}}
-2\varphi_{a,i,\mathbf{j}}
+i\chi_{a,i-1,\mathbf{j}}
-i\chi_{a,i+1,\mathbf{j}}
\right)
\nonumber\\
&+\frac{a_\mathrm{s}^3a_\tau}{2m}\,\Delta^\text{lat}\bigg(
\varphi_{a,i-1,\mathbf{j}}
+\varphi_{a,i+1,\mathbf{j}}
+i\chi_{a,i-1,\mathbf{j}}
-i\chi_{a,i+1,\mathbf{j}}
\bigg)
\nonumber\\
&+\mu a_\mathrm{s}^3a_\tau\,\bigg(
\varphi_{a,i-1,\mathbf{j}}
+\varphi_{a,i+1,\mathbf{j}}
+i\chi_{a,i-1,\mathbf{j}}
-i\chi_{a,i+1,\mathbf{j}}
\bigg)
\nonumber\\
&-ga_\mathrm{s}^3a_\tau\,\bigg(
\Psi_{i-1,\mathbf{j}}\,\varphi_{a,i-1,\mathbf{j}}
+\Psi_{i,\mathbf{j}}\,\varphi_{a,i+1,\mathbf{j}}
+i\Psi_{i-1,\mathbf{j}}\,\chi_{a,i-1,\mathbf{j}}
-i\Psi_{i,j,k,l}\,\chi_{a,i+1,\mathbf{j}}
\bigg)+\eta_{a,i,\mathbf{j},\varphi}(\vartheta)
\end{align}
\begin{align}
\label{eq:deriv2}
\nonumber
\frac{\partial\chi_{a,i,\mathbf{j}}}{\partial \vartheta}
=&\ a_\mathrm{s}^3\,\left(
\chi_{a,i+1,\mathbf{j}}
+\chi_{a,i-1,\mathbf{j}}
-2\chi_{a,i,\mathbf{j}}
-i\varphi_{a,i-1,\mathbf{j}}
+i\varphi_{a,i+1,\mathbf{j}}
\right)
\nonumber\\
&+\frac{a_\mathrm{s}^3a_\tau}{2m}\,\Delta^\text{lat}\bigg(
\chi_{a,i-1,\mathbf{j}}
+\chi_{a,i+1,\mathbf{j}}
-i\varphi_{a,i-1,\mathbf{j}}
+i\varphi_{a,i+1,\mathbf{j}}\bigg)
\nonumber\\
&+\mu a_\mathrm{s}^3a_\tau\,\bigg(
\chi_{a,i-1,\mathbf{j}}
+\chi_{a,i+1,\mathbf{j}}
-i\varphi_{a,i-1,\mathbf{j}}
+i\varphi_{a,i+1,\mathbf{j}}\bigg)
\nonumber\\
&-ga_\mathrm{s}^3a_\tau\,\bigg(
\Psi_{i-1,\mathbf{j}}\,\chi_{a,i-1,\mathbf{j}}
+\Psi_{i,\mathbf{j}}\,\chi_{a,i+1,\mathbf{j}}
-i\Psi_{i-1,\mathbf{j}}\,\varphi_{a,i-1,\mathbf{j}}
+i\Psi_{i,\mathbf{j}}\,\varphi_{a,i+1,\mathbf{j}}
\bigg)+\eta_{a,i,\mathbf{j},\chi}(\vartheta)\,,
\end{align}
where we have defined
\begin{align}
\Psi_{i,\mathbf{j}}
\equiv&\sum_a\Big(
\varphi_{a,i+1,\mathbf{j}}\,\varphi_{a,i,\mathbf{j}}
+\chi_{a,i+1,\mathbf{j}}\,\chi_{a,i,\mathbf{j}}
+i\varphi_{a,i+1,\mathbf{j}}\,\chi_{a,i,\mathbf{j}}
-i\varphi_{a,i,\mathbf{j}}\,\chi_{a,i+1,\mathbf{j}}
\Big)
\,,
\end{align}
and $\Delta^\text{lat}$ is the Laplacian on the lattice, cf. the discussion in chapter \ref{sec:constr_pi}. Taking $\psi$ and $\bar{\psi}$ to be the independent variables instead, the Langevin equations take the somewhat simpler form
\begin{align}
\nonumber\frac{\partial\psi_{a,i,\mathbf{j}}}{\partial\vartheta}=&a_\mathrm{s}^3\left(\psi_{a,i-1,\mathbf{j}}-\psi_{a,i,\mathbf{j}}\right)+\frac{a_\mathrm{s}^3a_\tau}{2m}\Delta^\text{lat}\psi_{a,i-1,\mathbf{j}}+\mu a_\mathrm{s}^3a_\tau\psi_{a,i-1,\mathbf{j}}\\\label{eq:CL_eq_bose1}
&-ga_\mathrm{s}^3a_\tau\left(\sum_b\bar{\psi}_{b,i,\mathbf{j}}\psi_{b,i-1,\mathbf{j}}\right)\psi_{a,i-1,\mathbf{j}}+\eta_{a,i,\mathbf{j}}(\vartheta)\\
\nonumber\frac{\partial\bar{\psi}_{a,i,\mathbf{j}}}{\partial\vartheta}=&a_\mathrm{s}^3\left(\bar{\psi}_{a,i+1,\mathbf{j}}-\bar{\psi}_{a,i,\mathbf{j}}\right)+\frac{a_\mathrm{s}^3a_\tau}{2m}\Delta^\text{lat}\bar{\psi}_{a,i+1,\mathbf{j}}+\mu a_\mathrm{s}^3a_\tau\bar{\psi}_{a,i+1,\mathbf{j}}\\\label{eq:CL_eq_bose2}
&-ga_\mathrm{s}^3a_\tau\left(\sum_b\bar{\psi}_{b,i+1,\mathbf{j}}\psi_{b,i,\mathbf{j}}\right)\bar{\psi}_{a,i+1,\mathbf{j}}+\eta_{a,i,\mathbf{j}}(\vartheta)^*\,.
\end{align}
Throughout this thesis, these stochastic differential equations are solved by a simple Euler-Maruyama scheme, as outlined in the previous section. For details of the numerical implementation and runtime benchmarks see appendix \sect{runtime}.

\clearpage

\thispagestyle{plain}
\section{Three-dimensional Bose gas\label{sec:3D}}
This chapter deals with the arguably simplest possible scenario for an ultracold bosonic gas: The three-dimensional, homogeneous Bose gas with $U(\mathcal{N})$-symmetric contact interaction. This system thus serves as an ideal test bed for the applicability and reliability of the complex Langevin method applied to gases of ultracold bosonic atoms. In particular, we will show that the results of well-known  approximate descriptions, Bogoliubov and Hartree-Fock theory, are well reproduced far below and above the Bose-Einstein transition, respectively. Furthermore, we will systematically discuss the effect of the discretization of imaginary time on the observables. We will also perform simulations close to the point of the Bose-Einstein transition and extract the shift of the Bose-Einstein transition temperature compared to the ideal value due to the interactions and compare the result to values from the literature.

\subsection{\label{sec:free}Ideal gas -- dependence on imaginary-time discretization}
As a first benchmark, we perform a complex Langevin simulation of the ideal Bose gas, i.e. at vanishing coupling, $g=0$. 
This serves, in particular, to discuss the errors induced by the discretization of imaginary time, corresponding to a truncation of the Fourier representations beyond a highest  Matsubara frequency. 

We consider a three-dimensional homogeneous gas, whose action is discretized on a $64^3\times N_\tau$ lattice with periodic boundary conditions and spatial lattice spacing $a_\mathrm{s}$, with the number of of lattice points $N_\tau$ chosen to be $N_\tau=16$. Setting $2ma_\mathrm{s}=1$ will permit us in the following to express all quantities in terms of $a_\mathrm{s}$, cf. appendix \sect{units}. The imaginary time lattice spacing is chosen as $a_\tau=0.05\,a_\mathrm{s}$. 

We tune the chemical potential to $\mu=-0.1\,a_\mathrm{s}^{-1}$, above the phase transition, where there is no macroscopic occupation of the $\mathbf{k}=0$ mode.
With the above choices for $N_\tau$ and $a_\tau$, the temperature results to be $T=1.25\,a_\mathrm{s}^{-1}$, which, together with $\mu$ corresponds to a total density $\rho=0.054\,a_s^{-3}$ and thus a critical temperature $T_\mathrm{c}=[\rho/\zeta(3/2)]^{2/3}2\pi/m=0.95\,a_\mathrm{s}^{-1}$ and a thermal de Broglie wave length of $\lambda_T= \sqrt{2\pi/mT} = 3.17 \,a_\mathrm{s}$. 

\begin{figure}[t]
	\centering\includegraphics[width=0.6\textwidth]{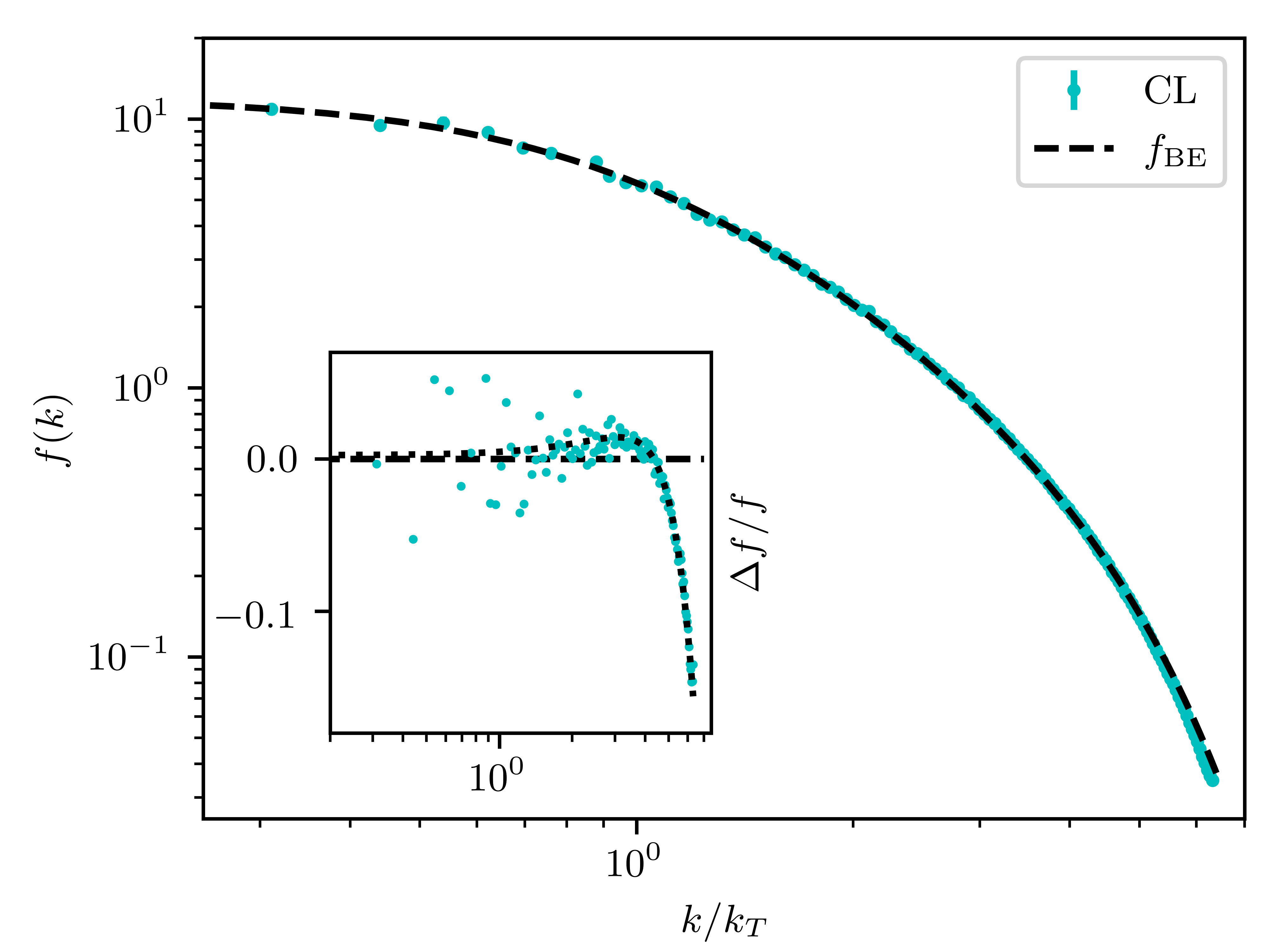}
	\caption{Angle-averaged momentum spectrum $f_\mathrm{CL}(k)$ of a single-component, non-interacting gas above the Bose-Einstein phase transition, at a temperature $T=1.25\,a_\text{s}^{-1}$ and chemical potential $\mu=-0.1\,a_\text{s}^{-1}$ (in units of the lattice constant $a_\text{s})$, as obtained by means of the complex Langevin method on a $64^{3}\times N_{\tau}$  lattice, with $N_\tau=16$ imaginary-time points.
		The momentum is measured in units of the inverse thermal de Broglie wave length $k_T=\lambda_T^{-1}=\sqrt{mT/2\pi}=0.315\,a_\mathrm{s}^{-1}$. 
		The black dashed line represents the Bose-Einstein distribution $f_\mathrm{BE}(k)$. 
		The inset shows the relative deviation $(\Delta f/f)^\mathrm{CL}_\mathrm{BE}=(f_\mathrm{CL}-f_\mathrm{BE})/f_\mathrm{BE}$ of the numerical and continuum analytic distributions. 
		The dotted black line in the inset represents the relative deviation as predicted from the analytical finite-$N_\tau$ computation.
	}
	\label{fig:freespectrum}
\end{figure}  

We simulate for a Langevin time of $\vartheta_\mathrm{max}=10^6\,a_\mathrm{s}^{-3}$, and begin the averaging after an equilibration time $\vartheta_0=10^5\,a_\mathrm{s}^{-3}$. Additionally, we average over 10 independent runs.
In each run, we initialize the field as $\psi(\tau,\mathbf{x})=0$ at each point on the lattice.

Let us consider the momentum distribution $f(\mathbf{k})=\langle a^\dagger_\mathbf{k}a_\mathbf{k}\rangle$. In figure \fig{freespectrum}, we show this momentum distribution extracted with complex Langevin, in angular-averaged form, i.e. $f_\mathrm{CL}(k)=f_\mathrm{CL}(|\mathbf{k}|)=\langle f_\mathrm{CL}(\mathbf{k})\rangle_{\Omega_\mathbf{k}}$.
As expected, the results for the distribution agree well with the Bose-Einstein distribution
\begin{align}
\label{eq:freeBED}
f_\mathrm{BE}(\mathbf{k})
&=\frac{1}{e^{\,\beta[\varepsilon(\mathbf{k})-\mu]}-1}
\,,
\end{align}
with free single-particle dispersion
\begin{align}
\label{eq:epsilonk}
\varepsilon(\mathbf{k})=\frac{\mathbf{k}^2}{2m}
\,,
\end{align}
over the whole range of momenta, while a weak deviation is seen at the smallest occupancies in the Boltzmann tail.
We amplify this discrepancy by showing, in the inset, the relative deviation $(\Delta f/f)^\mathrm{CL}_\mathrm{BE}=(f_\mathrm{CL}-f_\mathrm{BE})/f_\mathrm{BE}$ on the same logarithmic momentum scale. 

As we recall from section \sect{disc_eff}, the occupation number of the ideal gas, on a discrete imaginary-time lattice with $N_\tau$ lattice points, reads
\begin{align}
\label{eq:fFiniteNtau}
f_\mathrm{BE}(\mathbf{k};N_{\tau})
&=\sum_{n=0}^{N_{\tau}-1}\frac{1}{{N_\tau}(e^{2\pi i n/{N_\tau}}-1)+\beta(\varepsilon(\mathbf{k})-\mu)}
\,.
\end{align}
We show the relative deviation $(\Delta f/f)^{N_{\tau}}_\mathrm{BE}=(f_\mathrm{BE}(\mathbf{k};N_{\tau})-f_\mathrm{BE}(\mathbf{k}))/f_\mathrm{BE}(\mathbf{k})$ as a dotted line in the inset of \ref{fig:freespectrum}, which agrees on average with the deviation $(\Delta f/f)^\mathrm{CL}_\mathrm{BE}$ between the CL data and the exact result. 
This demonstrates that the systematic increase of the relative numerical error in the low-occupancy region at large momenta can be attributed to the truncation of the Matsubara series, to which the introduction of a finite lattice spacing of the imaginary time corresponds.

\begin{figure}	
	\centering\includegraphics[width=0.6\textwidth]{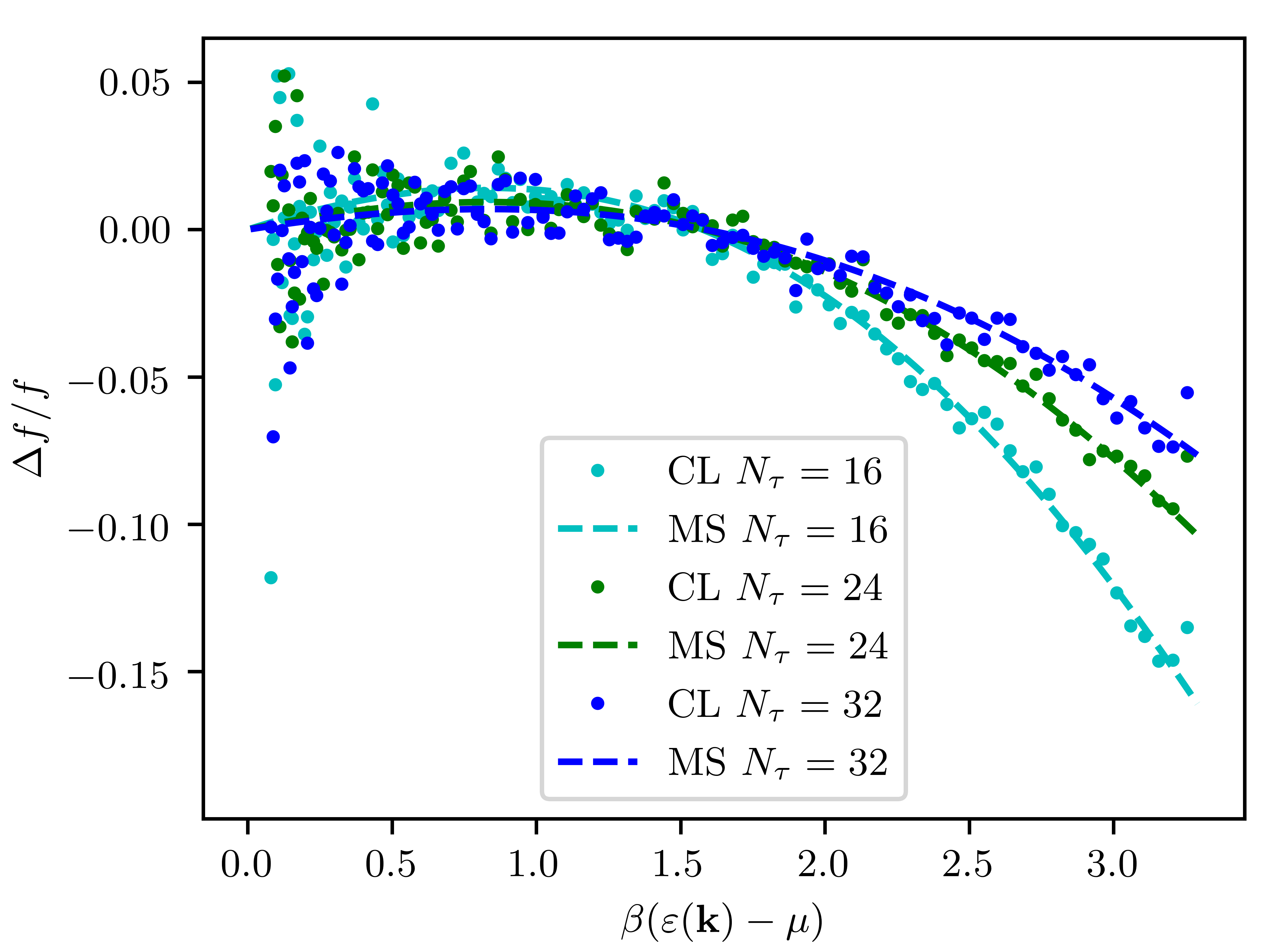}
	\caption{Comparison between the relative deviation $(\Delta f/f)^\mathrm{CL}_\mathrm{BE}=(f_\mathrm{CL}-f_\mathrm{BE})/f_\mathrm{BE}$ of the numerical and analytic, \eq{freeBED}, Bose-Einstein distributions (CL, points) and the finite-size deviation $(\Delta f/f)^{N_{\tau}}_\mathrm{BE}=[f_\mathrm{BE}(\mathbf{k};N_{\tau})-f_\mathrm{BE}(\mathbf{k})]/f_\mathrm{BE}(\mathbf{k})$ due to the truncation of the Matsubara series (MS, dashed lines), as a function of $\beta(\varepsilon(\mathbf{k})-\mu)$.
		The data is obtained for the same parameters as in \ref{fig:freespectrum}, but for three different $N_{\tau}\in\{16,24,32\}$.
	}
	\label{fig:comparison}
\end{figure}

In order to further elucidate this phenomenon, we perform runs with a different number of lattice points $N_{\tau}$ along the imaginary-time direction. 
The respective deviations $(\Delta f/f)^\mathrm{CL}_\mathrm{BE}$ and $(\Delta f/f)^{N_{\tau}}_\mathrm{BE}$ for $N_\tau=16$, $24$, $32$ are compared with each other in figure \fig{comparison}. 
As one can see, the deviation of the CL data from the exact Bose-Einstein distributions decreases with increasing $N_\tau$, as predicted by the analytical finite-$N_\tau$ results shown as dashed lines.

\begin{figure}
	\centering\includegraphics[width=0.6\textwidth]{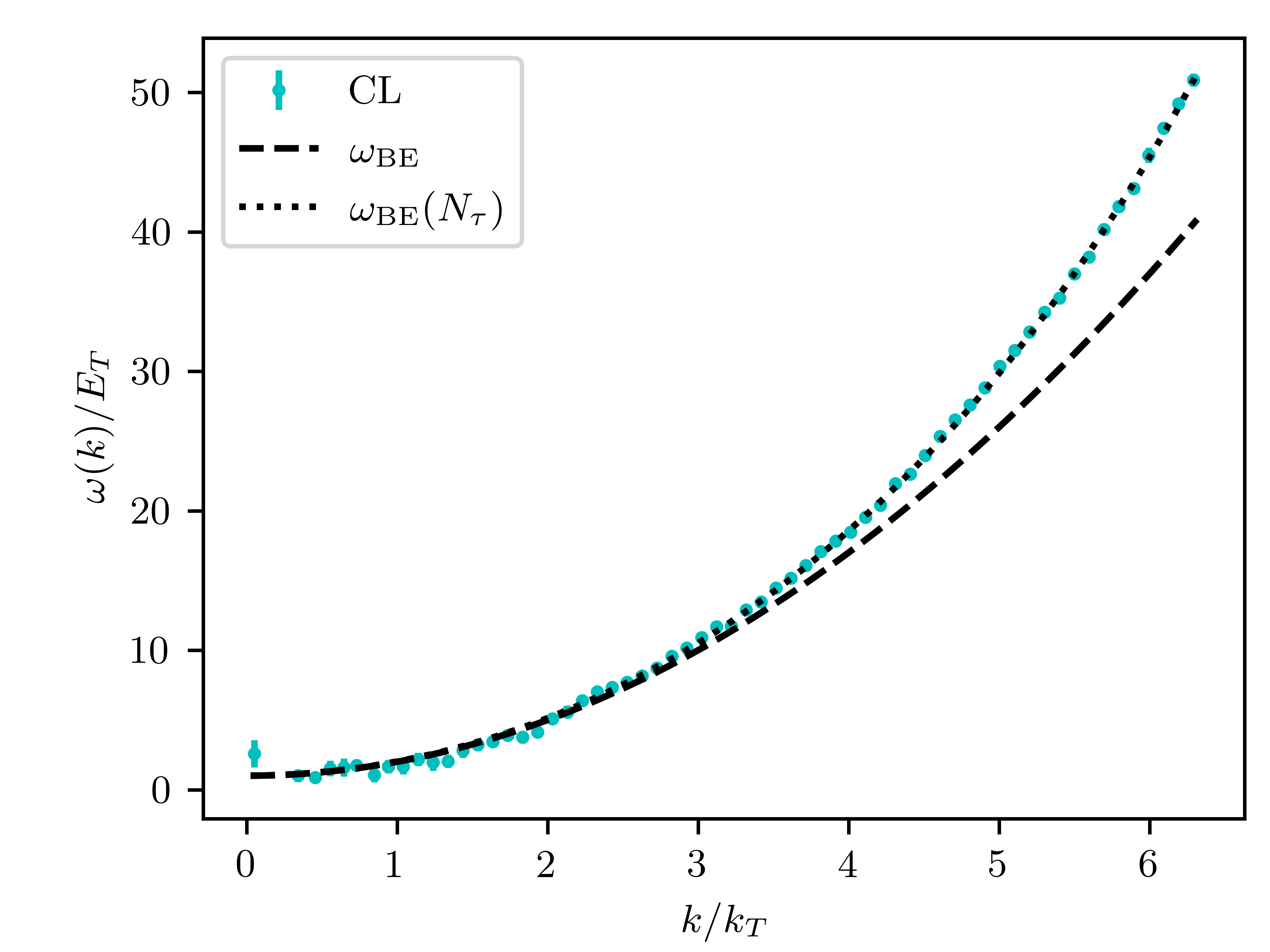}
	\caption{Dispersion $\omega_\mathrm{CL}(k)$ of the ideal Bose gas, as obtained according to \eq{disp} from the same data as used in \ref{fig:freespectrum} and averaged over the angular orientations of $\mathbf{k}$, in units of $E_T\equiv k_T^2/2m$. Note that for better visibility we doubled the width of the momentum bins in comparison to \ref{fig:freespectrum}.
		The black dashed line indicates the free-gas dispersion \eq{freegasdisp}, whereas the corresponding prediction \eq{freegasdispNtau} on a discrete temporal lattice is shown as a dotted black line.
		The momentum is measured in units of the inverse thermal de Broglie wave length $k_T=0.315\,a_\mathrm{s}^{-1}$.
	} 	
	\label{fig:freedisp}
\end{figure}

\begin{figure}	
	\centering\includegraphics[width=0.6\textwidth]{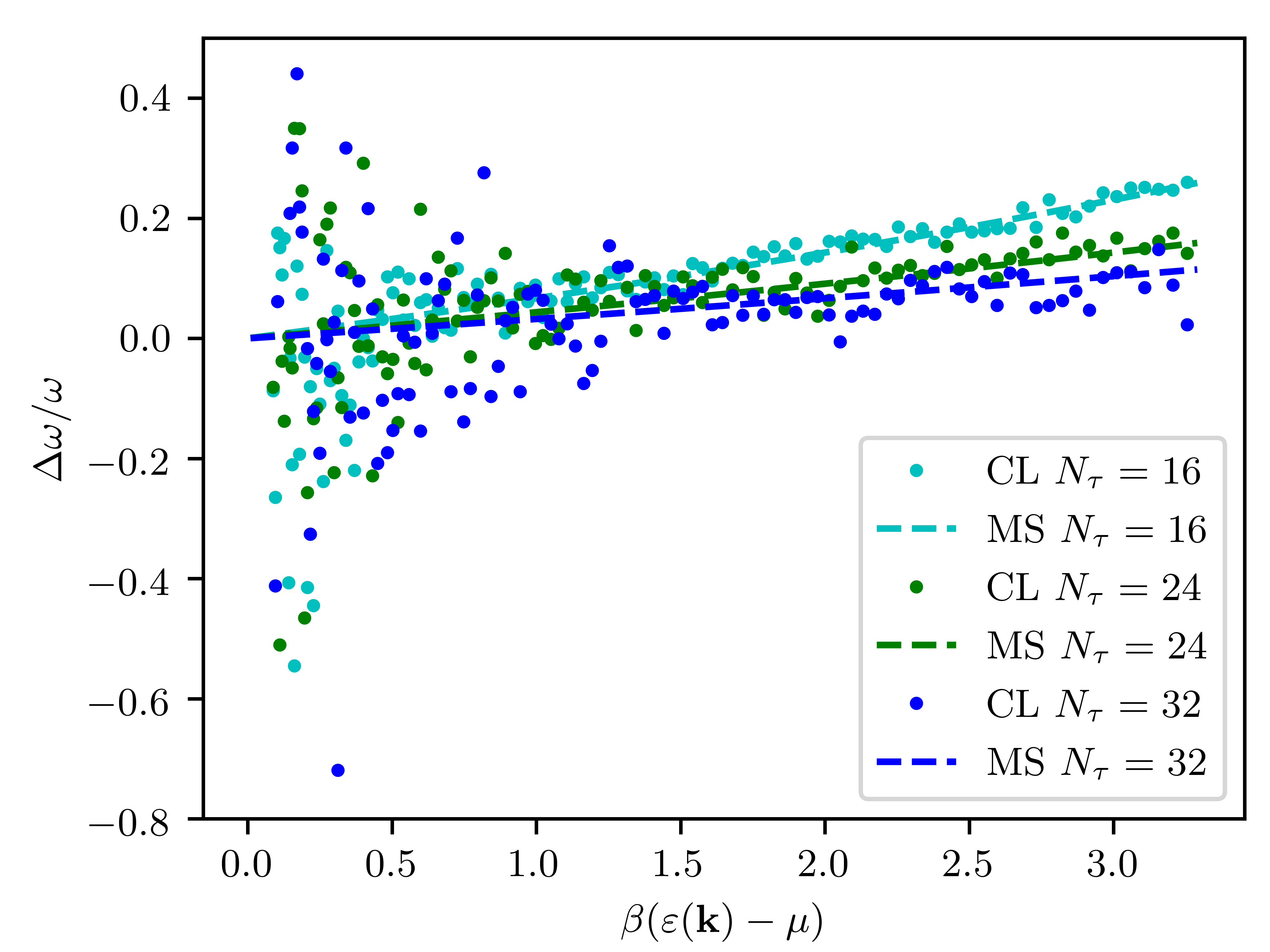}
	\caption{Comparison between the relative deviation $(\Delta \omega/\omega)^\mathrm{CL}_\mathrm{BE}=(\omega_\mathrm{CL}-\omega_\mathrm{BE})/\omega_\mathrm{BE}$ of the numerical and analytic, \eq{freegasdisp}, free gas dispersion and the finite-size deviation $(\Delta \omega/\omega)^{N_{\tau}}_\mathrm{BE}=[\omega_\mathrm{BE}(\mathbf{k};N_{\tau})-\omega_\mathrm{BE}(\mathbf{k})]/\omega_\mathrm{BE}(\mathbf{k})$ due to the truncation of the Matsubara series, as a function of $\beta(\varepsilon(\mathbf{k})-\mu)$.
		The data is obtained for the same parameters as in \ref{fig:freespectrum}, but for three different $N_{\tau}\in\{16,24,32\}$.
	}
	\label{fig:comparison2}
\end{figure}

As we recall from \sect{observ},  the dispersion relation $\omega(\mathbf{k})$ may be extracted as 
\begin{align}
\label{eq:disp2}
\omega_\mathbf{k}=\frac{1}{\beta}\int\limits_0^\beta d\tau\,\sqrt{-\frac{\partial_\tau\partial_{\tau'}\langle a^\dagger_{\mathbf{k}}(\tau)a_{\mathbf{k}}(\tau')\rangle|_{\tau=\tau'}}
{\langle a^\dagger_{\mathbf{k}}(\tau)a_{\mathbf{k}}(\tau')\rangle|_{\tau=\tau'}}}\,.
\end{align}  
The angular-averaged dispersion extracted via \eq{disp2} from the complex Langevin simulation,  $\omega_\mathrm{CL}({k})=\omega_\mathrm{CL}(|\mathbf{k}|)=\langle \omega_\mathrm{CL}(\mathbf{k})\rangle_{\Omega_\mathbf{k}}$, is shown for $N_\tau=16$ in figure \fig{freedisp}.
The exact dispersion is given by the free-gas kinetic energy, shifted by the chemical potential, 
\begin{align}
\label{eq:freegasdisp}
\omega_\mathrm{BE}(\mathbf{k})
&= \varepsilon(\mathbf{k})-\mu
\,,
\end{align}
while the finite-size dispersions result, as described in section \sect{disc_eff}, as
\begin{align}
\label{eq:freegasdispNtau}
\omega_\mathrm{BE}(\mathbf{k};N_\tau)
=\sqrt{\frac{-N_\tau^2T^{2}}{f(\mathbf{k};N_\tau)}
	\sum_{n=0}^{N_{\tau}-1}\frac{2e^{-2\pi i n/{N_\tau}}-e^{-4\pi i n/{N_\tau}}-1}
	{{N_\tau}(e^{2\pi i n/{N_\tau}}-1)+\beta[\varepsilon(\mathbf{k})-\mu]}}
\,.
\end{align}
%
The deviation between the CL and continuum analytic results is visibly larger, but can be attributed, as for the momentum spectrum $f$, to the finite lattice resolution in imaginary-time direction and thus to the corresponding truncation of the sum over the Matsubara frequencies. This is again corroborated by a comparison of the deviation for different $N_\tau$, as shown in figure \fig{comparison2}.

\subsection{Interacting gas above the BEC phase transition}
\label{sec:IntBECabovePT}
In the previous chapter, we have performed as a first benchmark complex Langevin simulations of the non-interacting, i.e. ideal Bose gas. This benchmark, however, is to a certain extent trivial, as in this case the Langevin equations are purely linear and can also be solved exactly. In the following, we will thus consider the interacting case, $g>0$, for which the Langevin equations become non-linear. 

\begin{figure}
	\centering\includegraphics[width=0.6\textwidth]{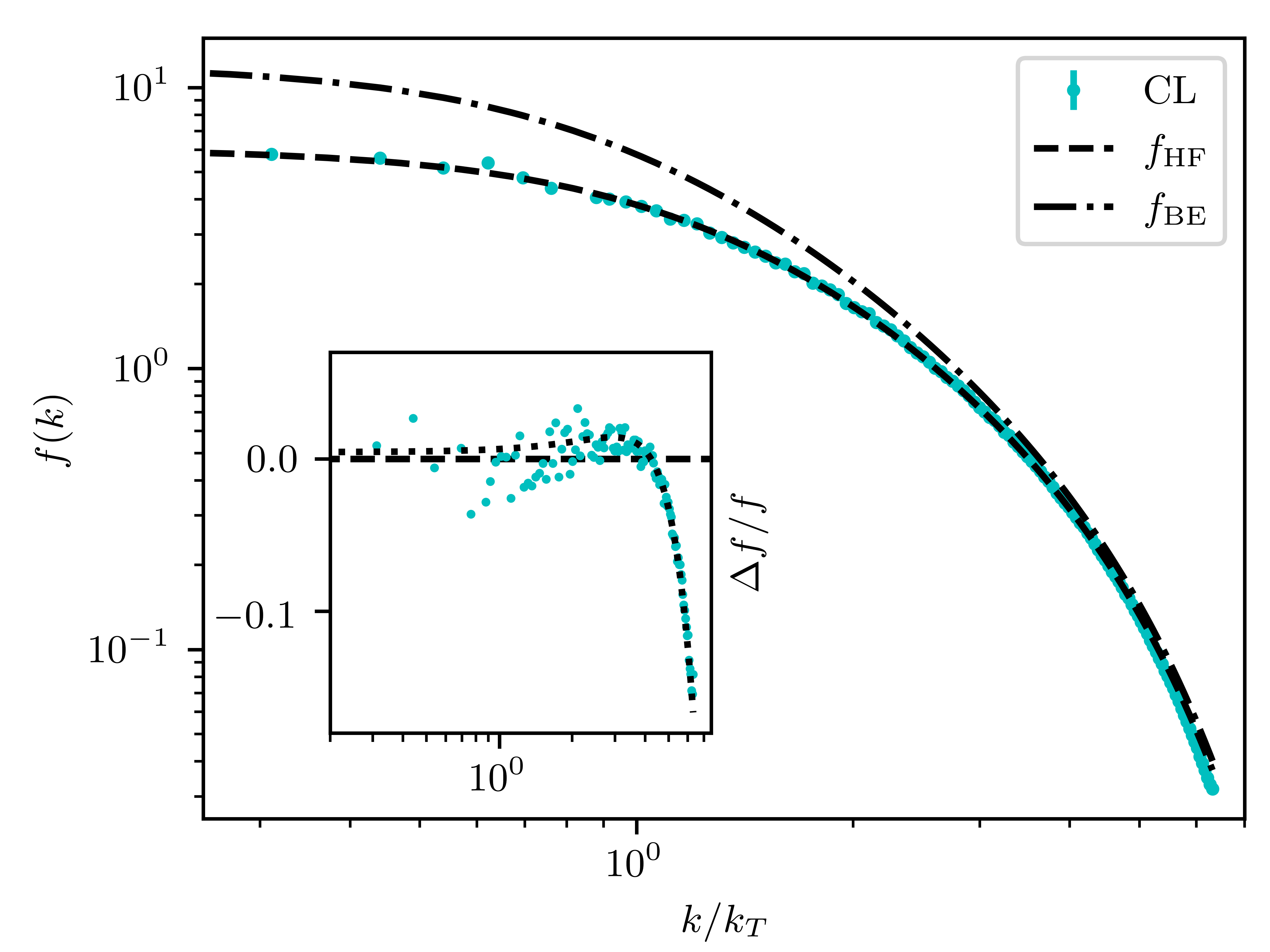}
	\caption{Angle-averaged momentum spectrum $f_\mathrm{CL}(k)$ of an interacting single-component gas above the BEC phase transition. 
		Temperature and chemical potential are chosen as in figure \fig{freespectrum}, the coupling strength is $g=1\,a_\mathrm{s}^2$, giving rise to a diluteness $\eta\approx 1.7\cdot 10^{-3}$.
		The momentum is measured in units of the inverse thermal de Broglie wave length $k_T=\lambda_T^{-1}=\sqrt{mT/2\pi}=0.315\,a_\mathrm{s}^{-1}$. 
		The CL data is obtained on a $64^{3}\times N_{\tau}$ lattice, with $N_{\tau}=16$ and averaged over $10$ runs.
		The spectrum fits well with the one calculated within the Hartree-Fock approximation (dashed black line). 
		For comparison we include the Bose-Einstein distribution of the corresponding non-interacting system (dash-dotted black line). 
		The inset shows the relative deviation $(\Delta f/f)^\mathrm{CL}_{\mathrm{HF}}=[f_\mathrm{CL}({k})-f_\mathrm{HF}({k})]/f_\mathrm{HF}({k})$. 
		The dotted black line in the inset represents the deviation $(\Delta f/f)^{N_{\tau}}_{\mathrm{HF}}=[f_\mathrm{HF}({k};N_{\tau})-f_\mathrm{HF}({k})]/f_\mathrm{HF}({k})$ of the finite-$N_{\tau}$ from the $N_{\tau}\to\infty$ version of the HF spectrum. 
	}
	\label{fig:abovetrans}
\end{figure}

In this section, we start with considering the case above the Bose-Einstein transition, where we can compare our numerical results to analytic predictions on the basis of the Hartree-Fock (HF) approximation. 
In the subsequent section, we will benchmark our CL results to Bogoliubov theory below the transition, for both one- and two-component systems.
Finally, we will explore the vicinity of the phase transition, where we can obtain a benchmark beyond perturbative approximations, by computing the relative shift of the critical temperature due to interactions, a quantity that is sensitive to non-perturbative fluctuations.

In the following, we will choose coupling strengths in the range from $g=0.1\,a_\mathrm{s}^2$ to $g=1\,a_\mathrm{s}^2$. According to the estimate \eq{diff_renorm_bare_coupling}, the corresponding difference between renormalized and bare coupling is then in the single-digit percentage range. For the purposes of this chapter, such deviation is negligible such that we will in the following not explicitly distinguish the bare and renormalized coupling constant. 

We choose the same values for temperature and chemical potential as in the previous section, $\mu=-0.1\,a_\mathrm{s}^{-1}$ and $T=1.25\,a_\mathrm{s}^{-1}$, but now turn on a non-vanishing, positive coupling constant $g=1\,a_\mathrm{s}^2$. 
Our simulation yields, for the chosen parameters, a density of $\rho= (0.0447\pm0.002)\,a_\mathrm{s}^{-3}$, where we extract $\rho$ as described in section \sect{observ}.
The corresponding diluteness parameter $\eta=\sqrt{\rho a^3}$ amounts to $\eta\approx 1.7\cdot 10^{-3}$, which is within the range of typical cold-gas experiments.
Figure \fig{abovetrans} shows the momentum spectrum of the dilute gas in analogy to the ideal-gas spectrum depicted in figure \fig{freespectrum}, on a $64^{3}\times N_{\tau}$ lattice, with $N_{\tau}=16$.

As we recall from section \sect{HF}, in three dimensions, the density obeys the following relationship in Hartree-Fock approximation:
\begin{align}
\label{eq:selfkons}
\rho=\lambda_{T}^{-3}\,g_{3/2}\left(e^{\,\beta(\mu-2g\rho)}\right)
\,.
\end{align}
To determine $\rho$, we numerically solve \eq{selfkons}.
We obtain $\rho_\mathrm{HF}=0.0454\,a_\mathrm{s}^{-3}$ which can be compared to the density obtained from our CL simulations as indicated above, with the small systematic deviation due to finite-size and discretization effects. 
The density yields an effective chemical potential $\mu'=\mu-2g\rho=-0.191\,a_\mathrm{s}^{-1}$, which gives the Hartree-Fock momentum distribution $f_\mathrm{HF}(k)=f_\mathrm{BE}(k,\mu')$. In figure \fig{abovetrans}, the Hartree-Fock spectrum is shown together with the free Bose-Einstein distribution, from which it substantially deviates, and with the data from the complex Langevin simulation. We observe good agreement between the CL data and the Hartree-Fock spectrum, the only deviations occurring at high momenta, as in the non-interacting case.

The inset of figure \fig{abovetrans} shows the relative deviation $(\Delta f/f)^\mathrm{CL}_{\mathrm{HF}}=[f_\mathrm{CL}({k})-f_\mathrm{HF}({k})]/f_\mathrm{HF}({k})$ from the expected Hartree-Fock result and compares it to the deviation $(\Delta f/f)^{N_{\tau}}_{\mathrm{HF}}=[f_\mathrm{HF}({k};N_{\tau})-f_\mathrm{HF}({k})]/f_\mathrm{HF}({k})$ of the finite-$N_{\tau}$ from the $N_{\tau}\to\infty$ version of the HF spectrum, for $N_{\tau}=16$ points. Again, the remaining discrepancy disappears when taking into account the discretization effects.

\subsection{Bose gas in the condensed phase}
In this section, we consider a Bose gas far below the Bose-Einstein transition, in the condensed phase. 
In a condensate, the pattern of possible collective excitations fundamentally changes due to spontaneous symmetry breaking. According to the Goldstone theorem, this symmetry breaking leads to massless elementary excitations, which in the case of a Bose gas are sound-like phase fluctuations. In the case of a $U(\mathcal{N})$-symmetric multi-component system, as described by Hamiltonian \eq{H_UN}, there are additional massive elementary excitations corresponding to density variations between the components. In order to explore the applicability of the CL algorithm also in the latter case, we will not only simulate a one-component system as in the previous sections, but also a $U(2)$-symmetric two-component gas.

We will benchmark our simulations to the predictions of Bogoliubov theory, which is the suitable theoretical description for the weakly interacting Bose gas in the condensed phase, compare section \sect{Bog_theo}.

\subsubsection{One-component system}
We start with the single-component system, $\mathcal{N}=1$, for which we now choose a temperature $T=0.625\,a_\mathrm{s}^{-1}$, chemical potential $\mu=0.5\,a_\mathrm{s}^{-1}$, and coupling constant $g=0.1\,a_\mathrm{s}^2$. 
This results in a total density of $\rho\approx\mu/g=5\,a_\mathrm{s}^{-3}$, corresponding to a diluteness $\eta = 0.56\cdot 10^{-3}$.
As before, we work on a $64^{3}\times N_{\tau}$  lattice, with $N_\tau=32$.

It turned out that initializing the field as $\psi(\tau,\mathbf{x};\vartheta=0)=0$ for the Langevin evolution, as we did for all runs above the phase transition, the Langevin process is not able to build up a condensate, i.e. a highly over-occupied $\mathbf{k}=0$ mode even at temperatures far below the transition, at least not within the limited simulation time available.
This could be expected for large systems as the relative size of the phase space covered by the condensate mode is inversely proportional to the system size.

\begin{figure}
	\centering\includegraphics[width=0.6\textwidth]{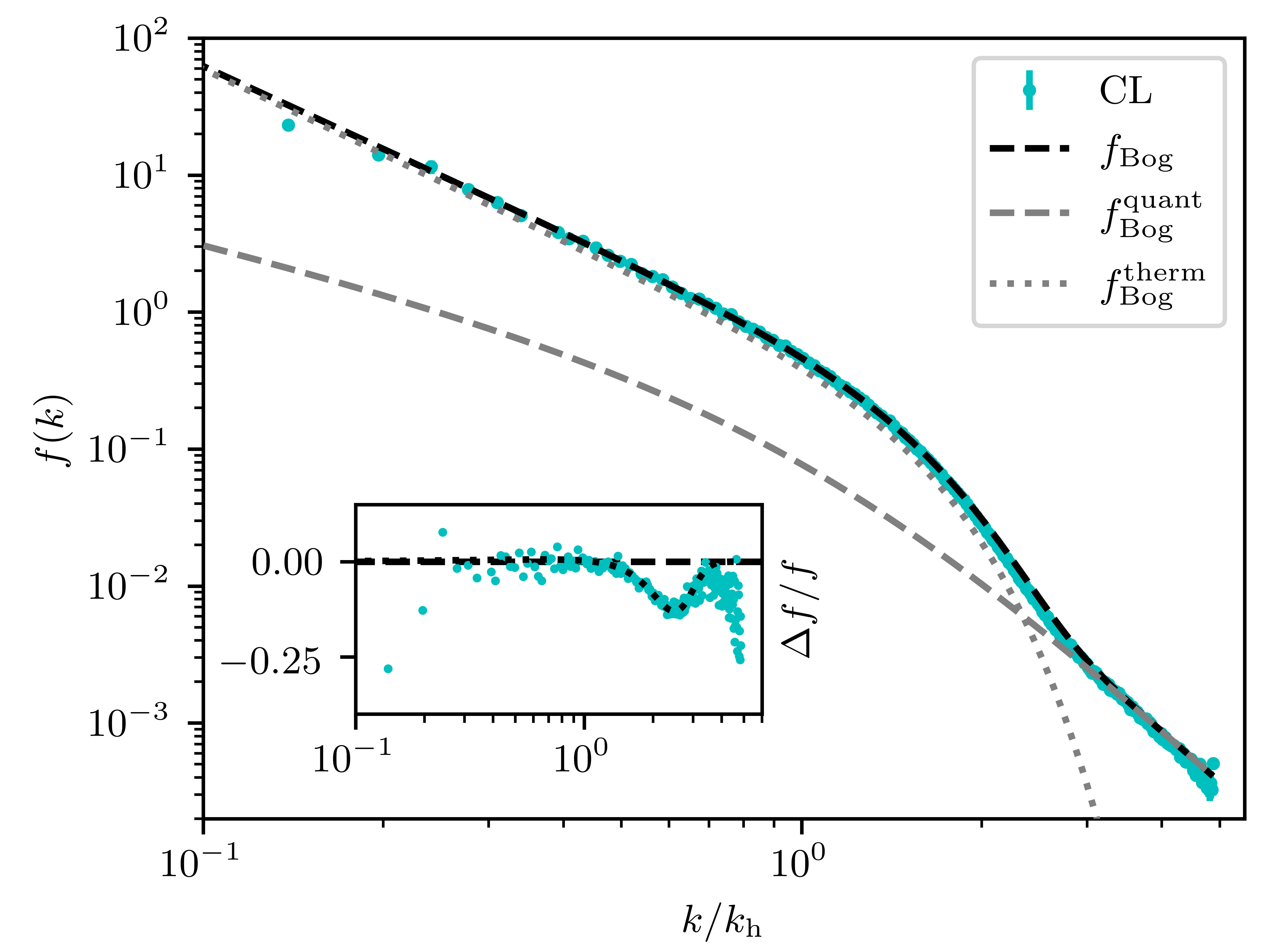}
	\caption{Momentum spectrum of the interacting single-component gas below the phase transition, at temperature  $T=0.625\,a_\mathrm{s}^{-1}$, chemical potential $\mu=0.5\,a_\mathrm{s}^{-1}$, and for a coupling $g=0.1\,a_\mathrm{s}^2$, corresponding to a diluteness $\eta\approx0.56\cdot10^{-3}$.
		It is obtained from CL dynamics on a $64^{3}\times (N_{\tau}=32)$ lattice, averaged over $4$ runs starting from a state with constant density $\rho_{0}=\mu/g=5\,a_\mathrm{s}^{-3}$. The CL process leads to a slightly larger condensate density  $\rho_{0}=f(0)/\mathcal{V}=(5.0309\pm0.0002)\,a_\mathrm{s}^{-3}$, corresponding to $f(0)\approx1.32\cdot10^{6}$, while the condensate depletion becomes $\rho'\equiv\rho-\rho_0=(0.0149\pm0.0001)\,a_\mathrm{s}^{-3}$.
		Momenta are given in units of the healing momentum $k_\mathrm{h}=\sqrt{2mg\rho_0}=0.71\,a_\mathrm{s}^{-1}$.
		The dashed black line represents the spectrum \eq{bogol_spectrum} predicted by Bogoliubov theory, consisting of a thermal (dotted gray line) and a quantum depletion (dashed grey line) part.
		The inset shows the relative deviation  $(\Delta f/f)^\mathrm{CL}_\mathrm{Bog}=(f_\mathrm{CL}-f_\mathrm{Bog})/f_\mathrm{Bog}$ of the CL data from the Bogoliubov spectrum. 
		The dotted black line indicates an estimate of the deviation $(\Delta f/f)^{N_{\tau}}_{\mathrm{Bog}}=[f_\mathrm{Bog}({k};N_{\tau})-f_\mathrm{Bog}({k})]/f_\mathrm{Bog}({k})$ of the finite-$N_{\tau}$ from the $N_{\tau}\to\infty$ version of the Bogoliubov spectrum, cf.~the main text.
	}
	\label{fig:belowtrans}
\end{figure}

\begin{figure}	
	\centering\includegraphics[width=0.6\textwidth]{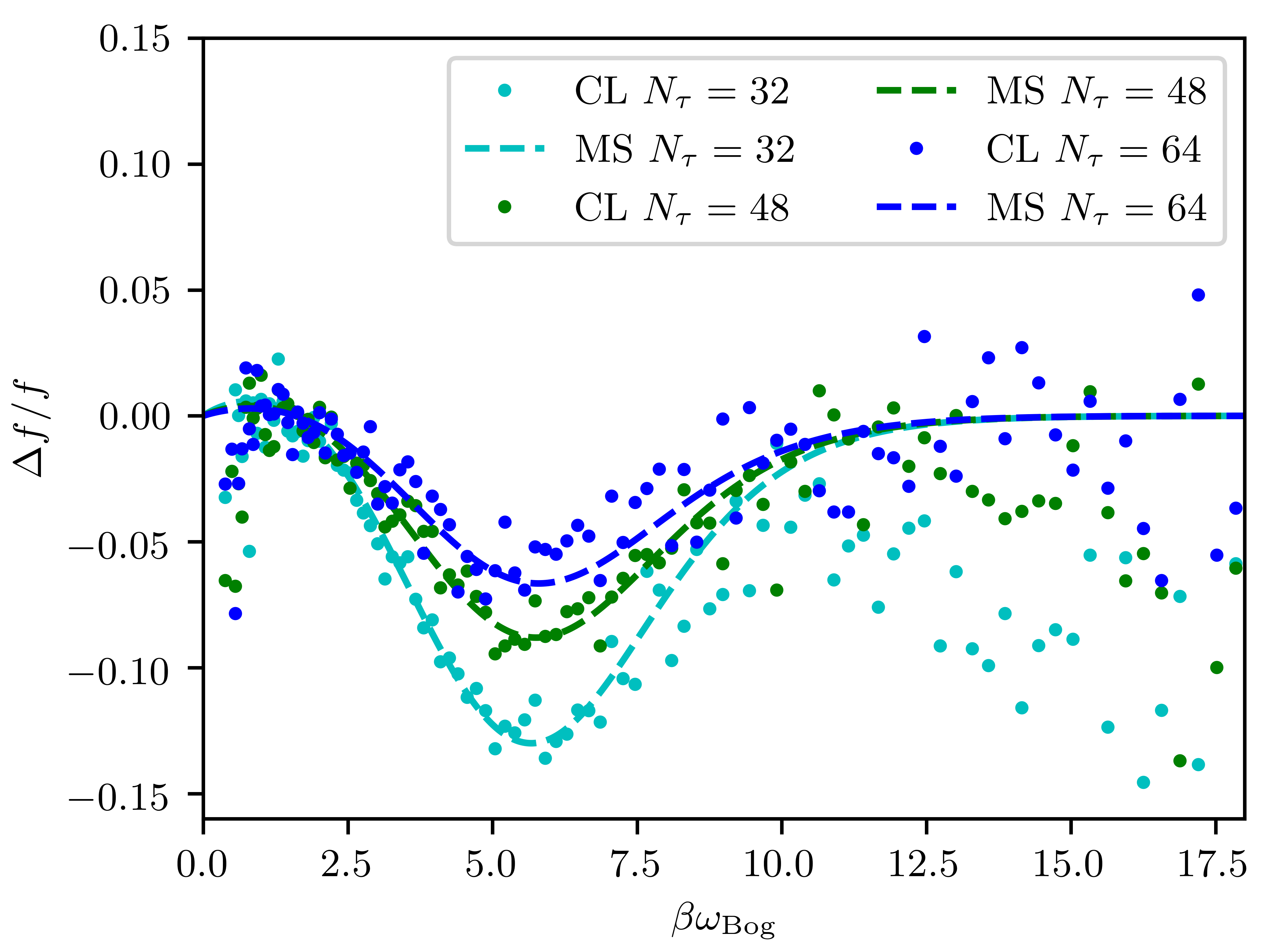}
	\caption{Comparison between the relative deviation $(\Delta f/f)^\mathrm{CL}_\mathrm{Bog}=(f_\mathrm{CL}-f_\mathrm{Bog})/f_\mathrm{Bog}$ of the numerical and Bogoliubov, \eq{bogol_spectrum}, distributions (CL, points) and the finite-size deviation $(\Delta f/f)^{N_{\tau}}_{\mathrm{Bog}}=[f_\mathrm{Bog}({k};N_{\tau})-f_\mathrm{Bog}({k})]/f_\mathrm{Bog}({k})$ due to the truncation of the Matsubara series (MS, dashed lines), as a function of $\beta\omega_{\mathrm{Bog}}$.
		The data is obtained for the same parameters as in figure \fig{belowtrans}, but for three different $N_{\tau}\in\{32,48,64\}$. Note that for better visibility we doubled the width of the momentum bins in comparison to figure \fig{belowtrans}.
	}
	\label{fig:comparisonBog}
\end{figure}

When starting instead from a spatially uniform, non-vanishing field configuration given by the mean-field value $\psi(\tau,\mathbf{x})\equiv\sqrt{\mu/g}$, the CL process leads to only a slight modification, which consists in a thermal and quantum depletion of the condensate mode and the buildup of the corresponding distribution of momentum excitations.

From Bogoliubov theory one expects the momentum distribution of the single-component gas to be, cf. section \sect{Bog_theo},
\begin{align}
\label{eq:bogol_spectrum}
f_{\mathrm{Bog}}(\mathbf{k})
&=\frac{1+2v_{\mathbf{k}}^{2}}{e^{\,\beta\omega_{\mathrm{Bog}}(\mathbf{k})}-1}
+v_{\mathbf{k}}^{2}
\,,
\end{align}
where
\begin{align}
\label{eq:bogol_v}
v_{\mathbf{k}}
&=\sqrt{\frac{1}{2}\left[\frac{\varepsilon(\mathbf{k})+g\rho_0}{\omega_{\mathrm{Bog}}(\mathbf{k})}-1\right]}
\,,
\end{align}
in which $\rho_{0}\approx\mu/g$ denotes the condensate density and the Bogoliubov dispersion of the elementary excitations is given by
\begin{align}
\label{eq:omegaBog3}
\omega_{\mathrm{Bog}}(\mathbf{k})
&=\sqrt{\varepsilon(\mathbf{k})\left[\varepsilon(\mathbf{k})+2g\rho_0\right]}
\,.
\end{align}
The second contribution to \eq{bogol_spectrum}, $v_{\mathrm{k}}^{2}$, gives rise to the quantum depletion, i.e. it accounts for all particles that are not in the condensate mode due to the interactions even at $T=0$.
The first term, which is proportional to the thermal occupancy of the quasiparticles, accounts for the fraction of particles which are non-condensed due to thermal excitations.

In figure \fig{belowtrans}, we show the analytically predicted spectrum \eq{bogol_spectrum} as a black dashed line together with the numerically computed one, which hardly deviate from each other, as reflected by the relative deviation $(\Delta f/f)^\mathrm{CL}_\mathrm{Bog}=(f_\mathrm{CL}-f_\mathrm{Bog})/f_\mathrm{Bog}$ depicted in the inset.
The grey dotted and dashed lines indicate the contributions to the analytic spectrum \eq{bogol_spectrum}, which account for the thermal and quantum depletion of the condensate, respectively.

We estimate the deviation caused by a finite $N_\tau$ by replacing the thermal distribution of non-interacting quasiparticles in \eq{bogol_spectrum} by the finite-$N_{\tau}$ expression, analogously to the ideal gas,
\begin{align}
\frac{1}{e^{\,\beta\omega_{\mathrm{Bog}}(\mathbf{k})}-1}
\to\sum_{n=0}^{N_{\tau}-1}\frac{1}{{N_\tau}(e^{2\pi i n/{N_\tau}}-1)+\beta\omega_{\mathrm{Bog}}(\mathbf{k})}
\,.
\end{align}
As can be inferred from the inset of figure \fig{belowtrans}, this captures the residual deviation between the CL and Bogoliubov distributions up to momenta above which the quantum depletion becomes more relevant than the thermal one.

In figure \fig{comparisonBog}, we compare the deviations of the distributions obtained with CL and the Bogoliubov distribution with the respective deviations of the finite-$N_{\tau}$ estimate from the full Bogoliubov distribution, for three different $N_{\tau}\in\{32,48,64\}$.
The results demonstrate that the Matsubara series truncation also affects the region of high momenta where $v_{k}^{2}$ dominates the spectrum.

\begin{figure}	
	\centering\includegraphics[width=0.6\textwidth]{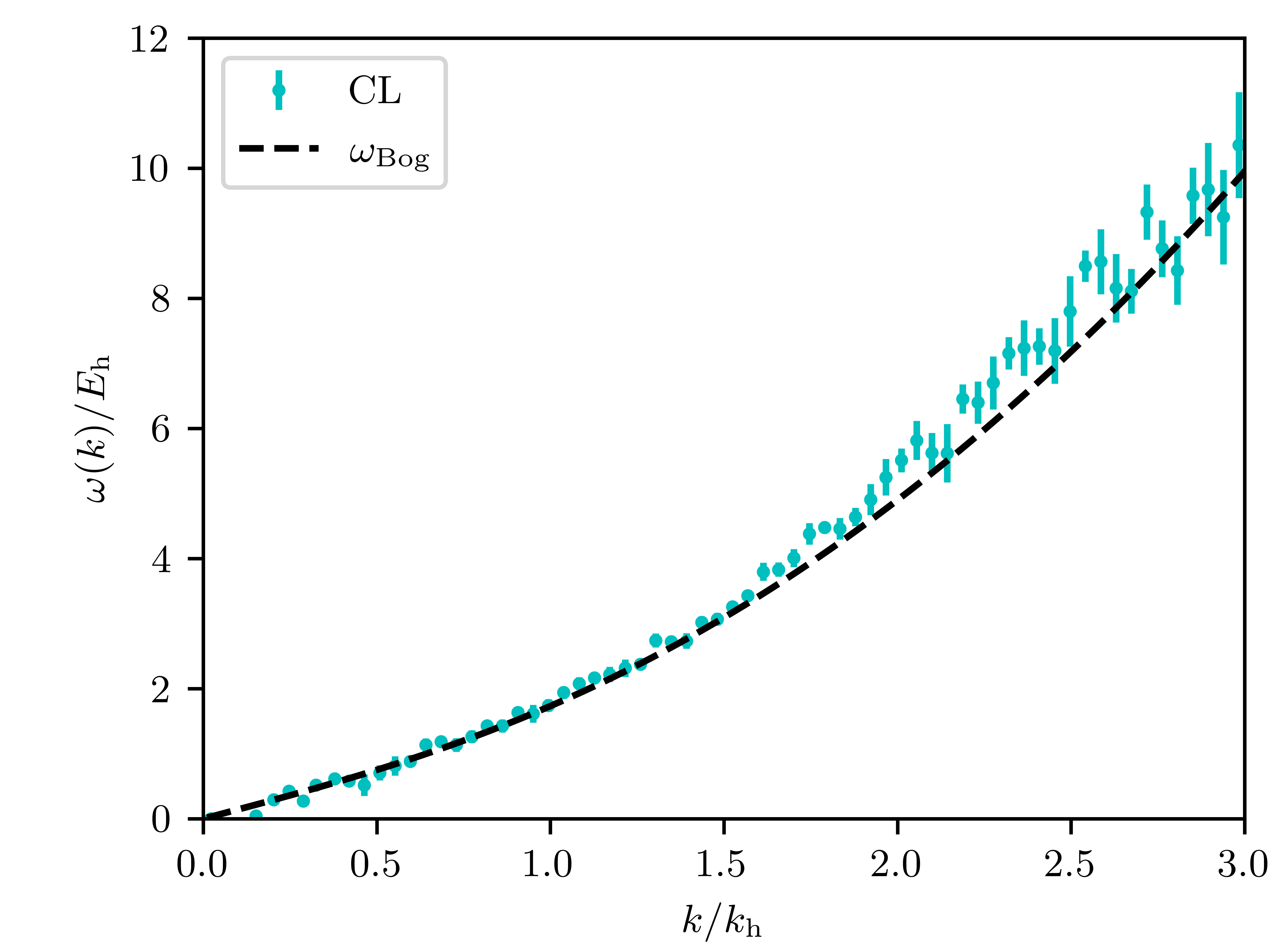}
	\caption{Dispersion of the interacting single-component gas below the phase transition in units of $E_\mathrm{h}\equiv k_\mathrm{h}^2/2m$, as obtained according to \eq{disp} from the same data as used in figure \fig{belowtrans} and averaged over the angular orientations of $\mathbf{k}$. Note that for better visibility we doubled the width of the momentum bins in comparison to figure \fig{belowtrans}.
		For comparison we show the dispersion predicted by Bogoliubov theory as a black dashed line.
	}
	\label{fig:belowtrans_disp}
\end{figure}

\begin{figure}
	\centering\includegraphics[width=0.6\textwidth]{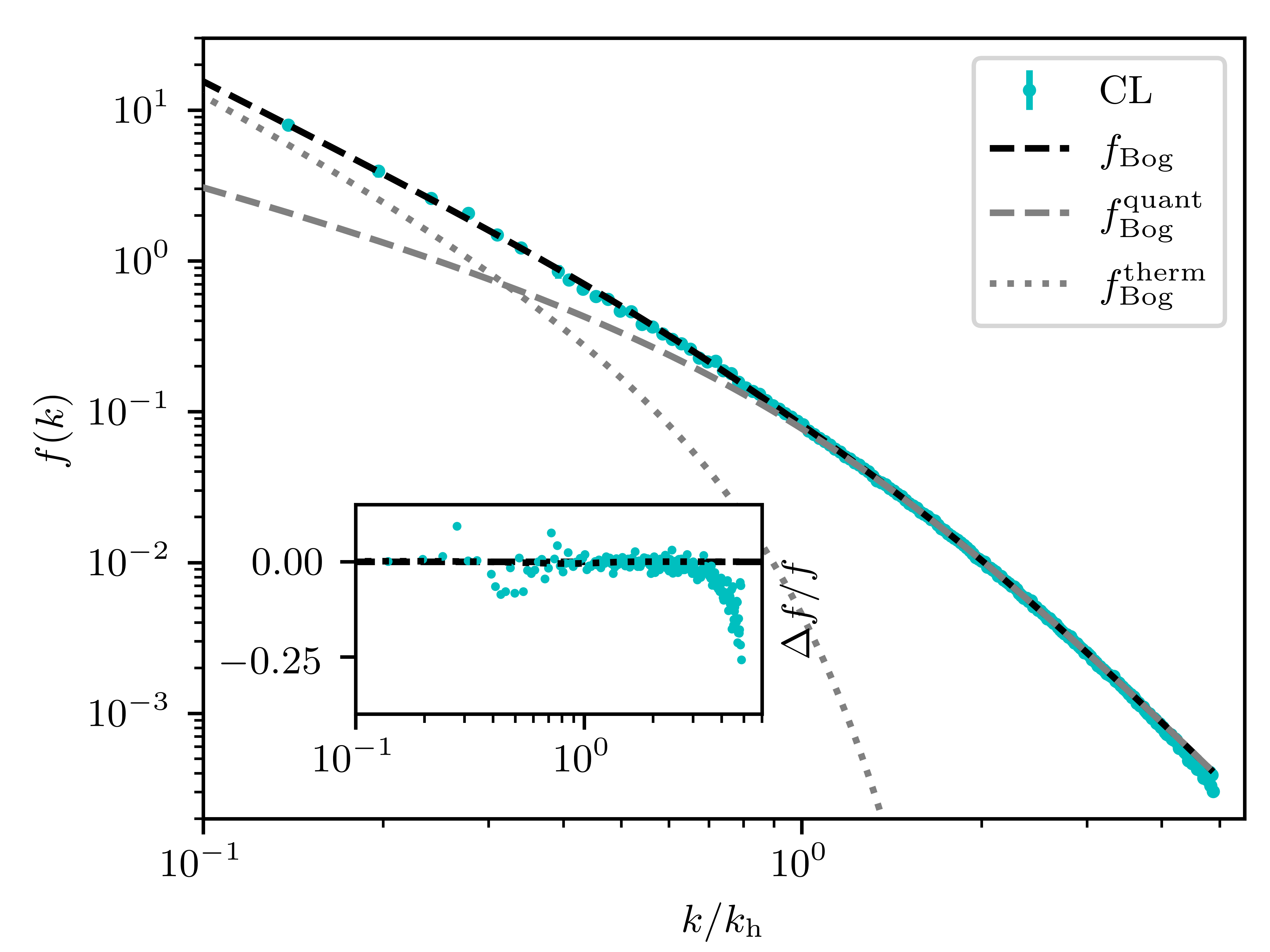}
	\caption{The same as in Fig. \ref{fig:belowtrans}, but for $T=0.15625\,a_\mathrm{s}^{-1}$ and $N_\tau=128$. At this low temperature, large parts of the spectrum are dominated by the quantum depletion, yet CL is still able to precisely describe the spectrum.
	}
	\label{fig:belowtrans_lowT}
\end{figure}
\begin{figure}
	\centering\includegraphics[width=0.6\textwidth]{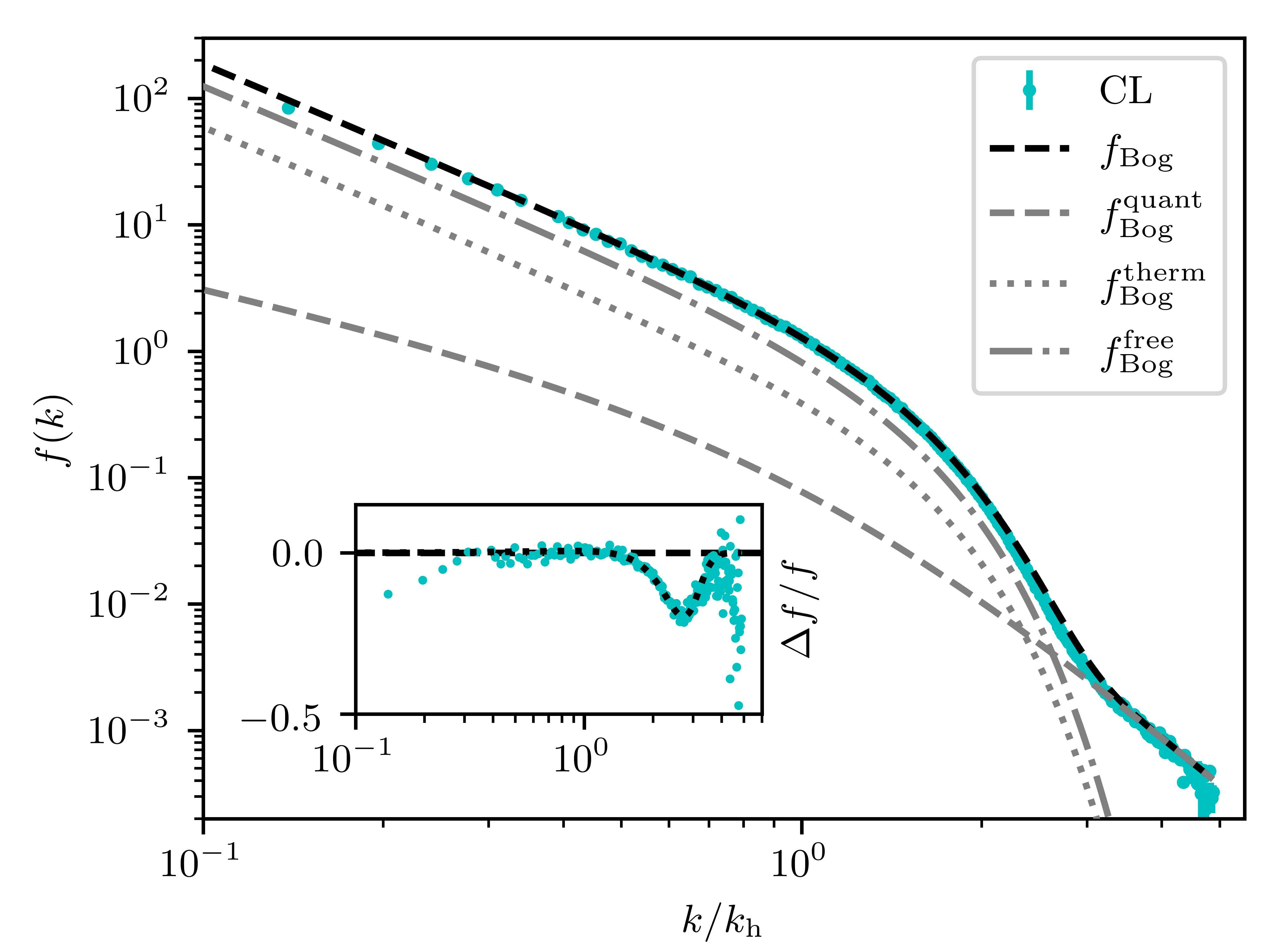}
	\caption{Momentum spectrum of the interacting two-component gas below the phase transition.
		The system is described by the U$(2)$-symmetric Hamiltonian \eq{H_UN}, i.e. $\mathcal{N}=2$.
		Parameters are, as before, $T=0.625\,a_\mathrm{s}^{-1}$, $\mu=0.5\,a_\mathrm{s}^{-1}$, and $g=0.1\,a_\mathrm{s}^2$, $N_{\tau}=32$, averaging over $3$ runs. 
		The spectrum \eq{bogol_spectrumON} predicted by $\mathcal{N}$-component Bogoliubov theory is represented by the black dashed line. 
		For comparison, we have also plotted the quantum (dashed gray line) and thermal (dotted gray line) contributions, as well as the contribution from the free relative number excitations between the two internal components (dashed-dotted line). 
		The inset shows the relative deviation $(\Delta f/f)^\mathrm{CL}_\mathrm{Bog,tot}=(f_\mathrm{CL}-f_\mathrm{Bog,tot})/f_\mathrm{Bog,tot}$ of the CL data from the Bogoliubov spectrum. 
		The dotted black line shows an estimate of the deviation $(\Delta f/f)^{N_{\tau}}_{\mathrm{Bog,tot}}=[f_\mathrm{Bog,tot}({k};N_{\tau})-f_\mathrm{Bog,tot}({k})]/f_\mathrm{Bog,tot}({k})$ of the finite-$N_{\tau}$ from the $N_{\tau}\to\infty$ version of the Bogoliubov spectrum, cf.~the main text.
	}
	\label{fig:belowtransU2}
\end{figure}

Also the Bogoliubov dispersion, which we show in figure \fig{belowtrans_disp} together with the numerical data, is well reproduced by the CL simulations.

Finally, it is important to note that the CL method is not only able to describe the part of the momentum spectrum which is dominated by thermal excitations but also to reproduce the occupation numbers at high momenta where the spectrum is dominated by the quantum contribution. In order to further corroborate this observation, we also performed a simulation for a much smaller temperature, $T=0.15625\,a_\mathrm{s}^{-1}$, i.e. we increased $N_\tau$ to $128$. For such low temperatures, the spectrum is almost entirely dominated by the quantum depletion. Still, the spectrum is accurately reproduced by CL, see figure \fig{belowtrans_lowT}. This gives hope that CL is a suitable tool for performing exact simulations also in the quantum regime.

\subsubsection{Two-component system}
We now repeat the above simulations for the case of a two-component system, with U$(2)$ symmetric Hamiltonian \eq{H_UN}, \eq{K_UN}.
From Bogoliubov theory (cf. section \sect{Bog_theo}), one obtains, as mentioned initially, besides the gapless elementary excitations \eq{omegaBog3}, an additional massive mode with dispersion given by \eq{epsilonk} for each of the remaining $\mathcal{N}-1$ degrees of freedom.
For the case of general $\mathcal{N}$, the total momentum distribution reads
\begin{align}
\label{eq:bogol_spectrumON}
f_{\mathrm{Bog},\mathrm{tot}}(\mathbf{k})
= \sum_{\alpha=1}^{\mathcal{N}}\langle a_{\mathbf{k},\alpha}^\dagger a_{\mathbf{k},\alpha}\rangle
=\frac{1+2v_{\mathbf{k},\mathcal{N}}^{2}}{e^{\,\beta\omega_{\mathrm{Bog}}(\mathbf{k})}-1}
+v_{\mathbf{k},\mathcal{N}}^{2}
+\frac{\mathcal{N}-1}{e^{\,\beta\varepsilon(\mathbf{k})}-1}
\,,
\end{align}
where
\begin{align}
\label{eq:bogol_vON}
v_{\mathbf{k},\mathcal{N}}
&=\sqrt{\frac{1}{2}\left[\frac{\varepsilon(\mathbf{k})+\mathcal{N}g\rho_0}{\omega_{\mathrm{Bog}}(\mathbf{k})}-1\right]}
\,,
\end{align}
and $\rho_0$ is the condensate density in every single component.

We show the results of our CL simulations in \ref{fig:belowtransU2} and compare them to the total Bogoliubov distribution \eq{bogol_spectrumON}, as well as to the quantum and thermal contributions from the Bogoliubov and free modes, respectively.

\subsection{Interacting system at the transition}
\label{sec:IntBGatCriticality}
The previous sections on the interacting Bose gas in the condensed and thermal phase demonstrated that the complex Langevin method works well in a non-trivial setting. However, the very fact that comparatively simple analytical approximations such as Bogoliubov and Hartree-Fock theory can well describe the resulting momentum spectra shows that these scenarios are not the most stringent test beds for the algorithm. In this section, we will thus consider the Bose gas close to the Bose-Einstein phase transition, which is subject to much stronger, non-perturbative fluctuations that are hard to capture by purely analytical approaches.  

We start by considering the occupation number spectrum to show a pure Rayleigh-Jeans power law in the IR as a signature of the phase transition. Thereafter, we will employ complex Langevin to determine the shift of the critical temperature due to interactions, a genuinely non-perturbative quantity.

\subsubsection{Rayleigh-Jeans scaling}
At the transition, we expect the occupation number to show Rayleigh-Jeans scaling,
\begin{align}
f(\mathbf{k})\sim \frac{1}{|\mathbf{k}|^2}
\,,
\end{align}
for small $|\mathbf{k}|$ in the infinite-volume limit. 
In principle, this could provide a method for determining the transition point, which, in practice, however, is suited only for obtaining a rough estimate of where the phase transition occurs. 
On the one hand, statistical errors are in general large for the relevant small momentum modes; 
on the other hand, in finite-size systems,  even at the transition, Rayleigh-Jeans scaling is not expected to be realized down to arbitrarily small momenta.

As reported in the previous section, the formation of a condensate by the Langevin process can take unrealistically long computation times. 
Hence, for simulations below the transition, a condensate must be seeded by choosing a non-vanishing zero-mode population in the initial configuration for the Langevin process. 
However, since we do not know a priori at which chemical potential condensation occurs, it is most convenient to approach the critical point from the non-condensed phase. 
Nonetheless, the fact that the Langevin equilibration time strongly increases for a condensed state gives us a further (numerical) indication for determining the transition point.

We choose a temperature $T=1.25\,a_\mathrm{s}^{-1}$ and a coupling $g=0.5\,a_\mathrm{s}^2$, resulting in a density $\rho\approx0.08\,a_\mathrm{s}^{-3}$ and thus a diluteness of $\eta\approx 0.8\cdot 10^{-3}$. 
From HF theory, one obtains the transition to occur at $\mu=2g\rho_\mathrm{c}^0$, with $\rho_\mathrm{c}^0$ the critical density in the free system. 
This amounts to $\mu_\mathrm{c}^{\text{HF}}/T=0.066$ for chosen temperature and coupling.

\begin{figure}[t]
	\centering\includegraphics[width=0.6\textwidth]{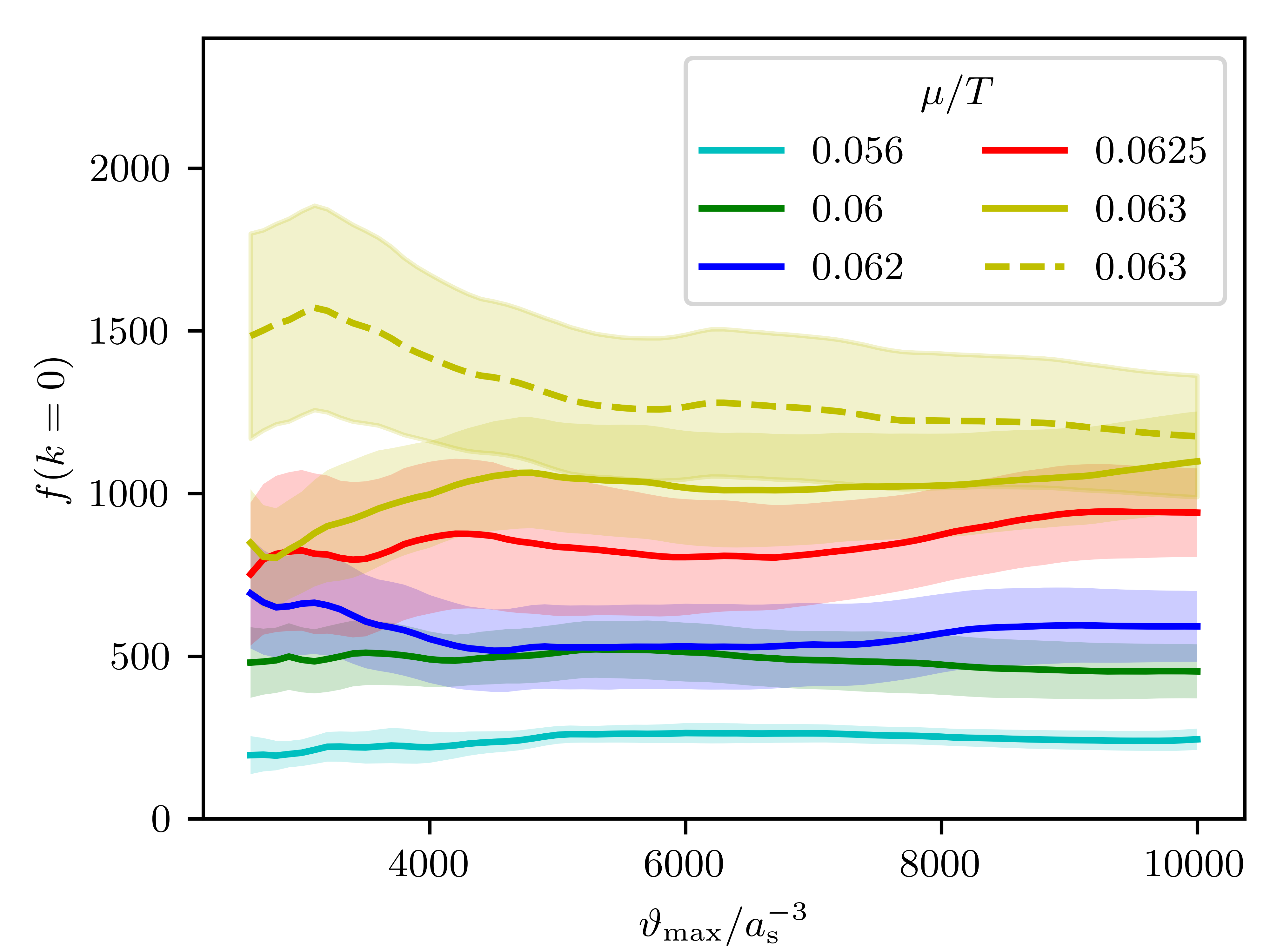}
	\caption{Langevin evolution of the zero mode occupancy $f(0)$ averaged over Langevin times $\vartheta=\vartheta_{0},\dots,\vartheta_{\mathrm{max}}$, cf.~\eq{f0CLaverage}, as a function of $\vartheta_{\mathrm{max}}$, with $\vartheta_0=2.5\cdot10^3\,a_\mathrm{s}^{-3}$, for five different chemical potentials $\mu$ close to the transition, with chemical potential decreasing from the uppermost to the lowermost curve, at a temperature  $T=1.25\,a_\mathrm{s}^{-1}$, coupling $g=0.5\,a_\mathrm{s}^2$.
		The data is obtained on a $64^{3}\times (N_{\tau}=16)$ lattice and the resulting densities $\rho\approx0.08\,a_\mathrm{s}^{-3}$ correspond to a diluteness $\eta\approx0.8\cdot10^{-3}$.
		For $\mu/T=0.056$, $0.06$, $0.062$, $0.0625$, no seed of the zero mode was given. 
		For $\mu/T=0.063$, we simulated with a seed of $f(0)=0$ (solid line) and  $f(0)\approx1475$ ($\psi=0.075\,a_\mathrm{s}^{-3/2}$) (dashed line). 
		Error bands are obtained from the variance of $10$ independent runs.
	}
	\label{fig:timeevol}
\end{figure}
We perform simulations for five values of the chemical potential between $\mu/T=0.056$ and $\mu/T=0.063$. 
The Langevin time evolution of the running average of the zero-mode occupancy, 
\begin{align}
\langle f(0,\vartheta)\rangle_{\vartheta_0}^{\vartheta_{\mathrm{max}}}
=\frac{1}{\vartheta_\text{max}-\vartheta_0}\int\limits_{\vartheta_0}^{\vartheta_\text{max}}d\vartheta \,f(k=0,\vartheta)
\,,
\label{eq:f0CLaverage}
\end{align}
is shown in figure \fig{timeevol}, as a function of the maximum Langevin time $\vartheta_{\mathrm{max}}$ and for $\vartheta_0=2.5\cdot10^3\,a_\mathrm{s}^{-3}$.
While we observe a fast convergence in $\vartheta_{\mathrm{max}}$ for $\mu/T=0.056$ and $0.06$, i.e.~still rather far away from the transition, the convergence of $f(0)$ appears to be more and more slowed down for $\mu/T=0.062$, $0.0625$, and $0.063$, especially for the latter value of the chemical potential, giving a first hint that the system approaches the transition.
In order to make sure that our results are reliable and not biased by insufficient simulation time, we performed additional runs with a non-zero seed of $\psi=0.075\,a_\mathrm{s}^{-3/2}$, corresponding to $f(0)\approx1475$, for the case of the largest chemical potential, $\mu/T=0.063$. 
The error bands indicate the variance over $10$ independent runs and thus illustrate the demand for statistics in the zero mode.
The long-time zero-mode occupancy appears to converge for all chemical potentials and especially the two simulations with different seed for $\mu/T=0.063$ are consistent with each other within the error bands. 
All runs used in the remainder of this chapter have been performed without a seed in the zero-mode. 

\begin{figure}
	\centering\includegraphics[width=0.6\textwidth]{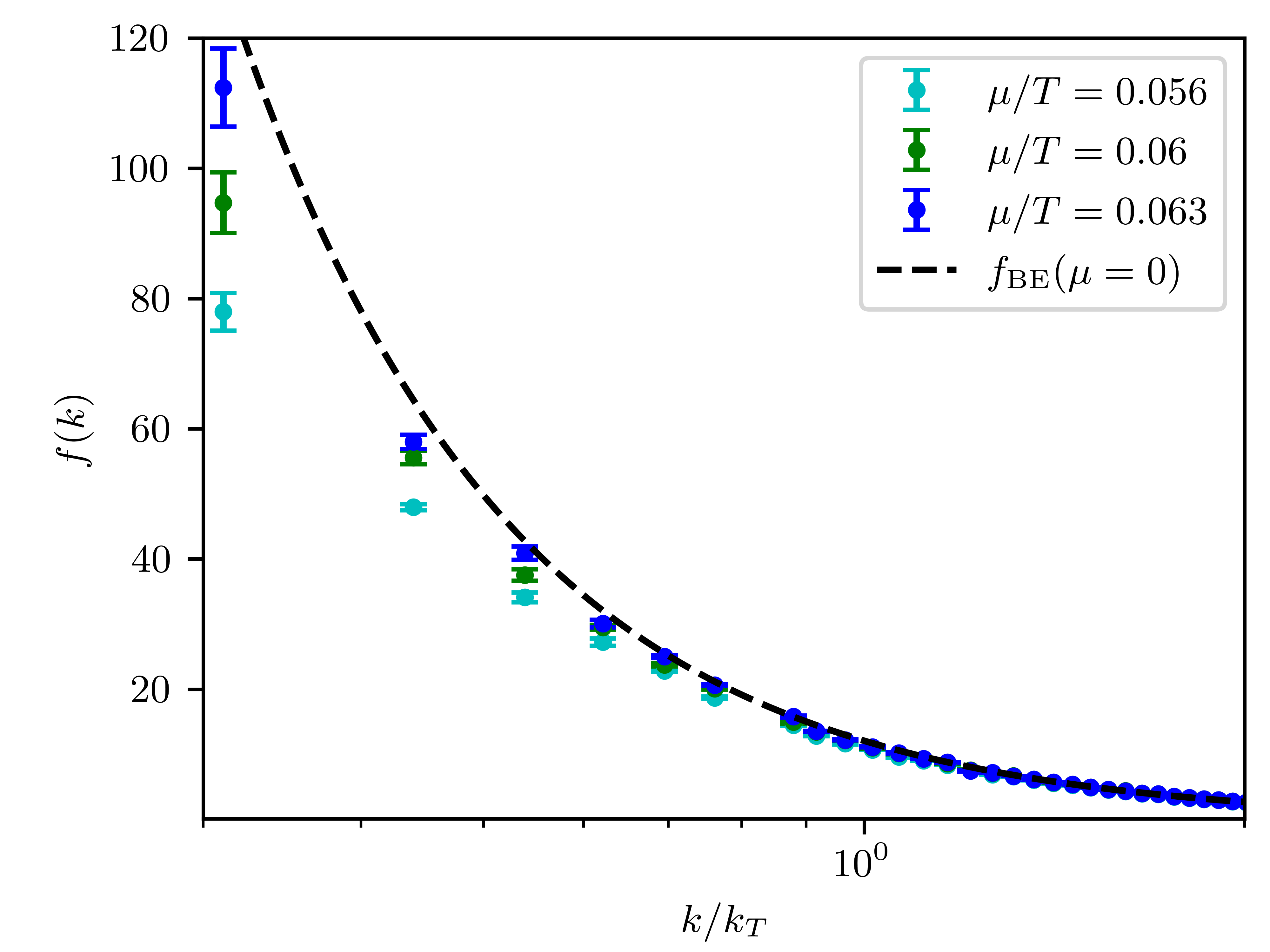}
	\caption{Infrared part of the momentum spectrum of the interacting single-component gas, for three different chemical potentials $\mu$ close to and below the transition, obtained by averaging up to $\vartheta_{\mathrm{max}}=10^{4}\,a_\mathrm{s}^{-3}$. 
		All other parameters are as in figure \fig{timeevol}. 
		Note that here only the $k$-axis carries a log scale for better visibility. 
	}
	\label{fig:acrosstrans}
\end{figure}
The full spectra $f(k)\equiv\langle f(k,\vartheta)\rangle_{\vartheta_0}^{\vartheta_{\mathrm{max}}}$ for three of the five configurations shown in figure \fig{timeevol} are depicted in figure \fig{acrosstrans}, as obtained from averaging up to $\vartheta_{\mathrm{max}}=10^{4}\,a_\mathrm{s}^{-3}$. 
They approach the Bose-Einstein distribution for $\mu=0$, while Rayleigh-Jeans scaling at low $k$ is not expected to be reached exactly in the finite-volume case.

\subsubsection{Shift of the critical temperature}
Let us now turn to the determination of the shift of the transition temperature due to interactions. 
At leading order in the dimensionless interaction strength, set by the diluteness $\eta=\sqrt{\rho a^3}$, the shift scales as $\eta^{2/3}$  \cite{baym1999transition}, and to next-to-leading order in $\eta$, it takes the form \cite{arnold2001tc}
\begin{align}
\label{eq:shiftc}
\frac{\Delta T_\mathrm{c}}{T_\mathrm{c}^0}=c \,\eta^{2/3}
+[c' \text{ln} ( \eta^{2/3} ) +c'' ]\,\eta^{4/3}\,,
\end{align}
where $c$, $c'$, and $c''$ are numerical constants, $T_\mathrm{c}^{0}=[\rho/\zeta(3/2)]^{2/3}2\pi/m$ is the critical temperature of the ideal Bose gas in $d=3$ dimensions and $\Delta T_\mathrm{c}\equiv T_\mathrm{c}-T_\mathrm{c}^0$ the shift of its value in the presence of interactions. 

\begin{figure}
	\centering\includegraphics[width=0.6\textwidth]{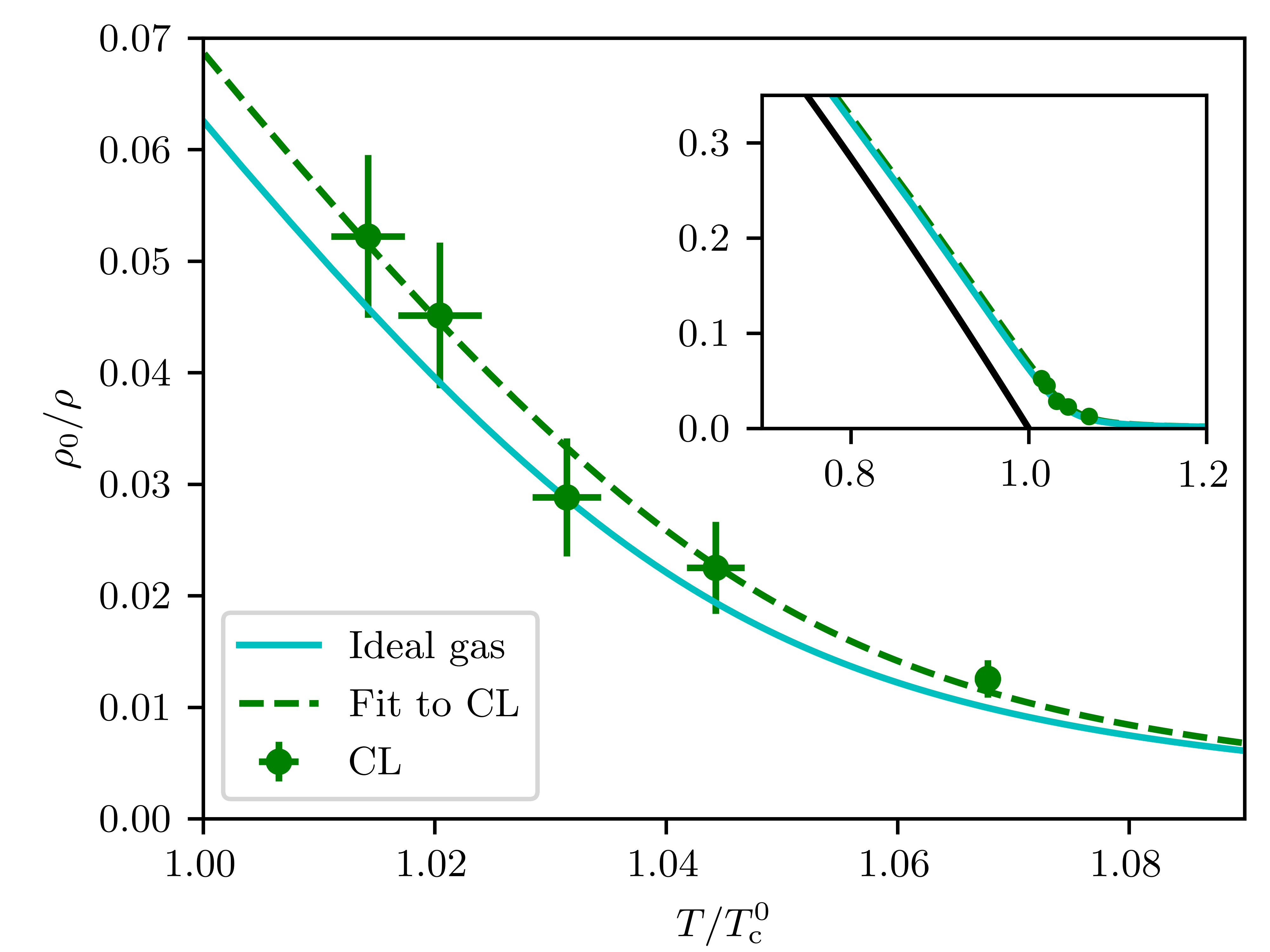}
	\caption{Condensate fraction $\rho_{0}/\rho$ as a function of $T/T_\mathrm{c}^{0}$, for a fixed temperature $T=1.25\,a_\mathrm{s}^{-1}$, coupling $g=0.5\,a_\mathrm{s}^2$ and system size $(64\,a_\mathrm{s})^{3}$. 
		The values obtained from the CL simulation for the five different chemical potentials $\mu$ from \ref{fig:timeevol} (green data points) are compared with the behavior of the ideal Bose gas (cyan solid line). 
		The dashed green line represents a fit of the ideal-gas behaviour, with shifted critical temperature,  to the simulation data as described in the main text, which gives $c=0.58\pm0.16$ quantifying the interaction-induced shift \eq{shiftc}.
		The inset depicts the same curves and data together with the near-critical condensate fraction in the thermodynamic limit (black solid line).
	}
	\label{fig:shift}
\end{figure}

Consider the condensate fraction $\rho_{0}/\rho$ in a free gas as a function of $T/T_\mathrm{c}^0$, for a fixed temperature $T$, and system size $\mathcal{V}=L^{3}$, and thus fixed ratio $\ell=L/\lambda_T$ of the system size $L$ and the thermal wave length $\lambda_{T}= \sqrt{2\pi/mT}$. 
Varying the chemical potential $\mu$ allows tuning the total density $\rho=N/\mathcal{V}$ as well as the zero-mode density $\rho_{0}=f(0)/\mathcal{V}$, and the critical temperature $T_\mathrm{c}^{0}$, such that
\begin{align}
{T}/{T_\mathrm{c}^{0}}
&=\ell^{2}\left(\frac{\zeta(3/2)}{N}\right)^{2/3}
\,,\\
{\rho_{0}}/{\rho}
&=\frac{z}{1-z}N^{-1}
\,,
\end{align}
where $z=\exp(\beta\mu)$ is the fugacity.
The total particle number as a function of $z$ in the continuum limit can be determined as the sum 
\begin{align}
N
=\sum_{m_{x}m_{y}m_{z}}
\frac{1}{z^{-1}\exp\left[{\pi}{\ell^{-2}}(m_{x}^2+m_{y}^2+m_{z}^2)\right]-1}
\,,
\end{align}
where the sums are performed over $m_{i}=-M_{i},\dots,M_{i}$, with $M_{x}=M_{y}=M_{z}$ chosen large enough to ensure convergence. 
The resulting dependence of the ideal-gas condensate fraction $\rho_{0}/\rho$ on $T/T_\mathrm{c}^{0}$ is shown in figure \fig{shift} as a solid cyan line.
For comparison, the inset shows the same curve over a wider range near the critical temperature, together with the condensate fraction $\rho_{0}/\rho=1-(T/T_\mathrm{c}^{0})^{3/2}$ that results in the thermodynamic limit $\mathcal{V}\to\infty$ (black solid line).

For the weakly interacting Bose gas we make use of the approximation that the functional form of the condensate fraction as a function of $T$ in units of the critical temperature is the same as in the non-interacting case, albeit with a shifted critical temperature, i.e.
\begin{align}
\left[{\rho_{0}}/{\rho}\right]({T}/{T_\mathrm{c}^{0}})
=\left[{\rho_{0}}/{\rho}\right]_\mathrm{free}(\alpha{T}/{T_\mathrm{c}^{0}})
\,,
\end{align} 
with $\alpha=T_c^0/T_c$ and thus $\Delta T_c/T_c^0=1/\alpha-1$. 
Within errors, this is confirmed by our data.
For each of the chemical potentials chosen in our CL simulations, we determine the total density $\rho$ from the sum over the occupation number spectrum over all points of the momentum-time lattice, multiplied with a Jacobi determinant to account for the sine-spacing of the momentum modes, as described in section \sect{observ}. We furthermore correct by the difference between the continuum and lattice densities determined for the ideal gas at the same temperature and chemical potential in order to account for the bias due to the finite momentum and Matsubara frequency cutoff, see appendix \sect{3D_supp} for details.
In this way, we determine condensate fractions and corresponding values of ${T}/{T_\mathrm{c}^{0}}$ for the five different near-critical chemical potentials $\mu\in\{0.063,\dots,0.056\}$ considered above, which we compare with the behavior of the ideal gas in the continuum in figure \ref{fig:shift}.

A least-squares fit of the ideal-gas curve to the simulation data (dashed green line) yields an interaction-induced shift of the critical temperature of
\begin{align}
\label{eq:Tshiftnum}
\Delta T_c/T_c^0
=1/\alpha-1
=0.00497\pm 0.00138\,.
\end{align}
Possible finite-size corrections to this result are discussed in appendix \sect{3D_supp}. For the density at $\mu/T=0.063$ the diluteness becomes $\eta=7.95\cdot10^{-4}$.
As a consequence, neglecting the $\mathcal{O}(\eta^{4/3})$-corrections in \eq{shiftc}, the constant $c$ in \eq{shiftc} takes the value
\begin{align}
\label{eq:cshift}
c=0.58\pm 0.16\,.
\end{align}

The determination of the constant $c$ has a long and controversial history \cite{stoof1992nucleation,gruter1997critical,holzmann1999bose,reppy2000density,arnold2000t,arnold2001bec,kastening2004bose,nho2004bose}, see \cite{andersen2004theory} for a review. 
Within the errors, our result \eq{cshift} is in good agreement with the value $c=0.7$ (no error provided) from \cite{holzmann1999bose}, while it is larger than $c=0.34\pm0.03$ from the PIMC simulations of \cite{gruter1997critical}  and  smaller than the average of more recent results which accumulate near $c\approx 1.3$, compare $c=1.32\pm0.02$ from \cite{arnold2001bec}, $c=1.29 \pm 0.05$ from \cite{kashurnikov2001critical} (Monte Carlo simulation of classical $O(2)$ field theory), $c=1.27\pm0.11$ from \cite{kastening2004bose} (variational perturbation theory),  and $c=1.32\pm0.14$ from \cite{nho2004bose} (PIMC simulation). 
Experiments with helium in Vycor glasses have found $c\approx5.1$ \cite{reppy2000density}, while this has been disputed in \cite{arnold2000t}. 

We also note that in \cite{arnold2001tc} the constants $c'$ and $c''$, which quantify the corrections to \eq{shiftc} of order $\eta^{4/3}$ were determined. 
The authors found $c=1.32\pm 0.02$, $c'=19.7518$ and $c''=75.7\pm0.4$, which for our diluteness $\eta=7.95\cdot10^{-4}$ yield an $\mathcal{O}(\eta^{4/3})$-correction to the shift $[c' \text{ln} ( \eta^{2/3} ) +c'' ]\,\eta^{4/3}=-0.00135\pm 0.00003$ and a total shift of $\Delta T_\mathrm{c}/T_\mathrm{c}^0=0.00998\pm 0.00017$, which is to be compared with \eq{Tshiftnum}.
In order to distinguish numerically the leading-order and next-to-leading order contributions would require an analysis for a range of different densities or interaction strengths and thus dilutenesses, which is beyond the scope of the present work.

In summary, our analysis demonstrates that the CL method gives access to a non-perturbative quantity (in the HF approximation, there is no shift of the transition temperature), consistent with previous results by order of magnitude.

\subsection{Conclusion}
In this chapter, we have shown that it is possible to evaluate the coherent-state path integral of the interacting Bose gas from first principles by means of the complex Langevin (CL) method in experimentally relevant coupling regimes and in three spatial dimensions. 

We found that the spectra and dispersions obtained via the Hartree-Fock and Bogoliubov approximations are well reproduced above and below the transition, respectively. 
By determining the shift of the critical temperature $\Delta T_\mathrm{c}$ due to interactions, we could show that the method is in principle capable of providing corrections beyond perturbative approximations. 

We have thereby had a first glimpse of the power of the complex Langevin method, which works almost flawlessly for a weakly interacting Bose gas and provides a comparatively easy access to exact quantum simulation thereof. This picture will be further consolidated in the subsequent chapters, where we will benchmark the method in more specialized settings (two-dimensional gas, external trapping potential, dipolar interactions).  

As the main limitation for the method we found the coupling strength: For the weakly interacting gas with diluteness of the order $\eta\sim 10^{-3}$ as in all simulations presented in this chapter, complex Langevin worked well, but started to get spoilt by runaway trajectories for higher diluteness as $\eta\sim 10^{-2}$. For the objective of providing a benchmark for ultracold atoms experiments, however, this is not too severe because it is also difficult experimentally to go beyond the weakly interacting regime.

\clearpage

\thispagestyle{plain}
\section{BKT transition in a two-dimensional Bose gas\label{sec:2D_gas}}
Experimentally, it is possible to create effectively two-dimensional ultracold Bose gases by making the harmonic confinement in one direction sufficiently strong, cf. section \sect{physics_bose}. This restriction of the gas to two spatial dimensions has a number of remarkable consequences. According to the Hohenberg-Mermin-Wagner theorem \cite{hohenberg1967existence,mermin1966absence}, there cannot be spontaneous breaking of a continuous symmetry and thus Bose-Einstein condensation in less than three spatial dimensions~\footnote{This applies to an infinitely extended system at nonzero temperature. In a finite-size two-dimensional Bose gas a considerable amount of the particles can still occupy the zero momentum mode.}. However, as shown by Berezinskii, Kosterlitz and Thouless \cite{berezinskii1972destruction_a,berezinskii1972destruction_b,kosterlitz1973ordering}, a topological phase transition characterized by the transition from a gas of free vortices to bound vortex-antivortex pairs occurs (Berezinskii-Kosterlitz-Thouless (BKT) transition). Two dimensions are also special in other respects. While in a three-dimensional Bose gas there is infinitely extended phase coherence below the Bose-Einstein transition temperature (the correlation function does not decay to zero but to a finite value at infinity) and in a one-dimensional system phase coherence is completely destroyed (the correlation function decays exponentially), a two-dimensional gas exhibits \textit{quasi-coherence}, characterized by an algebraically decaying correlation function, with an exponent that depends continuously on temperature, approaching zero for $T\to 0$. Furthermore, it has been predicted \cite{prokofev2002two} that the equation of state, i.e. the density as a function of chemical potential for fixed temperature, in two dimensions displays universal behavior in the weakly coupled regime. 

Due to these peculiarities, the two-dimensional Bose gas represents a very suitable system for a benchmark of the complex Langevin algorithm. In this chapter, we will thus explore the applicability of the method in this scenario. We will demonstrate that the known hallmarks of the BKT transition can be correctly reproduced, namely the vortex unbinding process, the algebraic decay of correlation functions and the universality of the equation of state. Furthermore, we will use the method to compute for the first time from first principles the non-universal BKT transition temperature in the weakly coupled regime. As it will turn out, the result deviates by around $5\%$ from the value that had previously been obtained from classical field theory simulations \cite{prokofev2001critical}. 

The organization of this chapter is as follows: In section \sect{BKT_theory}, we will review the physics of the BKT transition. We discuss first the behavior of correlation functions, superfluid density and the equation of state, before we move on to the topological characteristics across the transition. Finally, we discuss the finite-size scaling of critical quantities and their extrapolation to the thermodynamic limit, which is of crucial importance for the interpretation of numerical data. In section \sect{BKT_theory}, we present the results of the complex Langevin simulations, starting with the extraction of the critical density and chemical potential and then moving on to the evaluation of the equation of state, the momentum spectrum and the number of unbound vortices across the transition.
\subsection{Physics of a two-dimensional Bose gas \label{sec:BKT_theory}}
We briefly summarize a few of the most important characteristics of a Bose gas near the BKT transition that we will consider in the following. 
For general reviews of the theory cf., e.g., references \cite{altland2010condensed,jose2013years}. 

\subsubsection{Correlation functions, superfluid density and universality}
\label{sec:BKT}
The phase-ordered and disordered phases, which are separated by the BKT transition, can be distinguished by means of the first-order spatial correlation function $g(\mathbf{r})$, defined as
\begin{align}
g(\mathbf{r})=\langle\psi^\dagger(\mathbf{r}+\mathbf{x})\psi(\mathbf{x})\rangle
\,.
\end{align}
Considering a homogeneous system, in the disordered phase above the transition temperature $T_\text{BKT}$ the correlation function $g(\mathbf{r})$ falls off exponentially at large distances,
\begin{align}
g(\mathbf{r})\sim\exp\{-|\mathbf{r}|/\xi_\mathrm{c}\}\qquad (T>T_\text{BKT})\,,
\end{align}
with a temperature-dependent correlation length $\xi_\mathrm{c}$, which behaves as 
\begin{align}
\xi_\mathrm{c}(T)\sim \exp\left(\text{const}\cdot\sqrt{\frac{T_\text{BKT}}{T-T_\text{BKT}}}\right)\,.
\end{align} 
Below the transition, $g(\mathbf{r})$ shows algebraic behavior instead,
\begin{align}
g(\mathbf{r})\sim \left(|\mathbf{r}|/\xi_\mathrm{h}\right)^{-\alpha}
\qquad (T<T_\text{BKT})\,,
\label{eq:grbelowBKT}
\end{align}
on length scales $|\mathbf{r}|\gg\xi_\mathrm{h}$ larger than the zero-temperature healing length $\xi_\mathrm{h}=(2mg\rho)^{-1/2}\sim(2m\mu)^{-1/2}$.
The behavior of the momentum-space occupation number $f(\mathbf{k})$, defined as
\begin{align}
f(\mathbf{k})=\langle\psi^\dagger(\mathbf{k})\psi(\mathbf{k})\rangle
\,,
\label{eq:fk}
\end{align}
correspondingly scales as
\begin{align}
f(\mathbf{k})\sim {|\mathbf{k}|^{\alpha-2}}
\,,\qquad |\mathbf{k}|\to 0
\,.
\label{eq:fkscaling}
\end{align}
The scaling exponent $\alpha$ can be expressed in terms of the thermal de Broglie wave length $\lambda_T=\sqrt{2\pi/mT}$ and the superfluid density $\rho_\text{s}$ as
\begin{align}
\alpha=\frac{1}{\lambda_T^2\,\rho_\text{s}}
\,.
\label{eq:scalingexponent}
\end{align}
The superfluid density $\rho_\text{s}$ is defined as the fraction of the total particle density that upon an external boost of the system remains at rest. In $d$ spatial dimensions it may be computed as \cite{pollock1987path}
\begin{align}
\frac{\rho_\text{s}}{\rho}=1-\frac{\langle\mathbf{P}^2\rangle}{dmTN_\text{tot}}
\label{eq:rhos}
\,,
\end{align}
where $\mathbf{P}$ is the total momentum of the system, $N_\text{tot}$ the total particle number and $\rho=N_\text{tot}/\mathcal{V}$ the mean particle density. 
 
In the thermodynamic limit, decreasing the temperature through the transition, $\rho_\text{s}$, according to Nelson and Kosterlitz \cite{nelson1977universal}, makes a universal jump from $0$ to 
\begin{align}
\label{eq:nelson}
\rho_{\text{s},\text{c}}
=\frac{2mT}{\pi}
=\frac{4}{\lambda_{T}^{2}}
\,,
\end{align}
such that the scaling exponent approaches the critical value $\alpha=1/4$ for $T\to T_\text{BKT}^-$, while $\alpha$ goes to $0$ for $T\to 0$. Furthermore, at $T\to 0$ the superfluid density approaches the total density, $\rho_\text{s}\to\rho$, i.e. the entire gas becomes superfluid.

Close to the BKT transition, we may distinguish three momentum regimes.  Non-linear couplings between the momentum modes become important below the finite-temperature healing-length scale $k_\text{c}\sim\sqrt{mg\rho_\mathrm{s}}\sim m\sqrt{gT}\sim \sqrt{mg}\, k_{T}$ (where we have used \eq{nelson}), with thermal momentum $k_{T}\sim\sqrt{mT}$.
For momenta $k_\mathrm{c}\ll k\ll k_{T}$, the occupation number takes the Rayleigh-Jeans form $f(\mathbf{k})=2mT/|\mathbf{k}|^2$. For momenta above the thermal momentum $k\gg k_{T}$ the spectrum goes over to its Boltzmann tail, $f(\mathbf{k})=\exp(-|\mathbf{k}|^2/2mT)$. 

Whereas the \textit{superfluid} density \eq{nelson} takes a universal value at the transition, the \textit{total} density $\rho$ does not and is dependent on the microscopic details of the system instead.
It has been predicted in \cite{prokofev2001critical}, on the basis of the observation that the major contribution to the total density comes from the range of momenta $k_\text{c}\lesssim k\lesssim k_\text{UV}\sim k_{T}$, that, for $mg\ll1$, the critical density and chemical potential behave as~\footnote{In the following, we will consider the critical density and chemical potential at fixed temperature instead of the critical temperature at fixed density. While both points of view are physically equivalent, the former is technically more convenient as we work in the grand-canonical and not the canonical ensemble.}
\begin{align}
\label{eq:critical_dense}
\rho_\text{c}
&=\frac{1}{\lambda_T^2}\ln\left(\frac{\zeta_{\rho}}{mg}\right)
\,,
\\
\label{eq:critical_mu}
\mu_\text{c}
&=\frac{2g}{\lambda_T^2}\ln\left(\frac{\zeta_{\mu}}{mg}\right)
\,.
\end{align}
The numerical constants were given as $\zeta_{\rho}=380\pm3$ and $\zeta_{\mu}=13.2\pm0.4$~\footnote{In \cite {prokofev2001critical}, $\protect \zeta_{\rho}$ and $\protect \zeta_{\mu }$ were denoted as $\xi $ and $\xi _{\mu }$, respectively.}. 
As these were obtained from simulations of the classical field theory, which become exact in the limit $mg\to 0$, it will be interesting to check this prediction against the simulations of the full quantum model at nonzero $mg$. 

From the above results, one finds that the weakly coupled 2D Bose gas exhibits universal behavior across the BKT transition:
As was argued in \cite{prokofev2002two}, both $\rho_\text{c}$ and $\mu_\text{c}$ are non-universal quantities that depend on the ultraviolet (UV) cutoff scale $k_\text{UV}$, while the cutoff dependence cancels out when 
subtracting the chemical potential from the HF-type expression $2g\rho$, as well as in the differences $\rho-\rho_\text{c}$ and $\mu-\mu_\text{c}$. It was thus suggested that, upon introducing the dimensionless control parameter $X=(\mu-\mu_\text{c})/mgT$, one can write the equation of state in the universal form $(2g\rho-\mu)/(mgT)=\theta(X)$ with some function $\theta(X)$, which is valid within the so-called fluctuation region where $X$ varies on the order of unity.
Evaluating the difference of this relation to the critical one leads to the equation of state in the form 
\begin{align}
\rho(\mu)-\rho_\text{c}
=\frac{1}{\lambda_T^2}F\left(\frac{\mu-\mu_\text{c}}{mgT}\right)
\label{eq:eosuniversal}
\,,
\end{align}
where the universal function $F(X)$ is related to $\theta(X)$ by
\begin{align}
\label{eq:EOS_mean_field}
F(X)=\pi\left[\theta(X)+X-\frac{1}{\pi}\ln\left(\frac{\zeta_{\rho}}{\zeta_{\mu}}\right)\right]\,.
\end{align}
In the Bogoliubov limit, $X\to\infty$, the function $\theta$ asymptotically obeys 
\begin{align}
\theta(X)-\frac{1}{\pi}\ln\left[\theta(X)\right]
\to X+\frac{1}{\pi}\ln(2\zeta_{\mu})
\label{eq:thetaMF}
\,,
\end{align}
whereas for $X\to-\infty$, a Hartree-Fock approximation yields
\begin{align}
\theta(X)+\frac{1}{\pi}\ln\left[\theta(X)\right]
\to |X|-\frac{1}{\pi}\ln(\zeta_{\mu})
\label{eq:thetaMF2}
\,,
\end{align}
and the resulting universal scaling form \eq{eosuniversal} of the equation of state was confirmed in this limit by means of classical Monte Carlo simulations \cite{prokofev2002two} for different values of the coupling $mg$.

Eventually, also the superfluid density, for chemical potentials below the BKT transition and in its vicinity of the transition, obeys a functional form generalizing the Nelson-Kosterlitz result \eq{nelson} \cite{prokofev2002two},
\begin{align}
\label{eq:nelsoninfluctreg}
\rho_{\text{s}}
=\frac{2mT}{\pi}\,f(X)
\,,
\end{align}
with a universal function $f(X)$ obeying $f(X\to0^{+})=1$ and $f(X<0)\equiv0$.

\subsubsection{Topological characteristics}
\label{sec:TopoBKT}
A further important characteristic of the BKT transition is topological in nature: 
Whereas above $T_\text{BKT}$ one considers the gas to be dominated by free quantum vortices, which destroy the large-distance phase coherence, these mutually annihilate as pairs of vortices and anti-vortices and thus disappear below the transition, giving rise to an only algebraically slow decoherence \eq{grbelowBKT}.

What is understood here by ``quantum vortex'' is a local minimum of the classical energy functional of the form 
\begin{align}
\psi(r,\varphi)=f(r)\,e^{iq\varphi}\,,
\end{align}
where $r$ and $\varphi$ are meant to be polar coordinates, $q\in\mathbb{Z}$ is the integer vortex charge~\footnote{For $q>0$, we speak of a vortex, for $q<0$ of an antivortex. However, ``vortex'' is also used as an umbrella term for both vortices and antivortices.} and $f(r)$ is a function with the property $f(0)=0$ and $f(r\gg \xi_\mathrm{h})=\sqrt{\rho}=\text{const}$, with healing length $\xi_\mathrm{h}=\sqrt{2mg\rho}$. For our purposes, it suffices to assume
\begin{align}
f(r)\approx \sqrt{\rho}\,\Theta(r-\xi_\mathrm{h})\,.
\end{align}
The current density
\begin{align}
\mathbf{j}
=\frac{1}{2mi}\left[\psi^*\nabla\psi-\psi\nabla\psi^*\right]
\label{eq:current}
\end{align}
created by a single vortex reads
\begin{align}
\mathbf{j}_\text{v}(r,\varphi)=\frac{q\rho}{mr}\,\Theta(r-\xi_\mathrm{h})\,\mathbf{e}_\varphi\,.
\end{align}
The integrated squared current density of one vortex reads
\begin{align}
\int d^2r\,|\mathbf{j}_\text{v}|^2=\frac{2\pi q^2\rho^2}{m^2}\,\ln\left(\frac{L}{\xi_\mathrm{h}}\right)\,,
\end{align}
where $L$ is the system size. Let us now consider an ensemble of singly-quantized vortices (i.e. with $q=\pm 1$), neglecting higher-quantized vortices. The total current density produced by these vortices is the sum of the current densities produced by the single vortices. If the positions of the vortices are uncorrelated and positive and negative charges are well mixed, the cross terms in $|\mathbf{j}_\text{tot}|^2=|\sum_i \mathbf{j}_{\text{v},i}|^2$ vanish and every vortex contributes $2\pi\rho^2\ln(L/\xi_\mathrm{h})/m^2$ to the integral over $|\mathbf{j}_\text{tot}|^2$. Bound vortex-antivortex pairs instead will not contribute, as the current density created by both of them cancels out to zero at large distances. $|\mathbf{j}(\mathbf{r})|^2$ may thus be considered a measure for the mean density of unbound, free vortices. 

In a realistic Bose gas, it is not only the vortices that produce current density, but there will also be a background of sound waves whose current density superposes that of the vortices. We can isolate the contribution from the vortices by means of a Helmholtz decomposition of $\mathbf{j}$ into its irrotational and rotational part 
\begin{align}
\mathbf{j}
=\mathbf{j}_\text{irr}+\mathbf{j}_\text{rot}
\,.
\end{align}
The mean density of free, unbound vortices $\rho_{\mathrm{v,free}}$ is then approximately given by
\begin{align}
\label{eq:defrhovfree}
\rho_{\mathrm{v,free}}(\mathbf{r}) 
&\simeq \frac{m^2}{2\pi \ln(L/\xi_\text{h})\langle\rho\rangle^2}\,|\mathbf{j}_\text{rot}(\mathbf{r})|^2
\,.
\end{align}
A decrease of $\rho_{\mathrm{v,free}}$ across the transition reflects the vortex-anti-vortex recombination process, when going over from the disordered above to the ordered phase below the transition.

It should be noted that the full quantum theory in principle does not know anything about vortices, which are classical objects after all, despite the fact that as local minima they contribute significantly to the quantum path integral. All that can be computed in CL simulations and measured in experiment are expectation values of observables. However, we can construct observables motivated by classical reasoning. Defining the current density operator as $\hat{\mathbf{j}}=\left[\psi^\dagger\nabla\psi-(\nabla\psi^\dagger)\psi\right]/2mi$, we can interpret $|\mathbf{j}_\text{rot}(\mathbf{r})|^2$ in \eq{defrhovfree} as the observable $\colon \hat{j}_x^2+\hat{j}_y^2+\hat{j}_z^2\colon$, where $\colon\colon$ denotes normal ordering.

Furthermore, it should be noted that the above definition of the number of free unbound vortices neglects the fact that in a realistic Bose gas, the bulk density is not flat but subject to fluctuations. For the latter reason, what is often considered is the velocity field $\mathbf{v}=\mathbf{j}/\rho$ and not the current density $\mathbf{j}$, i.e. expectation values of the form $\langle |\mathbf{j}|^2/\rho^2\rangle$ instead of $\langle |\mathbf{j}|^2\rangle/\langle\rho^2\rangle$ are evaluated, which are not affected by bulk density fluctuations. However, as discussed in section \sect{vortices}, such expectation values are difficult to evaluate in complex Langevin, as they are not holomorphic in the fundamental fields. Despite this limitation, \eq{defrhovfree} can be expected to give at least an estimate for the number of vortices present in a system.

\subsubsection{\label{sec:finitesize}Finite-size scaling}
Numerical simulations must necessarily be performed in a finite volume. For obtaining properties in the thermodynamic limit, some extrapolation is necessary since finite-size corrections vanish only logarithmically in a two-dimensional Bose gas. 
BKT renormalization group (RG) theory allows determining the finite-size scaling of the superfluid density. 
On the basis of general scaling arguments, the superfluid density as a function of system size $L$, $\rho_\mathrm{s}(L)$, close to criticality can be written as a function of $L/\xi_\mathrm{c}$ where $\xi_\mathrm{c}\sim 1/k_\mathrm{c}(T)\sim(m\sqrt{gT})^{-1}$ is the finite-temperature healing length acting as a microscopic length scale. 
For the exact functional form, one needs to solve the BKT RG equations.
Keeping the temperature $T$ fixed, one obtains, in leading order, the finite-size correction to \eq{nelson} at the infinite-size critical chemical potential $\mu_\text{c}(L\to\infty)$ \cite{weber1988monte},
\begin{align}
\label{eq:rhos_LO}
\rho_\mathrm{s}^\text{LO}(L)
=\frac{2mT}{\pi}\Bigg(1&+\frac{1}{2\ln (L/\xi_\mathrm{c})+C}\Bigg)
\,,
\end{align}
with some constant $C$, which we have explicitly written, other than usually found in the literature, in terms of the microscopic healing-length scale $\xi_\text{c}$~\footnote{%
	When integrating the BKT RG equations from the microscopic scale $\xi_\mathrm{c}$ to the volume scale $L$, the constant $C$ defines the initial value of the $L$-dependent function $f_{L}$, equation \eq{nelsoninfluctreg}, at the microscopic scale $\xi_{c}$.
	For example, in the approximation chosen in \eq{rhos_LO}, it is the inverse of the deviation of $f_{\xi_\mathrm{c}}$ from unity, $C=(f_{\xi_\mathrm{c}}-1)^{-1}$.
},
which later-on, for a fixed temperature, determines the scaling in terms of $mg$.
Higher-order expressions are also available \cite{hsieh2013finite}. 
For example, to next-to-leading order one has 
\begin{align}
\label{eq:rhos_NLO}
\rho_\mathrm{s}^\text{NLO}(L)
=\frac{2mT}{\pi}\Bigg(1&+\frac{1}{2\ln (L/\xi_\mathrm{c})+C+\ln\left[C/2+\ln(L/\xi_\mathrm{c})\right]}\Bigg)
\,.
\end{align}
Instead of extrapolating the superfluid density to the infinite-size limit for each single parameter set, we will follow the somewhat different procedure described in \cite{prokofev2001critical}. 
Namely, one defines a finite-size critical potential $\mu_\mathrm{c}(L)$ by demanding that the Nelson-Kosterlitz criterion \cite{nelson1977universal}, equation \eq{nelson}, be fulfilled at $\mu_\mathrm{c}(L)$ in the finite-size system.
Using then the universal form \eq{nelsoninfluctreg} away from criticality, one finds from the Kosterlitz-Thouless RG equations that, in the $L\to\infty$ limit, $f$ obeys the relation $f^{-1}+\ln f=1+\kappa'X$ near $X=0$, with a numerical constant $\kappa'=0.61(1)$  \cite{prokofev2002two}.
Expanding $f$ about unity this yields that, at $\mu$ closely above $\mu_\mathrm{c}(L)$ and sufficiently large $L$, the superfluid density obeys
\begin{align}
\frac{\rho_\mathrm{s}\pi}{2mT}-1
=f\left(\frac{\mu-\mu_\mathrm{c}(L)}{mgT}\right)-1
\simeq \sqrt{\frac{2\kappa'[\mu-\mu_\mathrm{c}(L)]}{mgT}}
\,.
\end{align}
Combining this with the leading-order approximation \eq{rhos_LO}, the finite-size scaling form of the difference between $\mu_\mathrm{c}(L\to\infty)$ and $\mu_\mathrm{c}(L)$ results as 
\begin{align}
\mu_\mathrm{c}(L\to\infty)-\mu_\mathrm{c}(L)
\overset{\text{LO}}{\simeq} \frac{mgT}{2\kappa'\left[2\ln (L/\xi_\mathrm{c})+C\right]^2}
\,.
\end{align}
Finally, since the function $F(X)$ defined in equation \eq{eosuniversal} is, in leading order, linear in $X$ close to criticality ($X=0$), one finds that $\rho-\rho_\mathrm{c}(L)\sim [\mu-\mu_\mathrm{c}(L)]/g$.
Inserting $1/\xi_\mathrm{c}= m\sqrt{gT}$ at the fixed near-critical temperature $T$, one obtains the finite-size scaling of the critical density $\rho_\text{c}(L)$,
\begin{align}
\label{eq:finitesize}
\rho_\text{c}^\text{LO}(L)=\rho_\text{c}(L\to\infty)-\frac{AmT}{\ln^2\left(BLm\sqrt{gT}\right)}
\,,
\end{align}
with dimensionless constants $A$ and $B$, as introduced by \cite{prokofev2001critical}.
Analogously, the next-to-leading-order expression \eq{rhos_NLO} yields
\begin{align}
\label{eq:finitesizeNLO}
\rho_\text{c}^\text{NLO}(L)
=\rho_\text{c}(L\to\infty)
 -\frac{AmT}{\left[\ln\left(BLm\sqrt{gT}\right)+\frac12\ln\ln\left(BLm\sqrt{gT}\right)\right]^2}
\,.
\end{align}
For the extraction of the critical chemical potential from finite-size data, similar relations apply. The leading order expression reads
\begin{align}
\label{eq:finitesizemu}
\mu_\text{c}^\text{LO}(L)=\mu_\text{c}(L\to\infty)-\frac{A_{\mu}mgT}{\ln^2\left(B_{\mu}Lm\sqrt{gT}\right)}
\,,
\end{align}
and the next-to-leading-order expression is given by
\begin{align}
\label{eq:finitesizemuNLO}
\mu_\text{c}^\text{NLO}(L)
=\mu_\text{c}(L\to\infty)
-\frac{A_\mu mgT}{\left[\ln\left(B_\mu Lm\sqrt{gT}\right)+\frac12\ln\ln\left(B_\mu Lm\sqrt{gT}\right)\right]^2}
\,.
\end{align}

\subsection{Results of the CL simulations}
\label{sec:results}
In the following we present the results of our CL simulations of properties of a dilute two-dimensional Bose gas close to the BKT transition. 
We begin with a determination of the critical density $\rho_\text{c}$ as a function of $mg$ in order to check the validity of the prediction \eq{critical_dense} in the quantum regime. 
Subsequently, we evaluate the equation of state and the single-particle momentum spectrum and their scaling properties near criticality, and we close with an evaluation of the vorticity across the BKT transition. 

We express all quantities in units of the spatial lattice constant $a_\text{s}$ and choose $2ma_\mathrm{s}=1$, compare appendix \sect{units}. 
Unless specified differently, the simulations in this chapter were performed on a grid with $256^2$ spatial $\times\, 32$ (simulations in section \sect{critdens}) or $\times\, 16$ (elsewhere) lattice points in the imaginary-time direction, with periodic boundary conditions imposed in both the temporal and spatial directions. 
We set the temporal lattice spacing to $a_\tau=0.05\,a_\text{s}$, resulting in a temperature $T=1.25\,a_\text{s}^{-1}$ and a thermal wave length $\lambda_T=\sqrt{2\pi/mT}=3.17\,a_\text{s}$. 
We choose the Langevin time step to be $\Delta\vartheta=5\cdot 10^{-3}\,a_\text{s}$. 

Parameters are chosen such that the effects of spatial and imaginary-time discretization are typically smaller than the statistical errors of the CL simulations. 
The former can be estimated by comparing the discretized and continuous versions of the non-interacting gas. 
Discretization errors only affect the high-momentum modes where the system is effectively non-interacting. 
For $N_\tau=32$ and $\lambda_T=\sqrt{2\pi/mT}=3.17\,a_\text{s}$, this yields, as estimates for the errors on the density and the superfluid density, $\lambda_T^2\delta \rho\equiv\lambda_T^2(\rho^\text{cont}-\rho^\text{latt})=4\cdot 10^{-5}$ and $\lambda_T^2\delta \rho_\text{s}/4\equiv\lambda_T^2(\rho^\text{cont}_\text{s}-\rho^\text{latt}_{s})/4=-6\cdot 10^{-3}$, which are negligible in comparison with typical statistical errors.

As we recall from section \sect{physics_bose}, choosing $\Lambda_0=\sqrt{\rho}$~\footnote{Of course, the choice of $\Lambda_0$ is not unique. However, since $\Lambda_0$ enters only logarithmically, other choices of the same order of magnitude yield only deviations in the few-percent range. E.g. the relative deviation in $g_R$ between $\Lambda_0=\sqrt{\rho}$ and another common choice, $\Lambda_0=k_\mathrm{h}=\sqrt{2mg\rho}$, is $mg\ln(2mg)/4\pi$, i.e. less than $2\%$ for $mg<0.5$.}, the renormalized coupling in terms of the bare one reads 
\begin{align}
\label{eq:renormalization}
g_R=\frac{ g}{1+\frac{mg}{2\pi}\,\log\left(\Lambda/\sqrt{\rho}\right)}
\,.
\end{align}
Typical densities result on the order of $\rho\sim 1\,a_\text{s}^{-2}$. Inserting the lattice cutoff for the UV cutoff, $\Lambda=\pi/a_\text{s}$, equation \eq{renormalization} yields a relative deviation between the bare coupling $g$ and the renormalized coupling $g_R$ in the range of $\sim2\cdot 10^{-3}$ to $\sim4\cdot 10^{-2}$ for the coupling strengths considered in the following ($mg=0.0125\dots 0.2$). 
Since these shifts are rather tiny, we can, in the following, avoid an explicit distinction between $g$ and $g_R$ and choose the bare coupling that enters the simulations as equal to the renormalized coupling $g_R$.

\begin{figure}
	\centering\includegraphics[width=0.6\textwidth]{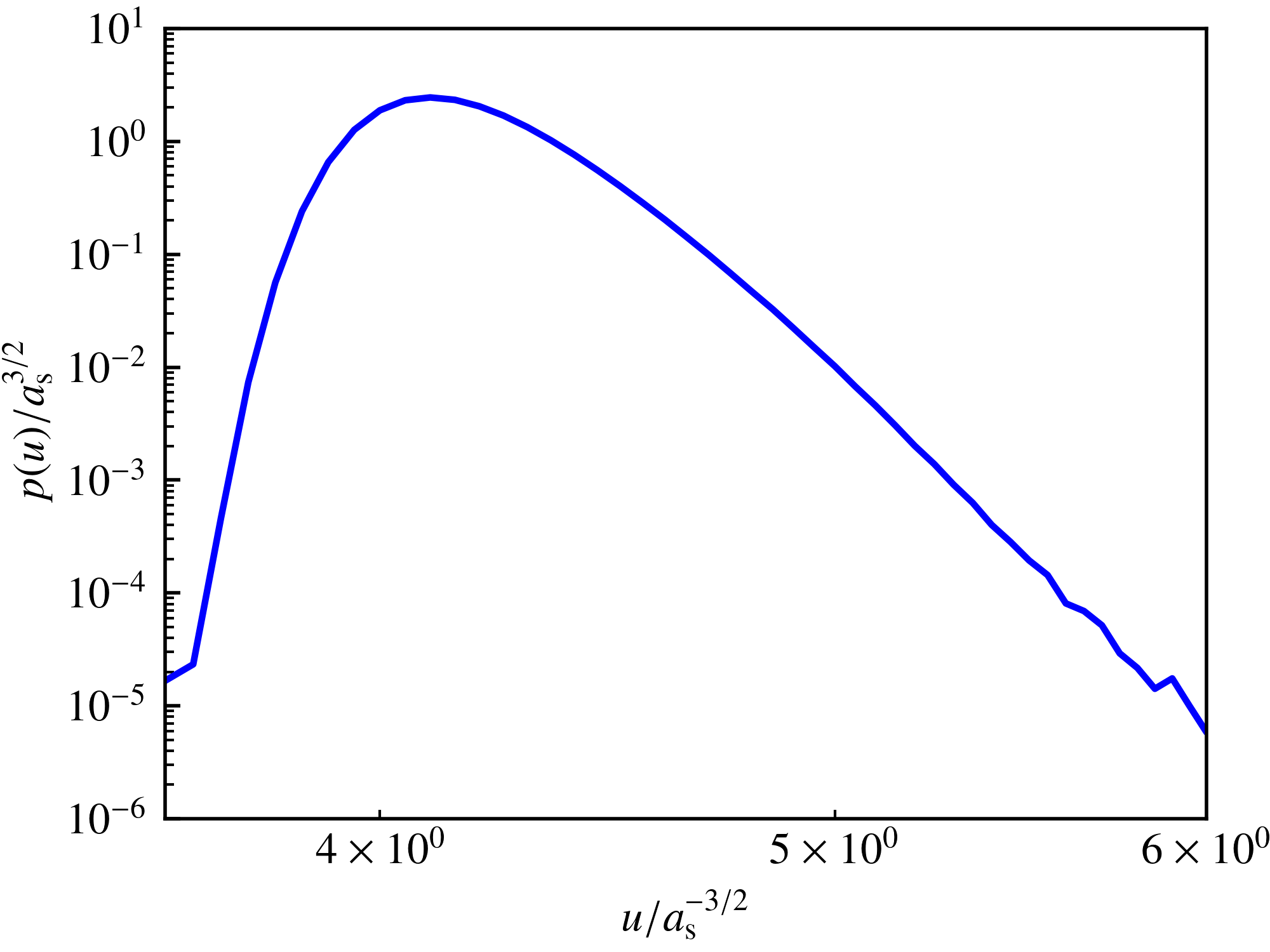}
	\caption{Distribution function of the drift magnitude $p(u)$ as defined in \eq{pu} for $mg=0.1$ at the BKT transition. In the double-logarithmic plot, $p(u)$ bends downwards, indicating that it decays faster than any polynomial and thus the results of the CL simulations can be trusted.
	}
	\label{fig:drift}
\end{figure}

As in the three-dimensional case, we find CL simulations to be limited to a region below some maximum coupling strength $mg$ beyond which the non-linearity in the evolution equations causes runaway trajectories which are impossible to average over in a reliable manner. 
The value of the coupling beyond which such effects dominate also depends on the proximity of the system to the BKT transition. 
Whereas far away from the transition we could simulate up to $mg\lesssim 0.4$, closer to criticality runaway trajectories frequently occur already for $mg\sim 0.2$.
These are no sharp thresholds and it is possible that larger coupling strengths can be reached by employing regulator techniques or adaptive step sizes \cite{aarts2010the,seiler2013gauge,attanasio2019dynamical,alvestad2021stable}. 
In probing the BKT transition we restrict ourselves to the regime $mg\leq0.1$, where runaway trajectories were found to be absent and (polynomial) observables appear to converge well.
When studying the equation of state away from the transition, we also include a coupling $mg=0.2$. 
We have checked from our simulations that the distribution function of the magnitude of the drift, $p(u)$, defined in equation \eq{pu}, decays faster than any polynomial, which has been identified as a necessary and sufficient criterion for the correctness of the complex Langevin method \cite{nagata2016argument}, cf. figure \fig{drift}.

\subsubsection{\label{sec:critdens}Critical density and chemical potential}
\begin{figure}
	\centering\includegraphics[width=0.495\textwidth]{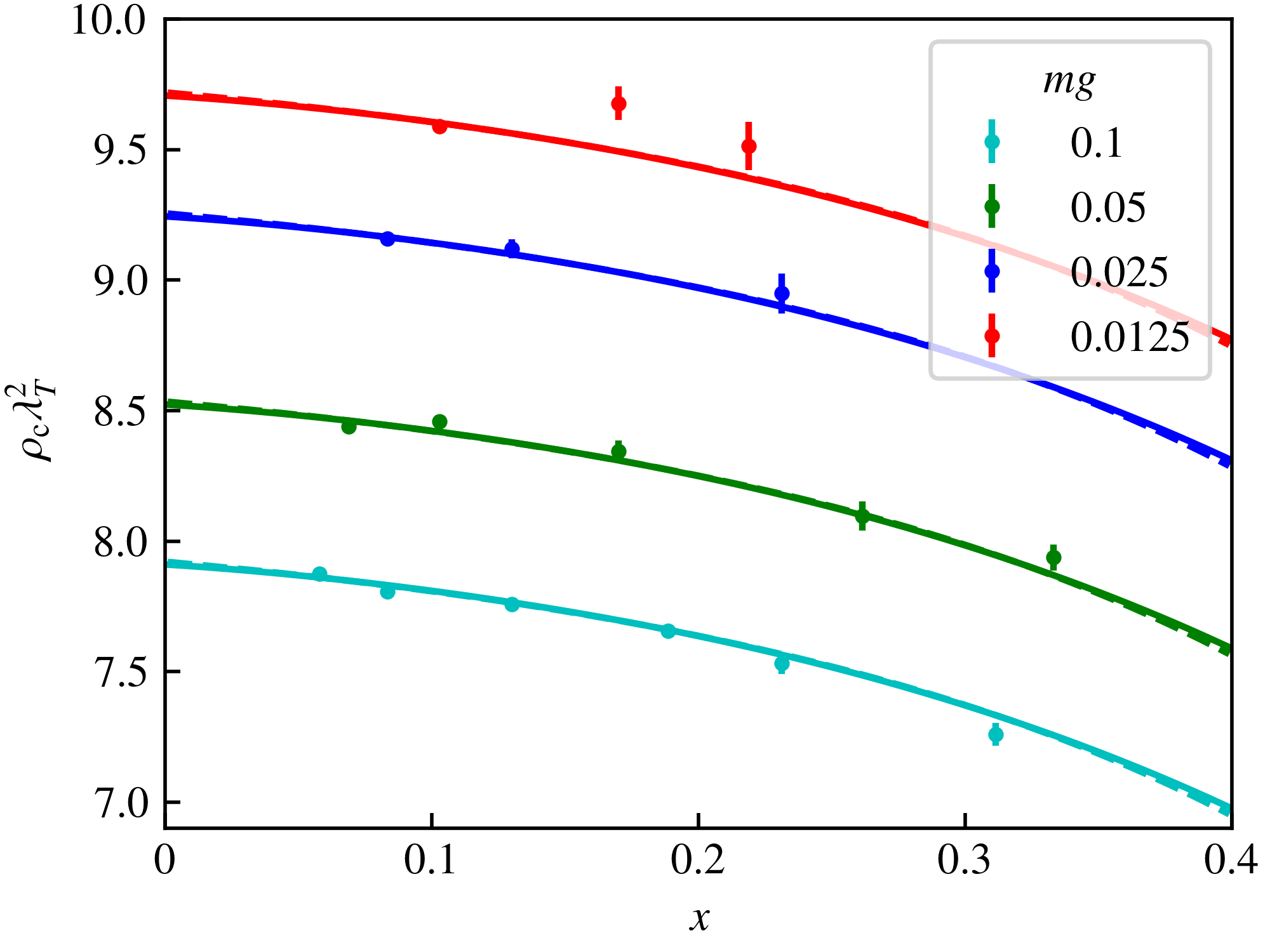}
	\centering\includegraphics[width=0.495\textwidth]{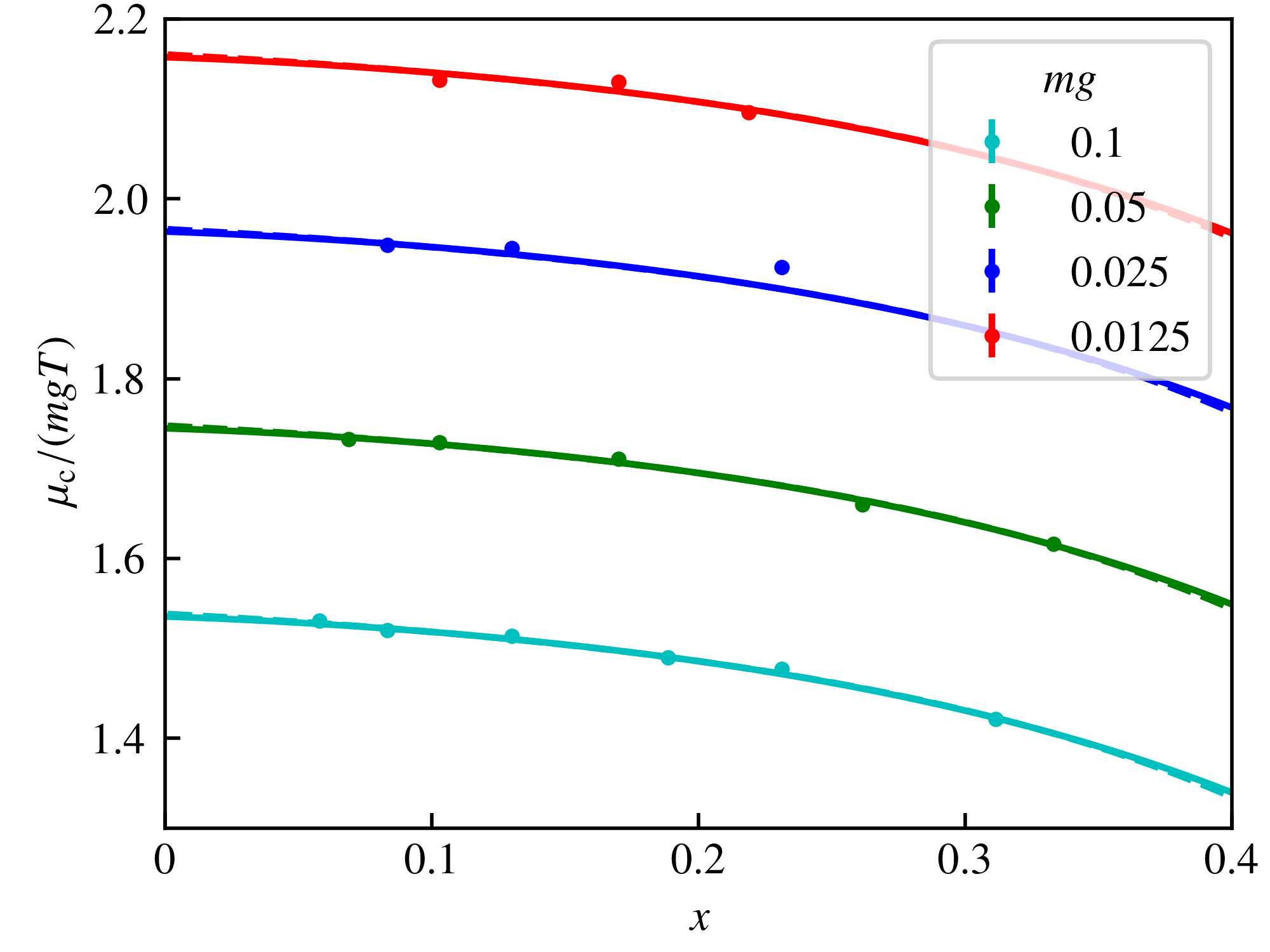}
	\caption{Critical density $\rho_\text{c}$ (left panel) and chemical potential $\mu_\text{c}$ (right panel) for various coupling strengths $mg$ as a function of the dimensionless parameter $x=\ln^{-2}\left(Lm\sqrt{gT}\right)$. Solid lines represent a fit of \eq{finitesize} and \eq{finitesizemu} to the data points. Extrapolating them for $x \to 0$ we can recover the critical density and chemical potential in the infinite-volume limit. Fits of the higher-order formulas \eq{finitesizeNLO} and \eq{finitesizemuNLO} are shown as dashed lines, which hardly deviate from the leading-order expressions. 
		Where no error bars are seen, they are shorter than the width of the data points.
	}
	\label{fig:finitesize}
\end{figure}
\begin{figure}
	\centering\includegraphics[width=0.495\textwidth]{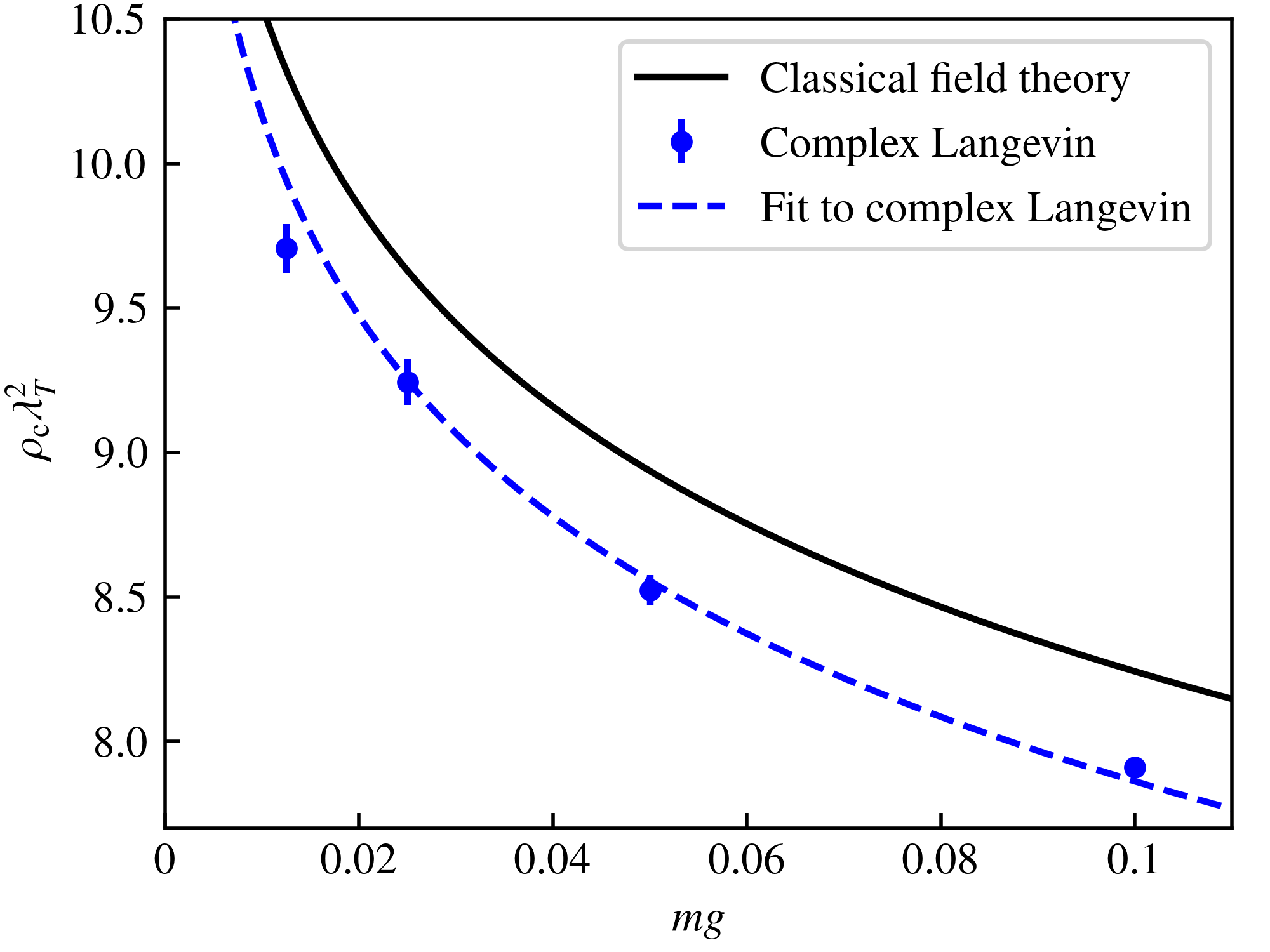}
	\centering\includegraphics[width=0.495\textwidth]{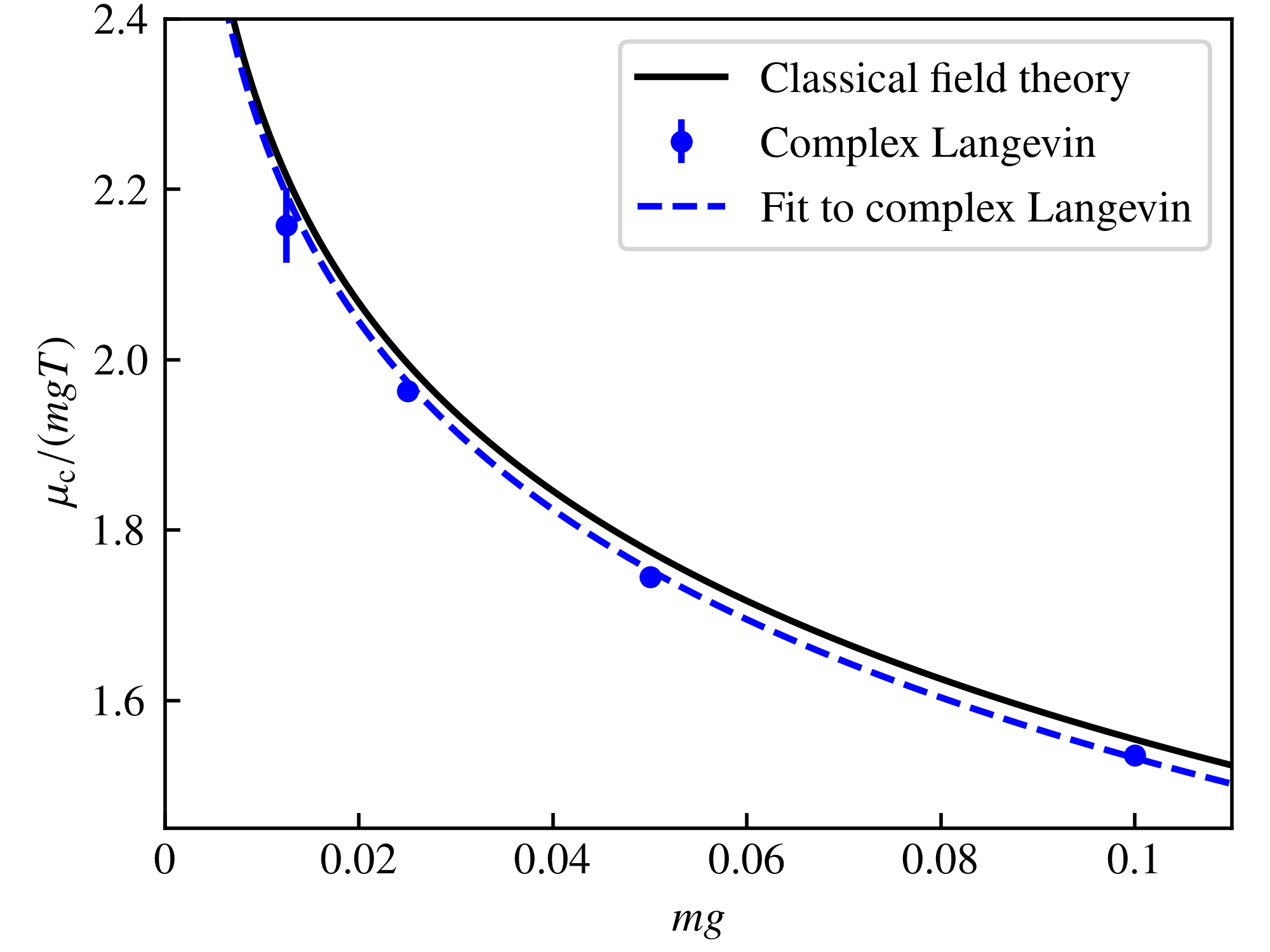}
	\caption{Critical density $\rho_\text{c}$ (left panel) and chemical potential $\mu_\text{c}$ (right panel) in the infinite-volume limit as functions of the coupling $mg$ computed with the complex Langevin simulation of the full quantum model (blue points), within a regime of couplings for which runaway processes are absent. The results are compared to the predictions \eq{critical_dense} with constant $\zeta_{\rho}=380$ and \eq{critical_mu} with $\zeta_{\mu}=13.2$ as obtained from the classical simulation of \cite{prokofev2001critical} (black lines). The critical quantities appear to be shifted downwards in the full quantum simulation by about $4\%$ ($2\%$). The blue dashed lines represent \eq{critical_dense} and \eq{critical_mu} with constants $\zeta_{\rho}=260\pm12$ and $\zeta_{\mu}=12.3\pm0.1$, as obtained from fits to our data.}
	\label{fig:rhoc}
\end{figure}

We start by extracting from our complex-Langevin data the critical density and chemical potential in the thermodynamic limit as functions of the coupling strength, in the weak-coupling regime.
To this end, we need to perform a finite-size scaling analysis as summarized in section \sect{finitesize}.
Our results can be compared with those from classical simulations, and, within limits, with quantum path-integral Monte Carlo as well as experimental results.

Equation \eq{rhos} allows us to extract the superfluid density $\rho_\text{s}$. Here, the total momentum operator $\mathbf{P}$ reads:
\begin{align}
\mathbf{P}=\sum_\mathbf{p} a_\mathbf{p}^\dagger a_\mathbf{p}\,\mathbf{p}\,.
\end{align}
The total momentum variance $\langle\mathbf{P}^2\rangle$ may then be computed as
\begin{align}
\nonumber\mathbf{P}^2
=\sum_{\mathbf{p}\mathbf{q}} a_\mathbf{p}^\dagger a_\mathbf{p} a_\mathbf{q}^\dagger a_\mathbf{q}\,\mathbf{p}
\cdot\mathbf{q}=\sum_{\mathbf{p}\mathbf{q}} a_\mathbf{p}^\dagger a_\mathbf{q}^\dagger a_\mathbf{p} a_\mathbf{q}\,\mathbf{p}
\cdot\mathbf{q}+\sum_{\mathbf{p}}a_\mathbf{p}^\dagger a_\mathbf{p}\,|\mathbf{p}|^2
\,, 
\end{align}
which in the path integral translates to 
\begin{align}
\label{eq:P2}
\langle\mathbf{P}^2\rangle
=\left\langle\frac{1}{N_\tau}\sum_i\left\{\left|\sum_\mathbf{p}\psi_{i+1,\mathbf{p}}^*\psi_{i,\mathbf{p}}\,\mathbf{p}\right|^2
+\sum_\mathbf{p}\psi_{i+1,\mathbf{p}}^*\psi_{i,\mathbf{p}}\,|\mathbf{p}|^2\right\}\right\rangle
\,,
\end{align}
where $|\cdot|^2$ has to be understood as the squared length of a vector, i.e. $|\mathbf{p}|^2=p_x^2+p_y^2+p_z^2$, and not as the squared modulus of a complex number. Disposing of a method to evaluate $\rho_\mathrm{s}$, we determine the finite-size critical point by matching $\rho_\mathrm{s}$ to the Nelson-Kosterlitz criterion \eq{nelson}. While we keep temperature fixed at $T=1.25\,a_\text{s}^{-1}$, we vary the chemical potential $\mu$, whereby both the total and the superfluid density change. As described in more detail in appendix \sect{secant}, we apply the secant algorithm to tune $\mu$ to the point where \eq{nelson} is fulfilled.

As outlined in section \sect{finitesize}, one must expect the corresponding critical (total) density $\rho_\mathrm{c}$ to be subject to finite-size effects.
We make use of the leading-order and next-to-leading order finite-size scaling forms \eq{finitesize} and \eq{finitesizeNLO}, respectively, which we found to be indistinguishable within our statistical errors.  

We computed the finite-size critical densities for four coupling strengths, $mg=0.1$, $0.05$, $0.025$, and $0.0125$, for several lattice sizes $L/a_\text{s}\in[24,\dots,256]$. 
These are shown in figure \fig{finitesize} vs.~the dimensionless parameter $x\equiv\ln^{-2}\left(Lm\sqrt{gT}\right)$. Error bars are obtained from the statistical variance of $12$ statistically independent runs.
Fitting equation \eq{finitesize} to our data points for all couplings and lattice sizes simultaneously we obtain $A=0.089\pm  0.015$ and $B=0.446\pm 0.048$. The parameters $A$ and $B$ yield the $x$-dependencies of the critical density shown as solid lines in the left panel of figure \fig{finitesize}.
By extrapolating these curves to $x=0$ we obtain our prediction for the infinite-size critical densities $\rho_\text{c}(mg)$ for the four different couplings $mg$. 
Our results of this are depicted in the left panel of figure \fig{rhoc}, in comparison with the functional dependence \eq{critical_dense} obtained in \cite{prokofev2001critical} from classical field theory simulations. 
Errors are the standard fit errors on the four fitting parameters $\rho_\text{c}(mg)$, which are defined as the width of the likelihood function in the direction of the respective parameter. 
This means, using a likelihood function $L(\mathbf{p})$ with parameter vector $\mathbf{p}$ and optimal parameters $\bar{\mathbf{p}}$, we define the error in $p_i$ as $\Delta p_i=(-L/\partial_i^2L)^{1/2}\,\big|_{\mathbf{p}=\bar{\mathbf{p}}}$.

One observes that the critical densities are shifted by around $4\%$ downwards as compared to the result from the classical simulations. 
If one assumes that the universal functional form of the critical density \eq{critical_dense} still holds in the full quantum model at (small) finite $mg$, albeit with a different non-universal constant $\zeta_{\rho}$, one may fit \eq{critical_dense} to our data, yielding
\begin{align}
\zeta_{\rho}
&=260\pm 12
\label{eq:zetarho}
\,,
\end{align}
as compared with $\zeta_{\rho}=380\pm3$ obtained in \cite{prokofev2001critical}. 
Note that the relative statistical error on $\zeta_{\rho}$ is larger than the relative shift between $g_R$ and $g$ even for the largest considered coupling, $(g-g_R)/g_R\sim 1.6\%$ for $mg=0.1$, which justifies our approximation $g_R\approx g$.

The above result for the critical density may be compared with the experimental findings of \cite{hung2011observation}, which gave a downward shift of $\sim 10\%$ in comparison to \eq{critical_dense} with $\zeta_{\rho}=380\pm3$.  
The experiment was conducted using an inhomogenous trapping potential, at couplings $mg=0.05,0.13,0.19,0.26$. Employing a local density approximation, the equation of state was determined from the density profile, from which in turn the authors extracted the critical densities. 
The effect of the local density approximation may need to be taken into account when comparing the by about a factor of two larger deviation with the present results. 

A different experimental approach was pursued in reference \cite{christodoulou2021observation}, where the authors employed the vanishing of the second-sound resonance as a criterion for the transition point in a uniform Bose gas. 
Similarly to \cite{hung2011observation}, the critical temperature was found to be shifted by $\sim10\%$ upwards, i.e.~the critical density by $\sim10\%$ downwards in comparison to the classical field theory result, though the statistical error is on the same order of magnitude as the shift itself. 
Additionally, the authors point to a possible bias due to finite-size effects. 
The results of further experiments \cite{ha2013strongly,fletcher2015connecting} are consistent, within their error bars, with either of the present results and the predictions of \cite{prokofev2001critical}.

In contrast to our findings, the PIMC simulations of \cite{pilati2008critical} gave a shift of the critical temperature to lower values, i.e., of the critical density $\rho_\mathrm{c}$ to values higher than the ones predicted in  \cite{prokofev2001critical}. 
These simulations were performed, however, in the strong-coupling regime $mg> 1$ ($mg=1.36,2.73,5.46$) that we did not find being accessible to CL. 

We performed an analogous finite-size analysis for the critical chemical potential, in which we fitted our data by means of the leading-order expression \eq{finitesizemu}.
This gave $A_{\mu}=0.087 \pm 0.003$ and $B_{\mu}=0.4 \pm 0.007$, and the corresponding fits are shown in the right panel of figure \fig{finitesize}.
(Note that one should have $B=B_{\mu}$, which is confirmed within errors.
The same does not apply to $A$ and $A_{\mu}$.)
The corresponding infinite-size $\mu_\text{c}$ is shown, again vs.~the coupling $mg$ and in comparison with the results obtained from the classical simulations of \cite{prokofev2001critical}, in the right panel of figure \fig{rhoc}.
Assuming again the functional form of the critical chemical potential \eq{critical_mu} to hold in the full quantum model at small  $mg$, a fit of \eq{critical_mu} to our data yields
\begin{align}
\zeta_{\mu}
&=12.3\pm 0.1
\label{eq:zetamu}
\,,
\end{align}
as compared with $\zeta_{\mu}=13.2\pm0.4$ obtained in \cite{prokofev2001critical}.

For comparison, we also performed simulations of the classical field theory ourselves, as we can straightforwardly turn off quantum effects in our simulation by setting $N_\tau=1$. The results are presented in appendix \sect{classsim}.

\subsubsection{\label{sec:scaleinv}Equation of state and scale invariance}
\begin{figure}
	\centering\includegraphics[width=0.495\textwidth]{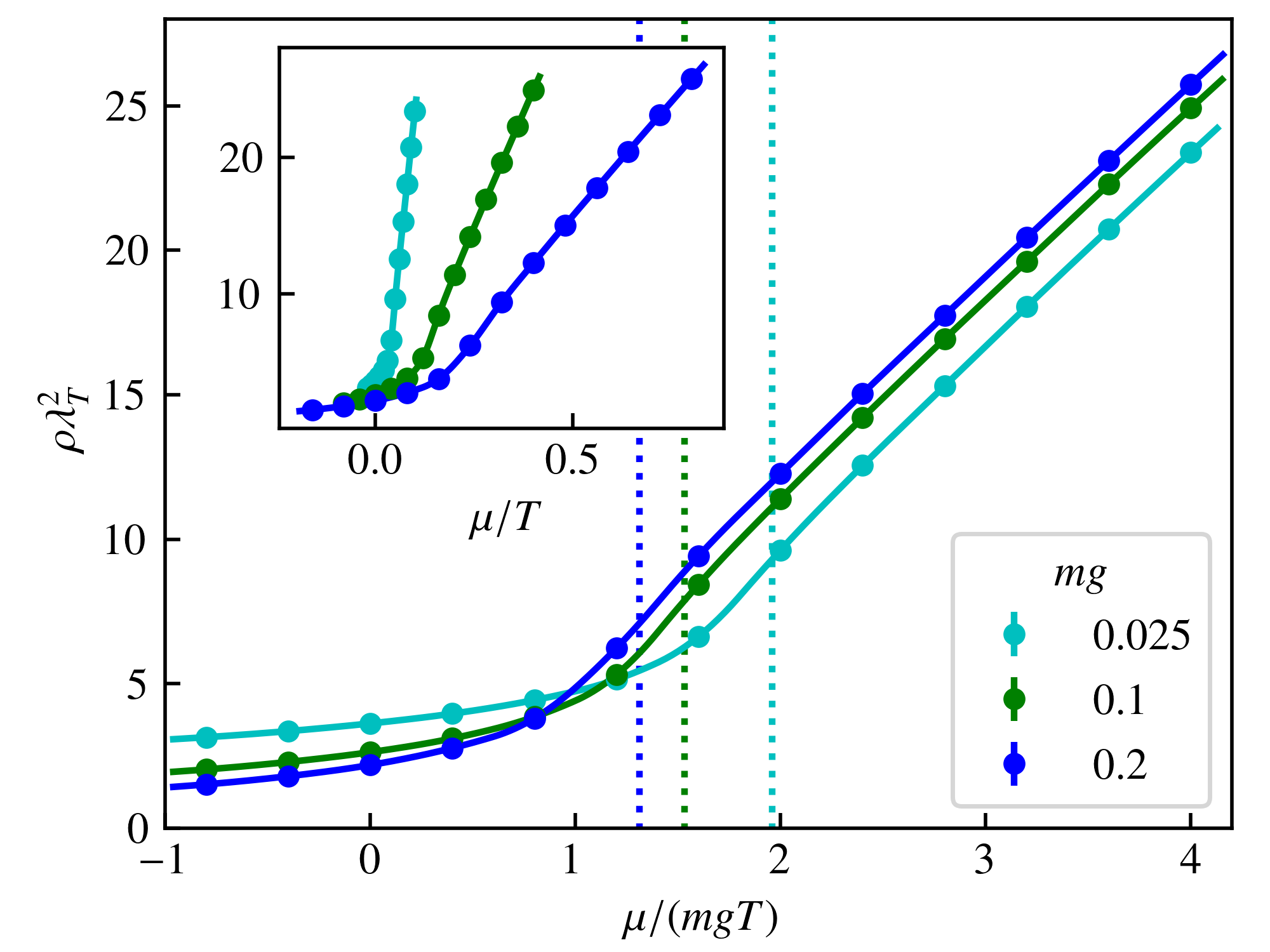}
	\centering\includegraphics[width=0.495\textwidth]{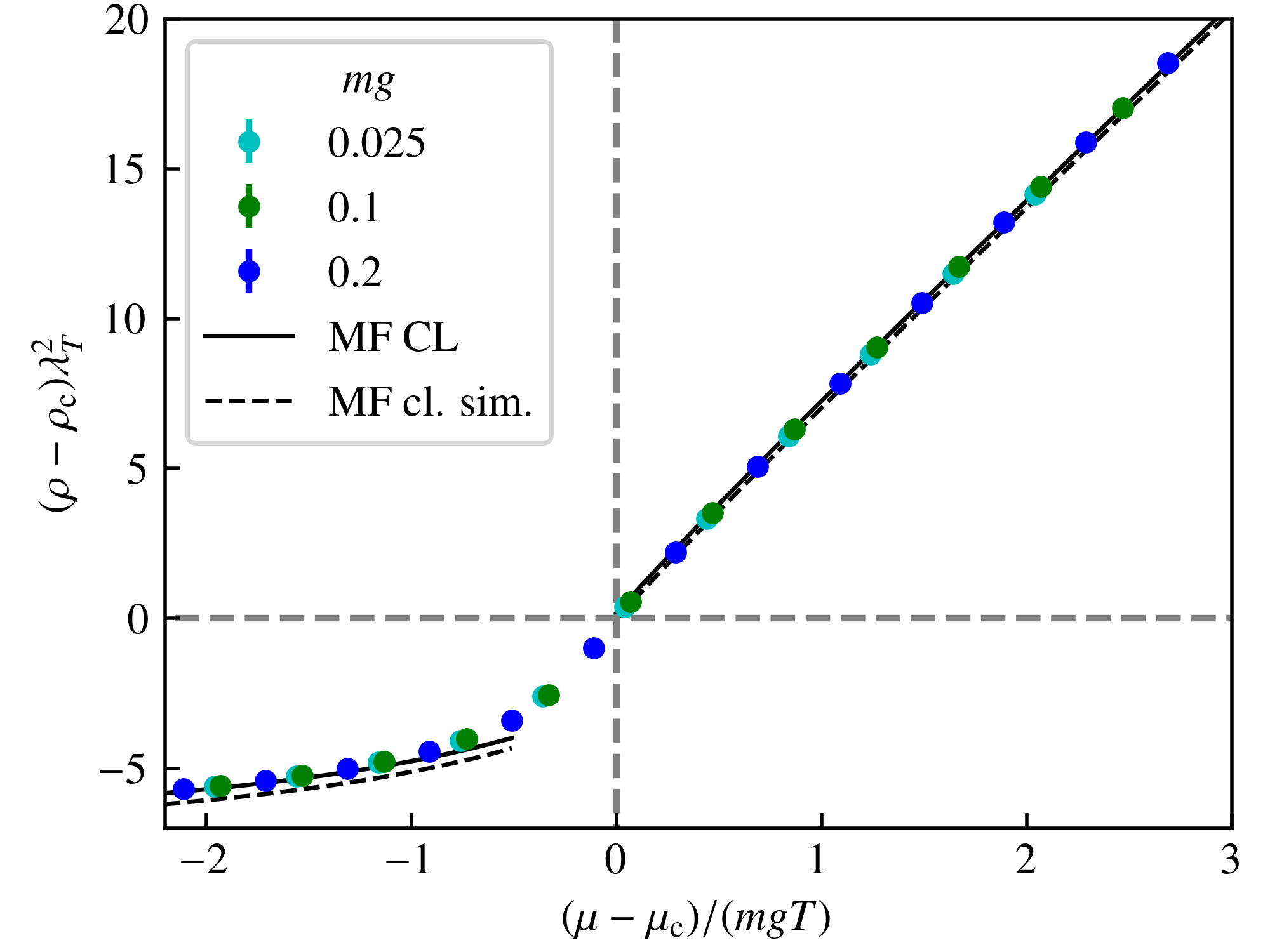}
	\caption{Equation of state, relating density $\rho$ and chemical potential $\mu$, for three values of the coupling $mg$. 
		The curves connecting the data points are obtained by spline interpolation. 
		Error bars are too small to be visible. 
		The inset of the left panel shows the densities as functions of $\mu/T$. 
		In the main part of the left panel, their arguments are rescaled with the coupling $mg$. The critical chemical potentials are shown as dotted vertical lines in the respective color.
		In the right panel, the resulting curves are shifted by the critical chemical potentials and densities, cf.~equation \eq{eosuniversal}. 
		The curves collapse for all three couplings considered, giving evidence of the universality of the 2D Bose gas near the BKT transition. 
		For the critical chemical potentials and densities $\rho_\text{c}$ and $\mu_\text{c}$ we took the result from  section \sect{critdens} for $mg=0.1$, whereas $\rho_\text{c}$ and $\mu_\text{c}$ for $mg=0.025$ and $mg=0.2$ are obtained from the differences $\mu_\text{c}(mg=0.025)-\mu_\text{c}(mg=0.1)$ and $\mu_\text{c}(mg=0.2)-\mu_\text{c}(mg=0.1)$ as well as $\rho_\text{c}(mg=0.025)-\rho_\text{c}(mg=0.1)$ and $\rho_\text{c}(mg=0.2)-\rho_\text{c}(mg=0.1)$, which by means of a least-squares fit were determined such that all three curves collapse. 
		The black lines in the right panel represent the approximations \eq{thetaMF} and \eq{thetaMF2}, valid for $X=(\mu-\mu_\text{c})/(mgT)\gg1$ and $X\ll-1$, respectively.  
		The dashed lines show them for $\zeta_{\rho}=380$ and $\zeta_{\mu}=13.2$ of \cite{prokofev2001critical}, the solid lines for the values \eq{zetarho} and \eq{zetamu}.
	}
	\label{fig:eos}
\end{figure}

As was outlined in section \sect{BKT}, the weakly interacting Bose gas is expected to feature universal behavior in the vicinity of the BKT transition. 
In order to analyze our data with respect to such universality, we depict, in the left panel of figure \fig{eos}, the equation of state $\rho(\mu)$ for a set of chemical potentials as obtained from our simulations for three different coupling strengths $mg\in\{0.025,0.1,0.2\}$, with $\mu$ given in units of $mgT$.
The data points are spline-interpolated for better visibility.
The respective critical $\mu_\mathrm{c}$ are indicated as thin vertical lines.
Since most data points are sufficiently far away from criticality, also for $mg=0.2$ runaway trajectories were mostly absent. However, for the two data points closest to the transition point, they started to occasionally occur, in which case we excluded the respective run, whereby we could still obtain a reasonable convergence of the total particle number. 
The inset shows the unscaled data as functions of $\mu/T$.

As is demonstrated in the right panel of figure \fig{eos}, shifting the curves by the respective $\rho_\text{c}\lambda_{T}^{2}$ downwards and $\mu_\text{c}/mgT$ to the left, with the $\rho_\text{c}$ and $\mu_\text{c}$ obtained as specified below, all curves collapse according to \eq{eosuniversal} to a single scaling function $F$.
Hence our data corroborates the predicted scaling within the depicted window around the critical point.

As similarly found for the classical simulations in \cite{prokofev2002two}, the approximate predictions \eq{thetaMF} and \eq{thetaMF2} (solid line) agree remarkably well with our CL results below and above the BKT transition, when choosing the non-universal constants $\zeta_\rho=260$ and $\zeta_{\mu}=12.3$ that we found from our simulations. Note that the non-universal parameter $\zeta_\mu$ enters the function $\theta(X)$ via \eq{thetaMF} and \eq{thetaMF2}. Additionally, $F(X)$ depends on $\theta_0=\ln(\zeta_\rho/\zeta_\mu)/\pi$, for which we found $\theta_{0}=\pi^{-1}\ln(\zeta_{\rho}/\zeta_{\mu})
=0.97 \pm 0.01$, as compared to $\theta_{0}=1.068\pm0.01$ reported in \cite{prokofev2002two}. Hence, the concrete functional form of $F(X)$ comes out slightly different from our simulation than predicted in \cite{prokofev2002two}. For comparison, we also show the scaling form $F(X)$ for the parameters $\zeta_{\rho}=380$ and $\zeta_{\mu}=13.2$ of \cite{prokofev2001critical} (dashed line). 

Under the approximation that the universal scaling relation \eq{eosuniversal} holds, at least for the not too large values $mg$ considered here, we can also determine the shift in the critical chemical potentials and densities between different $mg$, by fitting the curves to collapse onto each other.
This allows extracting the critical chemical potentials and densities independent of the procedure described in the previous section once $\mu_\mathrm{c}$ and $\rho_\mathrm{c}$ are known for one value of $mg$. 
The $\mu_\mathrm{c}$ and $\rho_\mathrm{c}$ employed in this section were obtained in this way, with $\mu_\mathrm{c}(mg=0.1)=(1.531\pm0.006)\,mgT$ and $\rho_\mathrm{c}(mg=0.1)=(7.874\pm0.008)\lambda_T^{-2}$ taken from the $256^2$ lattice results of section \sect{critdens}. 
For a comparison of the critical densities obtained in this way, with those from section \sect{critdens}, see appendix \sect{critdensscal}.

\subsubsection{\label{sec:corrfunc}Single-particle momentum spectra}
%
\begin{figure}
	\centering\includegraphics[width=0.495\textwidth]{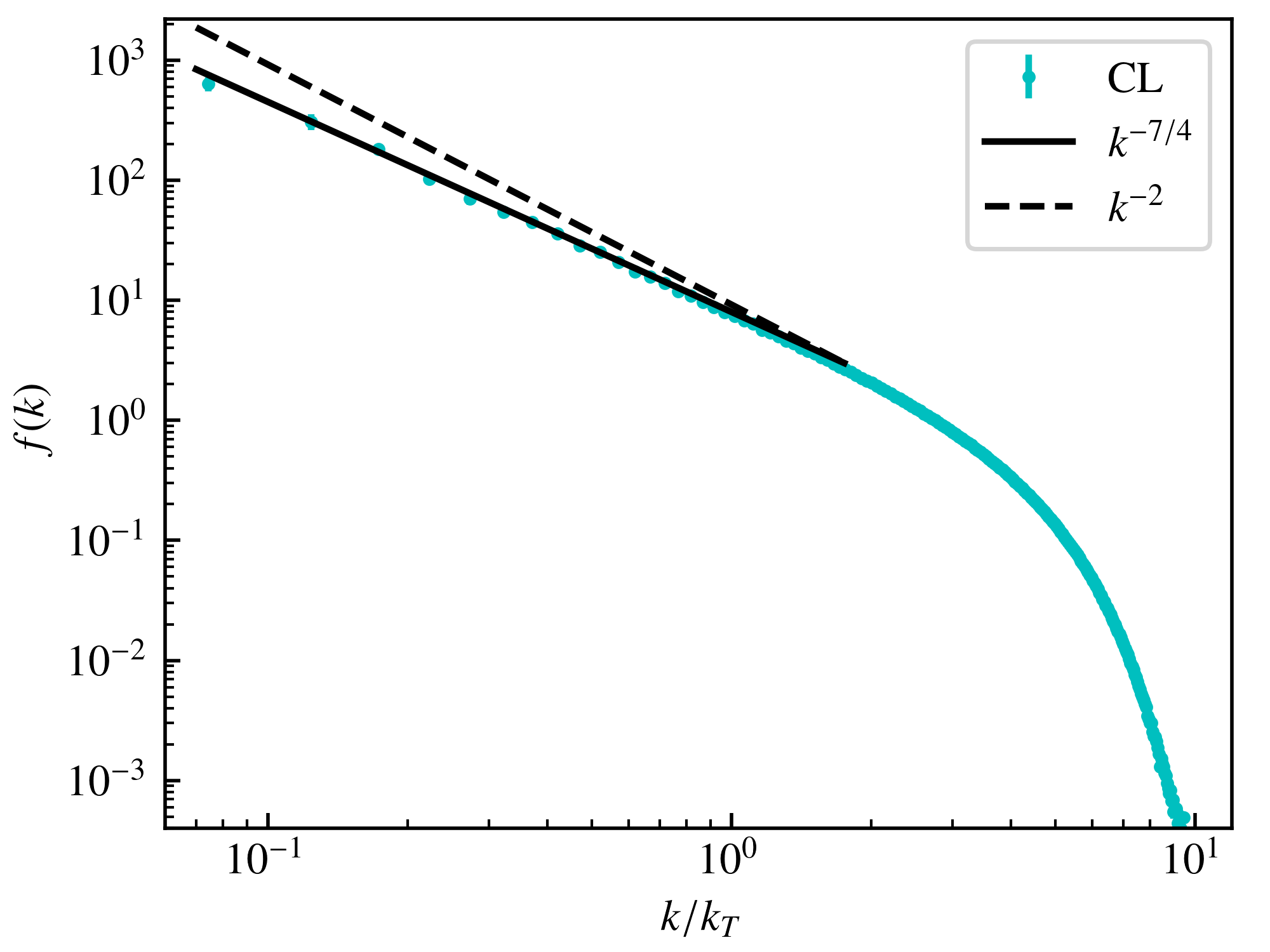}
	\centering\includegraphics[width=0.495\textwidth]{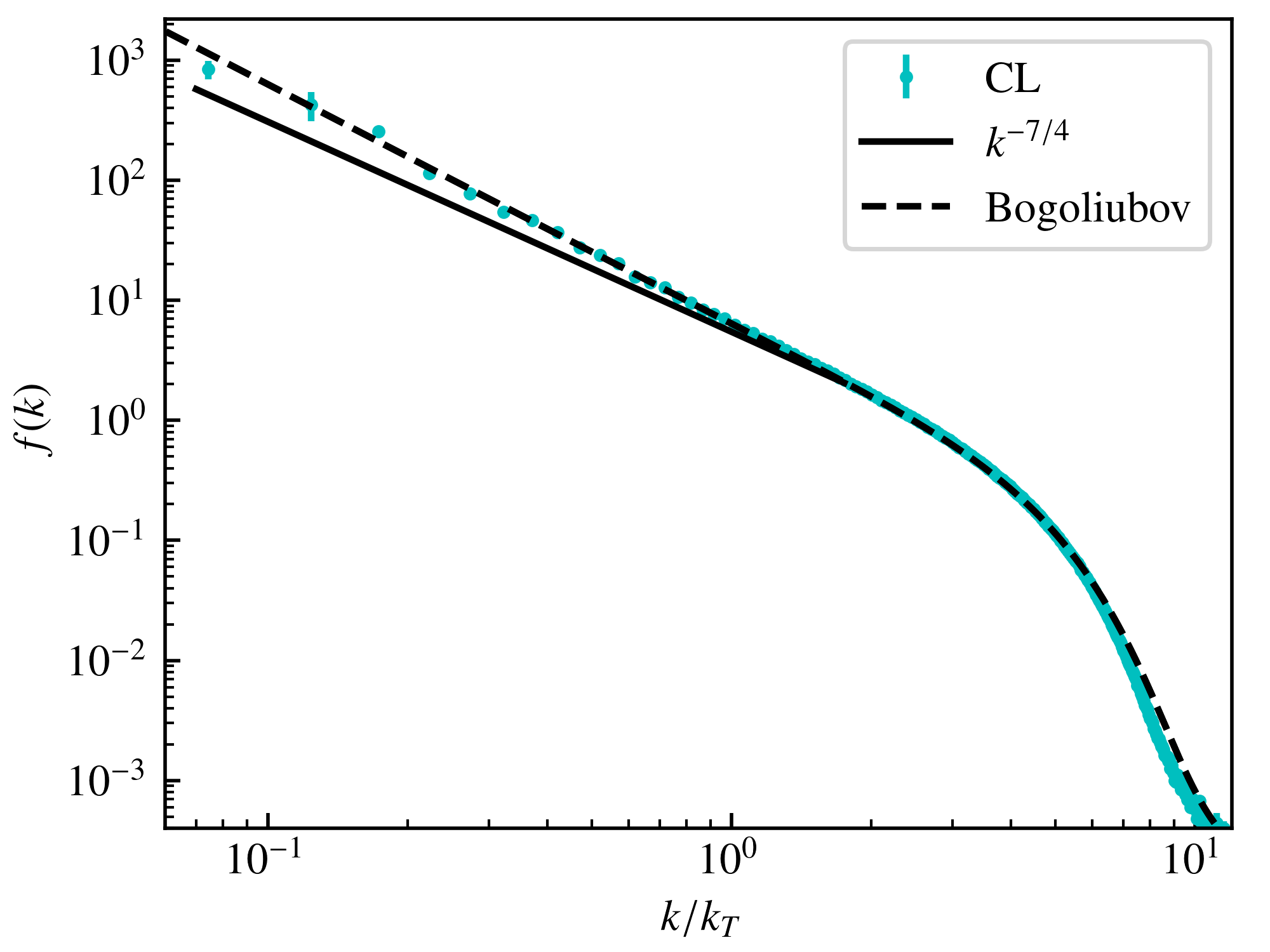}
	\caption{	
		Single-particle momentum spectra $f({k})=f(|\mathbf{k}|)$, equation \eq{fk}, for a coupling $mg=0.1$ and two different chemical potentials: $\mu/(mgT)=2$, slightly below the transition (left panel), and $\mu/(mgT)=4$, far below the transition (right panel). 
		The momentum is expressed in units of the thermal momentum $k_T=1/\lambda_T$. 
		Whereas slightly below the transition the spectrum approximately shows a $k^{-7/4}$ power law in the infrared, below the thermal momentum $k_{T}$, consistent with the near-critical scaling predicted by BKT theory, we find a Rayleigh-Jeans $\sim k^{-2}$ fall-off further below the transition, which forms part of the Bogoliubov distribution (dashed line). The condensate fractions $f(\mathbf{k}=0)/N_\text{tot}$ are $61\%$ and $82\%$, respectively, i.e. due to the finite extent of our system we find a macroscopic occupation of the condensate mode.
	}
	\label{fig:spectrum}
\end{figure}
\begin{figure}
	\centering\includegraphics[width=0.6\textwidth]{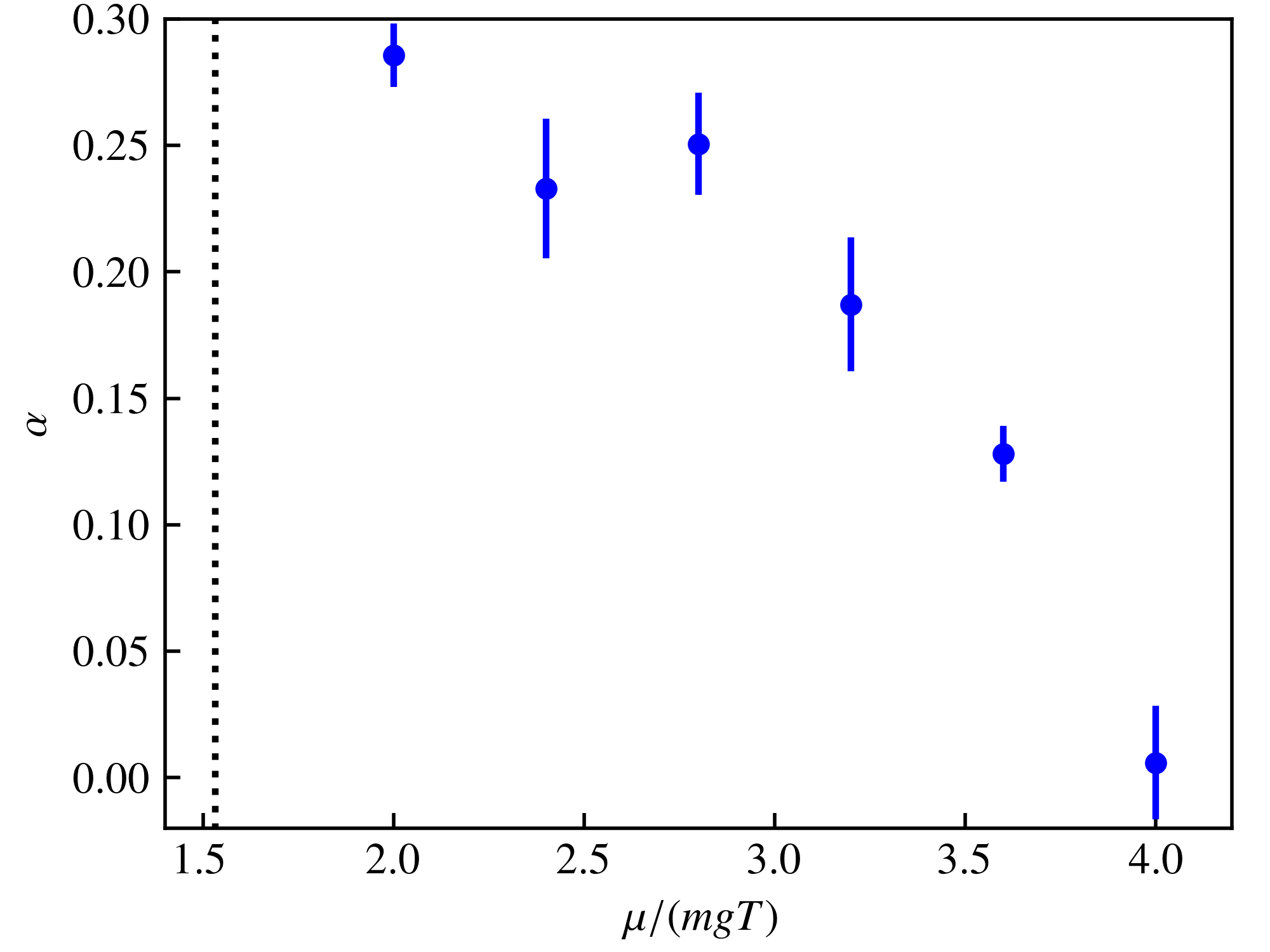}
	\caption{ 
		The power-law exponent $\alpha$ as a function of $\mu/(mgT)$, as obtained from a least-squares fit of a linear function to $\log f(k)$ vs. $\log k$, obtained from our simulations for $mg=0.1$. The critical chemical potential is marked by a dotted black line. Going from close to the transition to far below the transition, $\alpha$ decreases from $\simeq1/4$ to zero, in accordance with the prediction from BKT theory.
	}
	\label{fig:alphas}
\end{figure}

In this section, we use the CL data to evaluate the single-particle momentum-dependent occupation number distribution \eq{fk} for comparison with the infrared scaling \eq{fkscaling} expected close to and further below the BKT transition. 
We performed simulations for a coupling $mg=0.1$ at several chemical potentials ranging from $\mu/(mgT)=2$ to $\mu/(mgT)=4$, i.e. from slightly below to far below the BKT transition. 
The results for $\mu/(mgT)=2$ and $4$ are shown in figure \fig{spectrum}. 

As predicted by BKT theory, far below the transition, the distribution exhibits a $\sim k^{-2}$,  Rayleigh-Jeans power law in the infrared regime of momenta far below the temperature scale, while it falls off exponentially at larger momenta.
This is consistent with a thermal distribution of Bogoliubov quasiparticles, resulting in the dashed distribution in the right panel.
Close to the BKT transition, in contrast, the power law in the infrared is slightly reduced to $\sim k^{-7/4}$. 
In order to make this more systematic, we have extracted $\alpha$ for a range of chemical potentials by fitting a linear function to $\log f(k)$ vs.~$\log k$ in the power-law region. 
The results are shown in figure \fig{alphas}. 
As one can see, the value of $\alpha$ drops  from $\alpha\simeq1/4$ towards zero, between the chemical potentials slightly below and far below the transition, in accordance with the BKT prediction.

\subsubsection{\label{sec:vortices}Vortex unbinding across the BKT transition}
\begin{figure}
	\centering\includegraphics[width=0.495\textwidth]{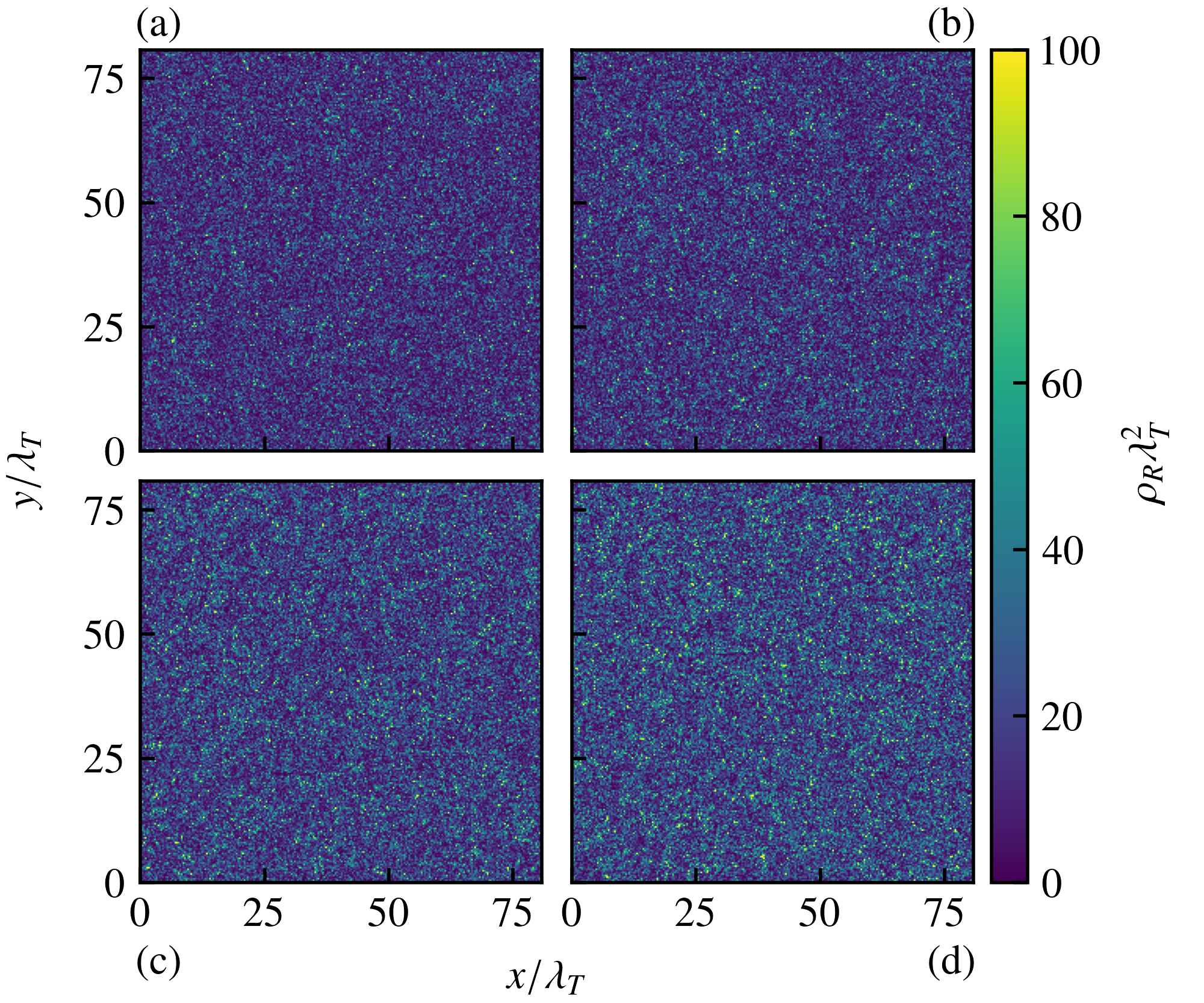}
	\centering\includegraphics[width=0.495\textwidth]{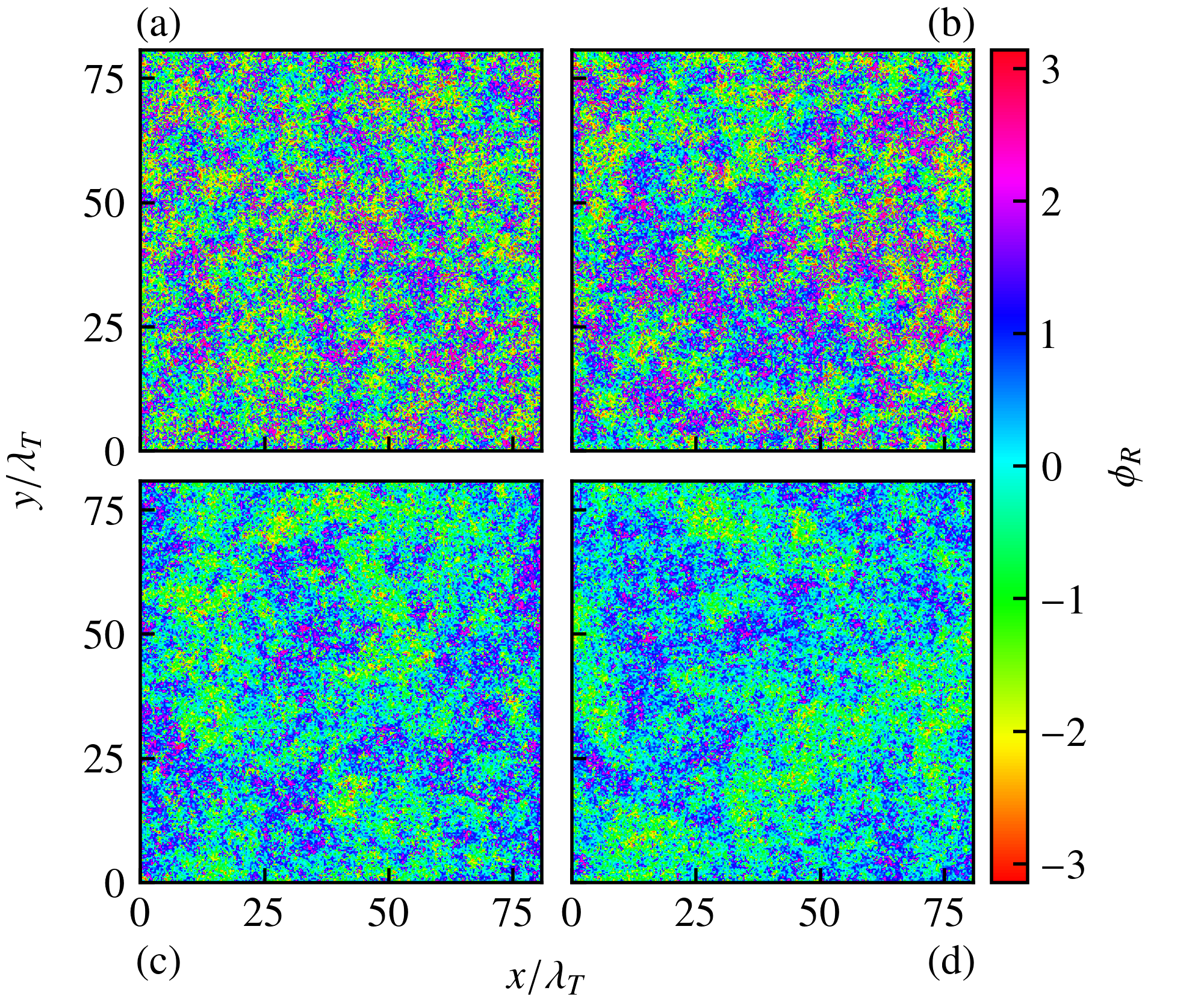}
	\caption{ 
		Snapshots of the position-space density $\rho_R\equiv \varphi_R^2+\chi_R^2$ at $\vartheta=4\cdot 10^3 \,a_\text{s}^{-3}$ for $mg=0.1$ and four different chemical potentials (a) $\mu/(mgT)=0.8$, (b) $1.2$, (c) $1.6$, and (d) $2.0$ (left panel). 
		The temperature is kept fixed while the chemical potential is varied, such that the healing length $\xi_\text{h}\equiv1/\sqrt{2m\mu}$ in units of the lattice spacing varies between $\xi_\text{h}=3.16\,a_\text{s}$ and $\xi_\text{h}=2\,a_\text{s}$. 
		The mean field densities $\lambda_T^2 \bar{\rho}\equiv \lambda_T^2 \mu/g$ are $\lambda_T^2 \bar{\rho}=5.02,7.54,10.05,12.56$. 
		As one can see, the position-space densities are completely dominated by fluctuations such that it is difficult to infer information about the topological phase transition from them. 
		The right panel shows the position-space phase $\phi_R\equiv \arg\left(\varphi_R+i\chi_R\right)$ for the same parameters, which indicates phase ordering across the transition.
	}
	\label{fig:realspace}
\end{figure}
One of the key features of the BKT transition is that it is characterized by the transition from a free vortex gas to bound vortex-antivortex pairs, which we here attempt to study within the CL framework. 
Let us first have a look at the configurations produced by the Langevin process in position space. 
We consider the density $\rho_R$ and phase $\phi_R$ characterizing the real parts of the complexified two-component field, defined as 
\begin{align}
\rho_R&\equiv\varphi_R^2+\chi_R^2
\\
\phi_R&\equiv\arg(\varphi_R+i\chi_R)
\,,
\end{align}
where we have used the notation introduced in section \sect{Langevin_eq}.
Their distributions across the spatial grid are shown in figure \fig{realspace}, for four values of the chemical potential $\mu$ in the vicinity of the BKT transition. 
While the density $\rho_R$ seems entirely dominated by noise, the phase $\phi_R$ reveals the phase ordering process across the transition.
However, one does not observe stable vortices in the position-space configurations. 
This is not a genuine property of complex Langevin simulations, but is also the case e.g. for ordinary Monte Carlo simulations of the XY model, cf. figure 1 in \cite{peled2019lectures}.

One may conclude that expectation values, i.e.~long-time averages along the Langevin trajectories, must be considered in order to gain insight into the topological phase transition. 
Typically, the analysis of topological properties of a Bose gas rests on the phase of the complex field $\psi$ or the velocity field $\mathbf{v}=\mathbf{j}/\rho$. For the specific case of a complex Langevin simulation, however, the problem is that both lead to observables that are non-holomorphic in the fields. 
In contrast, the CL algorithm requires the real-valued fields to be analytically continued, such that the action and observables should preferably be holomorphic. Non-holomorphic actions and observables lead to both computational problems (there can be numerical divergencies when the trajectories come close to the singularities) as well as conceptual ones (the correctness of the method is more difficult to establish) \cite{aarts2017complex}.
This represents a limitation to the applicability of the CL algorithm in this case. 

\begin{figure}
	\centering\includegraphics[width=0.6\textwidth]{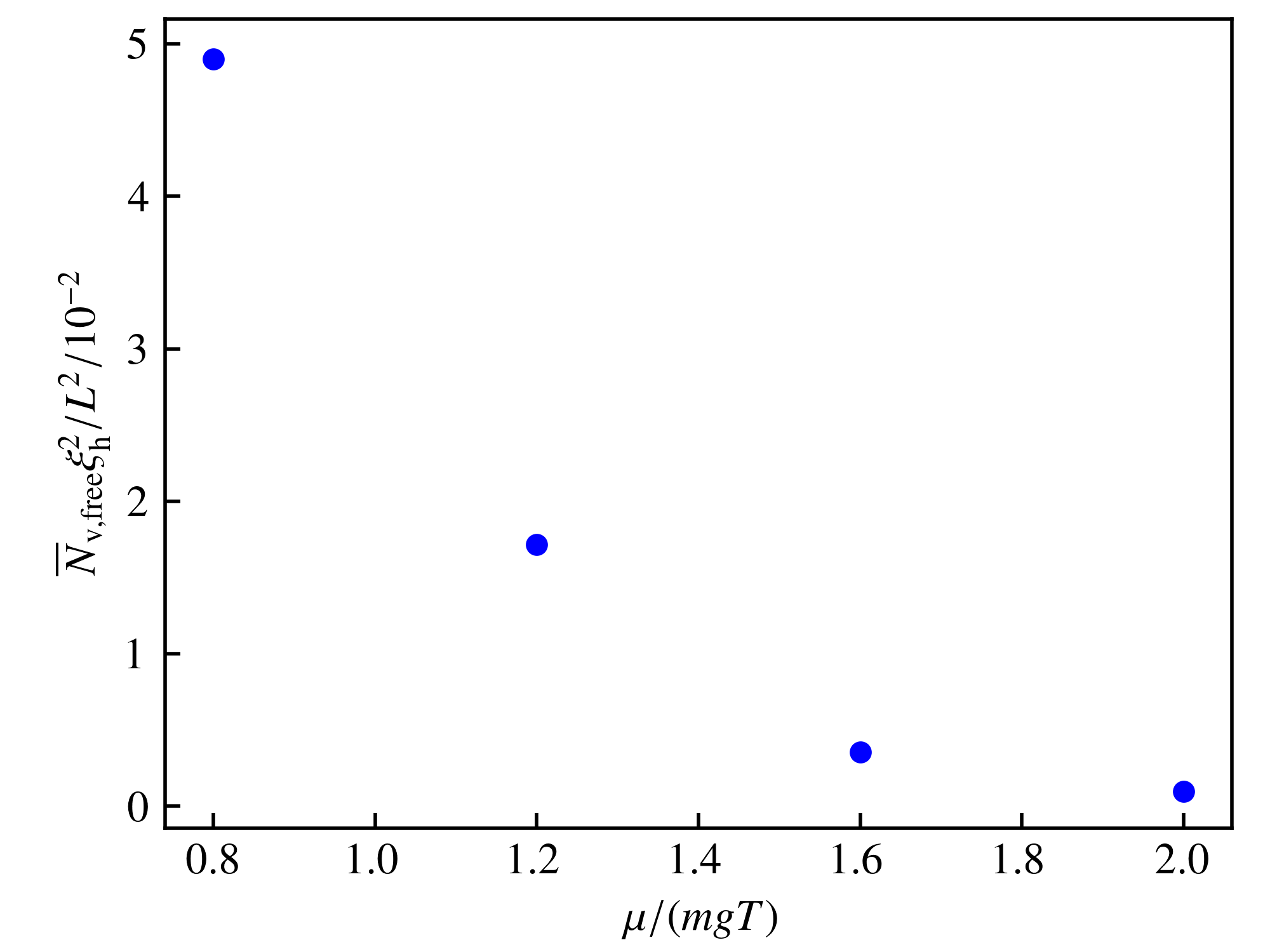}
	
	\caption{ 
		The mean density of free vortices, $\rho_{\mathrm{v,free}}\xi_\mathrm{h}^{2}$, as defined in \eq{defY}, as a function of $\mu/(mgT)$, for $mg=0.1$. 
		Error bars are too small to be visible.
	}
	\label{fig:jrot2}
\end{figure}

For this reason we resort to evaluating the current density instead of the velocity field,  as discussed in section \sect{TopoBKT}, which is perfectly holomorphic in $\varphi$ and $\chi$. Neglecting the effect of density fluctuation in the bulk, the mean free vortex density can be related to the rotational current density as $\rho_{\mathrm{v,free}}(\mathbf{r}) 
\simeq m^2|\mathbf{j}_\text{rot}(\mathbf{r})|^2/(2\pi \ln(L/\xi_\text{h})\langle\rho\rangle^2)$, equation \eq{defrhovfree}. $|\mathbf{j}_\text{rot}(\mathbf{r})|^2$ is most conveniently evaluated in momentum space. The momentum space current ${\mathbf{j}}_{i,\mathbf{p}}$ can be written as 
\begin{align}
\mathbf{j}_{i,\mathbf{p}}
=\frac{1}{2m}\sum_\mathbf{q}(\mathbf{p}+2\mathbf{q})\,\psi_{i+1,\mathbf{p}+\mathbf{q}}^*\psi_{i,\mathbf{q}}
\,.
\end{align}
The Helmholtz decomposition, in momentum space, reads 
\begin{align}
\mathbf{j}_{i,\mathbf{p}}
=\mathbf{j}_{i,\mathbf{p}}^\text{irr}+\mathbf{j}_{i,\mathbf{p}}^\text{rot}
=\mathbf{p}\frac{\mathbf{p}\cdot\mathbf{j}_{i,\mathbf{p}}}{|\mathbf{p}|^2}
-\frac{\mathbf{p}\times\mathbf{p}\times\mathbf{j}_{i,\mathbf{p}}}{|\mathbf{p}|^2}
\,.
\end{align}
The expectation value of the spatial integral over the squared current density, $\int d^2r\, |\mathbf{j}_\text{rot}(\mathbf{r})|^2$, can then be straightforwardly computed as 
\begin{align}
\int \mathrm{d}^2x\,\left\langle\,|\mathbf{j}_\text{rot}(\mathbf{x})|^2\right\rangle 
&=\left\langle\frac{1}{N_\tau}\sum_i\sum_\mathbf{p}
\,\mathbf{j}_{i,\mathbf{p}}^\text{rot}\cdot \mathbf{j}_{i,-\mathbf{p}}^\text{rot}\right\rangle
\,.
\end{align}
In figure \fig{jrot2}, we show the quantity
\begin{align}
\label{eq:defY}
\frac{\xi_\mathrm{h}^{2}}{L^2}\overline{N}_\text{v,free}
= \frac{\xi_\mathrm{h}^{2}}{L^2}\int \mathrm{d}^2x\,\left\langle\,\rho_{\mathrm{v,free}}(\mathbf{x})\right\rangle= \frac{\xi_\mathrm{h}^{2}}{L^2}\frac{m^2}{2\pi \ln(L/\xi_\text{h})\langle\rho\rangle^2}\int \mathrm{d}^2x\,\left\langle\,|\mathbf{j}_\text{rot}(\mathbf{x})|^2\right\rangle 
\,,
\end{align}
with healing length $\xi_\mathrm{h}=(2m\mu)^{-1/2}$, i.e., the mean number of free vortices per healing length squared, which can be considered to be a measure for the transition from a vortex gas to bound vortex-anti-vortex pairs, as a function of chemical potential. $\xi_\mathrm{h}^{2}\overline{N}_\text{v,free}/L^2$ is found to rapidly decay once the chemical potential is tuned across the transition.

\begin{figure}
	\centering\includegraphics[width=0.6\textwidth]{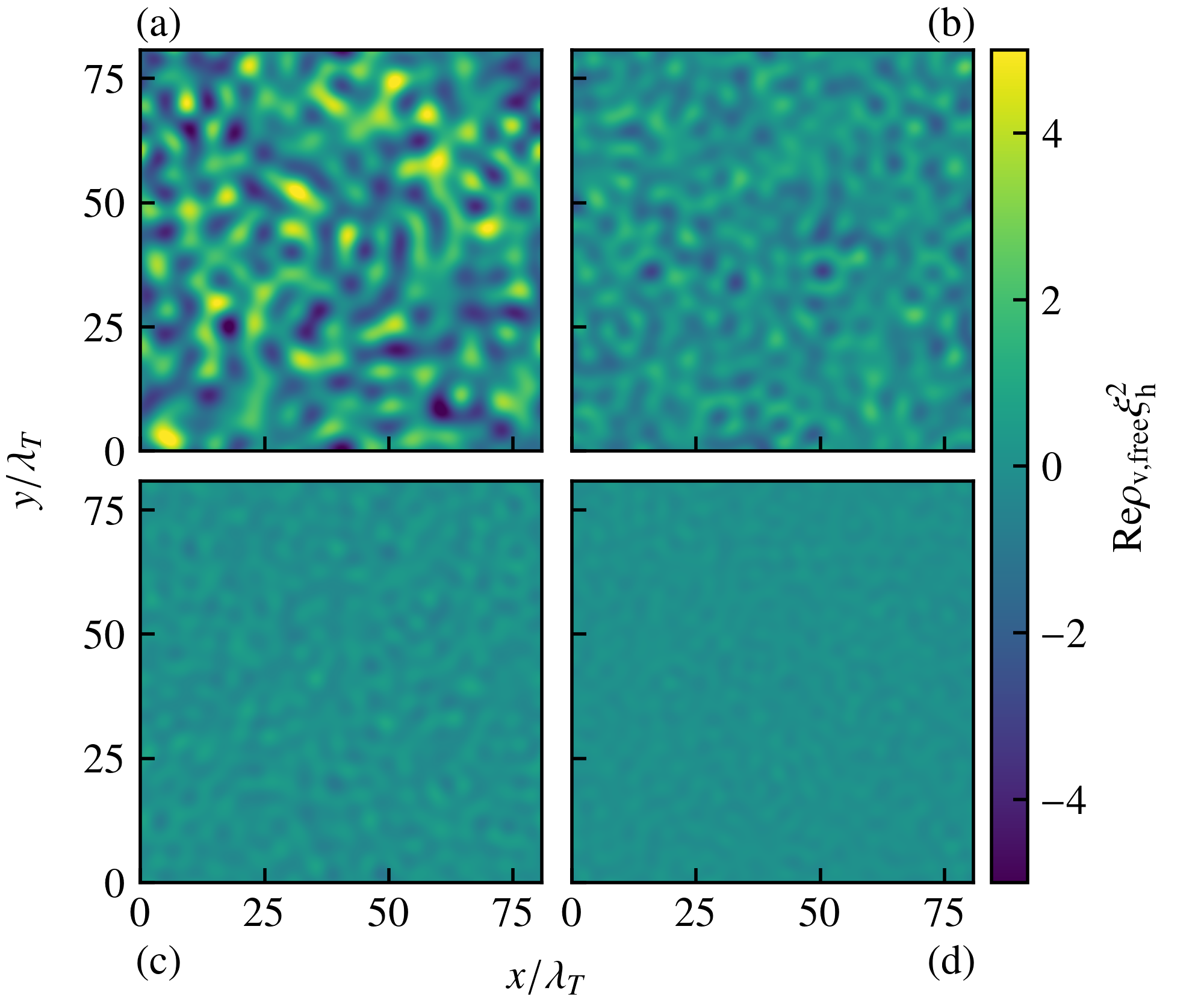}
	\caption{ 
		Snapshots of the real part of $\rho_{\mathrm{v,free}}(\mathbf{x})\xi_\mathrm{h}^2$, as defined in \eq{defrhovfree} for $mg=0.1$ and the same four different chemical potentials (a) $\mu/(mgT)=0.8$, (b) $1.2$, (c) $1.6$, and (d) $2.0$, corresponding to the data shown in figure \fig{realspace}. For a better visibility, fluctuations above the healing momentum $k_\mathrm{h}=\sqrt{2m\mu}$ are filtered out by a low-pass filter. Note that by virtue of the complexification prescription of the CL algorithm, $\rho_{\mathrm{v,free}}(\mathbf{x})\xi_\mathrm{h}^2$ will be in general complex in single realizations and its real part can become negative, while in the long-time average, $\rho_{\mathrm{v,free}}(\mathbf{x})\xi_\mathrm{h}^2$ will come out real and positive.
	}
	\label{fig:Ytilde}
\end{figure}

Snapshots of the spatial distribution of $\rho_{\mathrm{v,free}}(\mathbf{x})$, which can be computed by Fourier transforming ${\mathbf{j}}_{i,\mathbf{p}}$ back to real space and squaring it, are depicted in figure \fig{Ytilde}, for the same four chemical potentials as chosen in figure \fig{realspace}. Note that in snapshots of a CL simulation, this quantity will result complex in general, with the imaginary part averaging out only in the long-time average. In figure \fig{Ytilde}, this imaginary part is discarded.

The decrease of $\rho_{\mathrm{v,free}}$ across the transition corroborates the vortex-anti-vortex recombination process, when going over from the disordered above to the ordered phase below the transition.

\subsection{Conclusion}
Employing the complex Langevin algorithm, we have performed an ab-initio simulation of the interacting Bose gas in two spatial dimensions across the BKT phase transition. 
We have found that the CL method is able to successfully reproduce central characteristics of BKT physics, namely the universality in the equation of state, the algebraic decay of correlation functions and the vortex-unbinding mechanism. 
We have furthermore analyzed the dependence of the critical density on the interaction strength $mg$ and compared to  results from simulations of the classical field theory \cite{prokofev2001critical}. 
Our simulations of the full quantum model yield small but significant deviations from the classical-field-theory predictions. Unfortunately, the uncertainty of existing experimental results is slightly too large to definitely decide in favor of one of the two results for the critical density, as it is of the order of the shift itself. Furthermore, these experimental works were lacking a thorough finite-size analysis of their results. As it would suffice to reduce the error by a factor of $2$ to $3$, there is hope that future experimental measurements might eventually settle the question of the correct value of the critical density.

In conclusion, this work demonstrates that the CL algorithm can be a viable tool for performing full quantum simulations of the topological phase transition in a weakly interacting Bose gas, i.e., could be of use when quantum corrections beyond classical Gross-Pitaevskii simulation are of interest. 
Experimentally relevant applications include the BKT transition in external trapping potentials \cite{hadzibabic2006berezinskii,bisset2009quasicondensation} (see also chapter \sect{trap}), with long-range interactions \cite{fedichev2012bkt,giachetti2022berezinskii} or in multi-component gases \cite{kasamatsu2005vortices,kobayashi2019berezinskii,furutani2023berezinskii}.

\clearpage

\thispagestyle{plain}
\section{Density profiles in a harmonic trap\label{sec:trap}}
A very common experimental setting for ultracold Bose gases is a confinement in a harmonic trap, i.e. the atoms are subjected to an external potential of the form 
\begin{align}
V(\mathbf{r})=\frac{1}{2}m\left(\omega_x^2 x^2+\omega_y^2 y^2+\omega_z^2 z^2\right)\,,
\end{align}
such that the action in thermal equilibrium reads
\begin{align}
\nonumber S[\psi,\psi^*]=&\int \limits_0^\beta d\tau\int d^3r\Bigg\{\psi^*\partial_\tau\psi-\frac{1}{2m}\psi^*\Delta\psi\\&-\left[\mu-\frac{1}{2}m\left(\omega_x^2 x^2+\omega_y^2 y^2+\omega_z^2 z^2\right)\right]\psi^*\psi+\frac{g}{2}\left(\psi^*\psi\right)^2\Bigg\}\,.
\end{align}
As outlined in section \ref{sec:physics_bose}, if one trapping frequency, say $\omega_z$, is much larger than the chemical potential, $\omega_z\gg \mu$, we can integrate out one spatial dimension and arrive at an effective two-dimensional action:
\begin{align}
\nonumber S_{2D}[\psi,\psi^*]=&\int \limits_0^\beta d\tau\int d^2r\Bigg\{\psi^*\partial_\tau\psi-\frac{1}{2m}\psi^*\Delta\psi\\&-\left[\mu-\frac{1}{2}m\left(\omega_x^2 x^2+\omega_y^2 y^2\right)\right]\psi^*\psi+\frac{g_{2D}}{2}\left(\psi^*\psi\right)^2\Bigg\}\,,
\end{align}
where the effective two-dimensional coupling is given by
\begin{align}
\label{eq:g2D}
g_{2D}=\frac{\sqrt{8\pi}}{m}\frac{a}{l_z}
\end{align}
with the 3D scattering length $a$ that in first order is related to $g$ by $a=mg/(4\pi)$ and the effective extension in $z$-direction $l_z=1/\sqrt{m\omega_z}$. In the following, we will mainly consider the case of a two-dimensional gas in a trap and therefore drop again the subscript of $g_{2D}$.

A very natural question to ask is what will be the spatial density distribution $\rho(\mathbf{r})=\langle\psi(\mathbf{r})^\dagger\psi(\mathbf{r})\rangle$ in thermal equilibrium for given temperature $T$, chemical potential $\mu$ (or particle number $N_\text{tot}$), frequencies $\omega_{x,y,z}$ and coupling $g$~\footnote{The trapped thermal Bose gas can be uniquely described by three dimensionless parameters if all trapping frequencies are equal, $\omega_{x,y,z}\equiv\omega$ (otherwise the ratios between the frequencies add to the number): The ratio of chemical potential and trapping frequency $\mu/\omega$, the ratio of temperature and trapping frequency $T/\omega$ and the coupling strength $mg$ (2D) or  $a/l$ with $l=1/\sqrt{m\omega}$ (3D).}. Despite the apparent simplicity of this problem, it turns out that the interplay of thermal and quantum effects as well as the inhomogeneity of the external potential render the precise determination of $\rho(\mathbf{r})$ a non-trivial task. 

In lowest order, one can obtain an estimate for $\rho(\mathbf{r})$ in a mean field approximation, i.e. one searches for the function $\psi(\mathbf{r})$ that minimizes the classical energy functional
\begin{align}
E[\psi]=\int d^dr\left\{-\frac{1}{2m}\psi^*\Delta\psi-\mu |\psi|^2+V(\mathbf{r})|\psi|^2+\frac{g}{2}|\psi|^4\right\}\,.
\end{align}
The solution to this minimization problem can be easily obtained numerically by means of a steepest-decent algorithm, i.e. one evolves $\psi$ according to 
\begin{align}
\label{eq:steep_desc}
\frac{\partial\psi}{\partial\vartheta}=-\frac{\delta E}{\delta\psi^*}=-\frac{1}{2m}\Delta\psi-\mu \psi+V(\mathbf{r})\psi+g|\psi|^2\psi\,.
\end{align}
This corresponds exactly to performing a CL simulation with $N_\tau=1$ and the noise turned off~\footnote{Setting $N_\tau=1$ but keeping the noise amounts to a simulation of classical field theory instead, cf. section \ref{sec:class_field_theory}.}. An approximate analytical solution for the density profile can be obtained by neglecting the kinetic term $-\Delta\psi/2m$ in \eq{steep_desc}, which is possible for $\mu\gg \omega_{x,y,z}$ and sufficiently close to the center of the trap. Then, the density profile is simply given by
\begin{align}
\rho(\mathbf{r})=|\psi(\mathbf{r})|^2=\frac{\mu-V(\mathbf{r})}{g}\,.
\end{align}
This is known as the Thomas-Fermi (TF) density profile. As most good approximations, the Thomas-Fermi approximation signals the point of its own breakdown: For $V(\mathbf{r})>\mu$, the density would become negative, such that in this region $-\Delta\psi/2m$ cannot be negligible any more. For $\omega_x=\omega_y=\omega_z\equiv\omega$, this is the case for $r>r_\text{TF}$ with the Thomas-Fermi radius
\begin{align}
r_\text{TF}=\sqrt{\frac{2\mu}{m\omega^2}}\,.
\end{align}
The Thomas-Fermi radius gives an approximate measure for the size of the cloud of atoms in a trap. 

The most common simplifying assumption when considering corrections to the mean field density profile is the \textit{local density approximation (LDA)}. The idea is to treat the problem of finding the density at point $\mathbf{r}$ by computing the density in a homogeneous, infinitely extended system with constant external potential $V(\mathbf{r})\equiv\text{const}$. This approximation basically reduces the problem of finding $\rho(\mathbf{r})$ to finding the equation of state $\rho(\mu)$ in a homogeneous, infinitely extended system. In the center of the trap, the effective chemical potential $\mu'\equiv \mu-V(r)\gtrsim T$ is large and the system is in the condensed phase, while in the region outside the Thomas-Fermi radius, where mean-field theory predicts zero density, we have $\mu'<0$ and the particles are in the thermal phase. In general, the LDA is a good approximation as long as the potential varies sufficiently slowly in comparison to other important length scales such as the thermal wave length. For a harmonically trapped gas, this amounts to the condition $T\gg\omega$, which is typically well satisfied in experimental settings. The strongest deviations from the density profile obtained within the LDA can be expected to occur in the vicinity of the Thomas-Fermi radius $r_\mathrm{TF}$, where the LDA breaks down within mean-field theory.  

While the LDA substantially simplifies the problem of finding $\rho(\mathbf{r})$, there still remains the problem of finding $\rho(\mu)$. For this purpose, one can employ several approximate schemes such as the Hartree-Fock approximation, Bogoliubov theory, the Popov approximation and renormalization group theory \cite{lim2008correlation}. However, analytical approaches almost universally fail close to the highly non-perturbative BKT transition (or BE transition in a three-dimensional trap). As the BKT transition occurs around the Thomas-Fermi radius where also the deviations from the LDA itself are expected to be the strongest, it is precisely this region where the benefits from a full quantum simulation are the largest.  

Numerical approaches for determining the density profile apart from CL that do not rely on the local density approximation are the equally fully exact path-integral Monte Carlo (PIMC) method \cite{krauth1996quantum,heinrichs1998quantum,ruggeri2014quantum,yan2015incorporating} and the semi-classical stochastic Gross-Pitaevskii equation (SGPE) \cite{cockburn2012ab}.

In this short chapter, we will first present computations of the density profile for a generic parameter set. We study the effect of temperature on the density profile and examine the range of validity of the local density approximation. In the second part, we will compare to an actual experimental setting.

\subsection{Density profiles from CL}

\begin{figure}
	\centering\includegraphics[width=0.495\textwidth]{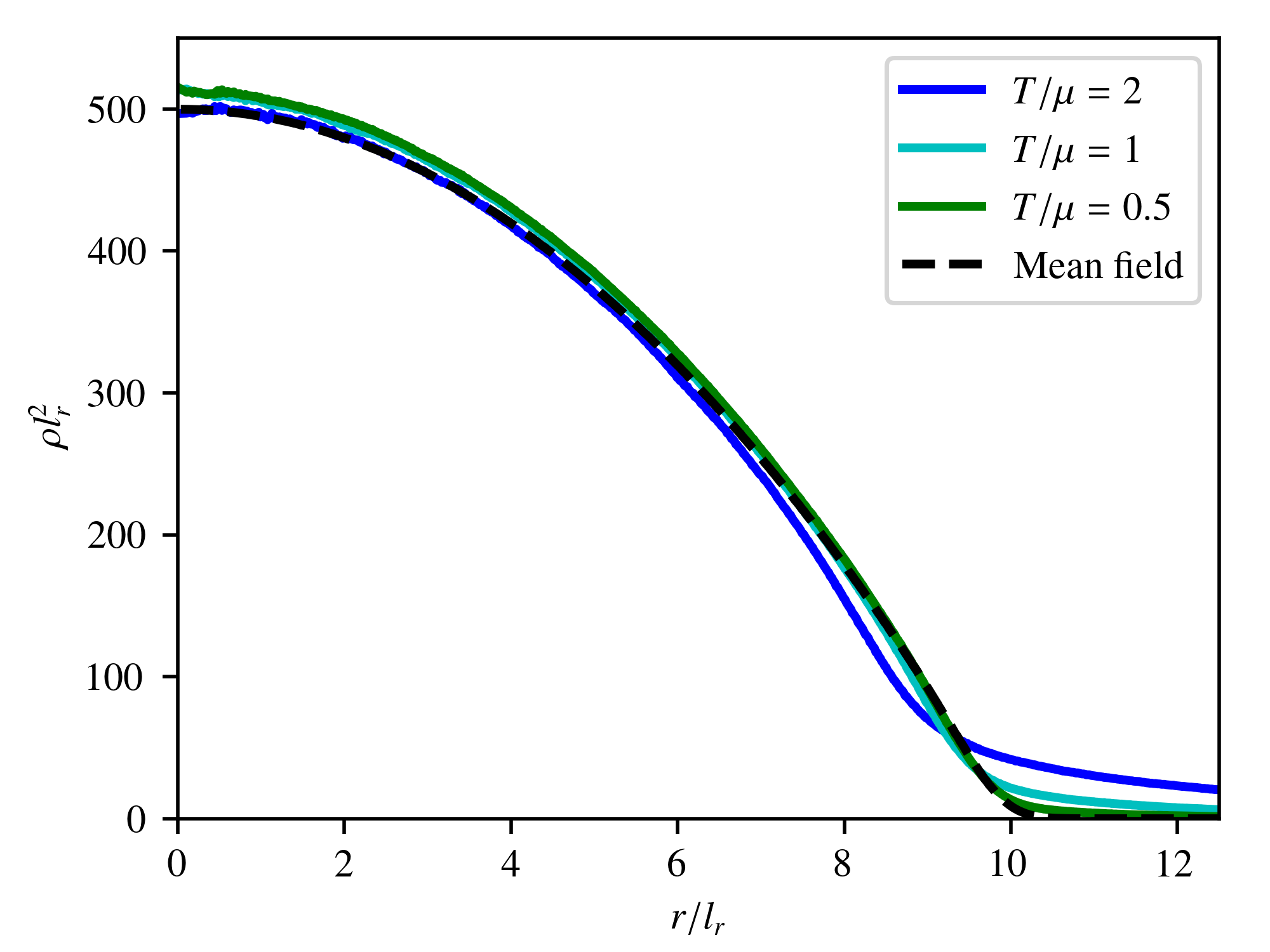}
	\centering\includegraphics[width=0.495\textwidth]{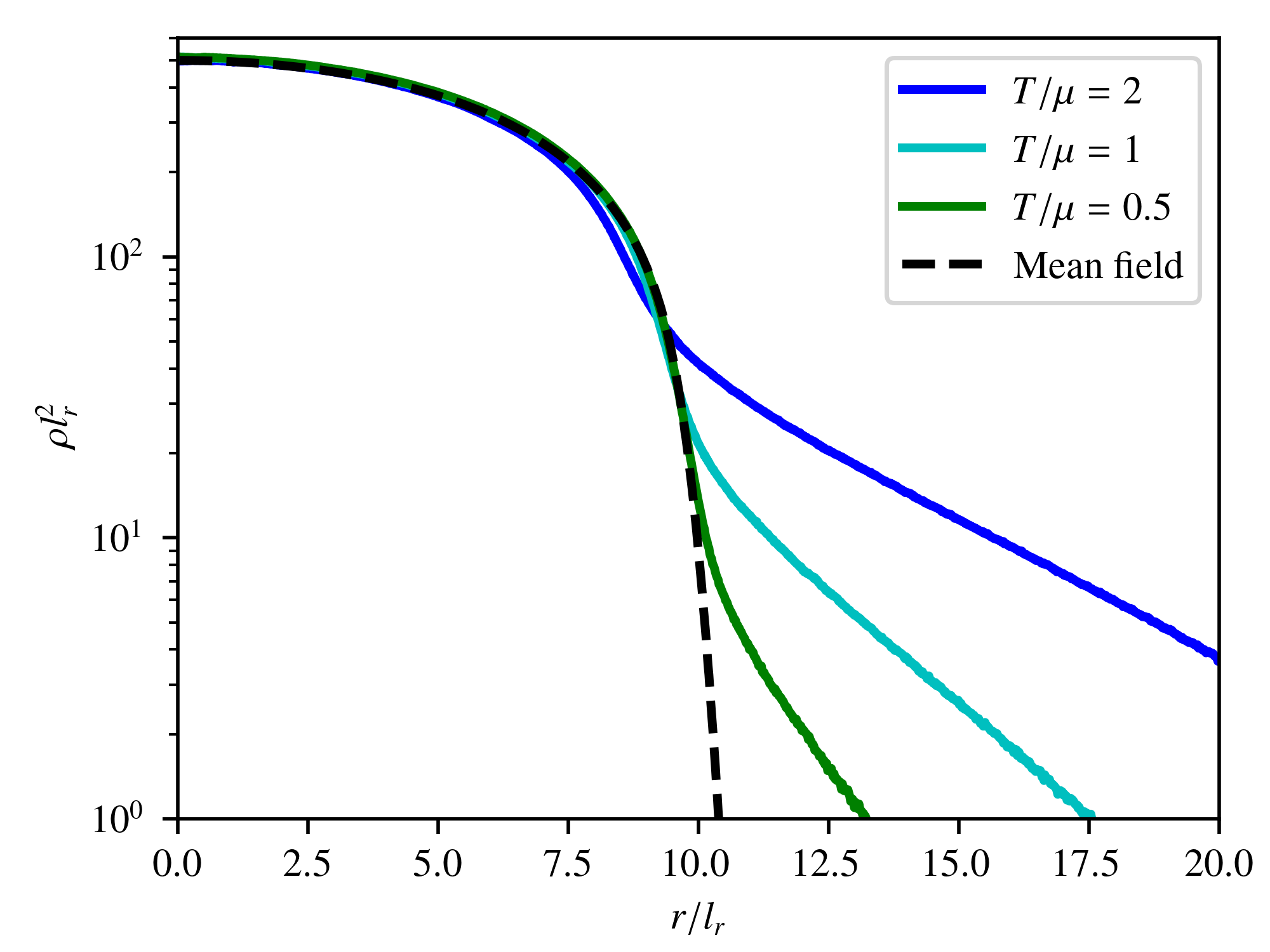}
	\caption{Radial density profile $\rho(r)$ of a two-dimensional harmonically confined Bose gas computed with the CL method. Parameters read $mg=0.1$, $\omega_x=\omega_y\equiv\omega_r$ and $\mu/\omega_r=50$. The temperature is varied between $\mu/T=0.5$ and $\mu/T=2$. The radial coordinate $r$ is expressed in units of the harmonic oscillator length $l_r\equiv\sqrt{1/m\omega_r}$ and the density in units of $l_r^{-2}$. The black dashed line represents the mean-field density profile, i.e. the minimum of the classical energy functional. The left panel shows a linear-linear plot while the right panel shows a linear-logarithmic plot of the same quantities. Note that the bare coupling strength for both the mean-field and full density profile is chosen to be the same and hence the renormalized one results slightly different: While the coupling plugged into the mean-field computation corresponds to a coupling defined at the scale of the healing momentum, the coupling plugged into the full simulation corresponds to a coupling defined at the lattice cutoff $\pi/ a_\mathrm{s}$.  
	}
	\label{fig:density_prof}
\end{figure}

Figure \fig{density_prof} depicts the density profile of a harmonically trapped, two-dimensional Bose gas computed with the CL method, for parameters $mg=0.1$, $\omega_x=\omega_y\equiv\omega_r$ and $\mu/\omega_r=50$ and three different temperatures. The action was discretized on a $512^2\times N_\tau$ lattice, with $N_\tau$ ranging from $20$ to $80$, $a_\tau=0.05\,a_\mathrm{s}$, $\mu=0.5\,a_\mathrm{s}^{-2}$ and equally spaced momenta. The Thomas-Fermi radius amounts to $r_\mathrm{TF}=141.4\,a_\mathrm{s}$ such that the extension of the computational lattice is sufficiently larger than the extension of the atomic cloud. In terms of the order of magnitude, these parameters correspond to typical experimental values, cf. the following section \sect{comp_densprof_exp}.

In the center of the trap, the density is slightly increased in comparison to the mean-field density, as a consequence of thermal and quantum depletion of the condensate. In the outer regions of the trap where the mean-field density is zero, the Langevin simulations produce thermally excited particles. Interestingly, in the intermediate region between the condensed and thermal phase, the density is \textit{lowered} in comparison to mean-field density. It is precisely this intermediate region where the BKT transition occurs, which renders the analytical description of this region most challenging.   

\begin{figure}
	\centering\includegraphics[width=0.495\textwidth]{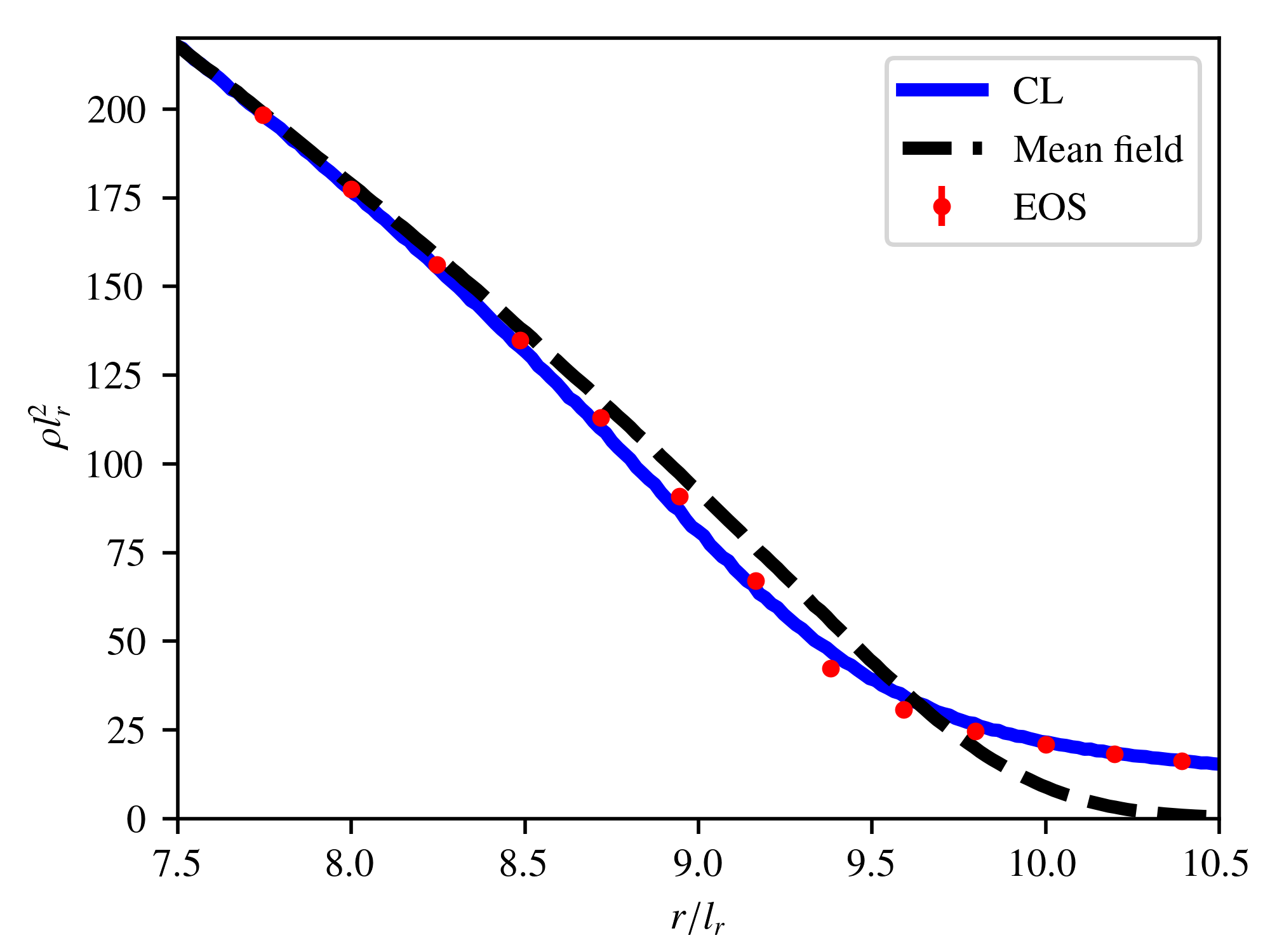}
	\centering\includegraphics[width=0.495\textwidth]{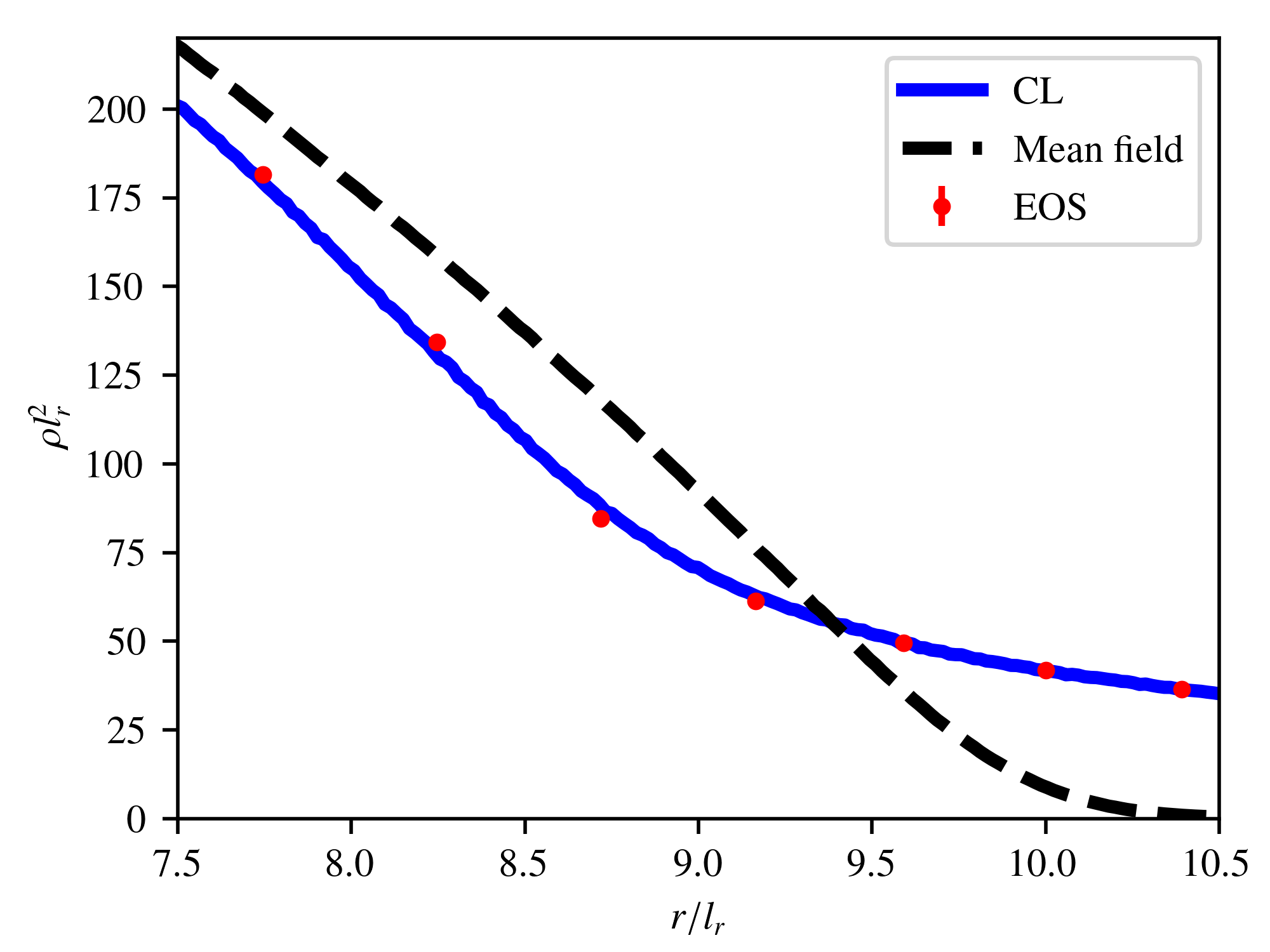}
	\caption{Full density profile from CL for the same parameters as in figure \fig{density_prof} and $T/\mu=1$ (left panel) and $T/\mu=2$ (right panel) around the Thomas-Fermi radius, in comparison to $\rho\left(\mu-\frac{1}{2}m\omega_r^2 r^2\right)$ (red dots), with the data for the equation of state $\rho(\mu)$ stemming from the calculations in chapter \sect{2D_gas}. For these large ratios of temperature and trapping frequency, $T/\omega_r=50$ (left panel) and $T/\omega_r=100$ (right panel), the deviations are tiny and are confined to a small region close to the Thomas-Fermi radius.
	}
	\label{fig:comparison_densprof_eos}
\end{figure}
\begin{figure}
	\centering\includegraphics[width=0.495\textwidth]{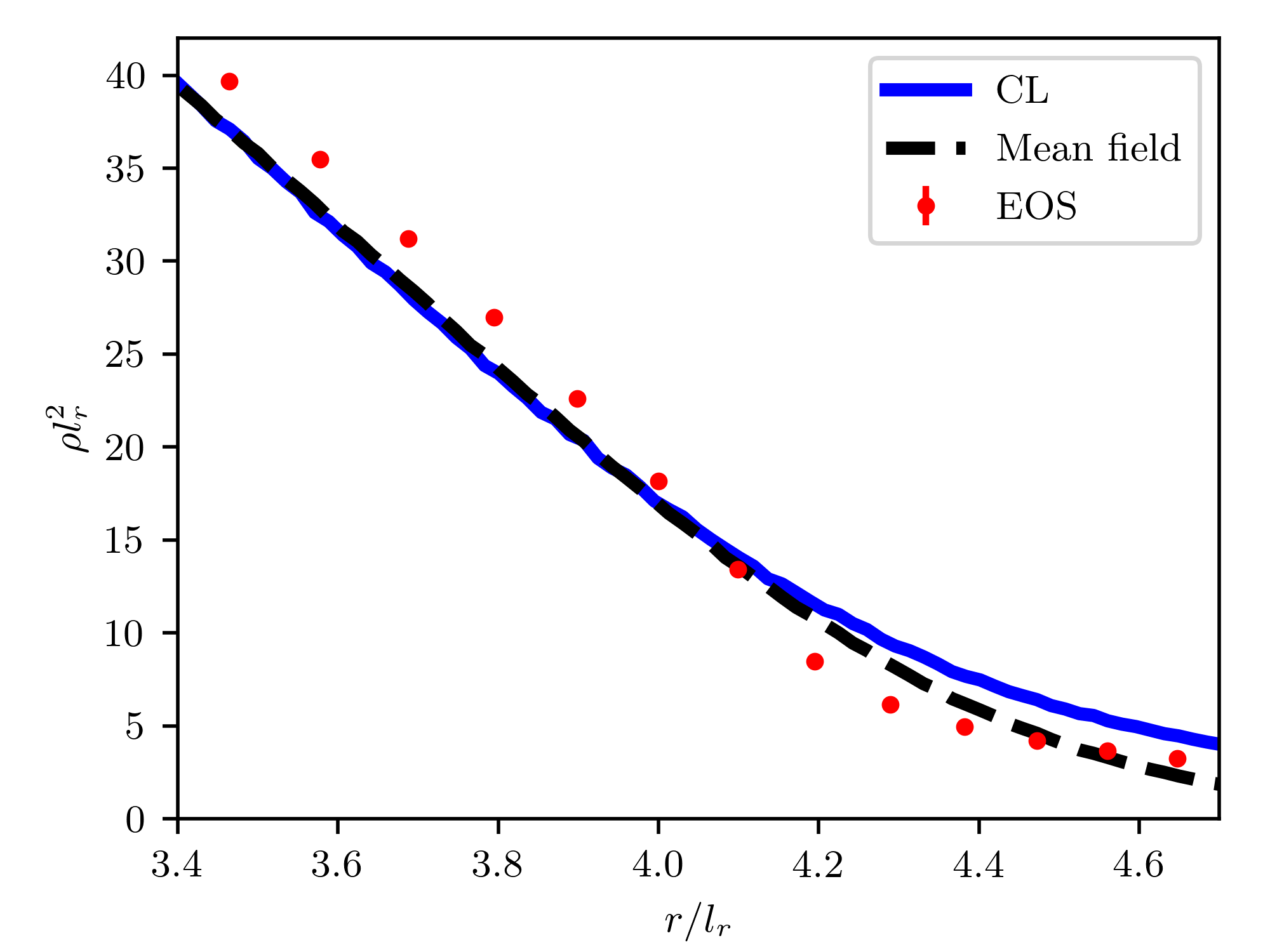}
	\centering\includegraphics[width=0.495\textwidth]{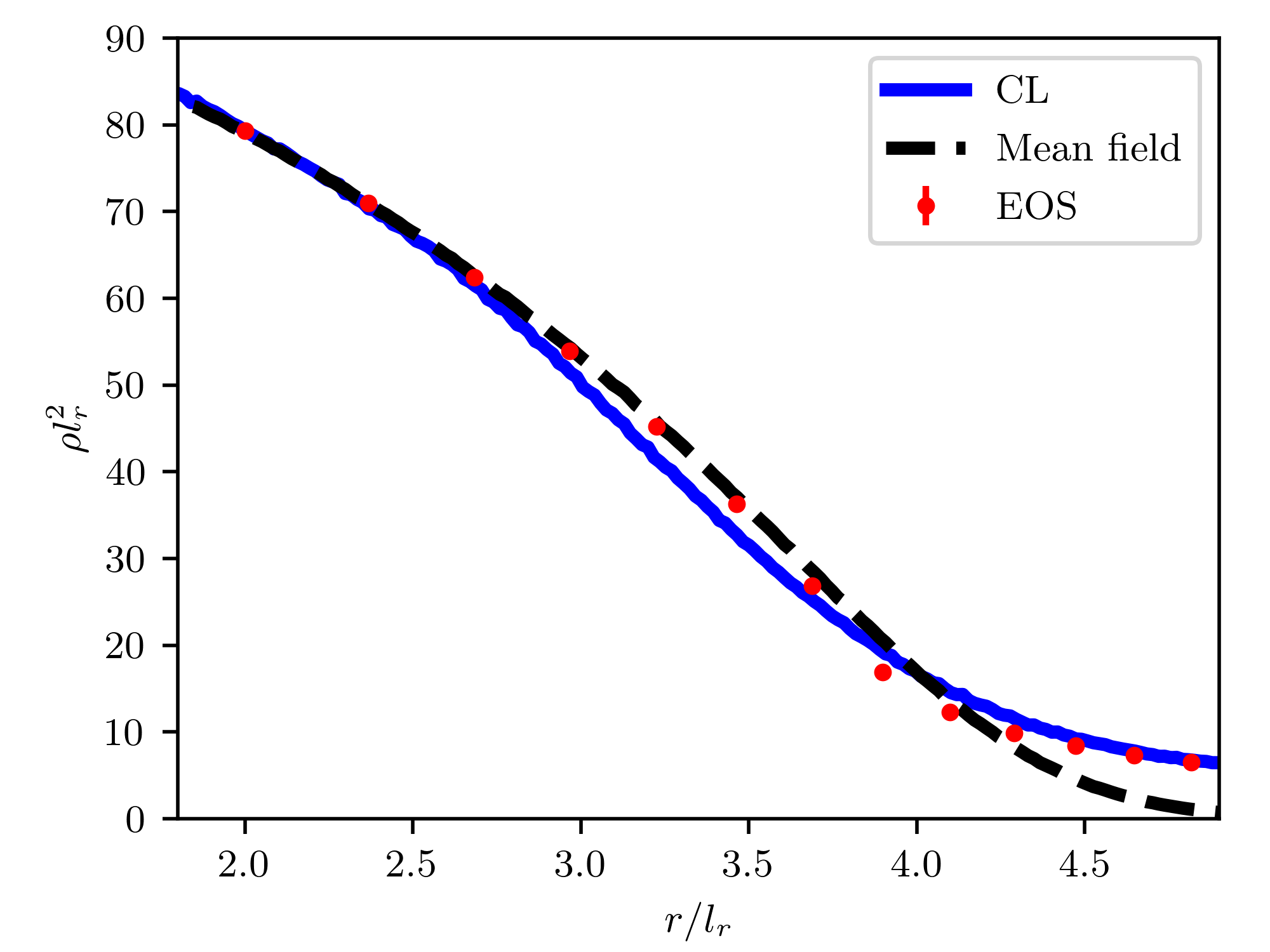}
	\caption{The same as in figure \fig{comparison_densprof_eos} but with $\mu/\omega_r=10$ and $T/\omega_r=10$ (left panel) and $T/\omega_r=20$ (right panel). The deviations from the LDA are substantially larger than in figure \fig{comparison_densprof_eos}, but still confined to the region around $r_\mathrm{TF}$.
	}
	\label{fig:comparison_densprof_eos2}
\end{figure} 

In figure \fig{comparison_densprof_eos}, we show a zoom into the region around the Thomas-Fermi radius for the same density profiles as in figure \fig{density_prof} (for $T/\mu=2$ and $T/\mu=1$) and compare them to the density profile that results from the local density approximation, i.e. $\rho\left(\mu-\frac{1}{2}m\omega_r^2 r^2\right)$. For the equation of state $\rho(\mu)$ we reuse the data computed with CL in chapter \sect{2D_gas} on a $256^2$ spatial lattice without external trapping potential. The deviations are small and confined to a region close to $r_\mathrm{TF}$, as it is expected for the comparatively large ratio of temperature and trapping frequency, $T/\omega_r=50$ and $T/\omega_r=100$, respectively, the LDA becoming good for $T\gg\omega$. For comparison, figure \fig{comparison_densprof_eos2} shows the full density profile and the LDA result for the substantially smaller temperatures $T/\omega_r=10$ and $T/\omega_r=20$, and the ratio of chemical potential and trapping frequency lowered to $\mu/\omega_r=10$ (in this way the ratio of temperature and chemical potential can be kept sufficiently large in order to make the thermal density profile substantially discernible from the mean field one while allowing for a way smaller ratio $T/\omega_r$). The deviations become considerably larger in this case but are still appearing only close to the Thomas-Fermi radius. 

These results justify the use of the LDA in the description of typical experimental settings. E.g. for the experiment that we want to describe in the following section, we have $T/\omega_r\sim 50$, for which the error of the LDA even close to $r_\mathrm{TF}$ is of the order of magnitude of $\sim 10\%$ and much smaller further away from $r_\mathrm{TF}$. However, they also indicate that for future high-precision comparisons to experiment, the deviations between the LDA and full density profiles might eventually play a role.

\subsection{Comparison to experiment\label{sec:comp_densprof_exp}}
Ultimately, one of the major applications of the numerically exact complex Langevin simulations of ultracold bosons is the high precision determination of physical quantities in comparison to the experiment. Density profiles in harmonic traps are very straightforward to extract experimentally, such that they offer themselves as a test bed for examining the potential of the CL method for such predictions of experimental quantities.

However, a substantial problem of a precision comparison with the experiment with regard to $\rho(\mathbf{r})$ is that there are almost no other precise ways of determining temperature $T$ and chemical potential $\mu$ of the atom cloud in the experiment than from $\rho(\mathbf{r})$ itself. Furthermore, there are experimental uncertainties on the total particle number and in the imaging procedure such that it is reasonable to introduce an unknown constant prefactor to the density $\rho(\mathbf{r})$. As the mass of the atoms $m$, the trapping frequency $\omega$ and the coupling $g$ are known independently, we remain with three unknown parameters, $T$, $\mu$ and a scaling factor $c$. Using the entire set of experimental data in order to fit three free parameters to the theoretical curve $\rho(\mathbf{r})$ obtained from CL simulations would greatly limit the predictive power of the theory, apart from the fact that large-scale Monte-Carlo simulations are not well suited as the input of an iterative fitting algorithm that requires numerous evaluations of the fitting function.

Thus we will employ the following procedure. We only take the thermal tail of the density profile, $\rho(r\gtrsim r_\mathrm{TF})$, as well as the density in the center of the trap, $\rho(r=0)$, as an input for determining the parameters $\mu$, $T$ and $c$. The rest of the profile is then predicted from the simulations, in particular the most crucial region close to $r_\mathrm{TF}$ where both BKT physics and the partial breakdown of the LDA make the prediction of $\rho(r)$ highly challenging for analytical descriptions. In principle, one may also take into account only the thermal tail. If the scaling factor $c$ is known independently, e.g. because the total particle number in the trap is known to a sufficient precision, the thermal tail is already sufficient to fit $\mu$ and $T$ well, as was e.g. done in \cite{hung2011observation}. 

As a fitting function for the thermal tail we employ the Hartree-Fock density in 2D, which is given by the solution of the equation (cf. section \sect{HF})
\begin{align}
\label{eq:therm_tail}
\rho(r;\mu,T)=-\lambda_{T}^{-2}\,\ln\left(1-\exp\left[\beta\left(\mu-\frac{1}{2}m\omega_r^2r^2-2g\rho\right)\right]\right)\,.
\end{align}
As shown in chapter \sect{3D}, this approximation provides an excellent description of the non-condensed phase. For the density in the center in the trap, one might simply assume the mean-field density $\rho(r=0)=\mu/g$.  However, we found this approximation not to be sufficient in practice, as there is a substantial deviation between the exact and mean-field density. Let us denote the ratio of the exact and mean-field density by $\alpha$, i.e. $\rho(r=0)=\alpha(\mu,T)\mu/g$, with $|\alpha(\mu,T)-1|\ll 1$. For $\alpha(\mu,T)$, one might also use some approximation scheme such as the (modified) Popov approximation. Here, we choose instead to determine $\alpha(\mu,T)$ from the CL simulations themselves. This requires an iterative approach: We first fit \eq{therm_tail} to the thermal tail of the experimental data with fitting parameters $\mu$ and $T$ and with the scaling factor $c$ fixed to $c=(\mu/g)/\rho_\text{exp}(r=0)$. We then let run the CL simulation with the resulting parameters and extract $\alpha(\mu,T)$ from this simulation. We then perform another fit to the experimental data with $c$ now fixed to $c=(\alpha\mu/g)/\rho_\text{exp}(r=0)$ and perform a new CL simulation with the resulting parameters. This procedure is iterated until convergence is reached. 

\begin{figure}
	\centering\includegraphics[width=0.495\textwidth]{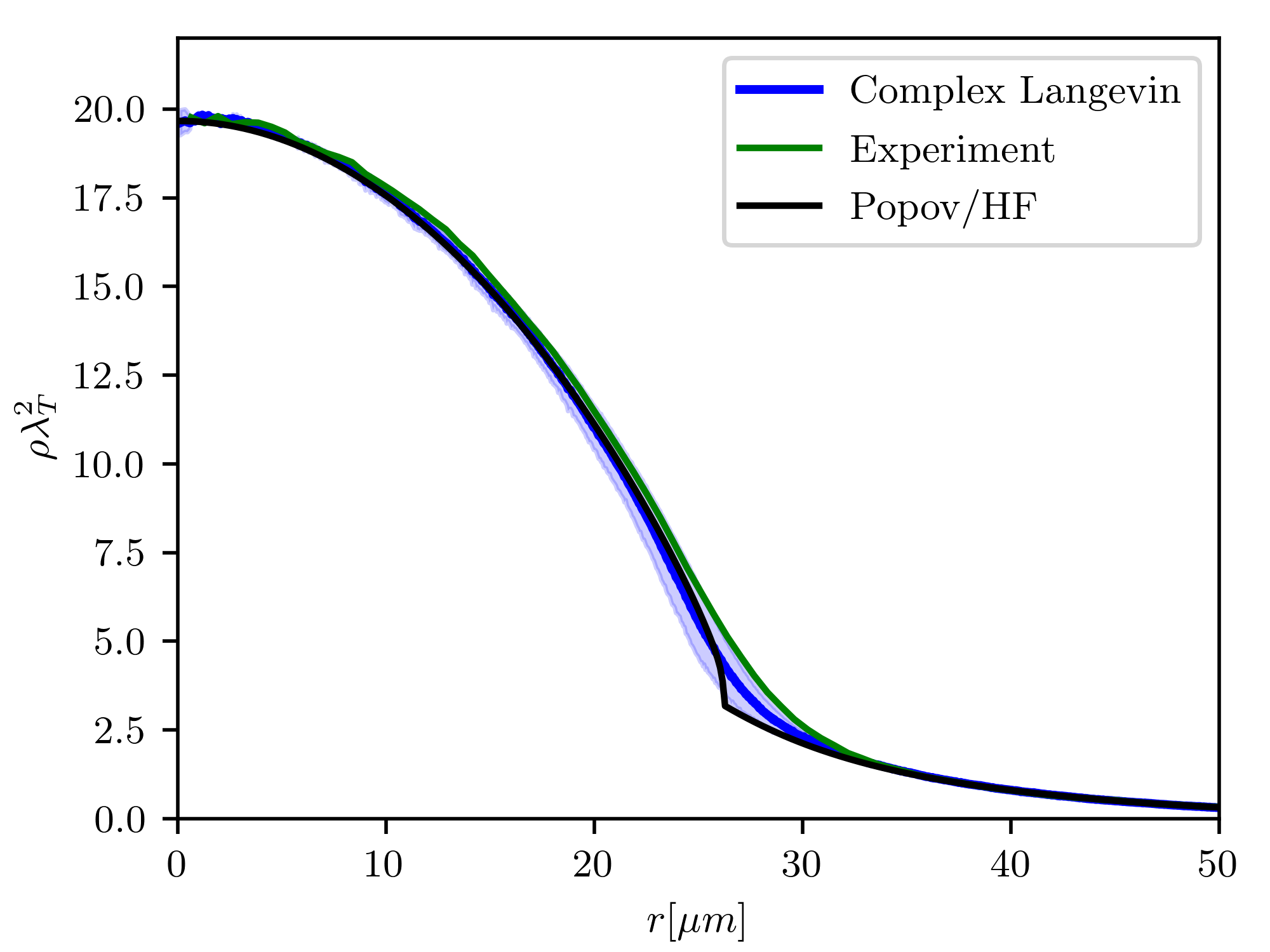}
	\centering\includegraphics[width=0.495\textwidth]{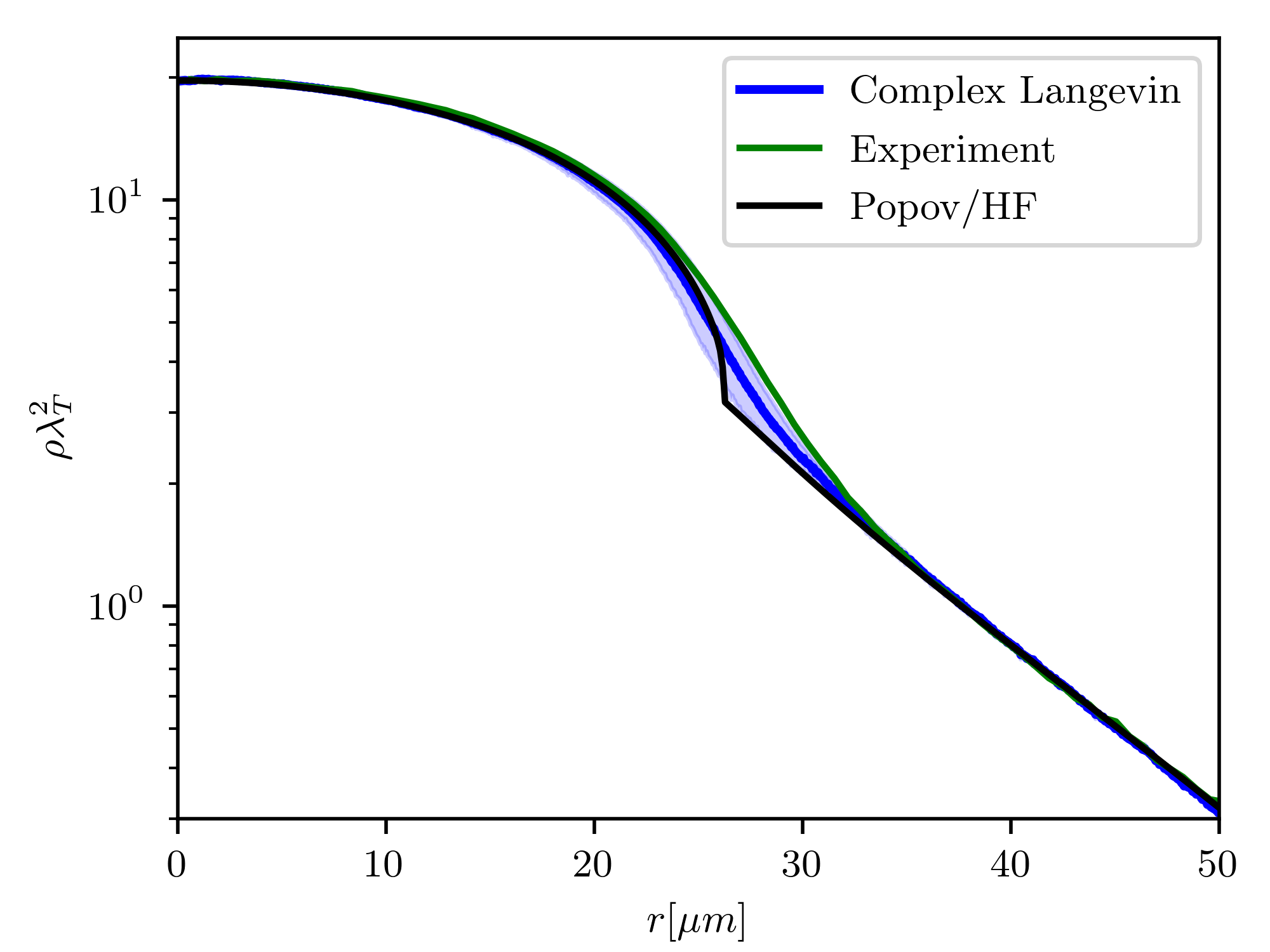}
	\caption{Density profile in a two-dimensional harmonically trapped Bose gas computed with complex Langevin (blue line) in comparison to the experimental one (green line). Left and right panel show the same data as a linear-linear and linear-logarithmic plot, respectively. The error band on the CL data represents the uncertainty in the chemical potential that describes the experimental setting. The black curve shows the prediction of the modified Popov approximation, \eq{mod_Popov1} and \eq{mod_Popov2}, which above the transition reduces to Hartree-Fock (HF) theory. Within the (comparatively large) error, the CL simulation agrees quantitatively with the experimental data. Qualitatively, it agrees better with the experimental shape of the density profile than the Popov approximation in so far as it does not feature the unphysical kink at the Thomas-Fermi radius that the latter produces.
	}
	\label{fig:comparison_exp}
\end{figure}

In the following, we will compare to data for the density profile in harmonic trap from the Heidelberg Potassium BEC (BECK) experiment \cite{hans2017an,viermann2022cosmological}, which allows for the preparation of ultracold atom clouds that are tightly confined in one direction, i.e. effectively two-dimensional. We consider a setting with the following parameters: The tight confinement trap frequency is $\omega_z=2\pi\times1.6\cdot 10^3\,\text{Hz}$ and the radial in-plane trapping frequency amounts to $\omega_r= 2\pi\times23.2\,\text{Hz}$. The mass of the potassium atoms is given by $m=39\,\mathrm{u}$. The three-dimensional scattering length is chosen to $a=400\,a_0$ with $a_0$ the Bohr radius. The data for the density profile that we compare with was produced as part of the study presented in \cite{viermann2022quantum}. The formula for the dimensionless coupling \eq{g2D} yields $mg=0.2645$. The full formula \eq{full_g2D} yields, upon inserting for $\Lambda_0$ the momentum cutoff of our simulation (in this way, the coupling of the simulation does not need to be further renormalized), a correction to this value of the order of only $\sim10\%$, which is also the order of magnitude of the experimental uncertainty to $a$. By applying the fitting procedure described above to the experimental data, we find a temperature $T=61\,\text{nK}$ and a chemical potential $\mu=6.7\cdot 10^{-31}\,\mathrm{J}$. While the fitting errors are comparatively small ($0.2\%$ on temperature and $2\%$ on chemical potential), we found the systematic error due to the choice of the window for fitting the thermal tail \eq{therm_tail} to be way larger. Estimating this error by varying the fitting window, we find $T=(61\pm 1)\,\text{nK}$ and $\mu=(6.7\pm0.4)\cdot 10^{-31}\,\mathrm{J}$. Figure \fig{comparison_exp} shows the result of a complex Langevin simulation on a $512^2\times 40$ lattice for these parameters in comparison to the experimental result. Unfortunately, the large uncertainty in the chemical potential leads also to a large uncertainty in the density profile~\footnote{A further source of uncertainty not considered here is the coupling constant whose value is subject to an experimental error of $5-10\%$.}. Within this uncertainty, the complex Langevin and experimental density profile agree with each other. For comparison, we also show the prediction of modified Popov (MP) theory in local density approximation as outlined in section \sect{popov}. Note that the coupling $g$ that enters equations \eq{mod_Popov1} and \eq{mod_Popov2} constituting the modified Popov approximation is a coupling defined at the scale of the healing momentum, whereas the coupling that enters the complex Langevin simulation is defined at the momentum cutoff of the simulation. To compute the correct coupling for the MP calculation, we employ the renormalization formula \eq{renormalization_general} improved by the finite-$a_\tau$ correction outlined in section \sect{coupling_renormalization}. Also note that above the transition point, the MP approximation reduces to the Hartree-Fock approximation. The complex Langevin result agrees qualitatively better with the experimental curve as it does not possess the unphysical kink at the Thomas-Fermi radius that the MP approximation features \footnote{Note, however, that such a kink, if it actually were physical, might also be washed out in the experimental averaging over multiple shots because the position of the trap center might shift between subsequent shots.}. A high-precision quantitative comparison, however, remains elusive due to the high uncertainty in the experimental parameters. Let us finally remark that the total particle number results as $N=2.6\cdot 10^4$. This may be compared to simulations of the two-dimensional trapped Bose gas by means of the path-integral Monte Carlo method \cite{heinrichs1998quantum,ruggeri2014quantum,ciardi2022finite} that reach only up to $N\sim1000$ atoms.

\subsection{Conclusion}
We have shown that the finite-temperature density profile in a two-dimensional trapped Bose gas may be computed  from first principles with the complex Langevin method and that in contrast to the other fully exact approach, path-integral Monte Carlo, the CL algorithm enables exact simulations for experimentally realistic particle numbers. We have used this to study the range of validity of the local density approximation. Furthermore we have compared the result of a complex Langevin simulation to the experimentally extracted density profile of the Heidelberg BECK experiment and found it to agree within errors. Unfortunately, however, the errors in the experimental parameters turned out to be too large for an actual precision comparison.

If these experimental uncertainties became smaller in the future, however, our approach could be of great usefulness for such high-precision comparisons between ultracold atoms experiments and theory, as to the best of our knowledge it is the only fully exact simulation method that is able to easily reach the comparatively high particle numbers of actual experiments. A scenario of particular interest is the regime of small $T/\omega$ where the local density approximation breaks down and thus most analytical descriptions fail.
\clearpage

\thispagestyle{plain}
\section{Dipolar gases\label{sec:dipolars}}
So far, we have only considered Bose gases composed of atoms possessing purely short-range interactions, i.e. with an interaction potential that can be approximated as $V(\mathbf{x}-\mathbf{y})=g\,\delta(\mathbf{x}-\mathbf{y})$, which is in very good approximation the case for alkali atoms, that were historically the first for which Bose-Einstein condensation was demonstrated experimentally and are still nowadays employed in most BEC experiments. Despite the numerous intriguing phenomena found in contact-interacting gases, adding a non-local interaction makes the physics even more interesting. In fact, there exist some atomic species which in their ground state possess a strong magnetic dipolar moment $\mu_\mathrm{d}$ that leads to long-range dipolar-dipolar interactions between the atoms~\footnote{The dipolar potential decreases as $1/r^3$ and is thus exactly at the edge of long-range  potentials, compare footnote \ref{shortrange}.}. The first Bose-Einstein condensate of strongly dipolar atoms was achieved with chromium \cite{griesmaier2005bose}. Later on, the even more magnetic elements  dysprosium and erbium could be condensed \cite{lu2011strongly,aikawa2012bose}. These experimental successes have triggered a large amount of theoretical work on dipolar BECs \cite{glaum2007critical,vanbijnen2007dynamical,lima2012beyond,zhang2019supersolidity,roccuzzo2020rotating,gallemi2022superfluid,smith2023supersolidity}.

One of the interesting effects of the dipolar interaction is that it can lead to a rotonic dispersion, i.e. one that features a local minimum at finite momentum. While such phenomenon is well known for the strongly interacting superfluid helium \cite{pitaevskii2016bose}, dipolar BECs offer the intriguing possibility to study roton excitations in a weakly-interacting setting. The stronger the dipolar interactions are in comparison to the usual contact interaction, the deeper the local minimum in the dispersion becomes, until it eventually touches zero, rendering the system unstable. The regime close to this instability is very susceptible to quantum and thermal fluctuations. Therefore, a method as complex Langevin that provides the possibility of a full quantum simulation is particularly beneficial here. Even more intriguing is the regime beyond this instability: While one would expect the gas to immediately collapse, experimentally one observes a surprisingly long-lived state of matter exhibiting both phase coherence and a periodic density modulation, which is known as supersolid. This has been theoretically explained in a semi-classical manner by adding a term derived from the LHY correction to the classical energy functional that accounts for the effect of quantum fluctuations \cite{wachtler2016quantum}, which (meta-)stabilize the system against collapse. Examining this mechanism in an ab initio simulation is of particular interest.  

This chapter is organized as follows: As a first check on the applicability of complex Langevin for the case of long-range interactions, we perform CL simulations in a three-dimensional box with periodic boundary conditions in the stable region and compare to Bogoliubov theory. Subsequently, as a more stringent benchmark, we show that the method is able to reproduce the analytically predicted rotonic dispersion relation of a dipolar gas that is harmonically confined in the polarization direction. In the next section, we perform a similar calculation but for a concrete experimental setting, namely for the Innsbruck Erbium experiment, and compare to the experimental results, attempting to settle the reason for the disagreement between experiment and previous Bogoliubov calculations. Finally, we will demonstrate the stabilizing effect of the quantum fluctuations and the formation of a supersolid state of matter in the classically unstable region of a dipolar gas within a first-principles CL simulation. 
\subsection{Physics of dipolar bosonic atoms}
In the typical experimental setting for dipolar atomic gases, all the dipoles are polarized along one direction by means of an external magnetic field. In the following, we will always take this direction to be the $z$-direction in our mathematical description. Under these assumptions, the dipolar interaction potential $V_\text{dip}(\mathbf{r}_1-\mathbf{r}_2)$ between two atoms located at positions $\mathbf{r}_1$ and $\mathbf{r}_2$ reads
\begin{align}
\label{eq:dip}
V_\text{dip}(\mathbf{r}_1-\mathbf{r}_2)=\frac{\mu_0\mu_\mathrm{d}^2}{4\pi}\frac{1-3(r_{z,1}-r_{z,2})^2/|\mathbf{r}_1-\mathbf{r}_2|^2}{|\mathbf{r}_1-\mathbf{r}_2|^3}\,,
\end{align}
where $\mu_0$ is the magnetic field constant. For convenience, one commonly defines a number of further quantities. We introduce the dipolar coupling constant
\begin{align}
C_\mathrm{dd}\equiv \mu_0\mu_\mathrm{d}^2\,,
\end{align}
the dipolar scattering length
\begin{align}
a_\mathrm{dd}\equiv \frac{mC_\mathrm{dd}}{12\pi}\,,
\end{align}
and the ratio of dipolar and contact interaction strength
\begin{align}
\epsilon_\mathrm{dd}\equiv \frac{a_\mathrm{dd}}{a}\,.
\end{align}
The Fourier transform of the dipolar potential reads
\begin{align}
\label{eq:FT_dip}
V_\text{dip}(\mathbf{k})=C_\mathrm{dd}\left(\frac{k_z^2}{|\mathbf{k}|^2}-\frac{1}{3}\right)\,.
\end{align}
In comparison to the standard contact interacting gas, the dipole interaction leads to an additional term in the Hamiltonian of the form
\begin{align}
H_\text{dip}=\frac{1}{2}\int d^3x\int d^3y\,\frac{C_\mathrm{dd}}{4\pi}\frac{1-3(r_{z,1}-r_{z,2})^2/|\mathbf{r}_1-\mathbf{r}_2|^2}{|\mathbf{r}_1-\mathbf{r}_2|^3}\psi^\dagger(\mathbf{r}_1)\psi^\dagger(\mathbf{r}_2)\psi(\mathbf{r}_1)\psi(\mathbf{r}_2)\,.
\end{align}
This corresponds to a classical Hamiltonian density 
\begin{align}
\nonumber\mathcal{H}_\mathrm{dip}\left[\psi,\psi^*\right](\mathbf{x})=\frac{1}{2}\int d^3y\,\psi^*(\mathbf{y})\psi(\mathbf{y})\frac{C_\mathrm{dd}}{4\pi}\frac{1-3(x_z-y_z)^2/|\mathbf{x}-\mathbf{y}|^2}{|\mathbf{x}-\mathbf{y}|^3}\psi^*(\mathbf{x})\psi(\mathbf{x})\,.
\end{align}
The complex Langevin equations for a dipolar Bose gas with external potential $V_\text{ext}$ read (compare the non-dipolar CL equations \eq{CL_eq_bose1} and \eq{CL_eq_bose2}):
\begin{align}
\nonumber\frac{\partial\psi_{i,\mathbf{j}}}{\partial\vartheta}=&a_\mathrm{s}^3\Bigg\{\psi_{i-1,\mathbf{j}}-\psi_{i,\mathbf{j}}+a_\tau\Bigg[\frac{1}{2m}\Delta^\text{lat}\psi_{i-1,\mathbf{j}}+\left(\mu-V^\text{ext}_\mathbf{j}\right)\psi_{i-1,\mathbf{j}}\\&-g\,\bar{\psi}_{i,\mathbf{j}}\psi_{i-1,\mathbf{j}}\psi_{i-1,\mathbf{j}}-U^\text{dip,lat}_{i-1,\mathbf{j}}\psi_{i-1,\mathbf{j}}\Bigg]\Bigg\}+\eta_{i,\mathbf{j}}\\
\nonumber\frac{\partial\bar{\psi}_{i,\mathbf{j}}}{\partial\vartheta}=&a_\mathrm{s}^3\Bigg\{\bar{\psi}_{i+1,\mathbf{j}}-\bar{\psi}_{i,\mathbf{j}}+a_\tau\Bigg[\frac{1}{2m}\Delta^\text{lat}\bar{\psi}_{i+1,\mathbf{j}}+\left(\mu-V^\text{ext}_\mathbf{j}\right)\bar{\psi}_{i+1,\mathbf{j}}\\&-g\,\bar{\psi}_{i+1,\mathbf{j}}\psi_{i,\mathbf{j}}\bar{\psi}_{i+1,\mathbf{j}}-U^\text{dip,lat}_{i,\mathbf{j}}\bar{\psi}_{i+1,\mathbf{j}}\Bigg]\Bigg\}+\eta_{i,\mathbf{j}}^*\,,
\end{align}
where 
\begin{align}
U^\text{dip,lat}_{i,\mathbf{j}}=&\frac{1}{N_xN_yN_z}\sum_{\mathbf{k}\mathbf{j}'}\exp\left[i\mathbf{k}\cdot(a_\mathrm{s}\mathbf{j}-a_\mathrm{s}\mathbf{j}')\right] \tilde{V}_\mathbf{k}^{\text{dip}}\,\bar{\psi}_{i+1,\mathbf{j}'}\psi_{i,\mathbf{j}'}\,.
\end{align}
Here, $\tilde{V}_\mathbf{k}^{\text{dip}}$ is the dipolar potential in Fourier space. If one inserts here the analytic expression for the Fourier transform, \eq{FT_dip}, this amounts to a dipolar potential that possesses the exact form \eq{dip} up to arbitrarily large distances. For a computational lattice with periodic boundary conditions, this has the effect that atoms interact with imaginary copies of themselves. This effect can sometimes be desired, as it allows for emulating an infinitely extended system on a finite lattice. In most cases, however, especially when describing a trapped atom cloud, it is rather a bug than a feature. In order to avoid this effect, one typically imposes a large-distance cutoff to the dipolar potential. For some specific choices of the cutoff, e.g. a spherical cutoff, there exists an analytical expression for the Fourier transform. For other forms of the cutoff, it is straightforward to perform a numerical FFT of the desired cut off potential once at the beginning of the simulation. Details on this can be found in appendix \sect{dip_cutoff}.

Interestingly, the dipolar interaction is attractive for atoms at positions such that that the angle between their connecting line and the polarization direction is smaller than $\bar{\theta}\equiv\arccos(1/\sqrt{3})\approx 0.3\pi$, while for larger angles it becomes repulsive. In general, attractive interactions in atomic gases can lead to an instability, i.e. collapse of the atomic cloud, unless they are counterbalanced by a different interaction or external potential. Dipolar atoms also possess a repulsive contact interaction that can counterbalance the attractive part of the dipolar interaction.  In a homogeneous system without external potential, the contact interaction prevents a collapse as long as $\epsilon_\mathrm{dd}<1$. This can be seen by considering the total interaction potential in Fourier space, which reads $g+C_\mathrm{dd}\left(k_z^2/|\mathbf{k}|^2-1/3\right)$ and hence can become negative as soon as $C_\mathrm{dd}/3>g$, i.e. $\epsilon_\mathrm{dd}>1$. If a trapping potential is added in the polarization direction, the critical value of $\epsilon_\mathrm{dd}$ increases, because the trapping potential acts against the atoms piling up in the polarization direction. For typical experimental values of the trapping frequency $\omega_z/\mu\sim 0.1$, $\epsilon_\mathrm{dd}^\text{crit}$ remains between $1$ and $2$.  

Of particular interest is the regime $1<\epsilon_\mathrm{dd}<\epsilon_\mathrm{dd}^\text{crit}$ because here the dispersion relation $\omega(\mathbf{k})$ can feature a local minimum \cite{santos2003roton}. The corresponding excitations are known as rotons, in analogy with the case of superfluid helium. Intuitively, this can be understood as follows: For low momenta, the atoms do not possess enough kinetic energy to overcome the trapping potential in $z$-direction and to leave the $x$-$y$-plane, where they feel only the repulsive part of the dipolar interaction. For larger momenta, they can escape the $x$-$y$-plane such that they feel also the attractive part, which effectively lowers their total excitation energy. For even larger momentum, the single-particle regime is reached such that the excitation energy increases again. Together, these effects cause a local minimum in the dispersion relation. Such a local minimum has been observed experimentally \cite{petter2019probing}.

For $\epsilon_\mathrm{dd}>\epsilon_\mathrm{dd}^\text{crit}$, where the classical energy functional $E[\psi]=\int d^3r\,\mathcal{H}[\psi]$ is not bounded from below,
\begin{align}
\min_{\psi(\mathbf{r})}\,E[\psi]=-\infty\,,
\end{align}
one would expect a total collapse of the atoms to one point. Surprisingly, this is not observed experimentally, but instead the system stabilizes after an initial rapid growth of the roton mode occupancy \cite{chomaz2018observation}. One finds a long-lived meta-stable state, featuring supersolid properties, i.e. a periodic density modulation typical for solids combined with the long-range phase coherence typical of a superfluid \cite{tanzi2019observation,bottcher2019transitient,chomaz2019long}. For even larger $\epsilon_\mathrm{dd}$, the superfluidity is lost and a state of independent droplets of particles emerges \cite{roccuzzo2019supersolid,blakie2020supersolidity}.  

A theoretical description is offered by the extended Gross-Pitaevskii equation (EGPE) formalism \cite{wachtler2016quantum}. Consider the ground state energy density in Bogoliubov theory, including the Lee-Huang-Yang (LHY) correction to the classical energy density, for the case of a dipolar gas \cite{lima2012beyond}:
\begin{align}
\label{eq:LHY}
\mathcal{E}_0(\rho)=\frac{1}{2}\frac{4\pi a}{m} \rho^2\left[1+\frac{128}{15\sqrt{\pi}}\gamma(\epsilon_\mathrm{dd})\,a^{3/2}\rho^{1/2}\right]\,,
\end{align}
where 
\begin{align}
\label{eq:gamma}
\gamma(\varepsilon_\mathrm{dd})=\frac{1}{2}\int \limits_0^\pi d\theta\,\sin\theta \left[1+\epsilon_\mathrm{dd}\left(3\cos^2\theta-1\right)\right]^{5/2}\,.
\end{align}
Remarkably, the LHY correction scales as $\sim\rho^{5/2}$ in the particle density, while all interaction terms (also the dipolar interaction term) scale as $\sim\rho^2$. I.e. in the case of a collapse, the repulsive LHY term eventually outweighs the attractive dipolar term and leads to a stabilization. A caveat in this line of reasoning is that, in the case of interest, $\epsilon_\mathrm{dd}>1$, the prefactor $\gamma(\varepsilon_\mathrm{dd})$ turns complex. However, since the imaginary part is very small in the experimentally relevant regime $1<\epsilon_\mathrm{dd}<2$ (e.g. for $\epsilon_\mathrm{dd}=1.5$, the imaginary part is $160$ times smaller than the real part), one can still argue for a metastabilizing effect of the LHY correction and discard the imaginary part in practical computations. Making furthermore a local-density approximation, i.e. assuming that the LHY correction that is derived for an infinitely extended, homogeneous condensate remains locally valid also in the presence of an external potential, we can replace the microscopic classical Hamiltonian density by an effective EGPE functional that perturbatively captures also the effect of quantum fluctuations, i.e. 
\begin{align}
\mathcal{H}\left[\psi\right]\to \mathcal{H}_\text{EGPE}\left[\psi\right]=\mathcal{H}\left[\psi\right]+\frac{8\,\text{Re}\gamma(\epsilon_\mathrm{dd})}{15\pi^2}\,m^{3/2}g^{5/2}|\psi|^5\,.
\end{align}
The approach to simulate the extended classical Hamiltonian density $\mathcal{H}_\text{EGPE}\left[\psi\right]$ instead of the microscopic one $\mathcal{H}\left[\psi\right]$ in the unstable region has been able to reproduce the experimentally observed stabilization and occurrence of periodic density modulations \cite{wachtler2016quantum,ferrier2016observation}. Nonetheless, several caveats remain, namely the perturbative nature of the LHY correction, the local density approximation and the discarding of the imaginary part, such that an independent simulation by an exact method is desirable.  
\subsection{Bogoliubov spectra in a dipolar condensate}
As a first benchmark of the complex Langevin algorithm for dipolar atoms let us consider a three-dimensional homogeneous system with periodic boundary conditions in the stable phase, i.e. for $\epsilon_\mathrm{dd}<1$. We consider low temperatures $T\ll\mu$ such that we can compare to Bogoliubov theory and the quantum contribution to the Bogoliubov spectrum dominates over the thermal one. This regime is of particular interest because the effect of quantum fluctuations plays a crucial role in dipolar gases. 

\begin{figure}
	\centering\includegraphics[width=0.495\textwidth]{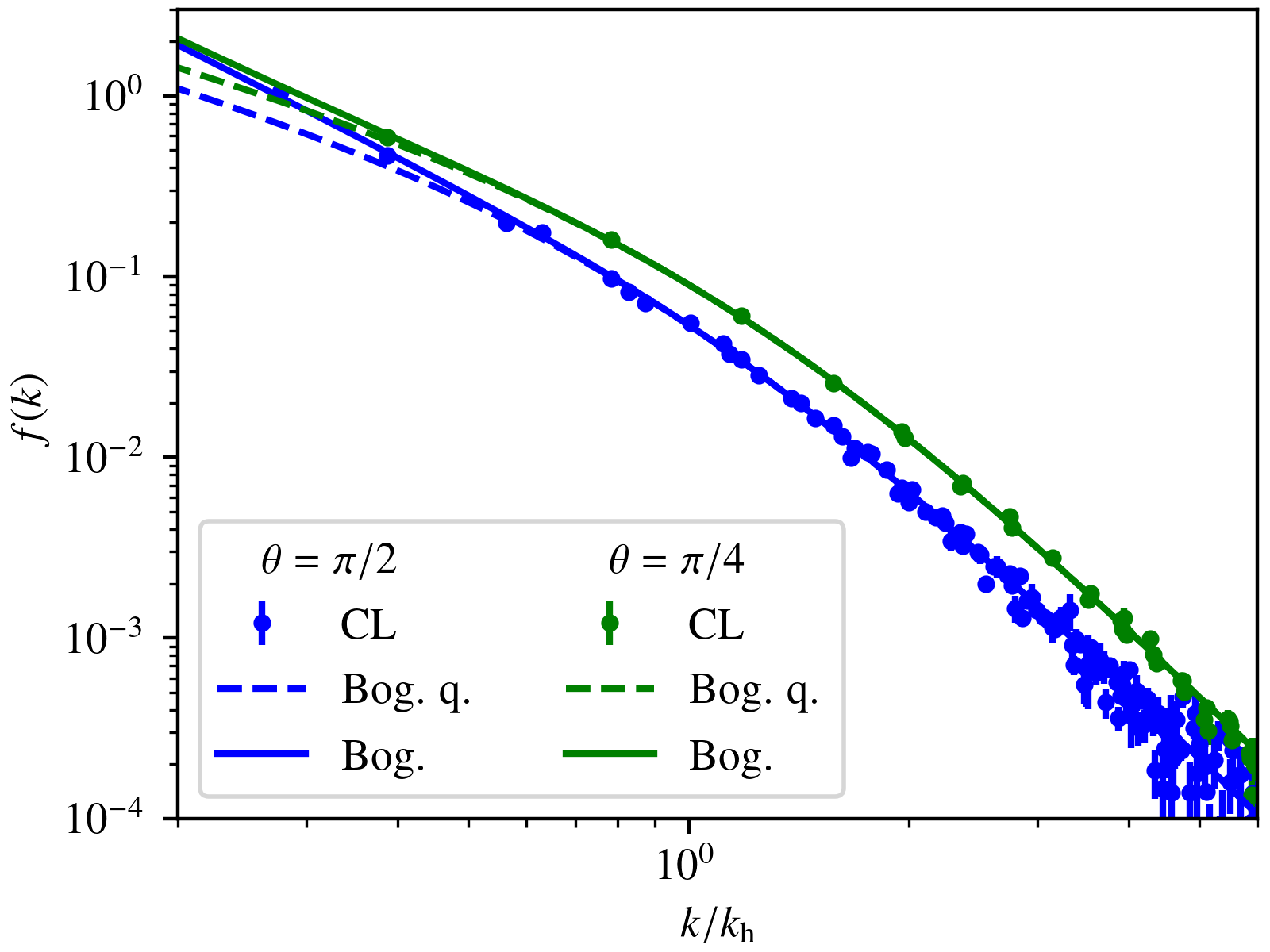}
	\centering\includegraphics[width=0.495\textwidth]{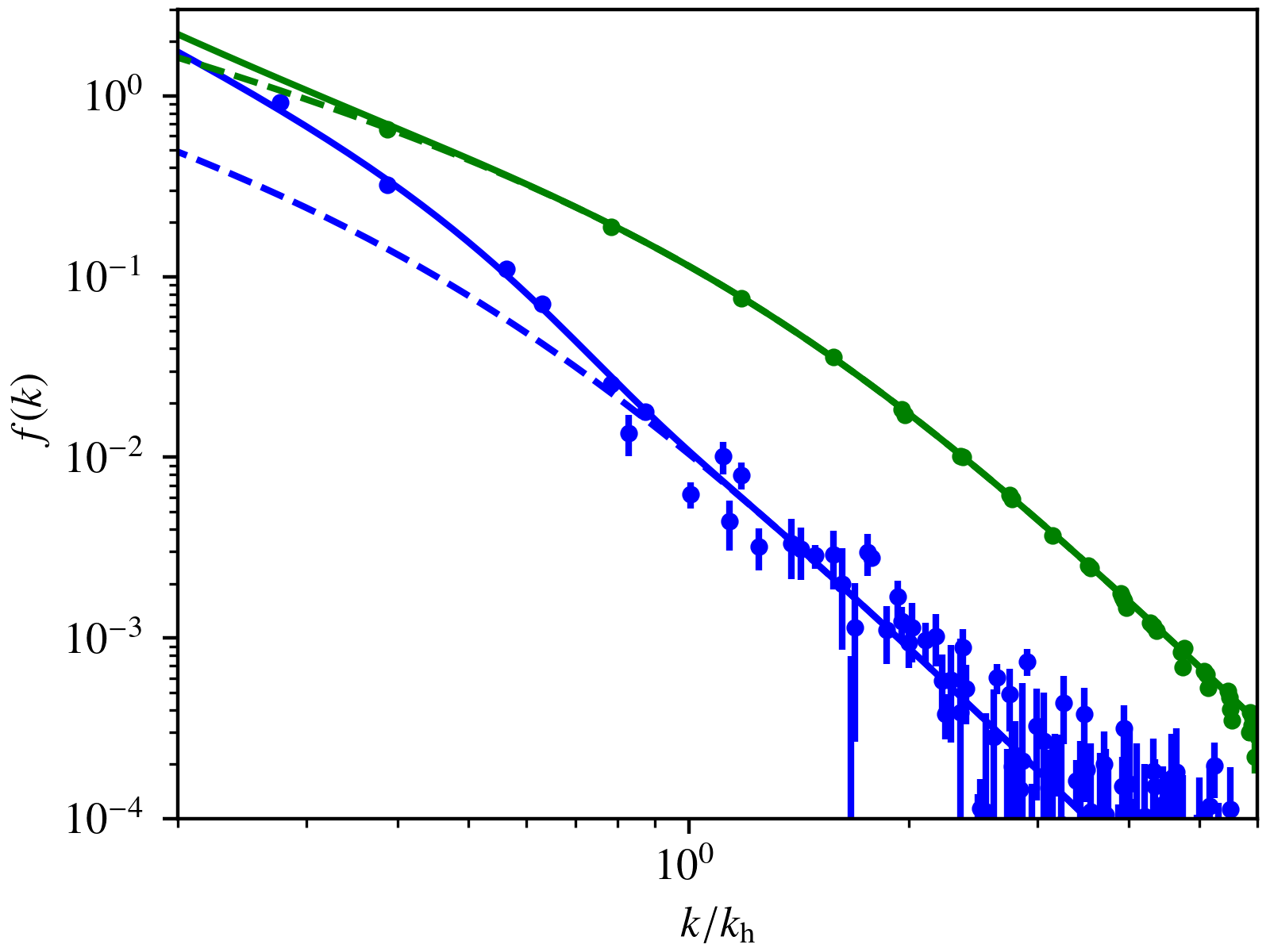}
	\caption{Occupation number $f$ as a function of radial momentum $k$, averaged over the azimuthal angle $\phi$, for polar angle $\theta=\pi/2$ (blue curves and data points) and $\theta=\pi/4$ (green curves and data points). The data points represent the occupation number as obtained from a CL simulation, the solid line represents the full Bogoliubov prediction for finite temperature and the dashed line the quantum part thereof, i.e. the Bogoliubov prediction at zero temperature. Parameters are $g=0.1\,a_\mathrm{s}^2$, $\mu=0.5\,a_\mathrm{s}^{-1}$ and $T/\mu=0.15625$. The ratio of dipolar and contact interaction is either $\epsilon_\mathrm{dd}=0.25$ (left panel) or $\epsilon_\mathrm{dd}=0.75$ (right panel).
	}
	\label{fig:dip_bog}
\end{figure}

In contrast to the purely contact interacting gas, the non-isotropic dipolar interaction has as its consequence that dispersion relations and momentum spectra are no longer functions of the modulus of the momentum only, $f(\mathbf{k})\neq f(k)$, but also of the polar angle $\theta$, i.e. $f(\mathbf{k})= f(k,\theta)$. For a three-dimensional dipolar gas enclosed in a box with periodic boundary conditions, the interaction potential in momentum space reads:
\begin{align}
\label{eq:full_pot_k}
V_\mathbf{k}=\begin{cases}g+C_\mathrm{dd}\left(k_z^2/|\mathbf{k}|^2-1/3\right)\,,\quad \mathbf{k}\neq 0\\g\,,\quad \mathbf{k}= 0\end{cases}\,.
\end{align} 
Using the definition of the polar angle $\theta$, $\cos^2\theta=k_z^2/|\mathbf{k}|^2$, and the ratio of dipolar and contact interaction $\epsilon_\mathrm{dd}=C_\mathrm{dd}/3g$, we can write the Bogoliubov dispersion in the dipolar case as 
\begin{align}
\omega(k,\theta)=\sqrt{\frac{k^2}{2m}\left[\frac{k^2}{2m}+2g\rho_0\left[1+\epsilon_\mathrm{dd}\left(3\cos^2\theta-1\right)\right]\right]}\,.
\end{align}
The occupation number $f(k,\theta)$ reads
\begin{align}
\label{eq:bogol_spectrum_dipolar}
f(k,\theta)
&=\frac{1+2v(k,\theta)^{2}}{e^{\beta\omega(k,\theta)}-1}
+v(k,\theta)^2
\,,
\end{align}
with 
\begin{align}
v(k,\theta)^2=\frac{1}{2}\left[\frac{k^2/2m+g\rho_0\left[1+\epsilon_\mathrm{dd}\left(3\cos^2\theta-1\right)\right]}{\omega(k,\theta)}-1\right]\,.
\end{align}
We perform simulations on a $32^3\times 512$ lattice, with $g=0.1\,a_\mathrm{s}^2$, $\mu=0.5\,a_\mathrm{s}^{-1}$ and $T/\mu=0.15625$, with $a_\mathrm{s}$ the lattice spacing. The spectra $f(\mathbf{k})$ are averaged over the azimuthal angle $\phi$ and binned in $k$ and $\theta$, with the angular binning width chosen as $\Delta\theta=0.01$. The resulting spectrum is shown in figure \fig{dip_bog} for $\epsilon_\mathrm{dd}=0.25$ and $\epsilon_\mathrm{dd}=0.75$, and for polar angles $\theta=\pi/2$ and $\theta=\pi/4$. As one can see, $f(k,\theta)$ changes substantially when varying $\theta$, and even more so for larger $\epsilon_\mathrm{dd}$. Furthermore, the temperature is chosen so low that thermal excitations are strongly suppressed and the spectrum is dominated by its purely quantum contribution. The complex Langevin simulation is able to accurately reproduce this spectrum. This is of particular importance because the quantum contribution plays a crucial role in the physics of dipolar atoms. In conclusion, we have performed a first sanity check on the application of complex Langevin to the dipolar Bose gas and found that the non-local, anisotropic interaction presents no particular challenge for the method. 

\subsection{Rotonic dispersion\label{sec:rot_disp}}
\begin{figure}
	\centering\includegraphics[width=0.495\textwidth]{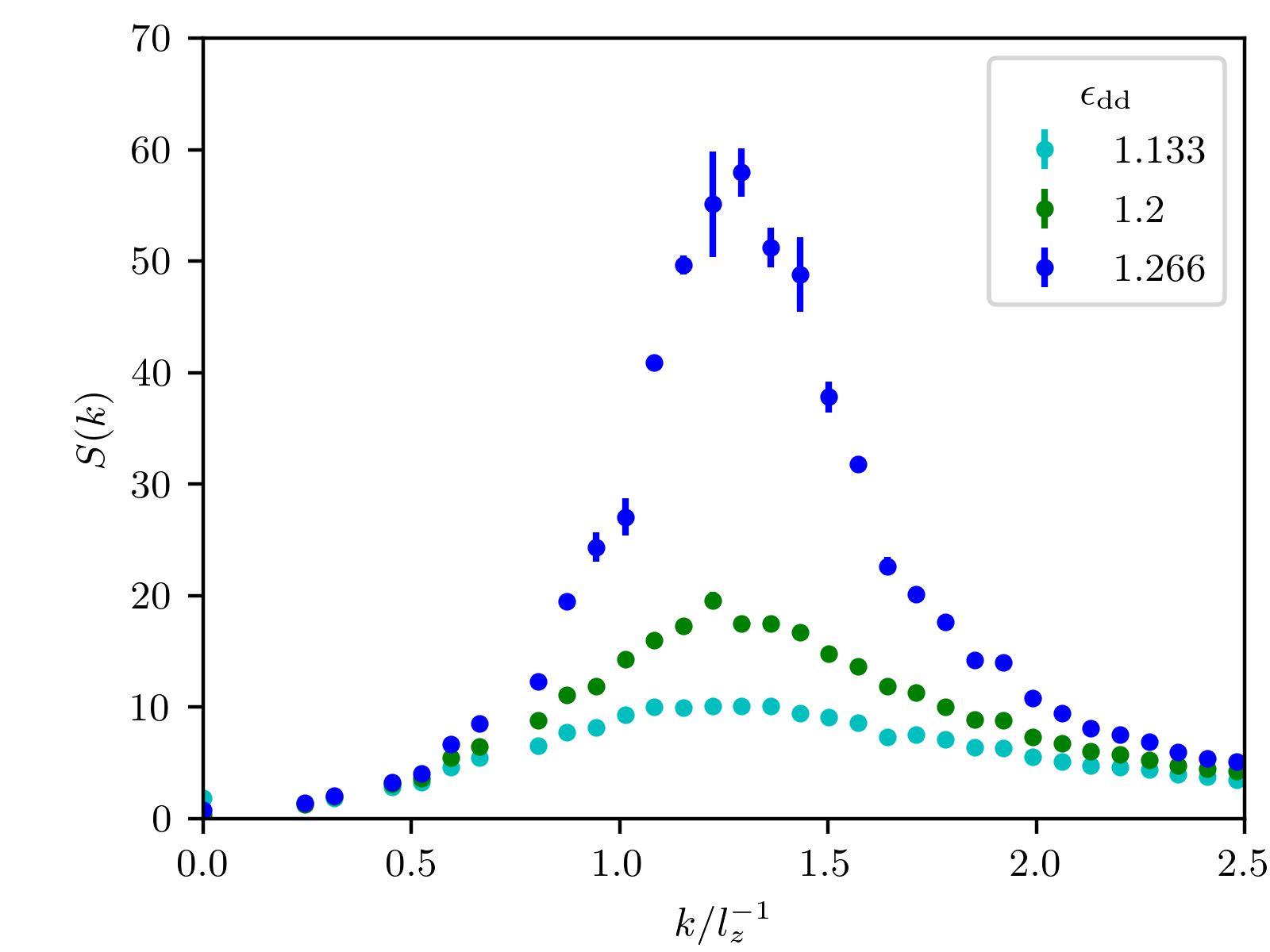}
	\centering\includegraphics[width=0.495\textwidth]{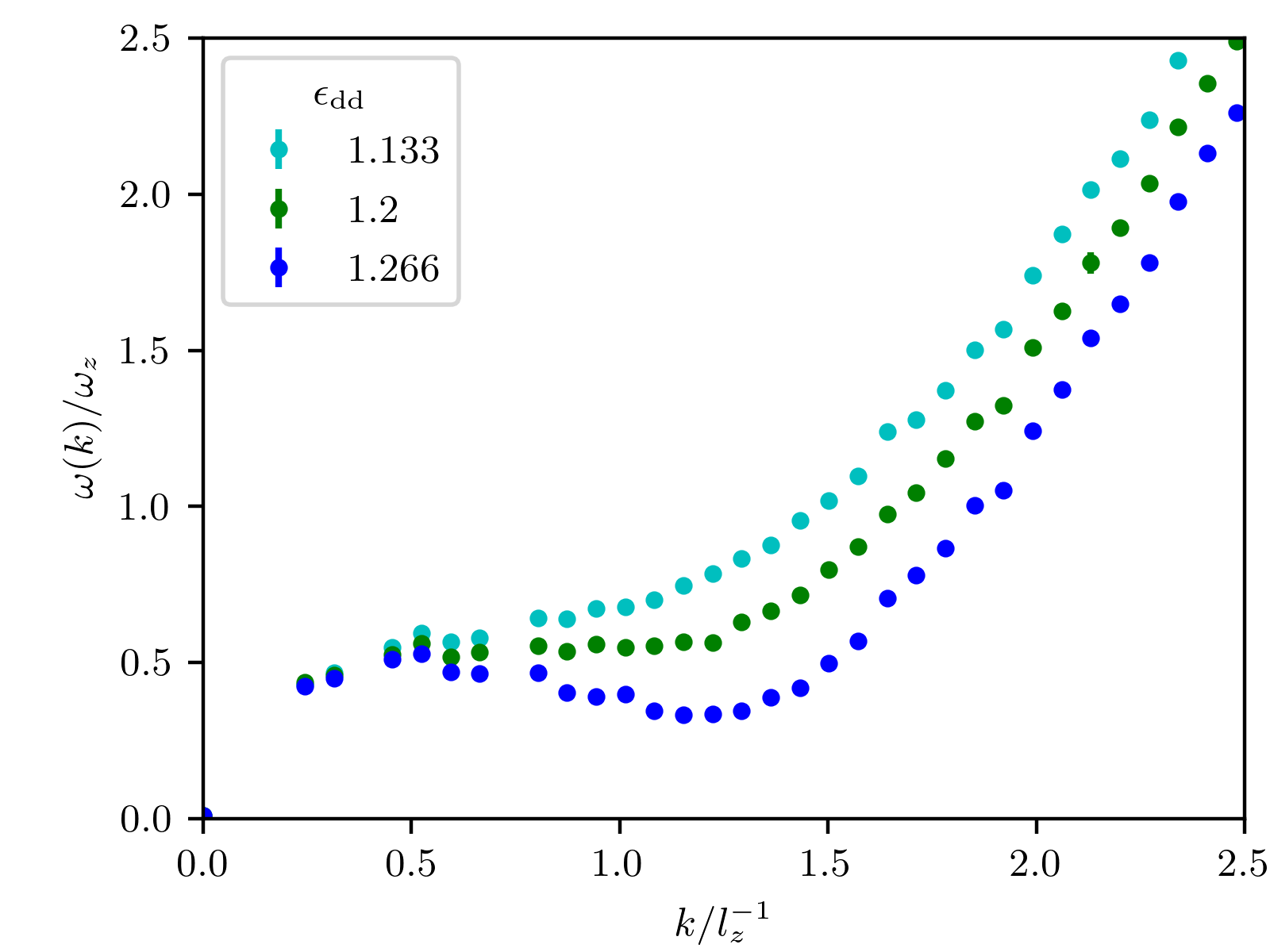}
	\caption{Static structure factor $S(k)$ (left panel) and dispersion $\omega(k)$ (right panel), angular-averaged over momenta in the $x$-$y$-plane, for a dipolar gas with periodic boundary conditions in $x$- and $y$-direction and a harmonic confinement in $z$-direction. $\omega(k)$ is given in units of the trapping frequency $\omega_z$ and momenta are expressed in units of the inverse of the harmonic oscillator length $l_z=\sqrt{1/m\omega_z}$. Parameters are $\mu/\omega_z=10$, $T/\mu=0.41667$ and $\mu m^3 g^2=1/800$. $\epsilon_\mathrm{dd}$ is varied between $1.133$ and $1.266$. Only for $\epsilon_\mathrm{dd}=1.266$ does the peak in the static structure factor become sharp enough to cause a pronounced local minimum in the dispersion.
	}
	\label{fig:disp_S_roton}
\end{figure} 
In this section, we want to demonstrate that we can achieve a rotonic dispersion relation in a gas of dipolar atoms within a complex Langevin simulation. In contrast to the spectrum of a homogeneous, periodic dipolar gas, which we reproduced in the previous section, a rotonic dispersion is expensive to compute even within the Bogoliubov formalism. Namely, the occurrence of rotonic excitations requires a trapping potential in the polarization direction of the dipolars and the application of the Bogoliubov formalism to an inhomogeneous system with external potential amounts to the diagonalization of a large matrix.

While we could in principle directly extract the dispersion $\omega(\mathbf{k})$ via \eq{disp}, we found that in practice the demand in statistics was impractically high in the case of dipolar atoms in the rotonic regime, in particular because an angular average as performed in chapter \sect{3D} is not possible due to the anisotropic nature of the dipolar interaction. Instead, we resorted to extract $\omega(\mathbf{k})$ indirectly from the \textit{static structure factor} (SSF), defined as  
\begin{align}
\label{eq:SSF}
S(\mathbf{k})=1+\frac{1}{N_\text{tot}}\int d^3r_1\,d^3r_2\,e^{i\mathbf{k}\cdot(\mathbf{r}_1-\mathbf{r}_2)}\langle \psi^\dagger(\mathbf{r}_1) \psi^\dagger(\mathbf{r}_2)\psi(\mathbf{r}_1) \psi(\mathbf{r}_2)\rangle\,,
\end{align}
with $N_\text{tot}$ the total particle number. Via the Feynman relation, the static structure factor is related to the dispersion as \cite{pitaevskii2016bose,klawunn2011local}
\begin{align}
\label{eq:Feynman_relation}
S(\mathbf{k})=\frac{\mathbf{k}^2}{2m\omega(\mathbf{k})}\,\coth\left(\frac{\omega(\mathbf{k})}{2T}\right)\,,
\end{align}
which is valid in the superfluid phase. The correlator \eq{SSF} can be straightforwardly evaluated in a CL simulation. Then, by numerically solving \eq{Feynman_relation} for $\omega(\mathbf{k})$, we can extract the dispersion relation.  

We simulate a simple scenario that allows for the occurrence of a rotonic dispersion, choosing periodic boundary conditions in the $x$-$y$-plane and a harmonic trapping potential with frequency $\omega_z$ in $z$-direction, i.e. the polarization direction. We choose as parameters $g=0.1\,a_\mathrm{s}^2$, $\mu=1\,a_\mathrm{s}^{-1}$, $\mu/\omega_z=10$ and $T/\mu=0.41667$, and we simulate on a $128^2\times 96\times 48$ lattice. The dipolar potential is cut off only in $z$-direction at half the extension of the computational lattice, while in $x$- and $y$-direction we keep it infinitely extended, in accordance with the choice of periodic boundary conditions. The total particle number results on the order of $N_\text{tot}\sim 2\cdot 10^6$, which is not only out of reach for PIMC simulations but also one order of magnitude larger than what can be typically achieved in experiment. We vary $\epsilon_\mathrm{dd}$ between $1.133$ and $1.266$. The resulting static structure factor is shown in the left panel of figure \fig{disp_S_roton}, while the right panel shows the dispersion relation computed according to \eq{Feynman_relation}. The SSF features pronounced peaks at a momentum of $k\sim 1.25\,l_z^{-1}$ (with $l_z=1/\sqrt{m\omega_z}$ the harmonic oscillator length) for all considered values of $\epsilon_\mathrm{dd}$. However, at finite temperature, this does not necessarily imply a local minimum in the dispersion but only a deviation from the Bogoliubov dispersion. It is only for $\epsilon_\mathrm{dd}=1.266$ that the peak is sharp enough to cause such a local minimum.

\subsection{Comparison to experiment}
\begin{figure}
	\centering\includegraphics[width=0.495\textwidth]{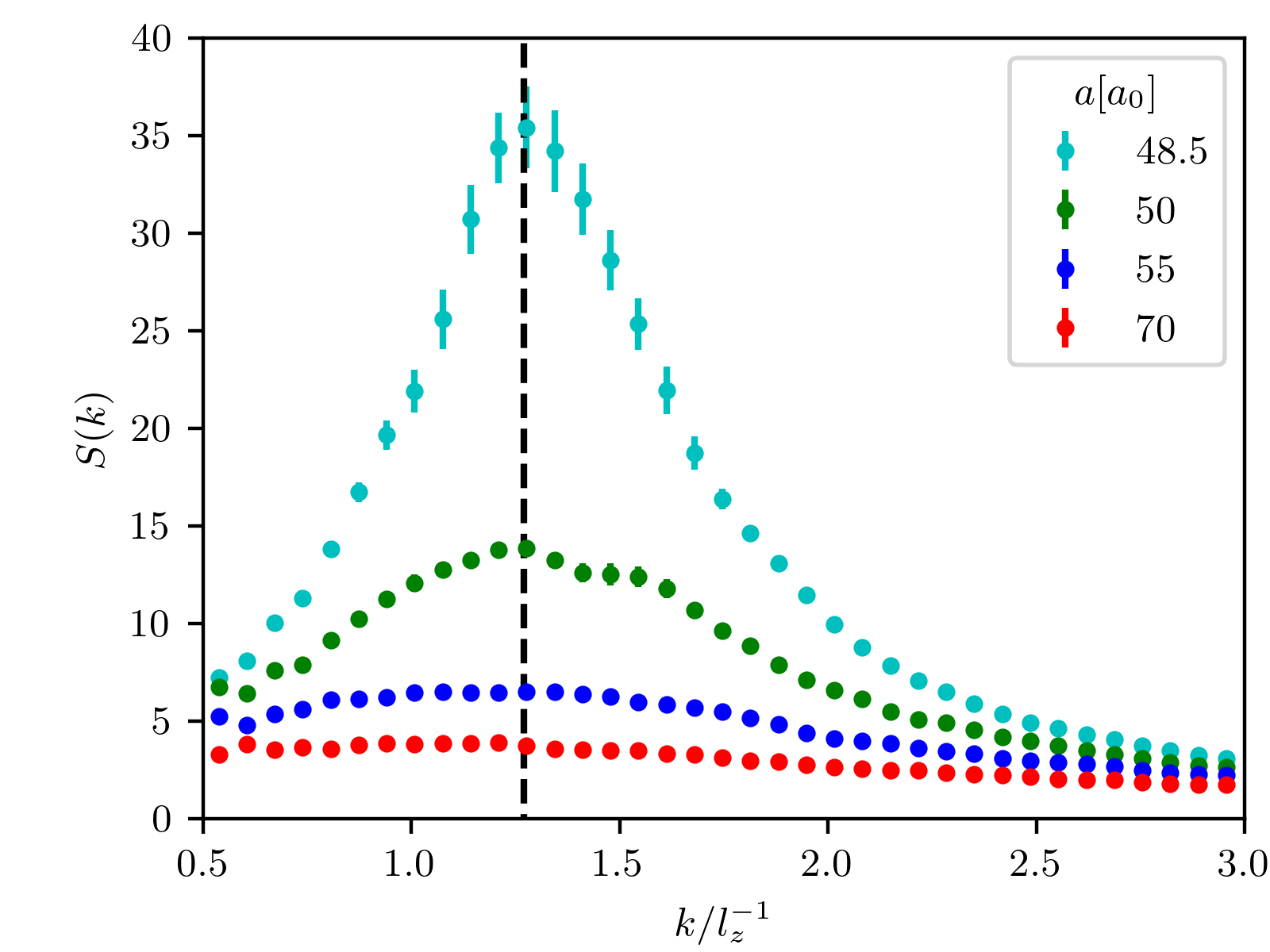}
	\centering\includegraphics[width=0.495\textwidth]{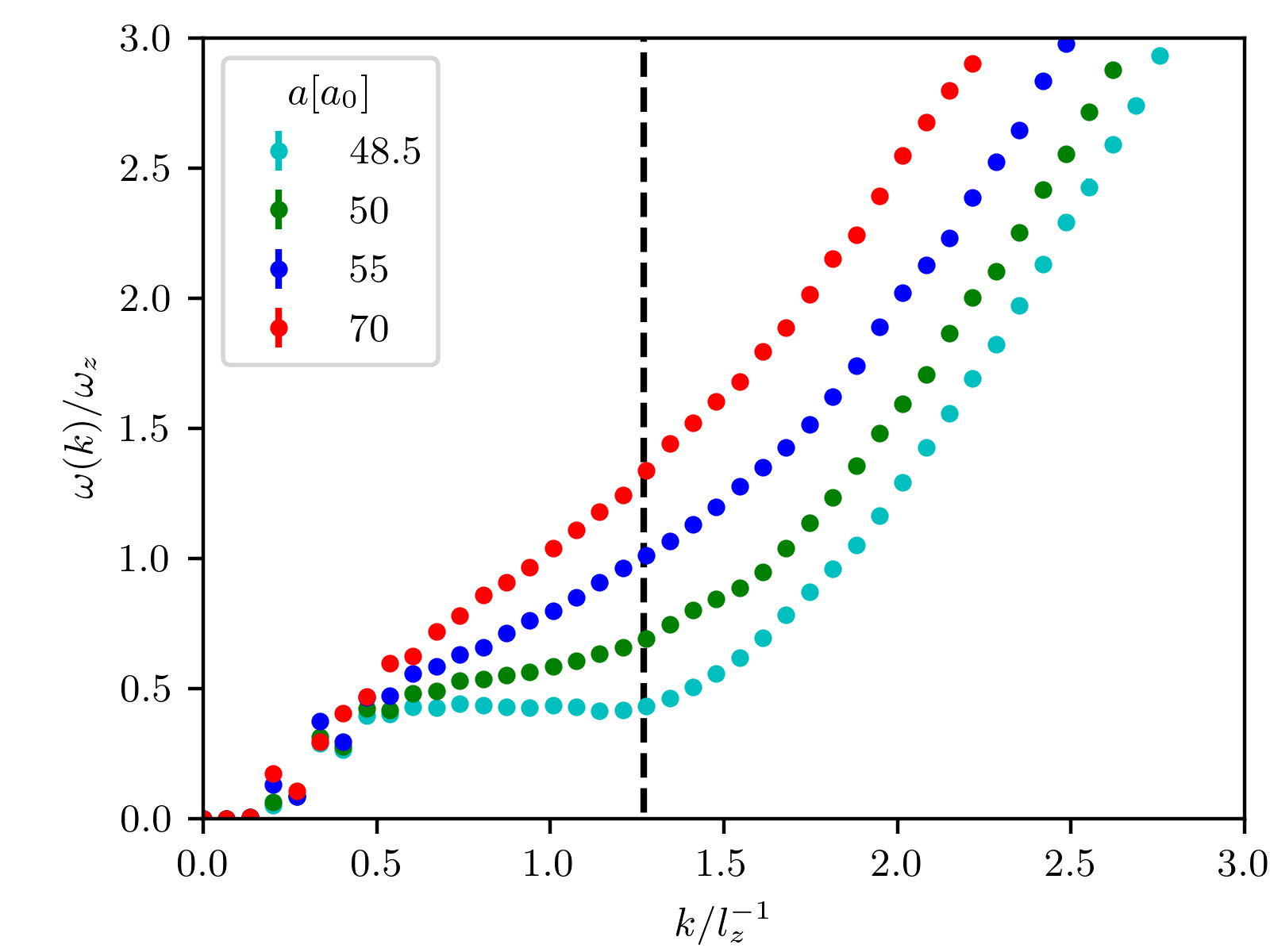}
	\caption{Static structure factor $S(k)\equiv S(k\mathbf{e}_y)$ (left panel) and dispersion $\omega(k)\equiv \omega(k\mathbf{e}_y)$(right panel) as extracted from a CL simulation with $N_\text{tot}=2.4\cdot 10^4$ particles and temperature $T= 50\,\text{nK}$ for four different scattering lengths $a$. The dashed black line marks the momentum $k_\text{rot}=1.27\,l_z^{-1}$ where $\omega_\text{rot}$ is extracted.
	}
	\label{fig:disp_different_a}
\end{figure}
\begin{figure}
	\centering\includegraphics[width=0.6\textwidth]{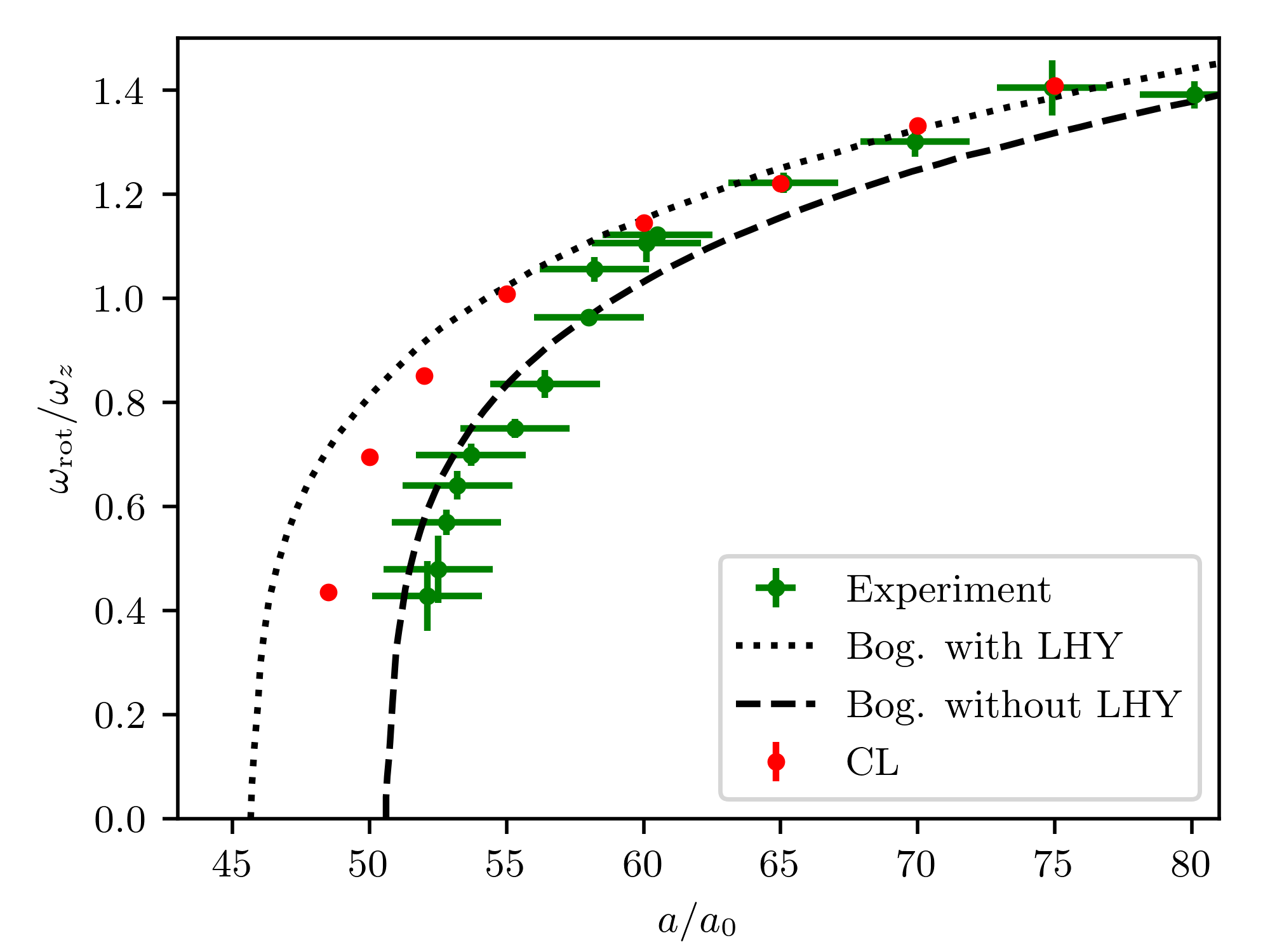}
	\caption{Roton gap $\omega_\text{rot}\equiv \omega(1.27\,l_z^{-1}\,\mathbf{e}_y)$ as a function of the scattering length $a$ in units of the Bohr radius $a_0$ as extracted from a CL simulation with $N_\text{tot}=2.4\cdot 10^4$ particles and temperature $T= 50\,\text{nK}$ (red points). Green points represent the experimental values, the dotted black line the Bogoliubov prediction with the LHY correction and the dashed black line the Bogoliubov prediction without the LHY correction. The experimental and Bogoliubov data is taken from \cite{petter2019probing}. 
	}
	\label{fig:exp_vs_theo_dip}
\end{figure}
In this section we compare the results of ab initio CL simulations for the roton excitation energy to an actual experimental setting. In an experiment conducted by the Ferlaino group at the University of Innsbruck \cite{petter2019probing}, the dispersion relation of a stable gas of erbium ($^{166}\text{Er}$) atoms was measured with Bragg spectroscopy. $^{166}\text{Er}$ possesses a mass of $m=166\,\mathrm{u}$ and a dipolar scattering length $a_\mathrm{dd}=65.5\,a_0$ with $a_0$ the Bohr radius. The atoms were trapped in a harmonic confinement with frequencies $\omega_{x,y,z} = 2\pi \times (261, 27, 256)\,\text{Hz}$, leading to cigar-shaped atom cloud. The harmonic oscillator length in $z$-direction $l_z=\sqrt{\hbar/m\omega_z}$, which in the following will serve as a unit of length, results as $l_z=0.49\,\text{\textmu} \mathrm{m}$. In the measurements that we aim to reproduce, the number of atoms in the trap amounted to $N_\text{tot}=(2.4\pm0.2)\cdot 10^4$~\footnote{More precisely, the number of \textit{condensed} atoms was extracted to be $N_\text{cond}=(2.4\pm0.2)\cdot 10^4$. In order to make our results comparable to the theory predictions from Bogoliubov theory, we will in the following assume that the amount of non-condensed atoms is negligible (this is at least the case for the simulations as the chosen temperature is close to the zero-temperature limit) and perform CL simulations with the \textit{total} number of particles fixed to $N_\text{tot}=2.4\cdot 10^4$.}. The s-wave scattering length was varied between $\sim52\,a_0$ and $\sim80\,a_0$, such that $\epsilon_\mathrm{dd}$ varies between $1.26$ and $0.82$. The Bragg spectroscopy gives access to the dynamic structure factor $S(\omega,\mathbf{k})$, which in the superfluid phase is proportional to the spectral function \cite{pitaevskii2016bose}. The latter in turn features $\delta$-peaks at the dispersion $\omega(\mathbf{k})$ (in a realistic system, these appear as sharp but finite-width peaks), such that this procedure allows to extract the energy of the elementary excitations. The experimental study considered the dispersion $\omega(\mathbf{k})$ for momenta in $y$-direction, i.e. the elongated direction of the atom cloud. Of particular interest was the energy of the rotonic excitations, which was defined as $\omega_\text{rot}\equiv\omega(k_\text{rot}\mathbf{e}_y)$, where the roton momentum $k_\text{rot}$ was taken to be $k_\text{rot}=1.27\,l_z^{-1}$, i.e. the approximate location of the local minimum in the dispersion. $\omega_\text{rot}$ was extracted as a function of $a$ and compared to numerical Bogoliubov calculations~\footnote{For an inhomogeneous system, even in the Bogoliubov approximation that linearizes the Hamiltonian, one needs to diagonalize a large matrix in order to bring this linearized Hamiltonian also to diagonal form, which must be done numerically.} of the trapped geometry either in the standard Bogoliubov approximation or including the LHY term, i.e. inserting back a term $8\,\text{Re}\gamma(\epsilon_\mathrm{dd})/(15\pi^2)\,m^{3/2}g^{5/2}(\psi^\dagger\psi)^{5/2}$ into the Hamiltonian before making the Bogoliubov approximation. Both approaches substantially disagreed among each other and with the experiment, rendering an exact calculation desirable. 

We discretize the experimental system on a $48\times 384\times 64\times N_\tau$ lattice with $N_\tau$ ranging from $14$ to $84$. The dipolar length in units of the lattice spacing $a_\mathrm{s}$ is chosen to be $a_\mathrm{dd}=(1.1/12\pi)\,a_\mathrm{s}$. The dipolar interaction is cut off in each direction at half the lattice size in that respective direction and we make sure that the lattice is twice as large as the atom cloud in every direction, cf. appendix \sect{dip_cutoff}. For the majority of the simulations, we set the temperature to $T=50\,\text{nK}$, which corresponds to $N_\tau=42$. As will be argued below, this temperature can be considered close to the zero-temperature limit. We fix the total particle number to $N_\text{tot}=2.4\cdot 10^4$. As we work in a grand-canonical framework, we first have to find the chemical potential that produces this particle number. We do so by means of a standard secant algorithm, which we break off as soon as the expectation value $\langle N_\text{tot}\rangle$ deviates less than $1\%$ from $2.4\cdot 10^4$.

We determine the s-wave scattering length $a$ that corresponds to our bare coupling $g$ by employing the finite-$a_\tau$ modified Born series up to NNLO, cf. section \sect{coupling_renormalization}:
\begin{align}
\label{eq:Born}
\frac{4\pi a}{m}=V_0-m \sum_\mathbf{p}\frac{V_{\mathbf{p}}V_{-\mathbf{p}}}{\mathbf{p}^2\left(1-a_\tau\frac{\mathbf{p}^2}{4m}\right)}+m^2\sum_{\mathbf{p},\mathbf{q}}\frac{V_{-\mathbf{p}}V_{\mathbf{q}}V_{\mathbf{p}-\mathbf{q}}}{\mathbf{p}^2\mathbf{q}^2\left(1-a_\tau\frac{\mathbf{p}^2}{4m}\right)\left(1-a_\tau\frac{\mathbf{q}^2}{4m}\right)}\,.
\end{align}  
For $V_\mathbf{k}$ we insert \eq{full_pot_k}. The momentum sums over all lattice momenta are performed numerically. For our choice of the momentum cutoff and the experimental parameters, we find renormalization corrections to the leading order expression $a=mg/4\pi$ on the order of $\sim10\%$, which is definitely not negligible for the present purpose.

The SSF and dispersion relation from the CL simulation are shown in figure \fig{disp_different_a} for several values of the scattering length $a$. We extract $\omega_\text{rot}$ from this data by fitting a linear function to the numerical $\omega(k\mathbf{e}_y)$ in the interval $[1.27-0.1,1.27+0.1]\,l_z^{-1}$ and take the value of this linear function at $1.27\,l_z^{-1}$ to be $\omega_\text{rot}$. In figure \fig{exp_vs_theo_dip}, we show $\omega_\text{rot}$ as a function of $a$ extracted from our simulations together with the experimental and Bogoliubov data from \cite{petter2019probing}. At scattering lengths $a\gtrsim 55\,a_0$, the CL data agrees well with the prediction of Bogoliubov theory \textit{including} the LHY correction, as does also the experimental data. It is only close the critical point, for $a\lesssim 55\,a_0$, that small deviations become visible. Namely, the CL simulation predicts the critical scattering length to be slightly larger than the prediction of Bogoliubov theory including the LHY correction, albeit not as large as it is predicted by Bogoliubov theory without the LHY correction. The experimental data instead appears to agree better with the latter than the former close to the critical point. Hence, the simulation results alleviate the discrepancy between theory and experiment in comparison to the Bogoliubov calculation with the LHY term but do not fully remove it. Numerous reasons for this remaining discrepancy are possible, which are in part discussed in reference \cite{petter2019probing} itself. These include underestimated uncertainties in the particle number and the scattering length, as well as deviations from the zero-temperature limit due to a non-negligible thermal condensate depletion.

\begin{figure}
	\centering\includegraphics[width=0.495\textwidth]{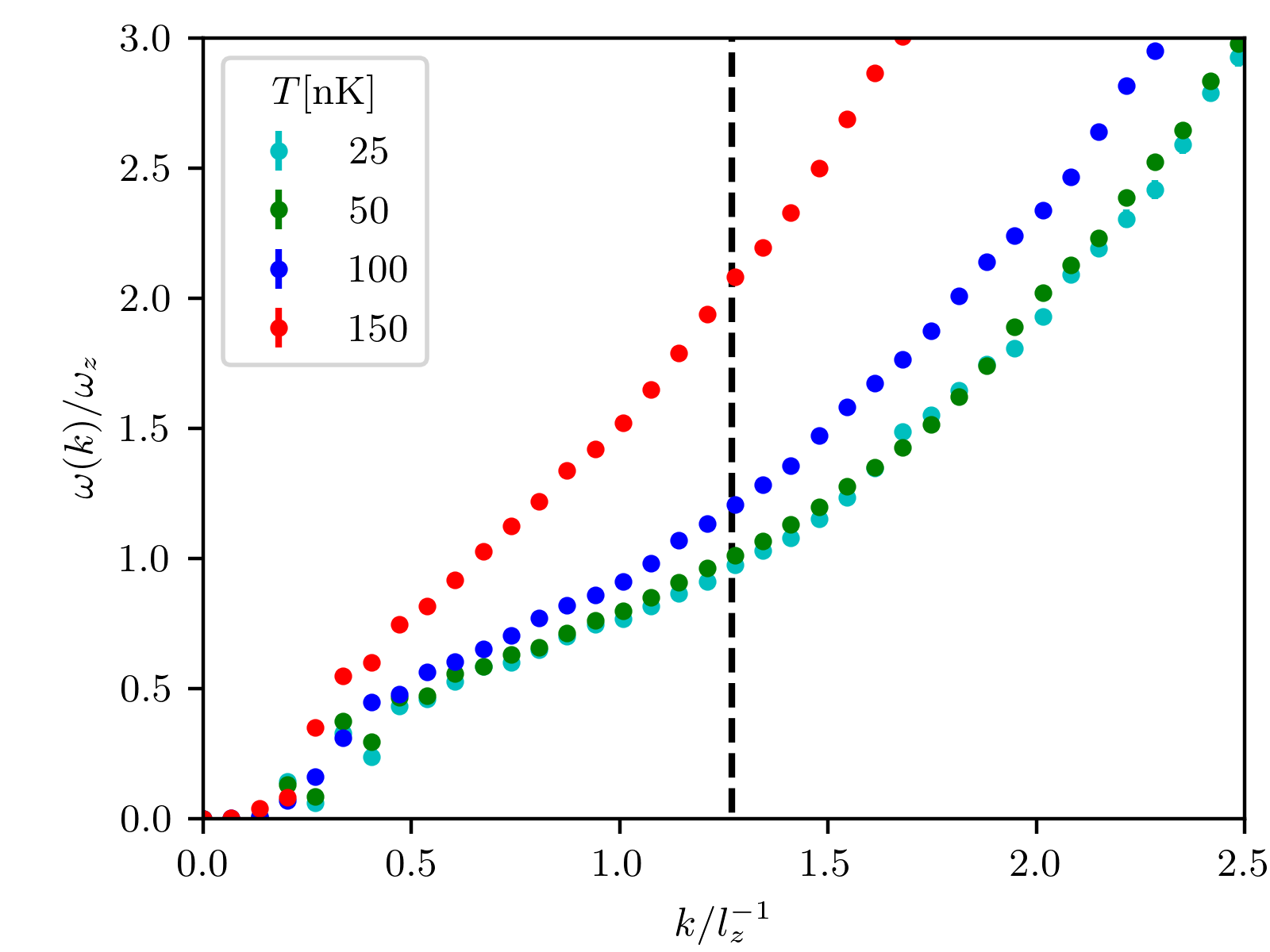}
	\centering\includegraphics[width=0.495\textwidth]{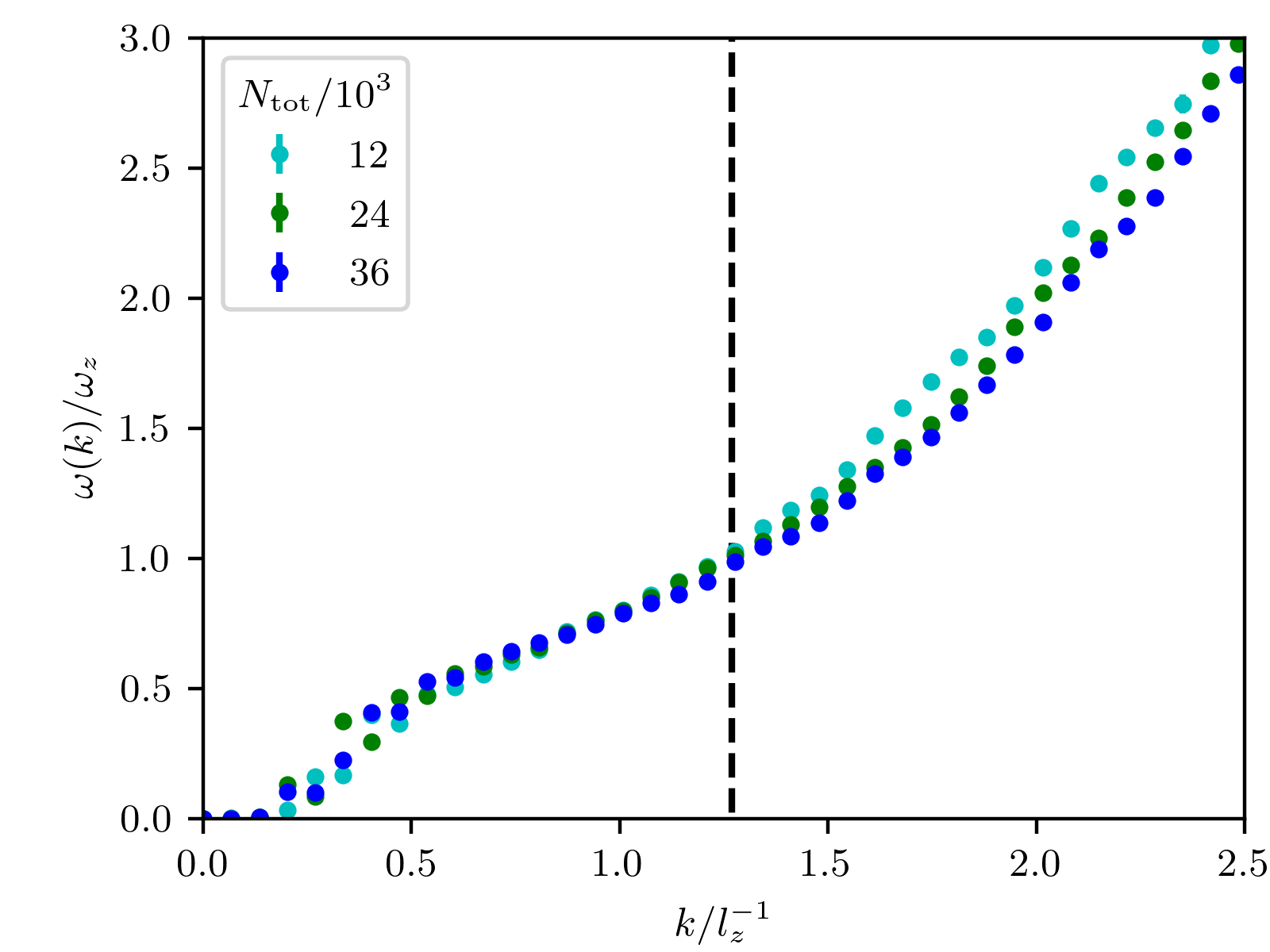}
	\caption{Dispersion relation at fixed scattering length $a=55\,a_0$ but varying temperature (left panel) and total particle number (right panel). The dispersion relation at $T=25\,\text{nK}$ and $T=50\,\text{nK}$ almost do not differ at all, indicating that at $T=50\,\text{nK}$ the zero-temperature limit is practically reached. The dispersion for $N_\text{tot}=1.2\cdot 10^4$ and $N_\text{tot}=3.6\cdot 10^4$ differs only slightly from that for $N_\text{tot}=24\cdot 10^3$, demonstrating a remarkable insensitivity of the dispersion to particle number. 
	}
	\label{fig:different_T_and_N}
\end{figure}

We finally want to explore how much a different temperature and particle number affect the dispersion relation. To that end we fix the scattering length to $a=55\,a_0$ and first vary the temperature (at fixed particle number $N_\text{tot}=2.4\cdot 10^4$), simulating, in comparison to the simulations presented so far, both at a lower temperature $T=25\,\text{nK}$ and at higher temperatures $T=100\,\text{nK}$ and $T=150\,\text{nK}$. The results are presented in the left panel of figure \fig{different_T_and_N}. Between the dispersion for $T=25\,\text{nK}$ and $T=50\,\text{nK}$ one observes almost no deviation beyond the statistical errors, suggesting that the simulations at $T=50\,\text{nK}$ whose results were presented so far have reached the zero-temperature limit with regard to the dispersion relation. For larger temperature, the excitation energy at the roton momentum appears to become larger. This result is in surprising contradiction with the findings of reference \cite{sanchez2023heating} where a \textit{decrease} of the roton energy at higher temperature was predicted. A possible explanation could be that we have fixed the \textit{total} particle number to $N_\text{tot}=2.4\cdot 10^4$ here (there is no other notion of particle number in an ab initio simulation), while the semi-classical approach outlined in \cite{sanchez2023heating} fixes the number of \textit{condensed} atoms. Fixing the condensate and increasing the temperature amounts to increasing the total number of particles, which in turn decreases the roton excitation energy, as the confining effect of the trapping in $z$-direction becomes weaker. However, it is also likely that the Feynman relation \eq{Feynman_relation} used here to compute the dispersion is not fully valid any more at these comparably high temperatures.

The right panel of figure \fig{different_T_and_N} shows the dispersion for fixed $a=55\,a_0$ and $T=50\,\text{nK}$ but varying $N_\text{tot}$. We decrease and increase $N_\text{tot}$ by $50\%$ in comparison to $N_\text{tot}=2.4\cdot 10^4$. Despite this large change, the dispersion seems to be only very slightly lowered and raised, respectively. This demonstrates a remarkable robustness of the roton gap with respect to particle number variations in experiment.  

\subsection{Supersolid from first principles}
In this section, we consider the same experimental scenario as in the previous one (erbium in a cigar-shaped trap), but examine what happens when we tune the scattering length beyond the critical point. As the complex Langevin algorithm enables exact simulations and thus includes the effect of quantum fluctuations automatically, which are believed to be of crucial importance in the formation of a supersolid, one should be able to observe the formation of such a state in a CL simulation even if only the microscopic action is simulated (i.e. without the LHY term). The density modulation characteristic of supersolids leads to comparatively high particle densities and hence high dilutenesses in the density peaks. This is problematic even in experiment because it causes strong three-body losses and therefore limits the lifetime of supersolids \cite{chomaz2019long}. For the complex Langevin algorithm this means a considerable limitation, as it breaks down for strong diluteness, the stronger non-linearity in the evolution equation causing runaways into the complex plane. We therefore restrict ourselves to the regime not too far below the transition point and also slightly lower the particle number in comparison to the previous section~\footnote{An alternative would be to consider a system with much smaller trapping frequencies or to decrease the dipolar scattering length in comparison to its actual physical value, but we here want to demonstrate the applicability of the algorithm to an experimentally realistic scenario.}. In this regime, runaway trajectories do already occur but still not so frequently as to spoil the convergence of observables.

\begin{figure}
	\centering\includegraphics[width=0.6\textwidth]{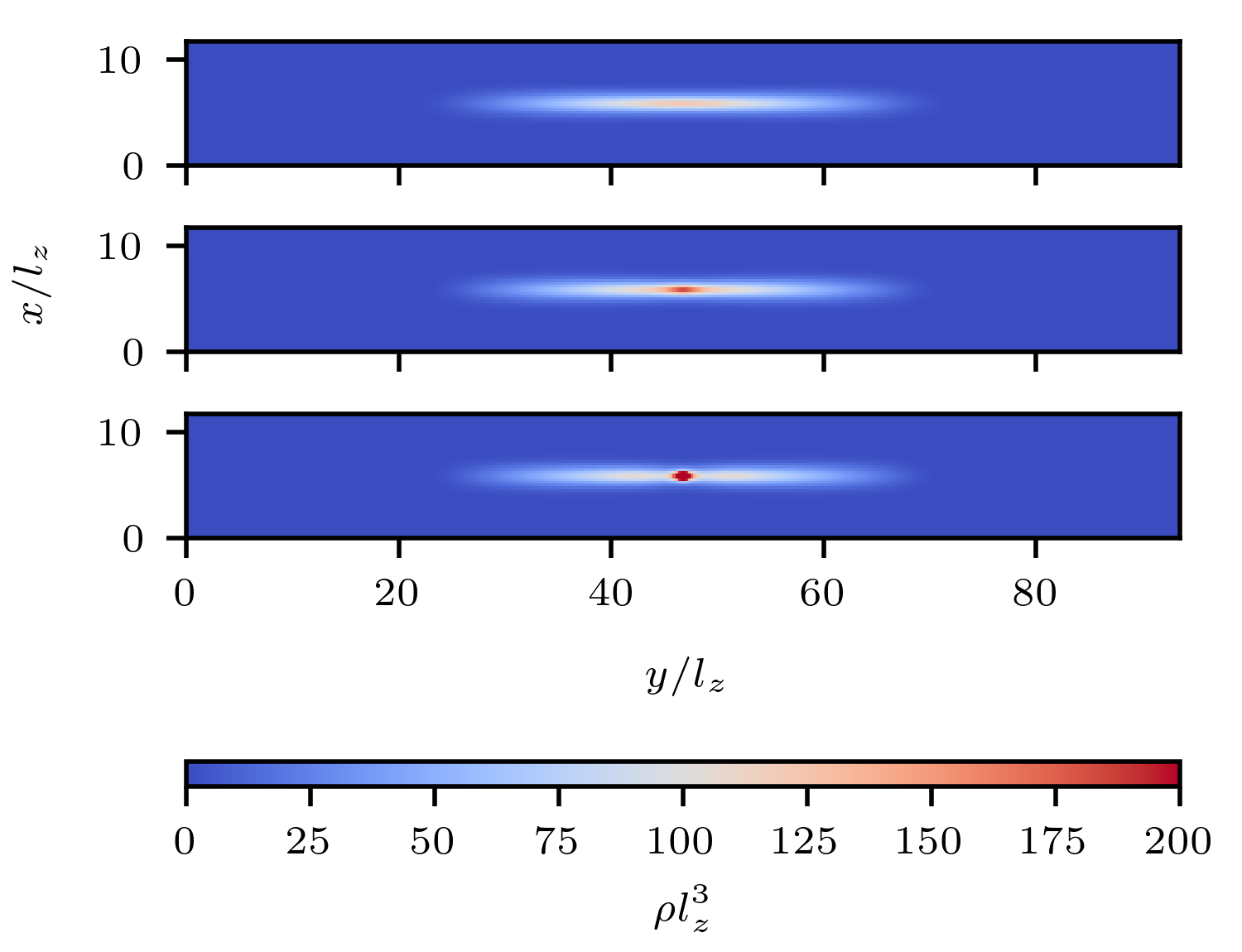}
	\caption{Auxiliary-time evolution of the classical field theory, for parameters $a=45\,a_0$, $\mu/\hbar\omega_z=4.07$, and the other parameters as specified in the main text. The panels show, from top to bottom, snapshots of the density in $x$-$y$-plane at auxiliary times $\vartheta=500,900,1000\,a_\mathrm{s}^{-3}$. The color scale is cut off at $\rho l_z^3=200$ such that the density distribution at the earlier times remains visible, with the actual density maximum at the latest time being $\rho_\text{max}l_z^3=1.7\cdot 10^3$. During the auxiliary time evolution, the atoms begin to pile up in the center of the trap as the dipolar interaction outweighs the repulsive contact interaction and the confining potential. Eventually, the system collapses into one point in the center of the trap.
	}
	\label{fig:coll}
\end{figure}

We choose a scattering length $a=45\,a_0$, i.e. $\epsilon_\mathrm{dd}=1.46$, $\mu/\hbar\omega_z=4.07$ and leave the other parameters (trapping frequencies, mass and dipolar scattering length of erbium) as in the previous section. Let us first demonstrate that this leads to a classically unstable system. To that end, we turn off the Langevin noise, i.e. perform a pure steepest descent evolution:
\begin{align}
\label{eq:steep_desc_dip}
\frac{\partial\psi}{\partial\vartheta}=-\frac{\delta E}{\delta\psi^*}\,,
\end{align}
where $E[\psi]$ is the classical energy functional. Furthermore, we set $g=4\pi a/m$ because in classical field theory, the coupling constant acquires no renormalization corrections. As in previous chapters, we choose as initial condition $\psi=\sqrt{\mu/g}=\text{const}$. Snapshots of the resulting evolution are shown in figure \fig{coll}. After a Thomas-Fermi profile has formed from the initially flat density, the atoms begin to pile up in the center of the trap due to the dipolar interaction and eventually collapse in one point.

\begin{figure}
	\centering\includegraphics[width=0.6\textwidth]{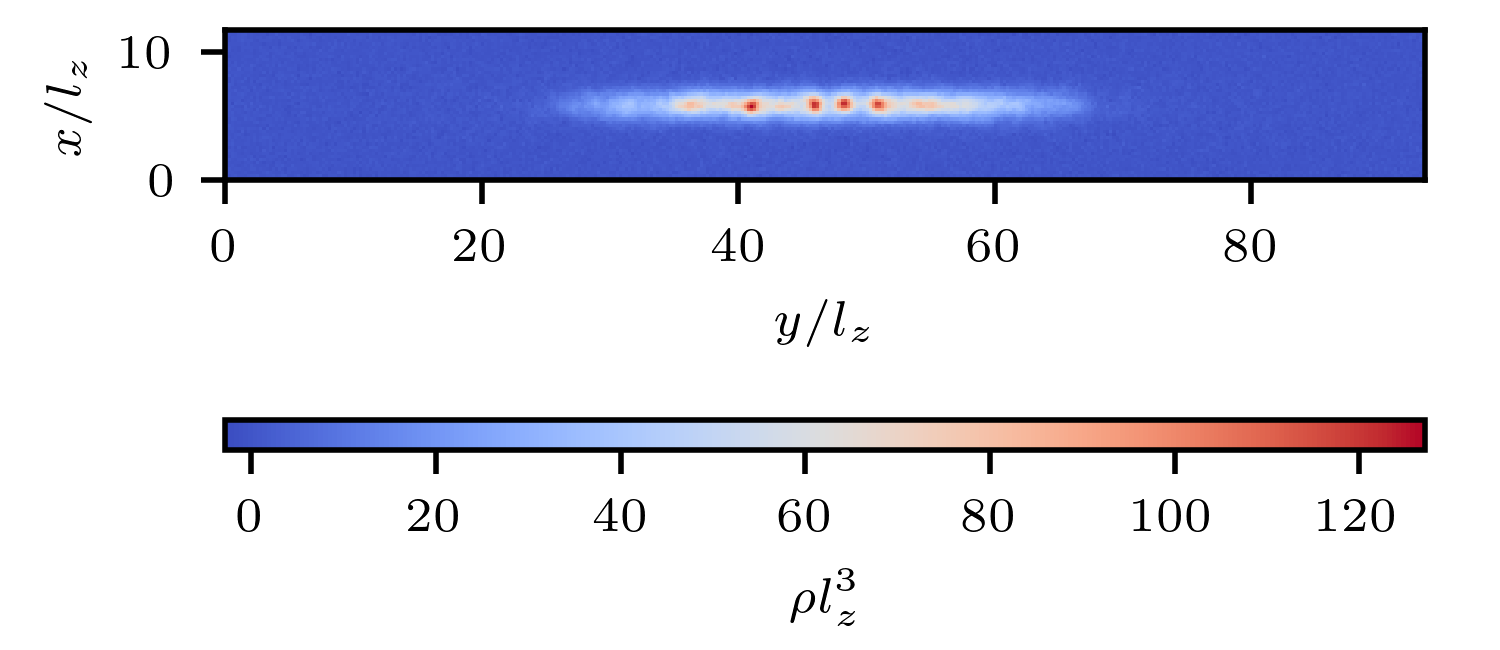}
	\caption{Snapshot of the density in real space, $\bar{\psi}(\mathbf{r})\psi(\mathbf{r})$, in the $x$-$y$ plane (in the center of the trapping potential in $z$-direction), averaged over the imaginary time $\tau$ and a short Langevin time of $\Theta=1000\,a_\mathrm{s}^{-3}$, in the full quantum simulation. Parameters are chosen exactly as in figure \fig{stab}, $a=45\,a_0$, $\mu/\hbar\omega_z=4.07$. Note, however, that $a$ is determined differently here than in the classical simulation because it acquires renormalization corrections in the full quantum theory, cf. the discussion in the main text. The total particle number results as $N_\text{tot}\sim1.8\cdot 10^4$. In contrast to the simulation of the classical field theory, the gas does not collapse any more, such that the quantum fluctuations inherently produced by the simulation appear to stabilize the gas against collapse. Furthermore, one observes spontaneous breaking of the translational symmetry.
	}
	\label{fig:stab}
\end{figure}

Next we turn on the Langevin noise for the exact same parameters (determining $g$, however, according to \eq{Born}). Now we observe no collapse of the atoms any more but the cloud maintains its shape for long simulation times. In figure \fig{stab}, we show a snapshot of the real part of the real-space density $\bar{\psi}(\mathbf{r})\psi(\mathbf{r})$, averaged over imaginary time $\tau$ and a short amount of Langevin time (actual snapshots without any averaging at all result very noisy). Not only does the system remain stable, but we can also observe in the single snapshot the spontaneous breaking of translational symmetry and the emergence of localized density peaks, i.e. the crucial property of a supersolid. 


\begin{figure}
	\centering\includegraphics[width=0.495\textwidth]{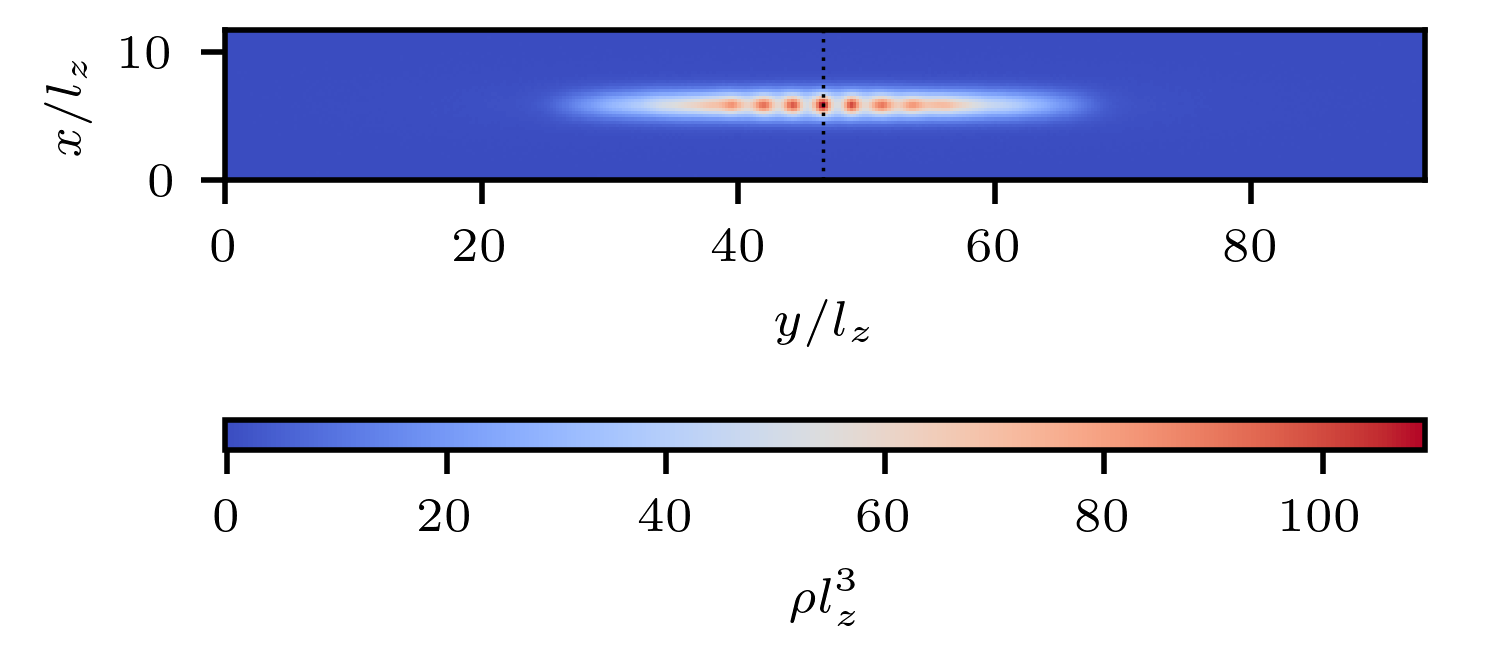}
	\centering\includegraphics[width=0.495\textwidth]{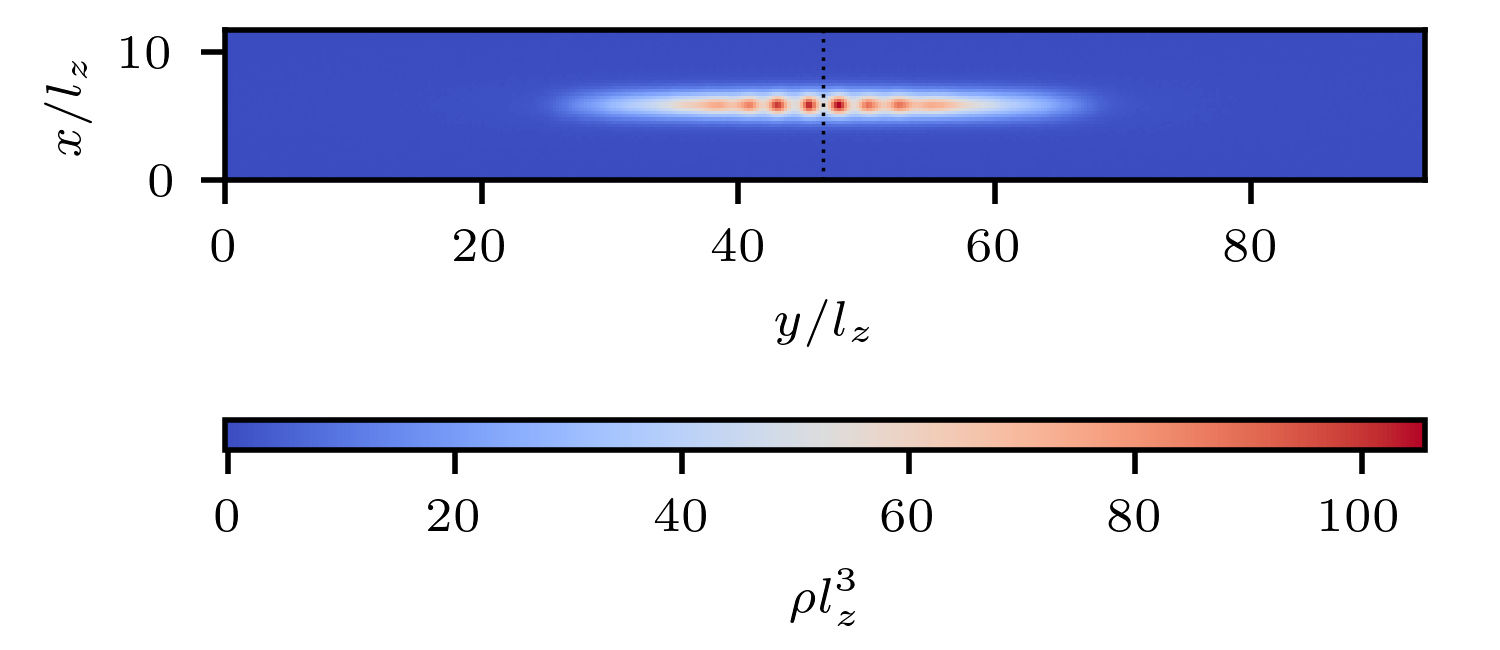}
	\caption{Expectation value of the density in the $x$-$y$ plane (in the center of the trapping potential in $z$-direction), i.e. $\bar{\psi}(\mathbf{r})\psi(\mathbf{r})$ in the long-time and multi-run average, with the run-average performed only among either seven-peak or eight-peak runs, cf. the main text. The left panel shows $\langle\rho\rangle_\text{7p}$, i.e. the density in the seven-peak quasi-ground-state and the right panel shows $\langle\rho\rangle_\text{8p}$, i.e. the density in the eight-peak quasi-ground-state. The dotted black line indicates the center of the trap. Parameters are the same as in figure \fig{stab}. 
	}
	\label{fig:density_average}
\end{figure}

\begin{figure}
	\centering\includegraphics[width=0.495\textwidth]{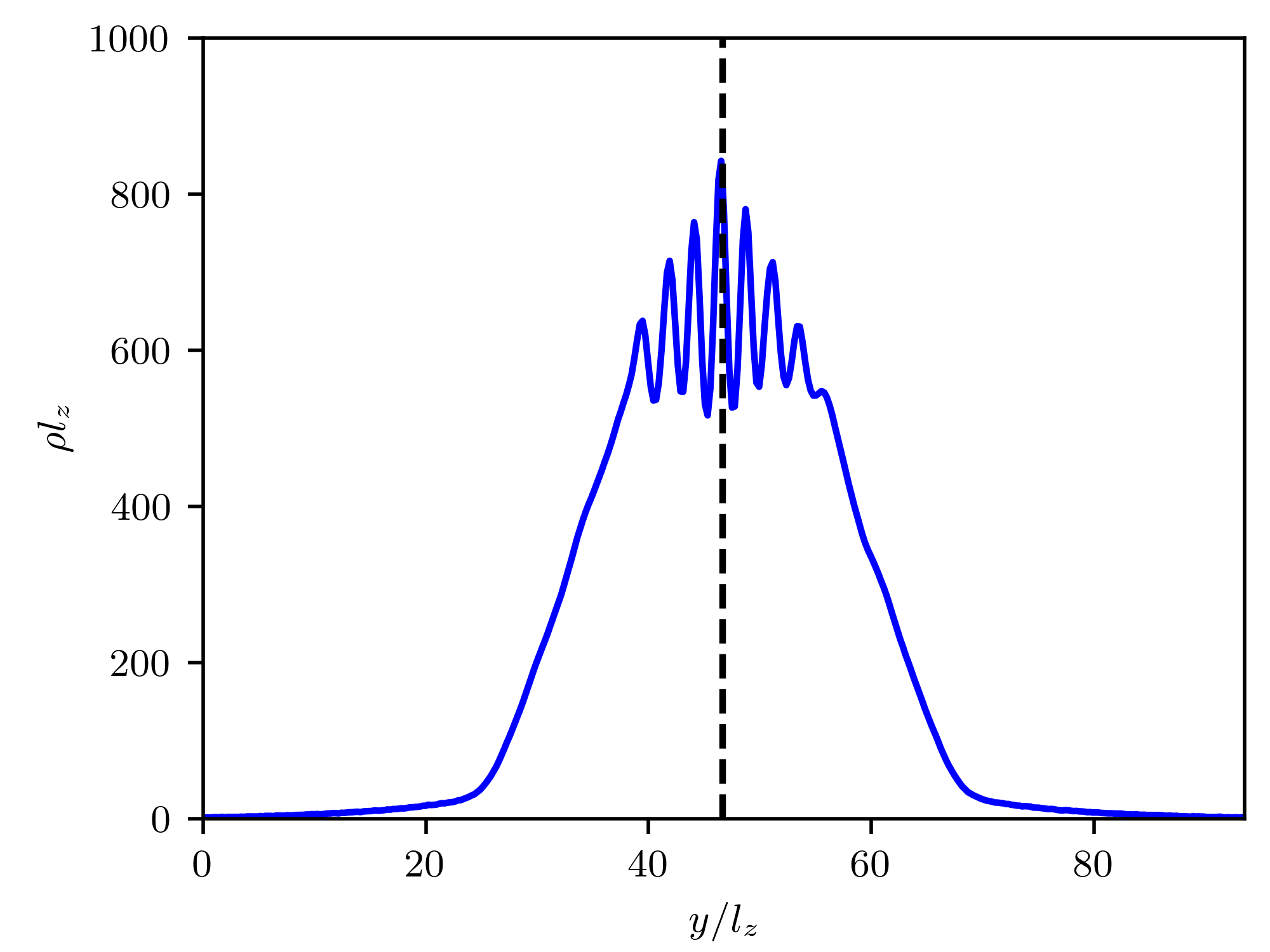}
	\centering\includegraphics[width=0.495\textwidth]{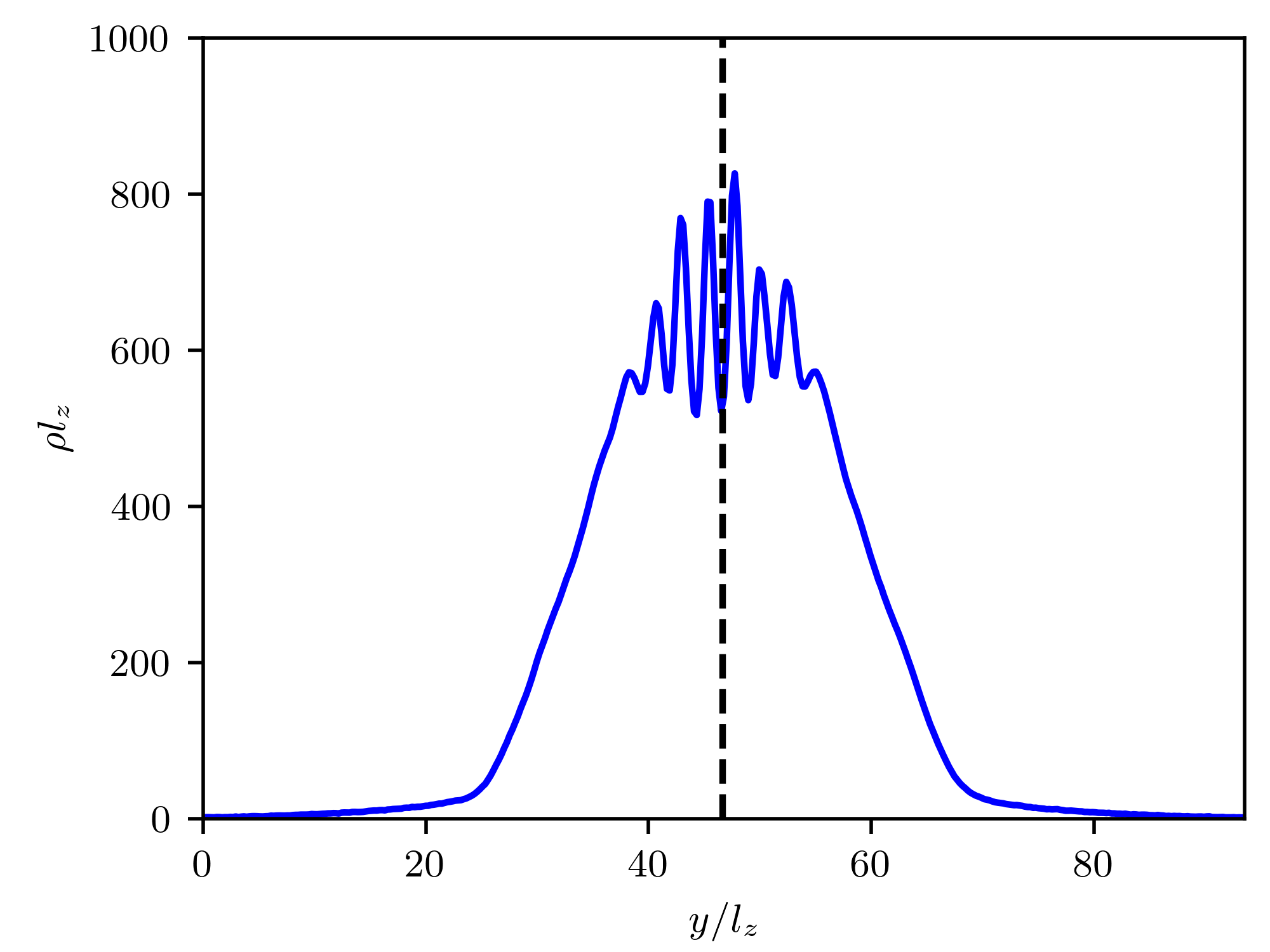}
	\caption{The same as in figure \fig{density_average}, i.e. the expectation value of the density $\rho$, but along the $y$-direction, integrated over the $x$- and $z$-direction. The left panel shows $\langle\rho\rangle_\text{7p}$, i.e. the density in the seven-peak quasi-ground-state and the right panel shows $\langle\rho\rangle_\text{8p}$, i.e. the density in the eight-peak quasi-ground-state. The dashed black line indicates the center of the trap. The persisting asymmetry in the peak heights is due to insufficient averaging, cf. the main text.
	}
	\label{fig:density_along_y}
\end{figure}

In order to make this more systematic, we must consider long-time and multi-run averages of the density, which in general provide the correct expectation values in complex Langevin simulations. In the particular scenario considered here, we find that in the long-time average, each single run converges to a stable, roughly symmetrical, modulated density profile with several maxima and minima. However, some runs produce a profile with a maximum in the center of the trap and in total seven peaks, while others produce a profile with a minimum in the center and in total eight peaks. This phenomenon suggests that in the energy landscape of the theory, there are two meta-stable local minima that are separated by a considerable energy barrier, such that ergodicity is broken to a certain extent. We will denote these as \textit{quasi-ground-states}. An experimental system is likely to randomly choose one of these quasi-ground-states in every realization by means of spontaneous symmetry breaking. In order to extract the density profile associated to each quasi-ground-state and not the superposition thereof, we choose to perform multi-run averages among seven-peak and eight-peak runs separately. The results are shown in figure \fig{density_average} (density in the $x$-$y$-plane in the center of the trapping potential in $z$-direction as a color plot) and \fig{density_along_y} (density along the $y$-direction integrated over $x$ and $z$). A periodic density modulation is visible~\footnote{Despite the multi-run and long-time average, the heights of the peaks are still not fully symmetric (which one would expect for the expectation value due to the symmetry of the problem). This is due to insufficient averaging. As the density profile converged rather slowly and these simulations were very expensive, we refrained from achieving full convergence, giving only a proof of concept.}. One also sees that the density by far does not drop down to zero at the minima, i.e. we are not in the independent quantum droplet phase (which would occur for even larger $\epsilon_\mathrm{dd}$) but in the supersolid phase.


While the CL simulation appears to stabilize the system against collapse in the simulation and is even capable of reproducing the periodic density modulation found in experiment and EGPE simulations, there are also some caveats about the application of the thermodynamic formalism in itself in this case.  It is easy to prove that in the classically unstable region where the classical energy functional $E[\psi]=\int d^3r\,\mathcal{H}[\psi]$ (with $\mathcal{H}[\psi]$ the classical Hamiltonian density) is not bounded from below, also the Hamiltonian operator, defining the full quantum theory, is not bounded from below, i.e.
\begin{align}
\min_{|\phi\rangle}\,\langle\phi|H|\phi\rangle=-\infty\,.
\end{align}
Namely, we can insert for the state $|\phi\rangle$ a coherent state product with parameters $\psi(\mathbf{r})$ in which case $\langle\phi|H|\phi\rangle=E[\psi]$. This means that $H$ possesses no real ground state and hence that the application of the standard Boltzmann-Gibbs thermodynamic formalism is not entirely well-defined here, as thermodynamic traces do not absolutely converge. Remarkably, there appears to be a suitable order of summation that makes them nonetheless converge (albeit not absolutely) to a physically reasonable ground state density profile, and even more remarkably, a complex Langevin simulation of the path integral appears to perform this ``correct'' summation automatically. 

\begin{figure}
	\centering\includegraphics[width=0.495\textwidth]{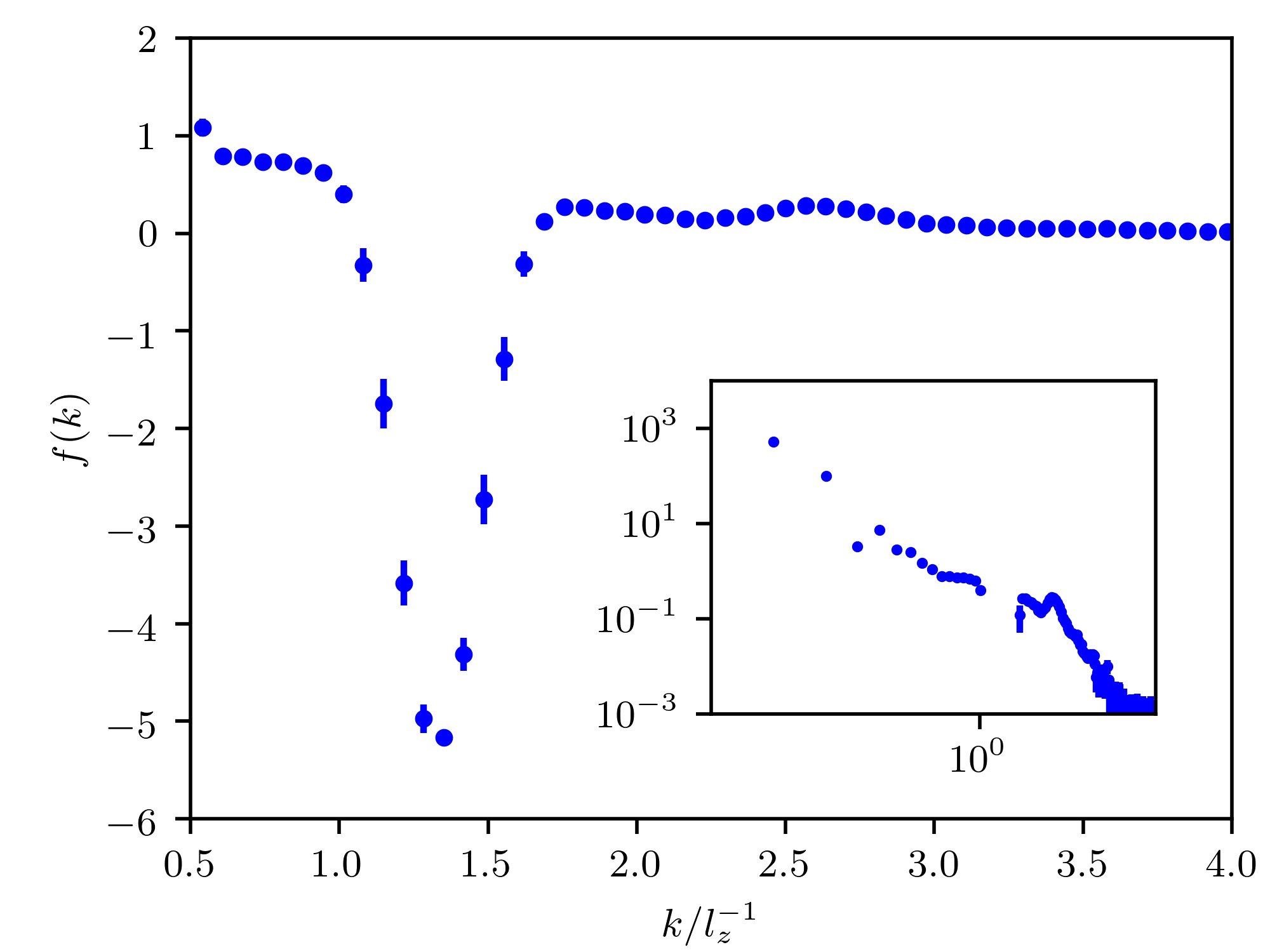}
	\centering\includegraphics[width=0.495\textwidth]{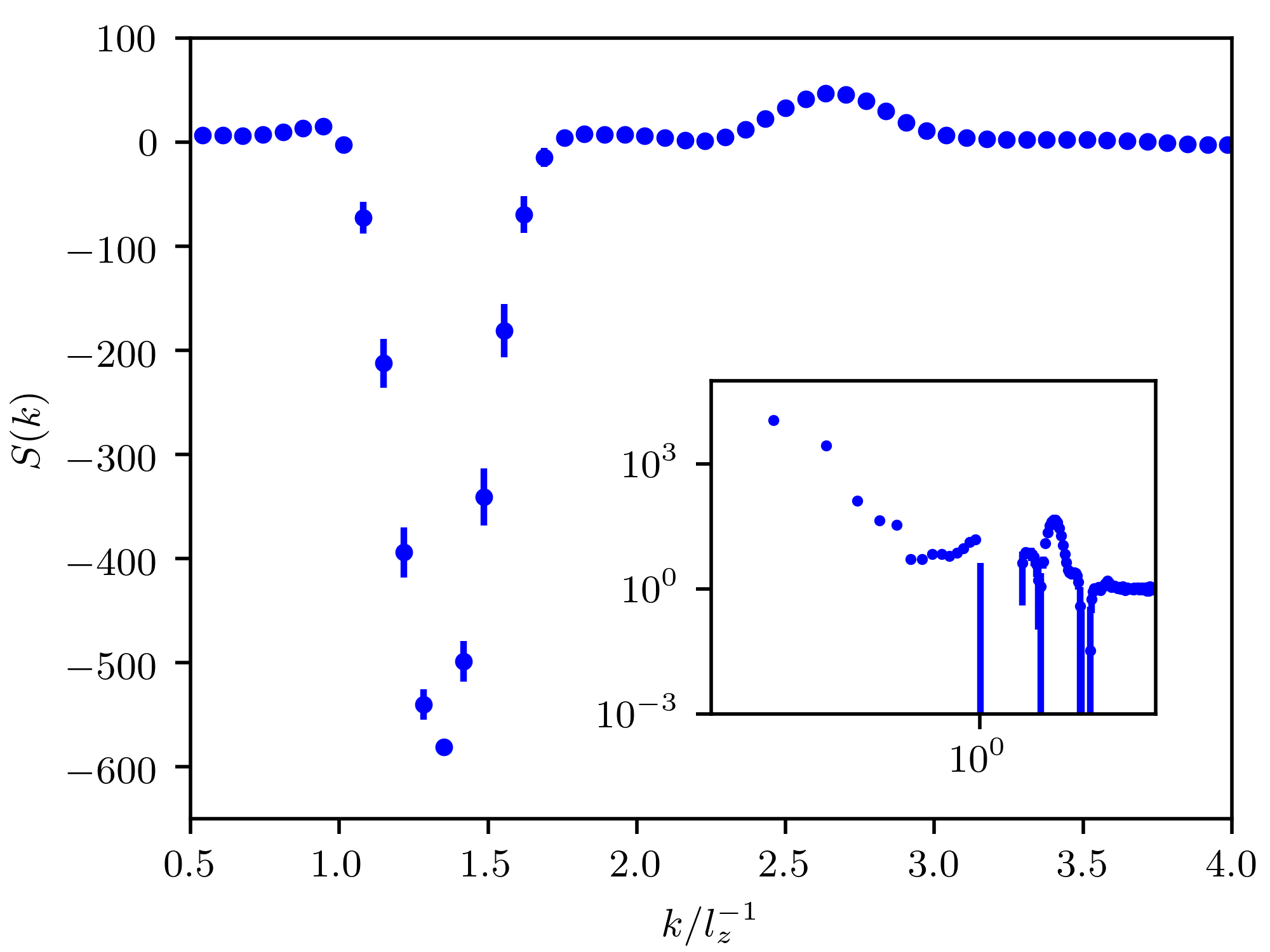}
	\caption{Momentum spectrum $f(k)=f(k\mathbf{e}_y)$ (left panel) and static structure factor $S(k)=S(k\mathbf{e}_y)$ (right panel) for parameters as in figure \fig{stab}, in the region around the roton momentum, where both feature a pronounced negative peak. The respective insets show $f(k)$ and $S(k)$ over the full momentum range in a double-logarithmic plot. 
	}
	\label{fig:Sk_and_fk}
\end{figure}

Despite this success of the thermal simulation, we encounter some unphysical results once we consider not only the ground-state density profile but also elementary excitations. In figure \fig{Sk_and_fk}, we show the SSF and the momentum spectrum $f(\mathbf{k})$ resulting from the simulations presented above. Both feature pronounced \textit{negative} peaks around the roton momentum. It is obvious that a negative $f(\mathbf{k})$ is unphysical, because in an experiment particle numbers can of course only result positive. But also a negative peak in the SSF is unphysical: In the real-space density-density correlator $\langle \rho(\mathbf{r}_0)\rho(\mathbf{r}_0+\mathbf{r})\rangle$, which is the Fourier transform of the SSF, it causes a periodic modulation, but with minima at $\mathbf{r}=0$ and multiples of the roton length, while for a system with solid-like periodic density modulation one would expect \textit{maxima} at these positions. 

Mathematically, one can interpret these results as a consequence of imaginary energies that are present in the system. In fact, analytically continuing the Bose-Einstein distribution or \eq{Feynman_relation} to imaginary $\omega$ results in negative $f$ and $S$. Such imaginary $\omega$ can be understood as stemming from the tiny but nonzero imaginary part of the LHY correction \eq{gamma}~\footnote{In the standard EGPE approach based on the local-density approximation, $\gamma(\epsilon_\mathrm{dd})$ becomes complex already in the stable region for $\epsilon_\mathrm{dd}>1$, but this is an artifact of the local-density approximation. We here have to imagine the LHY correction for some effective potential that turns complex only in the unstable region.}. Assuming that the Feynman relation \eq{Feynman_relation} remains valid in the unstable region and for $\omega^2<0$ may be analytically continued to $\omega=\pm i\Gamma$~\footnote{The validity of the Feynman relation rests on the validity of the Bogoliubov approximation that one can linearize the Hamiltonian and the density operator. This approximation will break down for vanishing $\omega$, but at the position of the peak in the SSF, $\omega$ acquires again a non-zero albeit imaginary value. In any case, this calculation provides at most an estimate of $\Gamma_\text{rot}$.}, we can extract the growth rate $\Gamma_\text{rot}$ of the roton mode from the height of the peak of the SSF, $S_\text{rot}$, as 
\begin{align}
\Gamma_\text{rot}\approx\sqrt{-\frac{k_\text{rot}^2T}{m S_\text{rot}}}\,,
\end{align}
where we have assumed $\Gamma_\text{rot} \ll T$ and hence approximated $\coth x\approx 1/x$. For the present scenario, we have $S_\text{rot}\sim 600$ and $k_\text{rot}\sim 1.35\,l_z^{-1}$, resulting in $\Gamma_\text{rot}=178\,\text{Hz}$. Note that in an experimental setting, this $\Gamma_\text{rot}$ would be a \textit{residual} growth rate, i.e. one that persists even after the repulsive LHY correction has brought to a halt the initial rapid population of the roton mode because the LHY correction possesses a small albeit non-zero imaginary part. Experimentally, roton growth rates have been measured \cite{chomaz2018observation}, but due to the quick onset of three-body effects \cite{chomaz2019long}, it would probably be hard if not unfeasible to measure this tiny residual growth rate that persist in the long-time limit.

\subsection{Conclusion}
We have demonstrated the viability of complex Langevin simulations of BECs featuring long-range, anisotropic dipolar interactions and have shown the power of ab initio simulations of dipolar quantum gases. First, we checked that in a homogeneous system with dipolar interactions in the stable phase ($\epsilon_\mathrm{dd}<1$) the spectra expected from dipolar Bogoliubov theory can be correctly reproduced. Further, we computed the dispersion relation in a simple scenario with periodic boundary conditions in $x$- and $y$-direction and a harmonic trapping potential in $z$-direction with complex Langevin and showed that it can feature a rotonic dispersion relation, as expected from analytic predictions and numerical solutions of the Bogoliubov-de-Gennes equations. As a practical application, we performed simulations in an actual experimental setting, comparing to experimental data for the roton gap, in order to settle the discrepancy between BdG calculations with and without an LHY corrections as well as experiment. Our results come out in between the different Bogoliubov computations but remain inconclusive with respect to the experiment. Finally, we have demonstrated the meta-stabilizing effect of quantum fluctuations in the classically unstable phase and have computed the density-modulated quasi-ground-state from first principles. Since the quantum fluctuations lead only to a meta-stabilization, our thermal simulation yielded unphysical results for some observables, namely the static structure factor and the particle occupation number at the roton momentum. We have argued how one can interpret these unphysical results when comparing to an actual experiment. 

In future studies, CL can serve to understand better the role and effect of the LHY correction that is currently widely used in the theoretical description of dipolar bosons. This includes both the stable phase, where we observed the CL results to lie between computations with and without this correction, as well as the unstable phase, where it appears to stabilize the system against collapse. Precision calculations of the critical point in various scenarios are also an important application, especially in comparison to the recently introduced ``improved LHY correction'' \cite{bombin2023quantum} that yields significant differences in comparison to the standard LHY term. Finally, the effect of dipolar interactions on thermal phase transitions (BE and BKT transition) is of particular interest \cite{glaum2007critical,filinov2010berezinskii,bombin2019berezinskii}. As in the non-dipolar case, complex Langevin simulations can be employed to compute non-universal quantities at the transition with high accuracy.

\clearpage

\thispagestyle{plain}
\section{Conclusion\label{sec:conclusion}}
In this thesis, we have demonstrated that the complex Langevin algorithm enables exact ab initio simulations of the interacting, thermal Bose gas in its field-theoretic description and have presented several exemplary applications. At first, we considered the three-dimensional, homogeneous case and benchmarked to known approximate analytical descriptions above and below the Bose-Einstein transition, Hartree-Fock and Bogoliubov theory, respectively. Furthermore, we employed CL to extract the non-perturbative shift of the Bose-Einstein transition temperature due to interactions and found the resulting value to lie within the range of values from the literature, albeit deviating substantially from the most recent results. We found that CL works well within the experimentally relevant weakly coupled regime of diluteness $\eta\sim 10^{-3}$ but that the Langevin evolution suffers from runaway trajectories and thus the algorithm breaks down for stronger coupling. Subsequently, we studied the two-dimensional (but still homogeneous) Bose gas, which is subject to much stronger fluctuations than the three-dimensional one and features a topological phase transition, thus providing a much more stringent benchmark for the complex Langevin method. We found CL to pass also this test, i.e.  it allowed a reliable extraction of observables for arbitrary temperatures above and below the phase transition as long as the the dimensionless coupling constant $mg$ remains in the regime of weak to intermediate coupling strength $mg\lesssim 0.2$, which as in the three-dimensional case roughly agrees with the experimentally feasible regime. In the third part of this work, we briefly looked at the two-dimensional Bose gas in a harmonic trap. We demonstrated how CL can be employed to calculate density profiles in such a system and examined the range of validity of the local density approximation. Furthermore, we compared to density profiles from experiment. In the last part, we considered dipolar Bose-Einstein condensates, where arguably quantum fluctuations play the most prominent role among the systems considered in this thesis. Apart from benchmark checks on the reliability of CL in this case (Bogoliubov theory, rotonic excitations), we attempted a precision calculation of the closing of the roton gap in comparison to experimental results. Furthermore, we demonstrated the stabilizing effect of quantum fluctuations in the classically unstable regime in an \textit{ab initio} simulation. 

All these promising results give hope that the complex Langevin approach to the description of ultracold bosonic gases can eventually establish itself alongside existing methods such as path integral Monte Carlo and the numerous varieties of semi-classical approaches (GPE, SGPE, TWA etc.). In particular, it is a fortunate accident that the method appears to work in typical experimental parameter regimes and breaks down only for coupling strengths that are also harder to achieve in experiment.

With its particular strengths and weaknesses, the CL approach appears as tailor-made to complement the existing methods. Compared to semi-classical methods, CL has the drawback of a substantially higher computational cost, because $d+1$-dimensional lattices must be evolved instead of $d$-dimensional ones. On the other hand, it is able to simulate the quantum many-body problem \textit{exactly} and thus captures also the effect of quantum fluctuations that semi-classical methods do not include. Additionally, it does not require a regularization in the UV since the ultraviolet catastrophe of classical field theory is avoided. Meanwhile, the CL approach has in common with the semi-classical approaches that it describes the gas of bosonic particles in terms of fields, which is the most suitable description for experimental settings with highly occupied modes and large numbers of particles. The ability to easily simulate arbitrary particle numbers without change in the computational cost is an advantage over the equally fully exact PIMC approach. On the other hand, the PIMC algorithm does not possess the limitation in terms of the coupling strength that we found for the CL algorithm, such that the methods nicely complement each other. 

For weakly interacting Bose gases, quantum fluctuations rarely change the behavior of a system dramatically and have qualitative effects. The only system studied in this thesis that violates this rule is the dipolar Bose in the classically unstable phase where quantum fluctuations stabilize the system against collapse. However, even this effect can be straightforwardly captured by mean-field simulations by simply adding a  term accounting for the quantum fluctuations to the classical energy functional. As already noted in the introduction, it is thus likely that the main field of application of CL lies in precision physics rather than the exploration of novel states of matter and the qualitative description of physical phenomena.

With this regard, two major directions can be pursued. On the one hand, as mentioned several times throughout this thesis, the CL approach appears as an ideal tool of the theory side for highly precise theory-experiment comparisons. Rendering the latter feasible, however, would require an increase in precision in comparison to most experimental results available at the time of this writing. On the other hand, CL can serve as an intra-theory benchmark as well. Novel as well as established approximations, e.g. Bogoliubov and Popov theory in all their varieties, classical field theory and the various renormalization group schemes, can be benchmarked against numerically exact CL simulations in order to establish their range of applicability and precision in a given physical scenario. Furthermore, CL might be useful in the study of some fundamental questions related to the field theoretic description of ultracold bosonic atoms, e.g. questions about the renormalization of the theory. 

There exist numerous physical scenarios that would be interesting to examine in future CL studies. First and foremost one must mention multi-component Bose-Einstein condensates, which have not been considered at all throughout this thesis apart from the brief study of the $U(2)$-symmetric Bose gas in chapter \sect{3D}. These include spinor gases \cite{kawaguchi2012spinor} with the special case of spin-orbit coupled gases \cite{lin2011spin}, where the components consist in atoms in different hyperfine levels; and mixtures of different atomic species \cite{modugno2002two}. What makes these systems particularly interesting is the fact that they typically  do not only possess an ordered and a disordered phase as one-component Bose-Einstein condensates. Rather, they can feature numerous ordered phases. E.g. the spin-1 spinor gas possesses a ferromagnetic, anti-ferromagnetic, polar and easy-plane phase \cite{kawaguchi2012spinor}, depending on the external parameters. Between these phases, the gas undergoes quantum phase transitions, which have been studied theoretically \cite{tomasz2013double,debelhoir2016first,mittal2020many,roy2023finite} and experimentally \cite{bookjans2011quantum,qiu2020observation,cominotti2023ferromagnetism}. The theoretic description of such transitions is in general hard due to their non-perturbative nature, but complex Langevin provides a possibility to simulate them from first principles. While some steps have been undertaken in the direction of studying multi-component gases with CL \cite{attanasio2020thermodynamics,mcgarrigle2023emergence,mayr2024}, the vast majority of scenarios remains unexplored. This includes, in particular, the interesting question of the interplay between the thermal phase transition and the transitions between the different ordered phases. Apart from multi-component BECs, there are still multiple intriguing questions related to dipolar gases in equilibrium that have not been explored here, e.g. how dipolar interactions affect BKT physics \cite{filinov2010berezinskii,bombin2019berezinskii}. Finally, it has nowadays become possible experimentally to engineer almost arbitrary external potentials for a gas of ultracold atoms. Choosing this potential on purpose such that the local density approximation breaks down provides a large playground for ab initio simulations with CL, in comparison to experiment. For example, a step potential makes it possible to bring different phases in contact with each other and to study the penetration depth of one phase into the other.  

Apart from this plethora of potential applications to systems of ultracold atoms, one might imagine that CL would be useful also for other systems of non-relativistic bosons. These include, inter alia, magnon condensates in (anti-)ferromagnets \cite{nikuni2000bose}, exciton-polariton systems \cite{kasprzak2006bose}, photon BECs \cite{klaers2010bose} and condensates of dark matter on galactic scales \cite{hu2000fuzzy}. To the best of our knowledge, none of these have been subjected so far to any CL simulation. 

Finally, a potential direction of future research is to go beyond purely thermal simulations. It is unlikely that complex Langevin will ever serve to simulate full real-time dynamics for anything but short-time evolution, although some progress has been made in this direction \cite{alvestad2023towards,boguslavski2023stabilizing}. Nonetheless, applications to sufficiently strongly dissipative Bose-Einstein condensates, for which the sign problem is less severe because of imaginary contributions in the Hamiltonian (resulting in real parts in the real-time action), might be feasible. This is almost certainly true for the case of infinite dissipation, which differs from the thermal case only by the different boundary conditions. It would thus be interesting to study how much the dissipation can be lowered before CL breaks down. A potential application could be to study the annihilation process of a vortex-antivortex pair in a dissipative superfluid, which has been studied in an approximate manner with the Gross-Pitaevskii equation and is conjectured to be dual a classical gravitational system \cite{wittmer2021vortex}.

We hope that the work presented here might stimulate further research in some of these promising directions.

\clearpage

\appendix

\thispagestyle{plain}
\section{Units \label{sec:units}}
Throughout this thesis, we employ three different types of units: SI units, natural units and numerical units. SI units are only rarely used when specifying experimental parameters. Instead, all theoretical formulas are formulated in \textit{non-relativistic natural units}, which are chosen such that 
\begin{align}
\hbar=k_B=1\,.
\end{align}
Notably, in contrast to the relativistic natural unit system, the speed of light is \textit{not} set to one (which would not make sense in a non-relativistic setting). In this unit system, temperature and energy have both the unit of inverse time, momentum has the unit of inverse length, but length and time do not have the same unit, nor does mass have the unit of inverse length (instead, it has the unit of time divided by length squared). 

As the speed of light is not set to one in non-relativistic natural units and length and time are two distinct quantities, there is still one degree of freedom left. For numerical computations, we employ a unit system where this degree of freedom is also eliminated by setting
\begin{align}
\label{eq:num_units}
2ma_\mathrm{s}=1\,,
\end{align}
where $a_\mathrm{s}$ is the lattice spacing. Now everything can be expressed in terms of $a_\mathrm{s}$~\footnote{In practice, $a_\mathrm{s}$ is set to $1$ in the actual numerical computation.}. The convention \eq{num_units} is chosen such that the dimension of physical quantities resembles those in the relativistic natural unit system, i.e. mass has now the dimension of inverse length and time has the dimension of length.

Matching numerical quantities to experimental ones is straightforwardly done by combining experimental quantities to a dimensionless quantity, which then has to agree with the respective quantity in the numerics, e.g.
\begin{align}
\frac{\mu^\text{num}}{\omega^\text{num}}&\overset{!}{=}\frac{\mu^\text{exp}}{\hbar\omega^\text{exp}}\\
\frac{T^\text{num}}{\omega^\text{num}}&\overset{!}{=}\frac{k_BT^\text{exp}}{\hbar\omega^\text{exp}}\\
\frac{a^\text{num}}{\sqrt{2a_\mathrm{s}/\omega^\text{num}}}&\overset{!}{=}\frac{a^\text{exp}}{\sqrt{\hbar/m^\text{exp}\omega^\text{exp}}}\,.
\end{align} 
At the same time, one must make sure that relevant physical quantities are properly resolved by $a_\mathrm{s}$. For example, in the condensed phase the momentum cutoff $\Lambda=\pi/a_\mathrm{s}$ should be much larger than the healing momentum in numerical units, i.e.
\begin{align}
\pi/a_\mathrm{s}\overset{!}{\gg} p_\mathrm{h}^\text{num}=\sqrt{\mu^\text{num}/a_\mathrm{s}}\,.
\end{align}
Throughout this thesis, we typically chose $\Lambda/p_\mathrm{h}^\text{num}\sim 5$. 
\clearpage

\thispagestyle{plain}
\section{Benchmarks of CL simulation runtimes\label{sec:runtime}}
The numerical evolution of the quantum fields according to the complex Langevin equations \eq{CL_eq_bose1} and \eq{CL_eq_bose2} was performed on NVIDIA graphic processing units (GPUs). Thanks to the extremely high parallelization that this hardware platform enables it is possible to perform in a reasonable amount of time the demanding simulation of four-dimensional lattices that the full quantum simulation requires. In particular, we assign to every single GPU thread the update of one lattice point. For the parallelized execution of the fast Fourier transform (necessary for evaluating the kinetic energy if spectral derivatives are employed, as well as the dipolar potential) and the generation of Gaussian random numbers, we rely on the existing cuFFT and cuRAND libraries. 

\begin{figure}[h]
	\centering\includegraphics[width=0.6\textwidth]{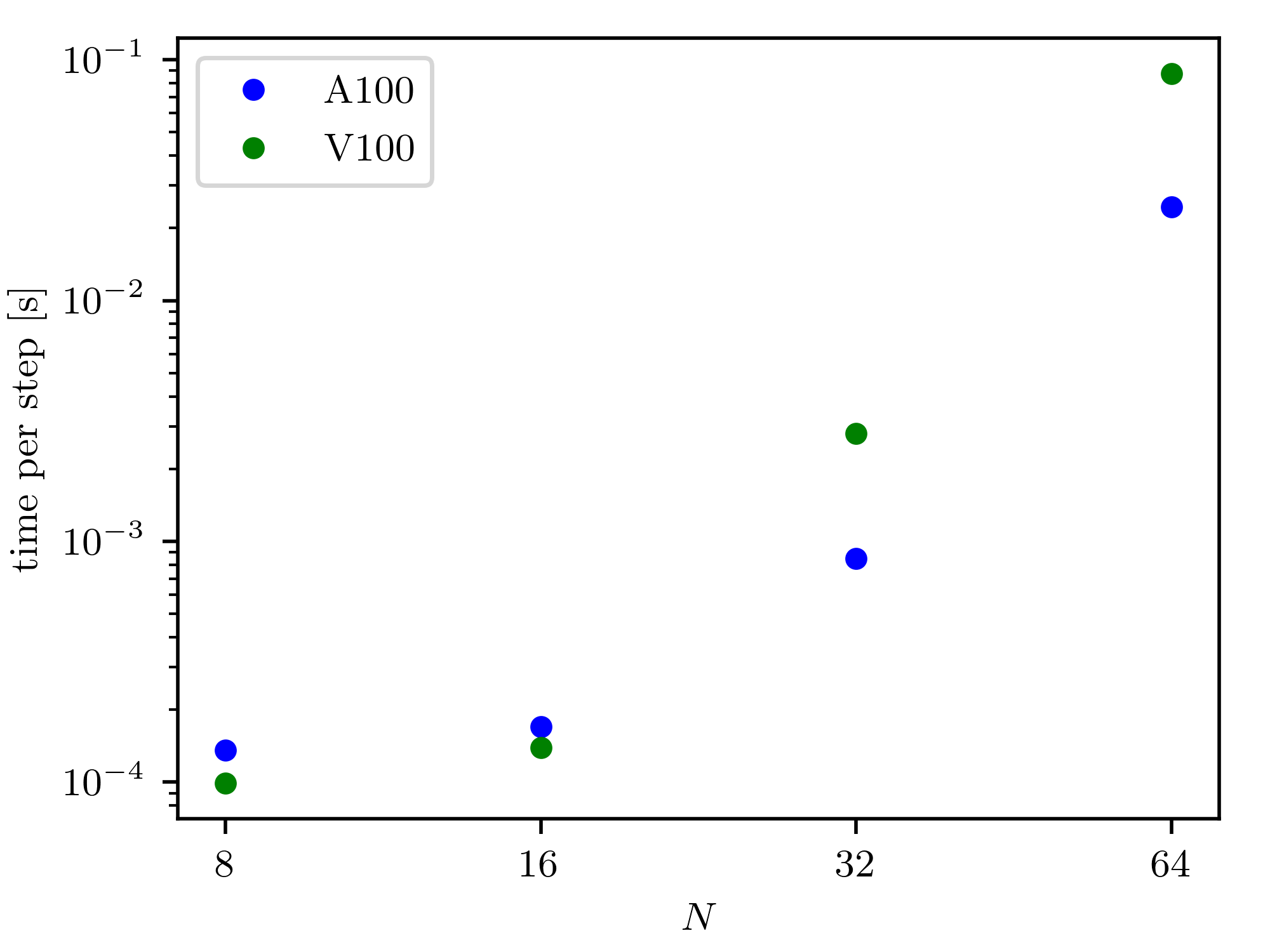}
	\caption{Computation time for one time step in seconds on a A100 and a V100, as a function of the lattice size $N$. Note that this time increases only by a small factor between $N=8$ and $N=16$ although the number of lattice points rises by a factor of $16$, a consequence of the massive parallelization of the GPU. Only when the number of lattice points exceeds the number of physically available threads does the computation time approximately scale with $N^4$.  
	}
	\label{fig:runtimes}
\end{figure}

Figure \fig{runtimes} shows the computation time per time step for an $N^4$-dimensional lattice (for a spectral derivative evaluation of the Laplacian and no dipole interaction), with $N$ ranging from $8$ to $64$, on two state-of-the-art GPUs, the NVIDIA V100 and A100 card. Throughout this work, the number of time steps for a single run was typically chosen on the order of $10^6$ to $10^7$, such that a run requires a simulation time of some hours to some days~\footnote{Additionally, we averaged over $\mathcal{O}(10)$ different runs. Given that enough resources are available, these can be executed in parallel on different machines.}. 

\vspace{1cm}

\clearpage

\thispagestyle{plain}
\section{Supplements to chapter 4\label{sec:3D_supp}}
\subsection{Correction of discretization effects}
On a computational lattice, systematic deviations of observables from their continuum counterparts are caused by both the finite number of Matsubara modes and the finite spatial lattice resolution. 
Both mainly affect the UV properties of the system and therefore, in describing the critical behavior near the phase transition, are expected to give a small correction only. 
Furthermore, for a weakly interacting gas above the transition, the spectrum is well approximated by an ideal-gas Bose-Einstein distribution with a shifted chemical potential, see section \sect{IntBECabovePT}. 
It is thus reasonable to account for the described systematic errors in the following way: 
For a free gas at $\mu\to 0$ in a box of fixed size $\mathcal{V}=L^{3}=(N_\mathrm{s}a_\mathrm{s})^3$, we numerically determine the deviation $\delta \rho=\rho(N_\mathrm{s}\to\infty,N_{\tau}\to\infty)-\rho(N_\mathrm{s},N_{\tau})$ between the density on the lattice and in the continuum (but for finite system size $\mathcal{V}$). 
$\delta\rho$ is then added to the density determined from numerical data for the interacting system. 
For the parameters as chosen in section \sect{IntBGatCriticality}, i.e. a $64^3\times 16$ lattice at $T=1.25\,a_\mathrm{s}^{-1}$, we obtain $\delta\rho=0.00154\,a_\mathrm{s}^{-3}$. 
This is a rather small correction, but it is significant in determining the constant $c$.
\subsection{Finite-size effects to the shift of the critical temperature}
In our extraction of the transition temperature shift due to interaction effects we refrained from performing a systematic finite-size analysis. Numerical results from reference \cite{kashurnikov2001critical} (whose authors performed such a systematic analysis and defined the shift in the finite system in a manner similar to ours) suggest that for the system size that was employed here ($20.2$ thermal wave lengths $\lambda_{T}$), finite size corrections are already significantly smaller than our statistical error. Namely, they find the shift of the critical temperature to scale with system size $L$ as
\begin{align}
\frac{\Delta T_\text{c}}{T_\text{c}^0}\sim \frac{1}{1+b_0 g (2m)^{3/2}T^{1/2}+\frac{a_1+b_1 g(2m)^{3/2}T^{1/2}}{(Lm^2Tg)^{1.038}}}
\end{align} 
with $a_1=1.29$, $b_0=0.123$, $b_1=0.744$~\footnote{Note that, in reference~\cite{kashurnikov2001critical}, constants were chosen  $2m=T=1$ in the factors multiplying $b_{0}$ and $b_{1}$.}.
For our parameters and system size, we obtain from this formula a deviation of $14.6\%$ between the finite-$L$ and the $L\to \infty$ value for the shift, which is only half our statistical error ($27.8\%$), which justifies our procedure. 

\begin{figure}[h]
	\includegraphics[width=0.495\textwidth]{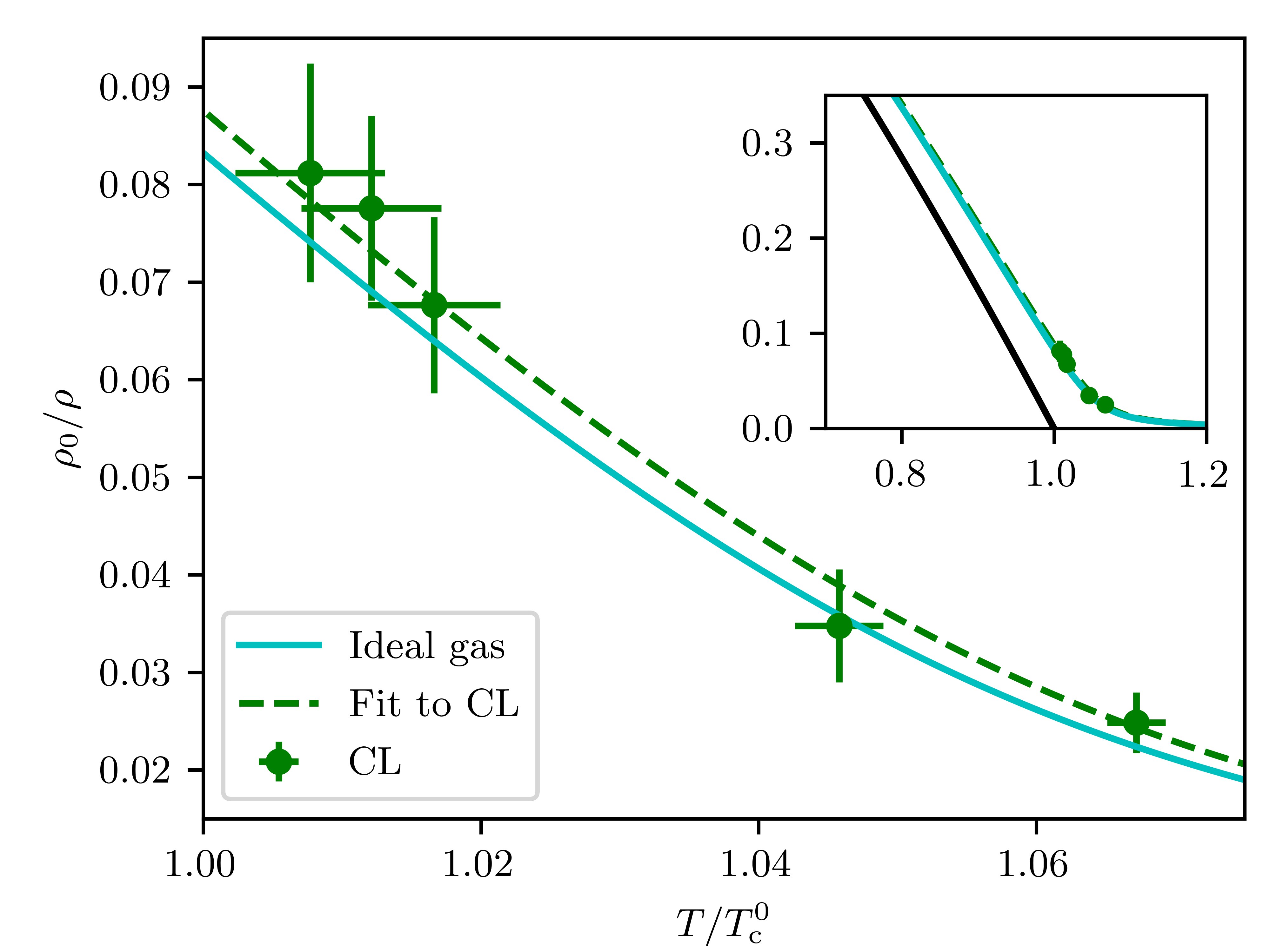}
	\includegraphics[width=0.495\textwidth]{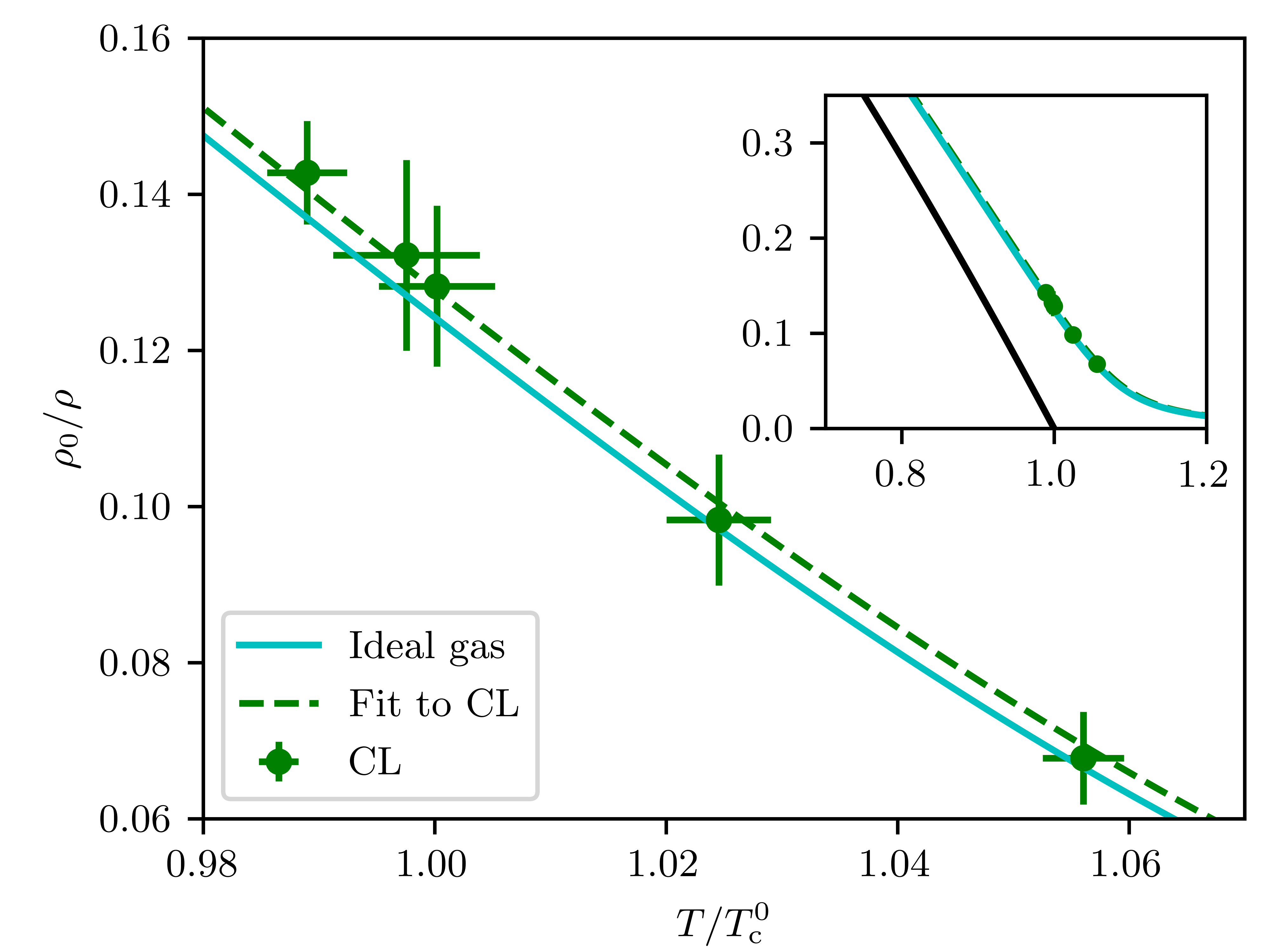}
	\caption{The same as in figure \fig{shift}, but on a $48^3$ (upper panel) and $32^3$ (lower panel) lattice. 
	}
	\label{fig:shift3248}
\end{figure}
Our own simulations corroborate this weak dependence on $L$. 
To demonstrate this, we perform the same analysis as in the main text but for the smaller $48^3$ and $32^3$  lattices, for the exact same chemical potentials. The result is shown in figure \fig{shift3248}. Here we obtain for the shift
\begin{align}
\Delta T_c/T_c^0&=0.0036\pm 0.0014\qquad (48^3\,\text{lattice})\\
\Delta T_c/T_c^0&=0.0031\pm 0.0008\qquad (32^3\,\text{lattice})\,.
\end{align}
Within the errors, this agrees with the result for the $64^3$ lattice, suggesting that finite-size effects already play a minor role. 

\clearpage

\thispagestyle{plain}
\section{Supplements to chapter 5\label{sec:2D_supp}}
\subsection{\label{sec:secant} Finding the critical chemical potential with the secant algorithm}
\begin{figure}
	\includegraphics[width=0.495\columnwidth]{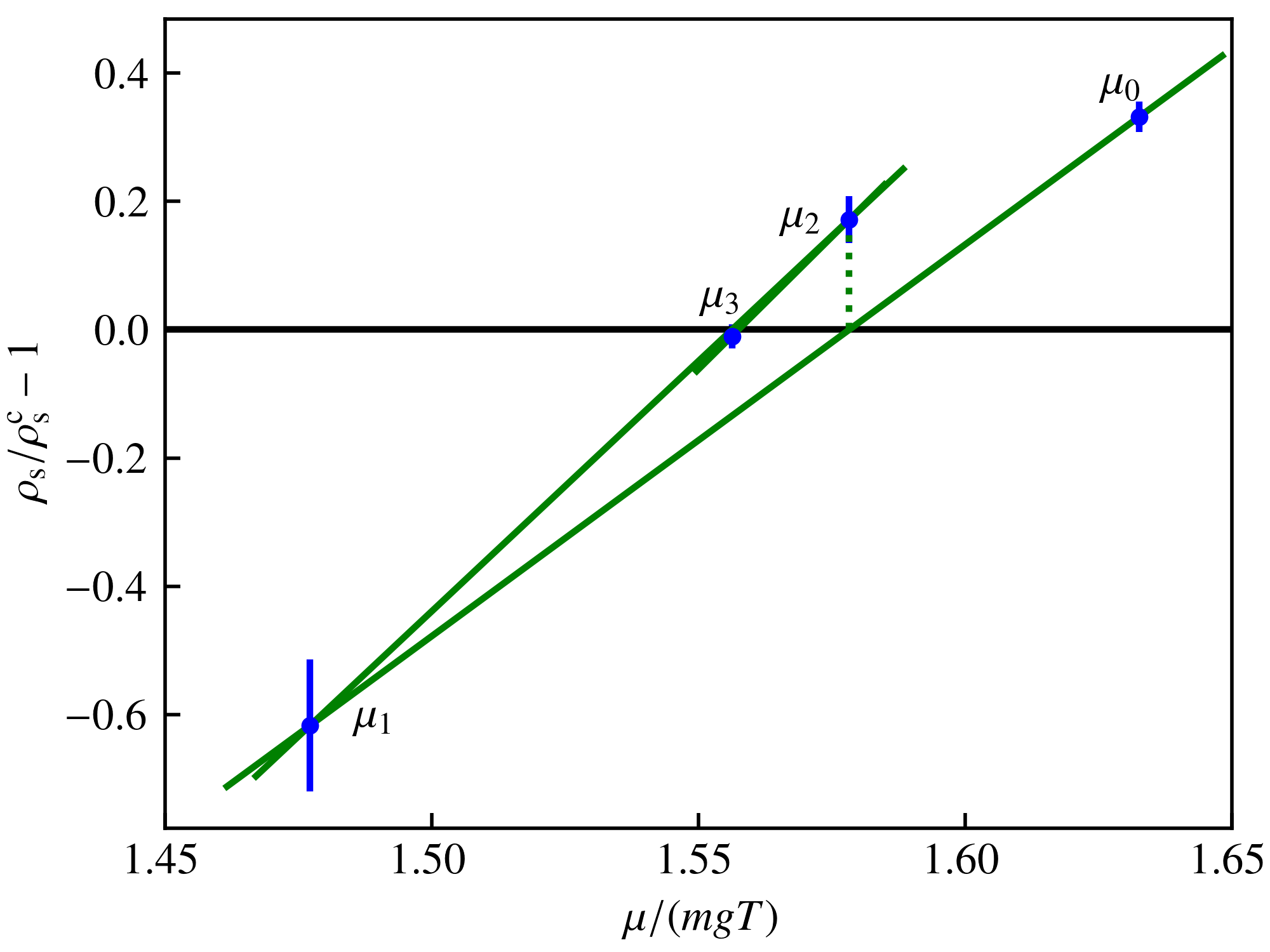}
	\includegraphics[width=0.495\columnwidth]{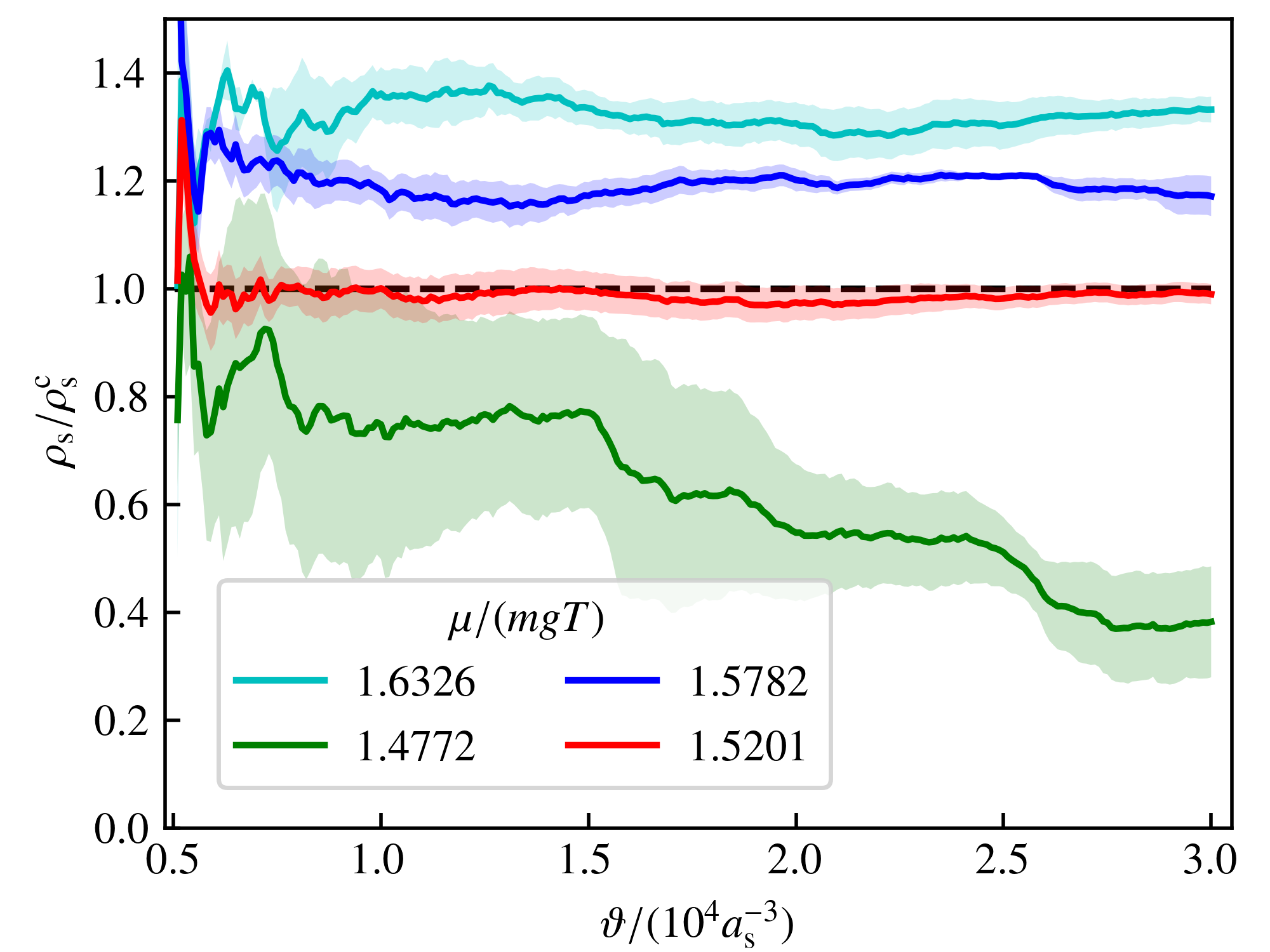}
	\caption{Determination of the BKT transition point by the secant method for $mg=0.1$ and $L/a_\mathrm{s}=128$. 
		Starting from two initial guesses for the critical chemical potential, $\mu_0/(mgT)=1.633$ and $\mu_1/(mgT)=1.477$, subsequent guesses are chosen as the intersection of the secant through the previous two data points and the $\mu/T$-axis (left panel). 
		The convergence is extremely fast. 
		The right panel shows the convergence of the  superfluid densities in Langevin time $\vartheta$, with error bands obtained from the variance of several statistically independent runs. 
	}	
	\label{fig:secant}
\end{figure}

We determined the chemical potential at the BKT transition point, i.e. the value $\mu_\mathrm{c}$ for which the Nelson criterion \eq{nelson} is fulfilled, by means of a secant algorithm, which converges faster than the common bisection algorithm. To this end, we start from two initial guesses for the critical chemical potential, $\mu_0$ and $\mu_1$, and determine subsequent estimates as 
\begin{align}
\mu_{i+2}
=\frac{\mu_{i}\tilde{\rho}_\mathrm{s}(\mu_{i+1})-\mu_{i+1}\tilde{\rho}_\mathrm{s}(\mu_{i})}
{\tilde{\rho}_\mathrm{s}(\mu_{i+1})-\tilde{\rho}_\mathrm{s}(\mu_{i})}
\,,
\end{align}
where $\tilde{\rho}_\mathrm{s}\equiv \rho_\mathrm{s}-\rho_\mathrm{s,c}$, and the critical superfluid density $\rho_\mathrm{s,c}$ is defined by \eq{nelson}. 
As initial guesses we chose $\mu_0=1.05\bar{\mu}$ and $\mu_1=0.95\bar{\mu}$ with $\bar{\mu}$ the estimate for the critical chemical potential from \cite{prokofev2001critical}, determined by \eq{critical_mu}. 
We stop the iteration once $\tilde{\rho}_\mathrm{s}$ is indistinguishable from $0$ within the statistical errors. This procedure is illustrated in figure \fig{secant} (left panel). 
As one can see, the convergence is extremely fast. 
In fact, we rarely had to perform more than 4 to 5 simulations for a given coupling and lattice size. 
Sometimes it occurred, however, that, due to statistical fluctuations, $\tilde{\rho}_\mathrm{s}(\mu_{i+1})$ and $\tilde{\rho}_\mathrm{s}(\mu_{i})$ resulted to be very close in magnitude, which could give rise to a divergence of the secant algorithm. 
In such cases we replaced $\tilde{\rho}_\mathrm{s}(\mu_{i})$ by hand by a suitable $\tilde{\rho}_\mathrm{s}(\mu_{j})$, with $j<i$, and thereafter continued the algorithm in the normal way.

\subsection{\label{sec:classsim} Classical field theory simulations}

Setting the number of lattice points in imaginary direction to one,  $N_\tau=1$, allows us to easily turn off quantum effects and to simulate a purely classical field theory. In this appendix, we briefly discuss the results of such simulations for comparison.

For such a theory, occupation numbers will follow a Rayleigh-Jeans law in the UV, $f(k)\sim \left(k^2/2m-\mu\right)^{-1}$, such that most quantities such as density or kinetic energy suffer from a UV divergence that is absent in the full quantum theory. In order to regulate said divergence, we follow the procedure of \cite{prokofev2001critical} and correct the densities and chemical potentials within classical field theory, $\rho^\text{class}$ and $\mu^\text{class}$, by subtracting the (non-interacting) Rayleigh-Jeans distribution and adding the (non-interacting) Bose-Einstein distribution, i.e.~we obtain the density and chemical potential $\rho$ and $\mu$ in the full quantum theory as
\begin{align}
\label{eq:rhomatch}
\rho&=\rho^\text{class}+\Delta\rho
\,,\\
\label{eq:mumatch}
\mu&=\mu^\text{class}-2g\Delta\rho
\,,
\end{align}
with 
\begin{align}
\Delta\rho
&\equiv\frac{1}{L^2}\sum_\mathbf{p}
\left[\frac{1}{\exp(\beta{\mathbf{p}^2}/{2m})-1}-\frac{2m}{\beta{\mathbf{p}^2}}\right]\,.
\end{align}
For the lattice spacing employed here, $\lambda_T=3.17\,a_\mathrm{s}$, this results in $\Delta \rho = 0.227\,a_\mathrm{s}^{-2}$. 
In \cite{prokofev2001critical}, the superfluid density was estimated by the number of windings in the systems, which are directly accessible within the employed worm algorithm. 
Since we have no access to winding numbers, we need, as in the quantum simulations in the main text, to resort to \eq{rhos}, which requires to deal with the UV divergence of $\langle\mathbf{P}^2\rangle$ that is absent in the quantum simulation. 
We here follow the same strategy as for the density and subtract the value of $\langle\mathbf{P}^2\rangle$ in a non-interacting classical field theory and add its value in a non-interacting quantum field theory. The latter reads
\begin{align}
\langle\mathbf{P}^2\rangle^\text{free}
&=\sum_\mathbf{p} \left\{\frac{1}{\exp(\beta\frac{\mathbf{p}^2}{2m})-1}
+\frac{1}{\left[\exp(\beta\frac{\mathbf{p}^2}{2m})-1\right]^2}\right\}\mathbf{p}^2\,,
\end{align}
such that we obtain the true $\langle\mathbf{P}^2\rangle$ from $\langle\mathbf{P}^2\rangle^\text{class}$ as 
\begin{align}
\langle\mathbf{P}^2\rangle
=\langle\mathbf{P}^2\rangle^\text{class}+\Delta\langle\mathbf{P}^2\rangle
\,,
\end{align}
with 
\begin{align}
\label{eq:deltaP2}
\nonumber\Delta\langle\mathbf{P}^2\rangle
\equiv \sum_\mathbf{p} &\left\{
\frac{1}{\exp(\beta{\mathbf{p}^2}/{2m})-1}
+\frac{1}{\left[\exp(\beta{\mathbf{p}^2}/{2m})-1\right]^2}-\ \frac{2m}{\beta{\mathbf{p}^2}}
-\left(\frac{2m}{\beta{\mathbf{p}^2}}\right)^2\right\}\mathbf{p}^2
\,,
\end{align}
which for our lattice gives $\Delta\langle\mathbf{P}^2\rangle/L^2=1.41\,a_\mathrm{s}^{-4}$.

\begin{figure}
	\includegraphics[width=0.495\columnwidth]{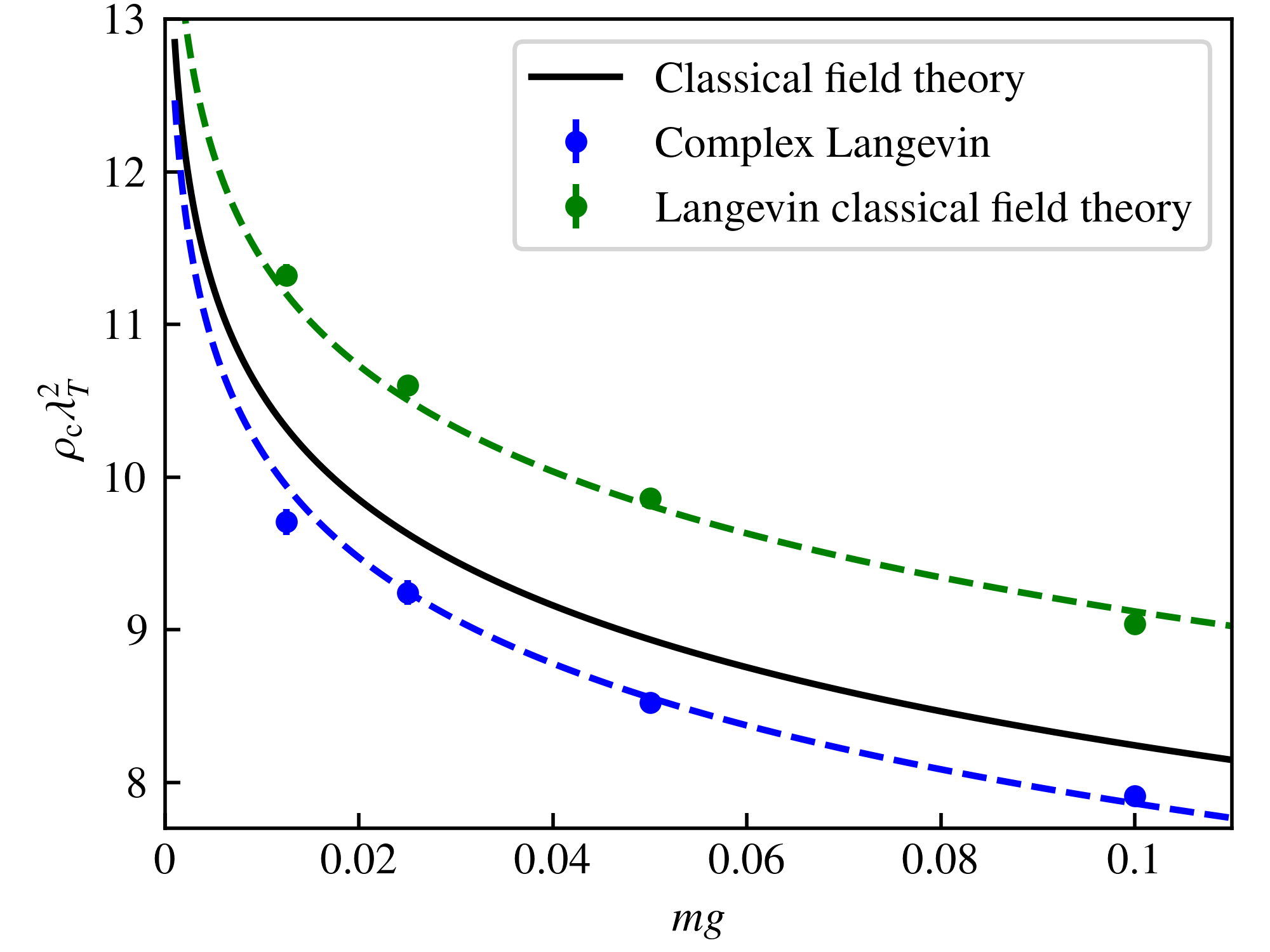}
	\includegraphics[width=0.495\columnwidth]{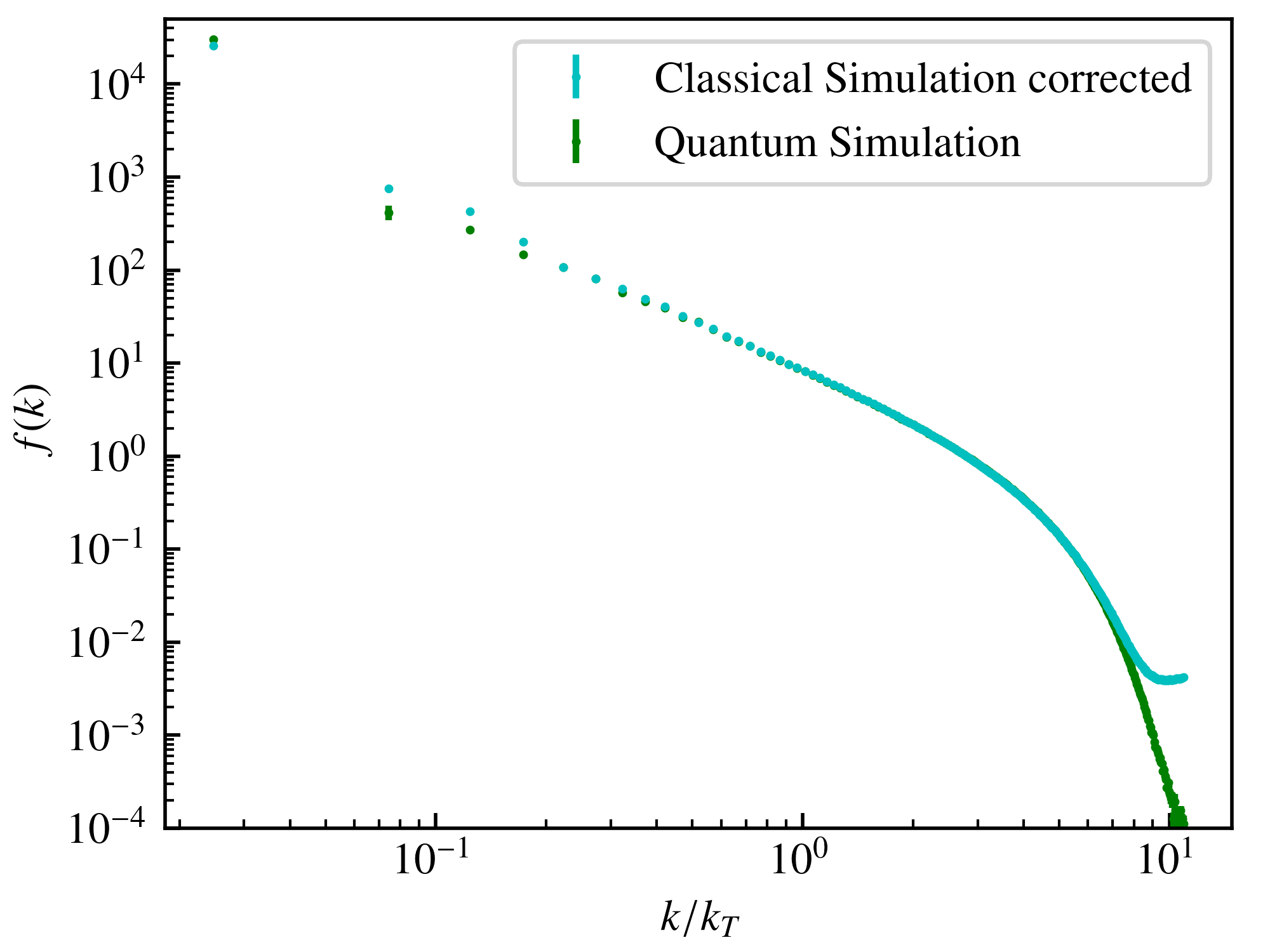}
	
	\caption{Left panel: Critical density $\rho_\text{c}$ as a function of the coupling $mg$ computed by Langevin simulations of the classical field theory, i.e. setting $N_\tau=1$ (green points) in comparison with the full quantum simulations (blue points) and the result from the classical field theory simulation of \cite{prokofev2001critical}. The critical densities from the classical Langevin simulation are substantially higher than those from the full quantum simulation, and they result as higher than those from the classical field theory simulation \cite{prokofev2001critical}.
		Right panel: Comparison of the momentum spectrum from the full quantum simulation (blue points) and the classical simulation (green points) for $mg=0.1$ in the critical region, where the latter has been corrected in the UV by subtracting a Rayleigh-Jeans distribution and adding a Bose-Einstein distribution at zero chemical potential. 
		Note that the zero mode has been shifted to a finite $k$ value in order to make it visible on the double-logarithmic scale. 
		The chemical potential $\mu$ of the quantum simulation is matched to the one of the classical simulation, $\mu^\text{class}$, according to $\mu=\mu^\text{class}-2g\Delta\rho$, equation \eq{mumatch}. 
		Apart from small deviations at very high momenta, the spectra agree well over almost the entire momentum range. However, deviations are visible in the strongly correlated IR-modes, with the quantum system being farther in the superfluid phase than the classical one. 
		The corresponding density of the quantum system results only by approximately $2\%$ larger than the corrected density of the classical system. }
	\label{fig:rhoclassical}
\end{figure}

In this way, we have repeated the extraction of the critical densities described in the main text for the classical field theory, results of which are shown in the left panel of figure \fig{rhoclassical}. 
The critical densities are by around $15\%$ larger than in the full quantum simulation and larger than the results from \cite{prokofev2001critical}, which could indicate that the different classical field theory simulations, in particular the different ways to extract the superfluid densities, are not entirely compatible with each other or some residual UV cutoff dependence remains.

In order to shed more light at the discrepancy between classical and quantum simulations, we compare, in the right panel of figure \fig{rhoclassical}, the momentum spectra obtained from the classical and quantum simulations, for $mg=0.1$, with the chemical potentials matched according to $\mu=\mu^\text{class}-2g\Delta\rho$. 
After subtracting the free Rayleigh-Jeans distribution and adding the Bose-Einstein distribution, the two spectra agree with each other for $k>k_\mathrm{c}$, apart from a small deviation in the far UV that is due to the Rayleigh-Jeans fall-off being slightly weaker than $k^{-2}$ and that, as we checked, has little effect on both the particle number and the superfluid density. 
For the strongly correlated IR modes, $k<k_\mathrm{c}$, however, one observes a substantial deviation, with the quantum system being already farther in the superfluid phase. 
This suggests that, while the effect of the density bias $\Delta \rho$ of the classical simulation can be accounted for in the description of the modes $k>k_\mathrm{c}$ by a shift of the effective chemical potential by $2g\Delta\rho$, its effect on the effective chemical potential governing the IR is more involved. 
The latter could be expected in view of the fact that at the transition point, the system is already substantially condensed, while the approximation $\mu_\text{eff}=\mu-2g\rho$ derives from the Hartree-Fock substitution of the interaction term, which is valid only in the non-condensed phase.

\subsection{\label{sec:critdensscal} Critical density from scaling}
The scaling relation \eq{eosuniversal} provides an independent method for extracting the differences of critical densities at different $mg$ (albeit no absolute values). Since applying the CL algorithm for determining the critical density for couplings $mg\gtrsim0.1$ was found to be increasingly impeded by runaway trajectories and thus we could not directly determine  $\rho_\mathrm{c}$ for this coupling regime, this extraction method proves useful in going beyond this limit. The latter is possible since only simulations further away from the transition point are required, for which CL breaks down later (i.e. at higher couplings). 

\begin{figure}
	\centering\includegraphics[width=0.6\columnwidth]{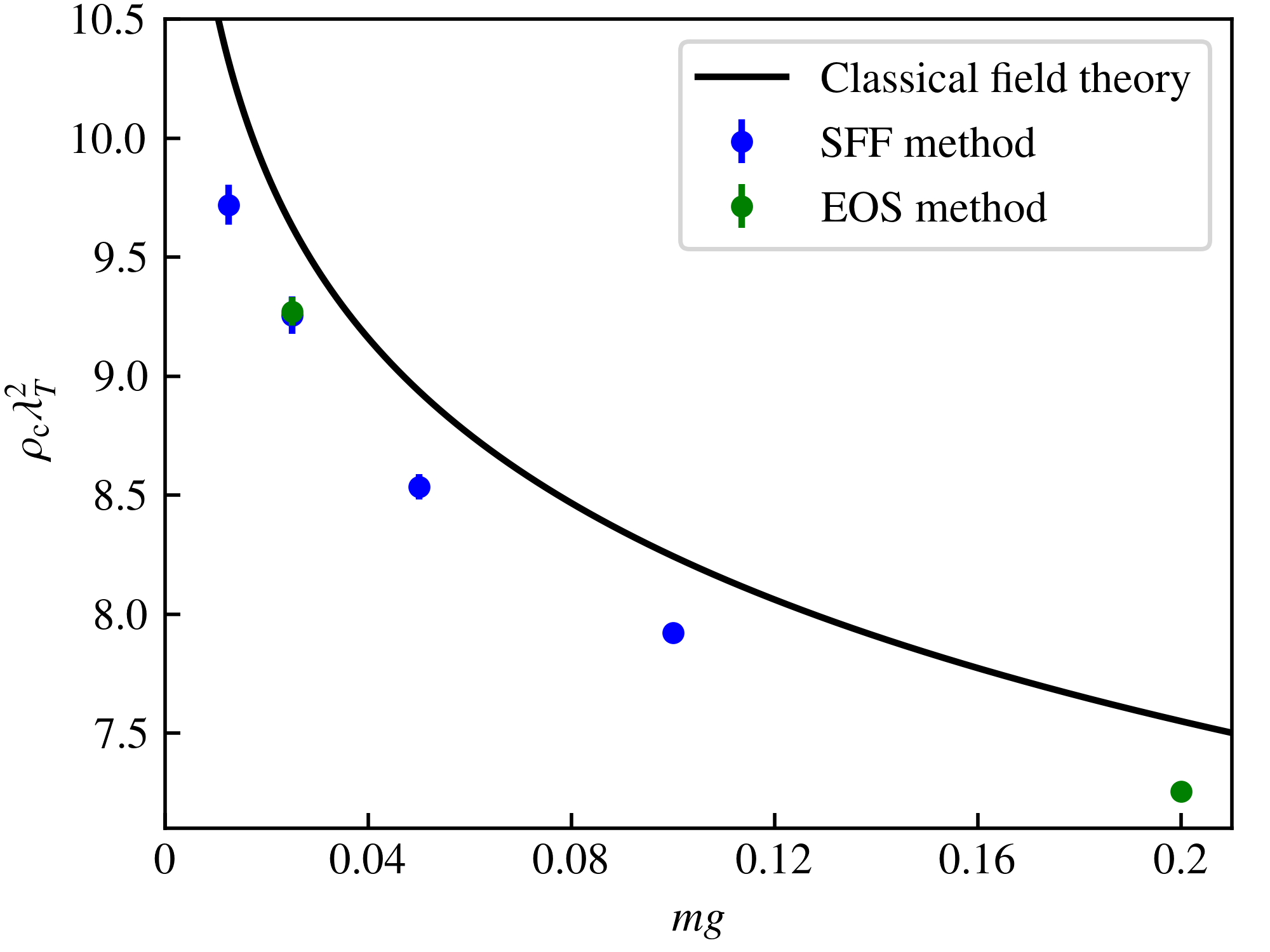}
	
	\caption{Critical density $\rho_\text{c}$ as a function of the coupling $mg$ computed by employing the Nelson criterion for the superfluid fraction (SFF), as described in \sect{critdens} (blue points), in comparison to the results from the rescaling of the equation of state (green points). Since the latter yields only differences of critical densities, we chose the critical density at $mg=0.1$ to be equal to the result from the superfluid fraction method. The black line represents \eq{critical_dense} with constant $\zeta_{\rho}=380$ obtained from the classical simulation of \cite{prokofev2001critical}.}
	\label{fig:rhocfromscaling}
\end{figure}

In \fig{rhocfromscaling}, we show the critical density, for $mg=0.025$ and $mg=0.2$ (green data points), as obtained by rescaling the equation of state for these couplings to the one for $mg=0.1$ and thus inferring it from the critical density $\lambda_T^2\rho_\mathrm{c}(mg=0.1)=7.921\pm0.032$.
The error bar is obtained as a combination of the error on the value for $mg=0.1$ and the fitting error obtained as the mean-square deviation of the rescaled curves from figure \fig{eos}. 
For comparison, the directly determined values from figure \fig{rhoc} are included as blue data points.

\clearpage

\thispagestyle{plain}
\section{Supplements to chapter 7}
\subsection{Dipolar interaction on a computational lattice \label{sec:dip_cutoff}}
The implementation of the long-range dipolar interaction on a computational lattice is somewhat tricky. In order to be able to benefit from the computational advantages of fast Fourier transforms (FFTs), we would like to impose periodic boundary conditions in the simulations. However, periodic boundary conditions in combination with the long-range nature of the dipolar interaction lead to the interaction of an atom with an infinite number of its imaginary copies. As mentioned in the main text, this can sometimes be desired, in which case it suffices to plug in for the numerical Fourier transform $\tilde{V}_\mathbf{k}^\text{dip}$ the analytical Fourier transform of the dipolar potential, i.e.
\begin{align}
\tilde{V}_\mathbf{k}^\text{dip}=\begin{cases}C_\mathrm{dd}\left(k_z^2/|\mathbf{k}|^2-1/3\right)\,,\quad \mathbf{k}\neq 0\\0\,,\quad \mathbf{k}= 0\end{cases}\,.
\end{align} 
If one wants to simulate an atom cloud trapped by a harmonic confinement, however, this interaction with imaginary copies is unphysical. Since the dipolar interaction decays as $\sim 1/r^3$, this bias will become negligible for sufficiently large lattices, but this would lead to unnecessarily large computational costs. A better choice is to cut off the dipolar potential at large distances, i.e. set it to $0$ beyond a certain cutoff. For some simple cutoff geometries, there exist analytical expressions for the Fourier transform. E.g. for a spherical cutoff (i.e. setting the interaction to zero for $|\mathbf{r}|>R$), one obtains ($\mathbf{k}\neq 0$) \cite{bisset2013theoretical}
\begin{align}
\tilde{V}_\mathbf{k}^\text{dip}=C_\mathrm{dd}\left(\cos^2\theta-1/3\right)\,\left[1+3\frac{\cos(Rk)}{R^2k^2}-3\frac{\sin(Rk)}{R^3k^3}\right]\,,
\end{align}
where $k=|\mathbf{k}|$ and $\cos^2\theta=k_z^2/|\mathbf{k}|^2$. However, often the atom cloud does not have a perfect spherical shape. The best solution in this case is to cut off the dipolar potential at half of the lattice extension in a particular direction (i.e. for $|x|>L_x/2$, $|y|>L_y/2$ or $|z|>L_z/2$) and to compute the Fourier transform numerically by a FFT once at the beginning of the simulation, i.e. to define: 
\begin{align}
\tilde{V}_\mathbf{k}^{\text{dip}}=a_\mathrm{s}^3\sum_{j_x=-N_x/2}^{N_x/2}\sum_{j_x=-N_y/2}^{N_y/2}\sum_{j_x=-N_z/2}^{N_z/2}\exp\left[i\mathbf{k}\cdot(a_\mathrm{s}\mathbf{j}-a_\mathrm{s}\mathbf{j}')\right]\frac{C_\mathrm{dd}}{4\pi}\frac{1-3(a_\mathrm{s}j_z)^2/|a_\mathrm{s}\mathbf{j}|^2}{|a_\mathrm{s}\mathbf{j}|^3}\,,
\end{align}
where $N_{x,y,z}$ is the number of lattice points in the respective direction. It is important to note that even with the cutoff to the dipolar potential, the lattice size must be taken at least twice as large as the atom cloud in order to avoid every interaction between imaginary copies. This is the price that is paid for the periodic boundary conditions, which make it possible to compute convolutions highly efficiently with FFTs.

In section \sect{rot_disp}, we consider a hybrid scenario where the system is trapped in $z$-direction and periodic in $x$- and $y$-direction. Then it makes sense to cut off the interaction in $z$-direction but not in $x$- and $y$-direction, i.e. to set the interaction to zero for $|z|>Z$. Also for this case there exists an analytical expression \cite{bisset2013theoretical}:
\begin{align}
\tilde{V}_\mathbf{k}^\text{dip}=C_\mathrm{dd}\left(\cos^2\theta-1/3\right)+C_\mathrm{dd}\,e^{-Zk_\rho} \left[\sin^2\theta \cos(Z k_z) - \sin\theta \cos\theta \sin(Zk_z)\right]\,,
\end{align}
where $k_\rho=\sqrt{k_x^2+k_y^2}$ . 

\clearpage

\thispagestyle{plain}

\clearpage

\thispagestyle{plain}
\section*{Acknowledgments}
I thank Thomas Gasenzer for the supervision of this PhD project and precious discussions and suggestions. I thank Martin Gärttner and Jan Pawlowski for being my co-supervisors. I thank Tilman Enss for being the second referee of this thesis. I thank Felipe Attanasio, Marc Bauer, Iacopo
Carusotto, Lauriane Chomaz, Francesca Ferlaino, Glenn Fredrickson, Christof Gattringer, Wyatt Kirkby, Stefan Lannig, Julian Mayr, Ethan McGarrigle, Aleksandr Mikheev, Markus Oberthaler, Duncan O'Dell, Andreea Oros, Jan Pawlowski, Axel Pelster, Thomas Pohl, Nikolay Prokof'ev, Davide Proment, Adam Ran\c{c}on, Niklas Rasch, Alessio Recati, Santo Roccuzzo, Christian-Marcel Schmied, Ido Siovitz, Marius Sparn, Helmut Strobel and Boris Svistunov for precious scientific discussions and exchange. I thank Francesca Ferlaino and Marius Sparn for providing me experimental data. I thank Alessio Recati for hosting me at the University of Trento. I thank my colleagues of the SynQS group Yannis Arck, Alexander Baum, Nils Becker, Alexandra Beikert, Brian Bostwick, Sören Breidenbach, Stefanie Czischek, Yannick Deller, Anton Eberhardt, David Feiz, Anna-Maria Glück, Paul Große-Bley, Maurus Hans, Lilo Höcker, Elinor Kath, Marcel Kern, Jan Kilinc, Wyatt Kirkby, Felix Klein, Hannes Köper, Stefan Lannig, Mai Le, Nikolas Liebster, Julian Mayr, Florian Meienburg, Aleksandr Mikheev, Andreea Oros, Niklas Rasch, Moritz Reh, Lisa Ringena, Julian Robertz, Santo Roccuzzo, Christian-Marcel Schmied, Florian Schmitt, Alexander Schmutz, Konstantinos Sfairopoulos, Ido Siovitz, Marius Sparn, Helmut Strobel, Celia Viermann, David Wachs, Magdalena Winkelvoß and Martin Zboron for the familiar atmosphere and the great time spent together. Finally, I thank my parents and Vittoria for their support during the course of my studies.

\end{document}